\documentclass[11pt]{article} 

\usepackage{textcomp, amssymb}  
\usepackage{anysize}            
\usepackage{amsfonts}
\usepackage{amsmath}
\usepackage{dsfont}
\usepackage{MnSymbol}
\usepackage{mathtools}
\usepackage{graphicx}
\usepackage{epsfig}
\usepackage{epstopdf}
\usepackage{lmerge}
\usepackage{tikz}
\usepackage{amsthm}
\usepackage{ifpdf}

\usetikzlibrary{calc,decorations.pathmorphing,shapes,graphs}

\newcounter{sarrow}
\newcommand\xrsquigarrow[1]{%
\stepcounter{sarrow}%
\mathrel{\begin{tikzpicture}[baseline= {( $ (current bounding box.south) + (0,-0.5ex) $ )}]
\node[inner sep=.5ex] (\thesarrow) {$\scriptstyle #1$};
\path[draw,<-,decorate,
  decoration={zigzag,amplitude=0.7pt,segment length=1.2mm,pre=lineto,pre length=4pt}]
    (\thesarrow.south east) -- (\thesarrow.south west);
    \end{tikzpicture}}%
}

\newcounter{sarrow1}

\newcommand\xnrsquigarrow[1]{%
\stepcounter{sarrow1}%
\mathrel{\begin{tikzpicture}[baseline= {( $ (current bounding box.south) + (0,-0.5ex) $ )}]
\node[inner sep=.5ex] (\thesarrow) {$\scriptstyle #1$};
\path[draw,<-,decorate,
  decoration={zigzag,amplitude=0.7pt,segment length=1.2mm,pre=lineto,pre length=4pt}]
    (\thesarrow1.south east) -- (\thesarrow1.south west);
    $\slashedarrowfill@\relbar\relbar/$
    \end{tikzpicture}}%
}

\makeatletter
\def\slashedarrowfill@#1#2#3#4#5{%
  $\m@th\thickmuskip0mu\medmuskip\thickmuskip\thinmuskip\thickmuskip
   \relax#5#1\mkern-7mu%
   \cleaders\hbox{$#5\mkern-2mu#2\mkern-2mu$}\hfill
   \mathclap{#3}\mathclap{#2}%
   \cleaders\hbox{$#5\mkern-2mu#2\mkern-2mu$}\hfill
   \mkern-7mu#4$%
}
\def\rightslashedarrowfillb@{%
  \slashedarrowfill@\relbar\relbar/\rightarrow}
\newcommand\xnrightarrow[2][]{%
  \ext@arrow 0055{\rightslashedarrowfillb@}{#1}{#2}}

\def\rightslashedarrowfille@{%
  \slashedarrowfill@\relbar\relbar/\twoheadrightarrow}
\newcommand\xntworightarrow[2][]{%
  \ext@arrow 0055{\rightslashedarrowfille@}{#1}{#2}}

\def\rightslashedarrowfillg@{%
  \slashedarrowfill@\relbar\relbar{\raisebox{.12em}{}}\twoheadrightarrow}
\newcommand\xtworightarrow[2][]{%
  \ext@arrow 0055{\rightslashedarrowfillg@}{#1}{#2}}

\def\rightslashedarrowfillx@{%
  \slashedarrowfill@\Relbar\Relbar/\rightrightarrows}
\newcommand\xnTworightarrow[2][]{%
  \ext@arrow 0055{\rightslashedarrowfillx@}{#1}{#2}}

\def\rightslashedarrowfilly@{%
  \slashedarrowfill@\Relbar\Relbar{\raisebox{.12em}{}}\rightrightarrows}
\newcommand\xTworightarrow[2][]{%
  \ext@arrow 0055{\rightslashedarrowfilly@}{#1}{#2}}

\pgfdeclareshape{slash underlined}
{
  \inheritsavedanchors[from=rectangle] 
  \inheritanchorborder[from=rectangle]
  \inheritanchor[from=rectangle]{north}
  \inheritanchor[from=rectangle]{north west}
  \inheritanchor[from=rectangle]{north east}
  \inheritanchor[from=rectangle]{center}
  \inheritanchor[from=rectangle]{west}
  \inheritanchor[from=rectangle]{east}
  \inheritanchor[from=rectangle]{mid}
  \inheritanchor[from=rectangle]{mid west}
  \inheritanchor[from=rectangle]{mid east}
  \inheritanchor[from=rectangle]{base}
  \inheritanchor[from=rectangle]{base west}
  \inheritanchor[from=rectangle]{base east}
  \inheritanchor[from=rectangle]{south}
  \inheritanchor[from=rectangle]{south west}
  \inheritanchor[from=rectangle]{south east}
  \inheritanchorborder[from=rectangle]
  \foregroundpath{
    \southwest \pgf@xa=\pgf@x \pgf@ya=\pgf@y
    \northeast \pgf@xb=\pgf@x \pgf@yb=\pgf@y
    \pgf@xc=\pgf@xa
    \advance\pgf@xc by .5\pgf@xb
    \pgf@yc=\pgf@ya
    \advance\pgf@xc by -1.3pt
    \advance\pgf@yc by -1.8pt
    \pgfpathmoveto{\pgfqpoint{\pgf@xc}{\pgf@yc}}
    \advance\pgf@xc by  2.6pt
    \advance\pgf@yc by  3.6pt
    \pgfpathlineto{\pgfqpoint{\pgf@xc}{\pgf@yc}}
    \pgfpathmoveto{\pgfqpoint{\pgf@xa}{\pgf@ya}}
    \pgfpathlineto{\pgfqpoint{\pgf@xb}{\pgf@ya}}
 }
}
\tikzset{nomorepostaction/.code=\let\tikz@postactions\pgfutil@empty}

\newtheorem{theorem}{Theorem}[section]
\newtheorem{definition}[theorem]{Definition}

\begin{document}

\begin{titlepage}
\thispagestyle{empty}

\hrule
\begin{center}
{\bf\LARGE Security Analyses of Quantum Communication Protocols}

\vspace{0.7cm}
--- Yong Wang ---

\vspace{2cm}
\begin{figure}[!htbp]
 \centering
 \includegraphics[width=1.0\textwidth]{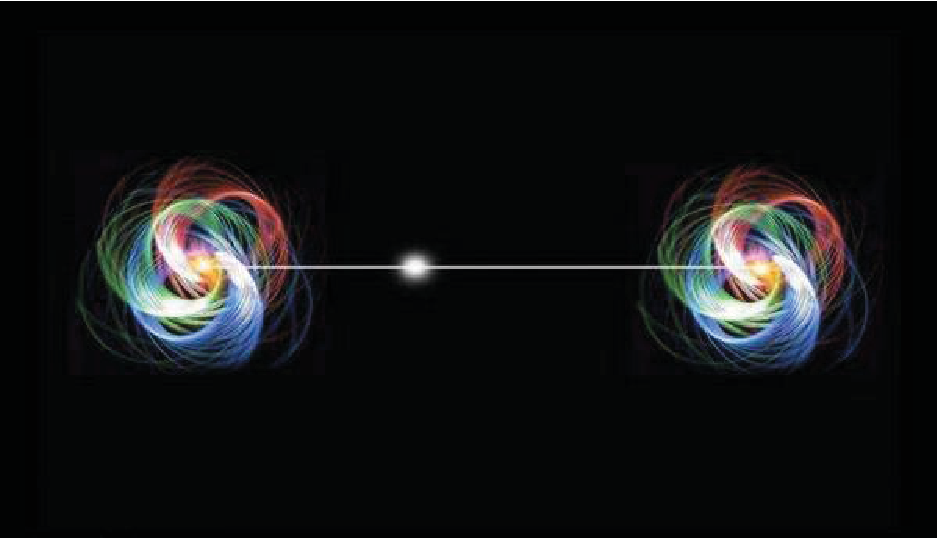}
\end{figure}

\end{center}
\end{titlepage}

\newpage 

\setcounter{page}{1}\pagenumbering{roman}

\tableofcontents

\newpage

\setcounter{page}{1}\pagenumbering{arabic}

        \section{Introduction}

Truly concurrent process algebras are generalizations to the traditional process algebras for true concurrency, CTC \cite{CTC} to CCS \cite{CC} \cite{CCS}, APTC \cite{ATC} to ACP \cite{ACP},
$\pi_{tc}$ \cite{PITC} to $\pi$ calculus \cite{PI1} \cite{PI2}, APPTC \cite{APPTC} to probabilistic process algebra \cite{PPA} \cite{PPA2} \cite{PPA3}.

In quantum process algebras, there are several well-known work \cite{PSQP} \cite{QPA} \cite{QPA2} \cite{CQP} \cite{CQP2} \cite{qCCS} \cite{BQP} \cite{PSQP} \cite{SBQP}, and we ever
did some work \cite{QPA11} \cite{QPA12} \cite{QPA13} to unify quantum and classical computing under the framework of ACP \cite{ACP} and probabilistic process algebra \cite{PPA}.

Now, it is the time to utilize truly concurrent process algebras APTC \cite{ATC} and APPTC \cite{APPTC} to model quantum computing and unify quantum and classical computing in this book.
This book is organized as follows. In chapter \ref{bg}, we introduce the preliminaries. In chapter \ref{qaptcg} and \ref{aqaptcg}, we introduce the utilization of APTC to unify quantum
and classical computing and its usage in verification of quantum communication protocols. In chapter \ref{qapptcg2} and \ref{aqapptcg}, we introduce the utilization of APPTC to unifying
quantum and classical computing and its usage in verification of quantum communication protocols.

\newpage\section{Backgrounds}\label{bg}

To make this book self-satisfied, we introduce some preliminaries in this chapter, including some introductions on operational semantics, proof techniques, truly concurrent process algebra
\cite{ATC} \cite{CTC} \cite{PITC} which is based on truly concurrent operational semantics, and also probabilistic truly concurrent process algebra and probabilistic truly concurrent
operational semantics, and also operational semantics for quantum computing.

\subsection{Operational Semantics}\label{OS}

The semantics of $ACP$ is based on bisimulation/rooted branching bisimulation equivalences, and the modularity of $ACP$ relies on the concept of conservative extension, for the
conveniences, we introduce some concepts and conclusions on them.

\begin{definition}[Bisimulation]
A bisimulation relation $R$ is a binary relation on processes such that: (1) if $p R q$ and $p\xrightarrow{a}p'$ then $q\xrightarrow{a}q'$ with $p' R q'$; (2) if $p R q$ and
$q\xrightarrow{a}q'$ then $p\xrightarrow{a}p'$ with $p' R q'$; (3) if $p R q$ and $pP$, then $qP$; (4) if $p R q$ and $qP$, then $pP$. Two processes $p$ and $q$ are bisimilar,
denoted by $p\sim_{HM} q$, if there is a bisimulation relation $R$ such that $p R q$.
\end{definition}

\begin{definition}[Congruence]
Let $\Sigma$ be a signature. An equivalence relation $R$ on $\mathcal{T}(\Sigma)$ is a congruence if for each $f\in\Sigma$, if $s_i R t_i$ for $i\in\{1,\cdots,ar(f)\}$, then
$f(s_1,\cdots,s_{ar(f)}) R f(t_1,\cdots,t_{ar(f)})$.
\end{definition}

\begin{definition}[Branching bisimulation]
A branching bisimulation relation $R$ is a binary relation on the collection of processes such that: (1) if $p R q$ and $p\xrightarrow{a}p'$ then either $a\equiv \tau$ and $p' R q$ or there is a sequence of (zero or more) $\tau$-transitions $q\xrightarrow{\tau}\cdots\xrightarrow{\tau}q_0$ such that $p R q_0$ and $q_0\xrightarrow{a}q'$ with $p' R q'$; (2) if $p R q$ and $q\xrightarrow{a}q'$ then either $a\equiv \tau$ and $p R q'$ or there is a sequence of (zero or more) $\tau$-transitions $p\xrightarrow{\tau}\cdots\xrightarrow{\tau}p_0$ such that $p_0 R q$ and $p_0\xrightarrow{a}p'$ with $p' R q'$; (3) if $p R q$ and $pP$, then there is a sequence of (zero or more) $\tau$-transitions $q\xrightarrow{\tau}\cdots\xrightarrow{\tau}q_0$ such that $p R q_0$ and $q_0P$; (4) if $p R q$ and $qP$, then there is a sequence of (zero or more) $\tau$-transitions $p\xrightarrow{\tau}\cdots\xrightarrow{\tau}p_0$ such that $p_0 R q$ and $p_0P$. Two processes $p$ and $q$ are branching bisimilar, denoted by $p\approx_{bHM} q$, if there is a branching bisimulation relation $R$ such that $p R q$.
\end{definition}

\begin{definition}[Rooted branching bisimulation]
A rooted branching bisimulation relation $R$ is a binary relation on processes such that: (1) if $p R q$ and $p\xrightarrow{a}p'$ then $q\xrightarrow{a}q'$ with $p'\approx_{bHM} q'$;
(2) if $p R q$ and $q\xrightarrow{a}q'$ then $p\xrightarrow{a}p'$ with $p'\approx_{bHM} q'$; (3) if $p R q$ and $pP$, then $qP$; (4) if $p R q$ and $qP$, then $pP$. Two processes $p$ and $q$ are rooted branching bisimilar, denoted by $p\approx_{rbHM} q$, if there is a rooted branching bisimulation relation $R$ such that $p R q$.
\end{definition}

\begin{definition}[Conservative extension]
Let $T_0$ and $T_1$ be TSSs (transition system specifications) over signatures $\Sigma_0$ and $\Sigma_1$, respectively. The TSS $T_0\oplus T_1$ is a conservative extension of $T_0$ if
the LTSs (labeled transition systems) generated by $T_0$ and $T_0\oplus T_1$ contain exactly the same transitions $t\xrightarrow{a}t'$ and $tP$ with $t\in \mathcal{T}(\Sigma_0)$.
\end{definition}

\begin{definition}[Source-dependency]
The source-dependent variables in a transition rule of $\rho$ are defined inductively as follows: (1) all variables in the source of $\rho$ are source-dependent; (2) if
$t\xrightarrow{a}t'$ is a premise of $\rho$ and all variables in $t$ are source-dependent, then all variables in $t'$ are source-dependent. A transition rule is source-dependent if
all its variables are. A TSS is source-dependent if all its rules are.
\end{definition}

\begin{definition}[Freshness]
Let $T_0$ and $T_1$ be TSSs over signatures $\Sigma_0$ and $\Sigma_1$, respectively. A term in $\mathbb{T}(T_0\oplus T_1)$ is said to be fresh if it contains a function symbol from
$\Sigma_1\setminus\Sigma_0$. Similarly, a transition label or predicate symbol in $T_1$ is fresh if it does not occur in $T_0$.
\end{definition}

\begin{theorem}[Conservative extension]\label{TCE}
Let $T_0$ and $T_1$ be TSSs over signatures $\Sigma_0$ and $\Sigma_1$, respectively, where $T_0$ and $T_0\oplus T_1$ are positive after reduction. Under the following conditions,
$T_0\oplus T_1$ is a conservative extension of $T_0$. (1) $T_0$ is source-dependent. (2) For each $\rho\in T_1$, either the source of $\rho$ is fresh, or $\rho$ has a premise of the
form $t\xrightarrow{a}t'$ or $tP$, where $t\in \mathbb{T}(\Sigma_0)$, all variables in $t$ occur in the source of $\rho$ and $t'$, $a$ or $P$ is fresh.
\end{theorem}

\subsection{Proof Techniques}\label{PT}

In this subsection, we introduce the concepts and conclusions about elimination, which is very important in the proof of completeness theorem.

\begin{definition}[Elimination property]
Let a process algebra with a defined set of basic terms as a subset of the set of closed terms over the process algebra. Then the process algebra has the elimination to basic terms
property if for every closed term $s$ of the algebra, there exists a basic term $t$ of the algebra such that the algebra$\vdash s=t$.
\end{definition}

\begin{definition}[Strongly normalizing]
A term $s_0$ is called strongly normalizing if does not an infinite series of reductions beginning in $s_0$.
\end{definition}

\begin{definition}
We write $s>_{lpo} t$ if $s\rightarrow^+ t$ where $\rightarrow^+$ is the transitive closure of the reduction relation defined by the transition rules of an algebra.
\end{definition}

\begin{theorem}[Strong normalization]\label{SN}
Let a term rewriting system (TRS) with finitely many rewriting rules and let $>$ be a well-founded ordering on the signature of the corresponding algebra. If $s>_{lpo} t$ for each
rewriting rule $s\rightarrow t$ in the TRS, then the term rewriting system is strongly normalizing.
\end{theorem}

\subsection{APTC with Guards -- $APTC_G$}

\begin{definition}[Prime event structure with silent event and empty event]
Let $\Lambda$ be a fixed set of labels, ranged over $a,b,c,\cdots$ and $\tau,\epsilon$. A ($\Lambda$-labelled) prime event structure with silent event $\tau$ and empty event $\epsilon$ is a tuple $\mathcal{E}=\langle \mathbb{E}, \leq, \sharp, \lambda\rangle$, where $\mathbb{E}$ is a denumerable set of events, including the silent event $\tau$ and empty event $\epsilon$. Let $\hat{\mathbb{E}}=\mathbb{E}\backslash\{\tau,\epsilon\}$, exactly excluding $\tau$ and $\epsilon$, it is obvious that $\hat{\tau^*}=\epsilon$. Let $\lambda:\mathbb{E}\rightarrow\Lambda$ be a labelling function and let $\lambda(\tau)=\tau$ and $\lambda(\epsilon)=\epsilon$. And $\leq$, $\sharp$ are binary relations on $\mathbb{E}$, called causality and conflict respectively, such that:

\begin{enumerate}
  \item $\leq$ is a partial order and $\lceil e \rceil = \{e'\in \mathbb{E}|e'\leq e\}$ is finite for all $e\in \mathbb{E}$. It is easy to see that $e\leq\tau^*\leq e'=e\leq\tau\leq\cdots\leq\tau\leq e'$, then $e\leq e'$.
  \item $\sharp$ is irreflexive, symmetric and hereditary with respect to $\leq$, that is, for all $e,e',e''\in \mathbb{E}$, if $e\sharp e'\leq e''$, then $e\sharp e''$.
\end{enumerate}

Then, the concepts of consistency and concurrency can be drawn from the above definition:

\begin{enumerate}
  \item $e,e'\in \mathbb{E}$ are consistent, denoted as $e\frown e'$, if $\neg(e\sharp e')$. A subset $X\subseteq \mathbb{E}$ is called consistent, if $e\frown e'$ for all $e,e'\in X$.
  \item $e,e'\in \mathbb{E}$ are concurrent, denoted as $e\parallel e'$, if $\neg(e\leq e')$, $\neg(e'\leq e)$, and $\neg(e\sharp e')$.
\end{enumerate}
\end{definition}

\begin{definition}[Configuration]
Let $\mathcal{E}$ be a PES. A (finite) configuration in $\mathcal{E}$ is a (finite) consistent subset of events $C\subseteq \mathcal{E}$, closed with respect to causality (i.e. $\lceil C\rceil=C$), and a data state $s\in S$ with $S$ the set of all data states, denoted $\langle C, s\rangle$. The set of finite configurations of $\mathcal{E}$ is denoted by $\langle\mathcal{C}(\mathcal{E}), S\rangle$. We let $\hat{C}=C\backslash\{\tau\}\cup\{\epsilon\}$.
\end{definition}

A consistent subset of $X\subseteq \mathbb{E}$ of events can be seen as a pomset. Given $X, Y\subseteq \mathbb{E}$, $\hat{X}\sim \hat{Y}$ if $\hat{X}$ and $\hat{Y}$ are isomorphic as pomsets. In the following of the paper, we say $C_1\sim C_2$, we mean $\hat{C_1}\sim\hat{C_2}$.

\begin{definition}[Pomset transitions and step]
Let $\mathcal{E}$ be a PES and let $C\in\mathcal{C}(\mathcal{E})$, and $\emptyset\neq X\subseteq \mathbb{E}$, if $C\cap X=\emptyset$ and $C'=C\cup X\in\mathcal{C}(\mathcal{E})$, then $\langle C,s\rangle\xrightarrow{X} \langle C',s'\rangle$ is called a pomset transition from $\langle C,s\rangle$ to $\langle C',s'\rangle$. When the events in $X$ are pairwise concurrent, we say that $\langle C,s\rangle\xrightarrow{X}\langle C',s'\rangle$ is a step. It is obvious that $\rightarrow^*\xrightarrow{X}\rightarrow^*=\xrightarrow{X}$ and $\rightarrow^*\xrightarrow{e}\rightarrow^*=\xrightarrow{e}$ for any $e\in\mathbb{E}$ and $X\subseteq\mathbb{E}$.
\end{definition}

\begin{definition}[Weak pomset transitions and weak step]
Let $\mathcal{E}$ be a PES and let $C\in\mathcal{C}(\mathcal{E})$, and $\emptyset\neq X\subseteq \hat{\mathbb{E}}$, if $C\cap X=\emptyset$ and $\hat{C'}=\hat{C}\cup X\in\mathcal{C}(\mathcal{E})$, then $\langle C,s\rangle\xRightarrow{X} \langle C',s'\rangle$ is called a weak pomset transition from $\langle C,s\rangle$ to $\langle C',s'\rangle$, where we define $\xRightarrow{e}\triangleq\xrightarrow{\tau^*}\xrightarrow{e}\xrightarrow{\tau^*}$. And $\xRightarrow{X}\triangleq\xrightarrow{\tau^*}\xrightarrow{e}\xrightarrow{\tau^*}$, for every $e\in X$. When the events in $X$ are pairwise concurrent, we say that $\langle C,s\rangle\xRightarrow{X}\langle C',s'\rangle$ is a weak step.
\end{definition}

We will also suppose that all the PESs in this paper are image finite, that is, for any PES $\mathcal{E}$ and $C\in \mathcal{C}(\mathcal{E})$ and $a\in \Lambda$, $\{e\in \mathbb{E}|\langle C,s\rangle\xrightarrow{e} \langle C',s'\rangle\wedge \lambda(e)=a\}$ and $\{e\in\hat{\mathbb{E}}|\langle C,s\rangle\xRightarrow{e} \langle C',s'\rangle\wedge \lambda(e)=a\}$ is finite.

\begin{definition}[Pomset, step bisimulation]
Let $\mathcal{E}_1$, $\mathcal{E}_2$ be PESs. A pomset bisimulation is a relation $R\subseteq\langle\mathcal{C}(\mathcal{E}_1),S\rangle\times\langle\mathcal{C}(\mathcal{E}_2),S\rangle$, such that if $(\langle C_1,s\rangle,\langle C_2,s\rangle)\in R$, and $\langle C_1,s\rangle\xrightarrow{X_1}\langle C_1',s'\rangle$ then $\langle C_2,s\rangle\xrightarrow{X_2}\langle C_2',s'\rangle$, with $X_1\subseteq \mathbb{E}_1$, $X_2\subseteq \mathbb{E}_2$, $X_1\sim X_2$ and $(\langle C_1',s'\rangle,\langle C_2',s'\rangle)\in R$ for all $s,s'\in S$, and vice-versa. We say that $\mathcal{E}_1$, $\mathcal{E}_2$ are pomset bisimilar, written $\mathcal{E}_1\sim_p\mathcal{E}_2$, if there exists a pomset bisimulation $R$, such that $(\langle\emptyset,\emptyset\rangle,\langle\emptyset,\emptyset\rangle)\in R$. By replacing pomset transitions with steps, we can get the definition of step bisimulation. When PESs $\mathcal{E}_1$ and $\mathcal{E}_2$ are step bisimilar, we write $\mathcal{E}_1\sim_s\mathcal{E}_2$.
\end{definition}

\begin{definition}[Weak pomset, step bisimulation]
Let $\mathcal{E}_1$, $\mathcal{E}_2$ be PESs. A weak pomset bisimulation is a relation $R\subseteq\langle\mathcal{C}(\mathcal{E}_1),S\rangle\times\langle\mathcal{C}(\mathcal{E}_2),S\rangle$, such that if $(\langle C_1,s\rangle,\langle C_2,s\rangle)\in R$, and $\langle C_1,s\rangle\xRightarrow{X_1}\langle C_1',s'\rangle$ then $\langle C_2,s\rangle\xRightarrow{X_2}\langle C_2',s'\rangle$, with $X_1\subseteq \hat{\mathbb{E}_1}$, $X_2\subseteq \hat{\mathbb{E}_2}$, $X_1\sim X_2$ and $(\langle C_1',s'\rangle,\langle C_2',s'\rangle)\in R$ for all $s,s'\in S$, and vice-versa. We say that $\mathcal{E}_1$, $\mathcal{E}_2$ are weak pomset bisimilar, written $\mathcal{E}_1\approx_p\mathcal{E}_2$, if there exists a weak pomset bisimulation $R$, such that $(\langle\emptyset,\emptyset\rangle,\langle\emptyset,\emptyset\rangle)\in R$. By replacing weak pomset transitions with weak steps, we can get the definition of weak step bisimulation. When PESs $\mathcal{E}_1$ and $\mathcal{E}_2$ are weak step bisimilar, we write $\mathcal{E}_1\approx_s\mathcal{E}_2$.
\end{definition}

\begin{definition}[Posetal product]
Given two PESs $\mathcal{E}_1$, $\mathcal{E}_2$, the posetal product of their configurations, denoted $\langle\mathcal{C}(\mathcal{E}_1),S\rangle\overline{\times}\langle\mathcal{C}(\mathcal{E}_2),S\rangle$, is defined as

$$\{(\langle C_1,s\rangle,f,\langle C_2,s\rangle)|C_1\in\mathcal{C}(\mathcal{E}_1),C_2\in\mathcal{C}(\mathcal{E}_2),f:C_1\rightarrow C_2 \textrm{ isomorphism}\}.$$

A subset $R\subseteq\langle\mathcal{C}(\mathcal{E}_1),S\rangle\overline{\times}\langle\mathcal{C}(\mathcal{E}_2),S\rangle$ is called a posetal relation. We say that $R$ is downward closed when for any $(\langle C_1,s\rangle,f,\langle C_2,s\rangle),(\langle C_1',s'\rangle,f',\langle C_2',s'\rangle)\in \langle\mathcal{C}(\mathcal{E}_1),S\rangle\overline{\times}\langle\mathcal{C}(\mathcal{E}_2),S\rangle$, if $(\langle C_1,s\rangle,f,\langle C_2,s\rangle)\subseteq (\langle C_1',s'\rangle,f',\langle C_2',s'\rangle)$ pointwise and $(\langle C_1',s'\rangle,f',\langle C_2',s'\rangle)\in R$, then $(\langle C_1,s\rangle,f,\langle C_2,s\rangle)\in R$.

For $f:X_1\rightarrow X_2$, we define $f[x_1\mapsto x_2]:X_1\cup\{x_1\}\rightarrow X_2\cup\{x_2\}$, $z\in X_1\cup\{x_1\}$,(1)$f[x_1\mapsto x_2](z)=
x_2$,if $z=x_1$;(2)$f[x_1\mapsto x_2](z)=f(z)$, otherwise. Where $X_1\subseteq \mathbb{E}_1$, $X_2\subseteq \mathbb{E}_2$, $x_1\in \mathbb{E}_1$, $x_2\in \mathbb{E}_2$.
\end{definition}

\begin{definition}[Weakly posetal product]
Given two PESs $\mathcal{E}_1$, $\mathcal{E}_2$, the weakly posetal product of their configurations, denoted $\langle\mathcal{C}(\mathcal{E}_1),S\rangle\overline{\times}\langle\mathcal{C}(\mathcal{E}_2),S\rangle$, is defined as

$$\{(\langle C_1,s\rangle,f,\langle C_2,s\rangle)|C_1\in\mathcal{C}(\mathcal{E}_1),C_2\in\mathcal{C}(\mathcal{E}_2),f:\hat{C_1}\rightarrow \hat{C_2} \textrm{ isomorphism}\}.$$

A subset $R\subseteq\langle\mathcal{C}(\mathcal{E}_1),S\rangle\overline{\times}\langle\mathcal{C}(\mathcal{E}_2),S\rangle$ is called a weakly posetal relation. We say that $R$ is downward closed when for any $(\langle C_1,s\rangle,f,\langle C_2,s\rangle),(\langle C_1',s'\rangle,f,\langle C_2',s'\rangle)\in \langle\mathcal{C}(\mathcal{E}_1),S\rangle\overline{\times}\langle\mathcal{C}(\mathcal{E}_2),S\rangle$, if $(\langle C_1,s\rangle,f,\langle C_2,s\rangle)\subseteq (\langle C_1',s'\rangle,f',\langle C_2',s'\rangle)$ pointwise and $(\langle C_1',s'\rangle,f',\langle C_2',s'\rangle)\in R$, then $(\langle C_1,s\rangle,f,\langle C_2,s\rangle)\in R$.

For $f:X_1\rightarrow X_2$, we define $f[x_1\mapsto x_2]:X_1\cup\{x_1\}\rightarrow X_2\cup\{x_2\}$, $z\in X_1\cup\{x_1\}$,(1)$f[x_1\mapsto x_2](z)=
x_2$,if $z=x_1$;(2)$f[x_1\mapsto x_2](z)=f(z)$, otherwise. Where $X_1\subseteq \hat{\mathbb{E}_1}$, $X_2\subseteq \hat{\mathbb{E}_2}$, $x_1\in \hat{\mathbb{E}}_1$, $x_2\in \hat{\mathbb{E}}_2$. Also, we define $f(\tau^*)=f(\tau^*)$.
\end{definition}

\begin{definition}[(Hereditary) history-preserving bisimulation]
A history-preserving (hp-) bisimulation is a posetal relation $R\subseteq\langle\mathcal{C}(\mathcal{E}_1),S\rangle\overline{\times}\langle\mathcal{C}(\mathcal{E}_2),S\rangle$ such that if $(\langle C_1,s\rangle,f,\langle C_2,s\rangle)\in R$, and $\langle C_1,s\rangle\xrightarrow{e_1} \langle C_1',s'\rangle$, then $\langle C_2,s\rangle\xrightarrow{e_2} \langle C_2',s'\rangle$, with $(\langle C_1',s'\rangle,f[e_1\mapsto e_2],\langle C_2',s'\rangle)\in R$ for all $s,s'\in S$, and vice-versa. $\mathcal{E}_1,\mathcal{E}_2$ are history-preserving (hp-)bisimilar and are written $\mathcal{E}_1\sim_{hp}\mathcal{E}_2$ if there exists a hp-bisimulation $R$ such that $(\langle\emptyset,\emptyset\rangle,\emptyset,\langle\emptyset,\emptyset\rangle)\in R$.

A hereditary history-preserving (hhp-)bisimulation is a downward closed hp-bisimulation. $\mathcal{E}_1,\mathcal{E}_2$ are hereditary history-preserving (hhp-)bisimilar and are written $\mathcal{E}_1\sim_{hhp}\mathcal{E}_2$.
\end{definition}

\begin{definition}[Weak (hereditary) history-preserving bisimulation]
A weak history-preserving (hp-) bisimulation is a weakly posetal relation $R\subseteq\langle\mathcal{C}(\mathcal{E}_1),S\rangle\overline{\times}\langle\mathcal{C}(\mathcal{E}_2),S\rangle$ such that if $(\langle C_1,s\rangle,f,\langle C_2,s\rangle)\in R$, and $\langle C_1,s\rangle\xRightarrow{e_1} \langle C_1',s'\rangle$, then $\langle C_2,s\rangle\xRightarrow{e_2} \langle C_2',s'\rangle$, with $(\langle C_1',s'\rangle,f[e_1\mapsto e_2],\langle C_2',s'\rangle)\in R$ for all $s,s'\in S$, and vice-versa. $\mathcal{E}_1,\mathcal{E}_2$ are weak history-preserving (hp-)bisimilar and are written $\mathcal{E}_1\approx_{hp}\mathcal{E}_2$ if there exists a weak hp-bisimulation $R$ such that $(\langle\emptyset,\emptyset\rangle,\emptyset,\langle\emptyset,\emptyset\rangle)\in R$.

A weakly hereditary history-preserving (hhp-)bisimulation is a downward closed weak hp-bisimulation. $\mathcal{E}_1,\mathcal{E}_2$ are weakly hereditary history-preserving (hhp-)bisimilar and are written $\mathcal{E}_1\approx_{hhp}\mathcal{E}_2$.
\end{definition}

\begin{definition}[Branching pomset, step bisimulation]
Assume a special termination predicate $\downarrow$, and let $\surd$ represent a state with $\surd\downarrow$. Let $\mathcal{E}_1$, $\mathcal{E}_2$ be PESs. A branching pomset bisimulation is a relation $R\subseteq\langle\mathcal{C}(\mathcal{E}_1),S\rangle\times\langle\mathcal{C}(\mathcal{E}_2),S\rangle$, such that:
 \begin{enumerate}
   \item if $(\langle C_1,s\rangle,\langle C_2,s\rangle)\in R$, and $\langle C_1,s\rangle\xrightarrow{X}\langle C_1',s'\rangle$ then
   \begin{itemize}
     \item either $X\equiv \tau^*$, and $(\langle C_1',s'\rangle,\langle C_2,s\rangle)\in R$ with $s'\in \tau(s)$;
     \item or there is a sequence of (zero or more) $\tau$-transitions $\langle C_2,s\rangle\xrightarrow{\tau^*} \langle C_2^0,s^0\rangle$, such that $(\langle C_1,s\rangle,\langle C_2^0,s^0\rangle)\in R$ and $\langle C_2^0,s^0\rangle\xRightarrow{X}\langle C_2',s'\rangle$ with $(\langle C_1',s'\rangle,\langle C_2',s'\rangle)\in R$;
   \end{itemize}
   \item if $(\langle C_1,s\rangle,\langle C_2,s\rangle)\in R$, and $\langle C_2,s\rangle\xrightarrow{X}\langle C_2',s'\rangle$ then
   \begin{itemize}
     \item either $X\equiv \tau^*$, and $(\langle C_1,s\rangle,\langle C_2',s'\rangle)\in R$;
     \item or there is a sequence of (zero or more) $\tau$-transitions $\langle C_1,s\rangle\xrightarrow{\tau^*} \langle C_1^0,s^0\rangle$, such that $(\langle C_1^0,s^0\rangle,\langle C_2,s\rangle)\in R$ and $\langle C_1^0,s^0\rangle\xRightarrow{X}\langle C_1',s'\rangle$ with $(\langle C_1',s'\rangle,\langle C_2',s'\rangle)\in R$;
   \end{itemize}
   \item if $(\langle C_1,s\rangle,\langle C_2,s\rangle)\in R$ and $\langle C_1,s\rangle\downarrow$, then there is a sequence of (zero or more) $\tau$-transitions $\langle C_2,s\rangle\xrightarrow{\tau^*}\langle C_2^0,s^0\rangle$ such that $(\langle C_1,s\rangle,\langle C_2^0,s^0\rangle)\in R$ and $\langle C_2^0,s^0\rangle\downarrow$;
   \item if $(\langle C_1,s\rangle,\langle C_2,s\rangle)\in R$ and $\langle C_2,s\rangle\downarrow$, then there is a sequence of (zero or more) $\tau$-transitions $\langle C_1,s\rangle\xrightarrow{\tau^*}\langle C_1^0,s^0\rangle$ such that $(\langle C_1^0,s^0\rangle,\langle C_2,s\rangle)\in R$ and $\langle C_1^0,s^0\rangle\downarrow$.
 \end{enumerate}

We say that $\mathcal{E}_1$, $\mathcal{E}_2$ are branching pomset bisimilar, written $\mathcal{E}_1\approx_{bp}\mathcal{E}_2$, if there exists a branching pomset bisimulation $R$, such that $(\langle\emptyset,\emptyset\rangle,\langle\emptyset,\emptyset\rangle)\in R$.

By replacing pomset transitions with steps, we can get the definition of branching step bisimulation. When PESs $\mathcal{E}_1$ and $\mathcal{E}_2$ are branching step bisimilar, we write $\mathcal{E}_1\approx_{bs}\mathcal{E}_2$.
\end{definition}

\begin{definition}[Rooted branching pomset, step bisimulation]
Assume a special termination predicate $\downarrow$, and let $\surd$ represent a state with $\surd\downarrow$. Let $\mathcal{E}_1$, $\mathcal{E}_2$ be PESs. A rooted branching pomset bisimulation is a relation $R\subseteq\langle\mathcal{C}(\mathcal{E}_1),S\rangle\times\langle\mathcal{C}(\mathcal{E}_2),S\rangle$, such that:
 \begin{enumerate}
   \item if $(\langle C_1,s\rangle,\langle C_2,s\rangle)\in R$, and $\langle C_1,s\rangle\xrightarrow{X}\langle C_1',s'\rangle$ then $\langle C_2,s\rangle\xrightarrow{X}\langle C_2',s'\rangle$ with $\langle C_1',s'\rangle\approx_{bp}\langle C_2',s'\rangle$;
   \item if $(\langle C_1,s\rangle,\langle C_2,s\rangle)\in R$, and $\langle C_2,s\rangle\xrightarrow{X}\langle C_2',s'\rangle$ then $\langle C_1,s\rangle\xrightarrow{X}\langle C_1',s'\rangle$ with $\langle C_1',s'\rangle\approx_{bp}\langle C_2',s'\rangle$;
   \item if $(\langle C_1,s\rangle,\langle C_2,s\rangle)\in R$ and $\langle C_1,s\rangle\downarrow$, then $\langle C_2,s\rangle\downarrow$;
   \item if $(\langle C_1,s\rangle,\langle C_2,s\rangle)\in R$ and $\langle C_2,s\rangle\downarrow$, then $\langle C_1,s\rangle\downarrow$.
 \end{enumerate}

We say that $\mathcal{E}_1$, $\mathcal{E}_2$ are rooted branching pomset bisimilar, written $\mathcal{E}_1\approx_{rbp}\mathcal{E}_2$, if there exists a rooted branching pomset bisimulation $R$, such that $(\langle\emptyset,\emptyset\rangle,\langle\emptyset,\emptyset\rangle)\in R$.

By replacing pomset transitions with steps, we can get the definition of rooted branching step bisimulation. When PESs $\mathcal{E}_1$ and $\mathcal{E}_2$ are rooted branching step bisimilar, we write $\mathcal{E}_1\approx_{rbs}\mathcal{E}_2$.
\end{definition}

\begin{definition}[Branching (hereditary) history-preserving bisimulation]
Assume a special termination predicate $\downarrow$, and let $\surd$ represent a state with $\surd\downarrow$. A branching history-preserving (hp-) bisimulation is a weakly posetal relation $R\subseteq\langle\mathcal{C}(\mathcal{E}_1),S\rangle\overline{\times}\langle\mathcal{C}(\mathcal{E}_2),S\rangle$ such that:

 \begin{enumerate}
   \item if $(\langle C_1,s\rangle,f,\langle C_2,s\rangle)\in R$, and $\langle C_1,s\rangle\xrightarrow{e_1}\langle C_1',s'\rangle$ then
   \begin{itemize}
     \item either $e_1\equiv \tau$, and $(\langle C_1',s'\rangle,f[e_1\mapsto \tau^{e_1}],\langle C_2,s\rangle)\in R$;
     \item or there is a sequence of (zero or more) $\tau$-transitions $\langle C_2,s\rangle\xrightarrow{\tau^*} \langle C_2^0,s^0\rangle$, such that $(\langle C_1,s\rangle,f,\langle C_2^0,s^0\rangle)\in R$ and $\langle C_2^0,s^0\rangle\xrightarrow{e_2}\langle C_2',s'\rangle$ with $(\langle C_1',s'\rangle,f[e_1\mapsto e_2],\langle C_2',s'\rangle)\in R$;
   \end{itemize}
   \item if $(\langle C_1,s\rangle,f,\langle C_2,s\rangle)\in R$, and $\langle C_2,s\rangle\xrightarrow{e_2}\langle C_2',s'\rangle$ then
   \begin{itemize}
     \item either $e_2\equiv \tau$, and $(\langle C_1,s\rangle,f[e_2\mapsto \tau^{e_2}],\langle C_2',s'\rangle)\in R$;
     \item or there is a sequence of (zero or more) $\tau$-transitions $\langle C_1,s\rangle\xrightarrow{\tau^*} \langle C_1^0,s^0\rangle$, such that $(\langle C_1^0,s^0\rangle,f,\langle C_2,s\rangle)\in R$ and $\langle C_1^0,s^0\rangle\xrightarrow{e_1}\langle C_1',s'\rangle$ with $(\langle C_1',s'\rangle,f[e_2\mapsto e_1],\langle C_2',s'\rangle)\in R$;
   \end{itemize}
   \item if $(\langle C_1,s\rangle,f,\langle C_2,s\rangle)\in R$ and $\langle C_1,s\rangle\downarrow$, then there is a sequence of (zero or more) $\tau$-transitions $\langle C_2,s\rangle\xrightarrow{\tau^*}\langle C_2^0,s^0\rangle$ such that $(\langle C_1,s\rangle,f,\langle C_2^0,s^0\rangle)\in R$ and $\langle C_2^0,s^0\rangle\downarrow$;
   \item if $(\langle C_1,s\rangle,f,\langle C_2,s\rangle)\in R$ and $\langle C_2,s\rangle\downarrow$, then there is a sequence of (zero or more) $\tau$-transitions $\langle C_1,s\rangle\xrightarrow{\tau^*}\langle C_1^0,s^0\rangle$ such that $(\langle C_1^0,s^0\rangle,f,\langle C_2,s\rangle)\in R$ and $\langle C_1^0,s^0\rangle\downarrow$.
 \end{enumerate}

$\mathcal{E}_1,\mathcal{E}_2$ are branching history-preserving (hp-)bisimilar and are written $\mathcal{E}_1\approx_{bhp}\mathcal{E}_2$ if there exists a branching hp-bisimulation $R$ such that $(\langle\emptyset,\emptyset\rangle,\emptyset,\langle\emptyset,\emptyset\rangle)\in R$.

A branching hereditary history-preserving (hhp-)bisimulation is a downward closed branching hp-bisimulation. $\mathcal{E}_1,\mathcal{E}_2$ are branching hereditary history-preserving (hhp-)bisimilar and are written $\mathcal{E}_1\approx_{bhhp}\mathcal{E}_2$.
\end{definition}

\begin{definition}[Rooted branching (hereditary) history-preserving bisimulation]
Assume a special termination predicate $\downarrow$, and let $\surd$ represent a state with $\surd\downarrow$. A rooted branching history-preserving (hp-) bisimulation is a weakly posetal relation $R\subseteq\langle\mathcal{C}(\mathcal{E}_1),S\rangle\overline{\times}\langle\mathcal{C}(\mathcal{E}_2),S\rangle$ such that:

 \begin{enumerate}
   \item if $(\langle C_1,s\rangle,f,\langle C_2,s\rangle)\in R$, and $\langle C_1,s\rangle\xrightarrow{e_1}\langle C_1',s'\rangle$, then $\langle C_2,s\rangle\xrightarrow{e_2}\langle C_2',s'\rangle$ with $\langle C_1',s'\rangle\approx_{bhp}\langle C_2',s'\rangle$;
   \item if $(\langle C_1,s\rangle,f,\langle C_2,s\rangle)\in R$, and $\langle C_2,s\rangle\xrightarrow{e_2}\langle C_2',s'\rangle$, then $\langle C_1,s\rangle\xrightarrow{e_1}\langle C_1',s'\rangle$ with $\langle C_1',s'\rangle\approx_{bhp}\langle C_2',s'\rangle$;
   \item if $(\langle C_1,s\rangle,f,\langle C_2,s\rangle)\in R$ and $\langle C_1,s\rangle\downarrow$, then $\langle C_2,s\rangle\downarrow$;
   \item if $(\langle C_1,s\rangle,f,\langle C_2,s\rangle)\in R$ and $\langle C_2,s\rangle\downarrow$, then $\langle C_1,s\rangle\downarrow$.
 \end{enumerate}

$\mathcal{E}_1,\mathcal{E}_2$ are rooted branching history-preserving (hp-)bisimilar and are written $\mathcal{E}_1\approx_{rbhp}\mathcal{E}_2$ if there exists a rooted branching hp-bisimulation $R$ such that $(\langle\emptyset,\emptyset\rangle,\emptyset,\langle\emptyset,\emptyset\rangle)\in R$.

A rooted branching hereditary history-preserving (hhp-)bisimulation is a downward closed rooted branching hp-bisimulation. $\mathcal{E}_1,\mathcal{E}_2$ are rooted branching hereditary history-preserving (hhp-)bisimilar and are written $\mathcal{E}_1\approx_{rbhhp}\mathcal{E}_2$.
\end{definition}

\subsubsection{$BATC$ with Guards}

In this subsection, we will discuss the guards for $BATC$, which is denoted as $BATC_G$. Let $\mathbb{E}$ be the set of atomic events (actions), $G_{at}$ be the set of atomic guards, $\delta$ be the deadlock constant, and $\epsilon$ be the empty event. We extend $G_{at}$ to the set of basic guards $G$ with element $\phi,\psi,\cdots$, which is generated by the following formation rules:

$$\phi::=\delta|\epsilon|\neg\phi|\psi\in G_{at}|\phi+\psi|\phi\cdot\psi$$

In the following, let $e_1, e_2, e_1', e_2'\in \mathbb{E}$, $\phi,\psi\in G$ and let variables $x,y,z$ range over the set of terms for true concurrency, $p,q,s$ range over the set of closed terms. The predicate $test(\phi,s)$ represents that $\phi$ holds in the state $s$, and $test(\epsilon,s)$ holds and $test(\delta,s)$ does not hold. $effect(e,s)\in S$ denotes $s'$ in $s\xrightarrow{e}s'$. The predicate weakest precondition $wp(e,\phi)$ denotes that $\forall s,s'\in S, test(\phi,effect(e,s))$ holds.

The set of axioms of $BATC_G$ consists of the laws given in Table \ref{AxiomsForBATCG21}.

\begin{center}
    \begin{table}
        \begin{tabular}{@{}ll@{}}
            \hline No. &Axiom\\
            $A1$ & $x+ y = y+ x$\\
            $A2$ & $(x+ y)+ z = x+ (y+ z)$\\
            $A3$ & $x+ x = x$\\
            $A4$ & $(x+ y)\cdot z = x\cdot z + y\cdot z$\\
            $A5$ & $(x\cdot y)\cdot z = x\cdot(y\cdot z)$\\
            $A6$ & $x+\delta = x$\\
            $A7$ & $\delta\cdot x = \delta$\\
            $A8$ & $\epsilon\cdot x = x$\\
            $A9$ & $x\cdot\epsilon = x$\\
            $G1$ & $\phi\cdot\neg\phi = \delta$\\
            $G2$ & $\phi+\neg\phi = \epsilon$\\
            $G3$ & $\phi\delta = \delta$\\
            $G4$ & $\phi(x+y)=\phi x+\phi y$\\
            $G5$ & $\phi(x\cdot y)= \phi x\cdot y$\\
            $G6$ & $(\phi+\psi)x = \phi x + \psi x$\\
            $G7$ & $(\phi\cdot \psi)\cdot x = \phi\cdot(\psi\cdot x)$\\
            $G8$ & $\phi=\epsilon$ if $\forall s\in S.test(\phi,s)$\\
            $G9$ & $\phi_0\cdot\cdots\cdot\phi_n = \delta$ if $\forall s\in S,\exists i\leq n.test(\neg\phi_i,s)$\\
            $G10$ & $wp(e,\phi)e\phi=wp(e,\phi)e$\\
            $G11$ & $\neg wp(e,\phi)e\neg\phi=\neg wp(e,\phi)e$\\
        \end{tabular}
        \caption{Axioms of $BATC_G$}
        \label{AxiomsForBATCG21}
    \end{table}
\end{center}

Note that, by eliminating atomic event from the process terms, the axioms in Table \ref{AxiomsForBATCG21} will lead to a Boolean Algebra. And $G9$ is a precondition of $e$ and $\phi$, $G10$ is the weakest precondition of $e$ and $\phi$. A data environment with $effect$ function is sufficiently deterministic, and it is obvious that if the weakest precondition is expressible and $G9$, $G10$ are sound, then the related data environment is sufficiently deterministic.

\begin{definition}[Basic terms of $BATC_G$]
The set of basic terms of $BATC_G$, $\mathcal{B}(BATC_G)$, is inductively defined as follows:
\begin{enumerate}
  \item $\mathbb{E}\subset\mathcal{B}(BATC_G)$;
  \item $G\subset\mathcal{B}(BATC_G)$;
  \item if $e\in \mathbb{E}, t\in\mathcal{B}(BATC_G)$ then $e\cdot t\in\mathcal{B}(BATC_G)$;
  \item if $\phi\in G, t\in\mathcal{B}(BATC_G)$ then $\phi\cdot t\in\mathcal{B}(BATC_G)$;
  \item if $t,s\in\mathcal{B}(BATC_G)$ then $t+ s\in\mathcal{B}(BATC_G)$.
\end{enumerate}
\end{definition}

\begin{theorem}[Elimination theorem of $BATC_G$]
Let $p$ be a closed $BATC_G$ term. Then there is a basic $BATC_G$ term $q$ such that $BATC_G\vdash p=q$.
\end{theorem}

We will define a term-deduction system which gives the operational semantics of $BATC_G$. We give the operational transition rules for $\epsilon$, atomic guard $\phi\in G_{at}$, atomic 
event $e\in\mathbb{E}$, operators $\cdot$ and $+$ as Table \ref{SETRForBATCG21} shows. And the predicate $\xrightarrow{e}\surd$ represents successful termination after execution of the event $e$.

\begin{center}
    \begin{table}
        $$\frac{}{\langle\epsilon,s\rangle\rightarrow\langle\surd,s\rangle}$$
        $$\frac{}{\langle e,s\rangle\xrightarrow{e}\langle\surd,s'\rangle}\textrm{ if }s'\in effect(e,s)$$
        $$\frac{}{\langle\phi,s\rangle\rightarrow\langle\surd,s\rangle}\textrm{ if }test(\phi,s)$$
        $$\frac{\langle x,s\rangle\xrightarrow{e}\langle\surd,s'\rangle}{\langle x+ y,s\rangle\xrightarrow{e}\langle\surd,s'\rangle} \quad\frac{\langle x,s\rangle\xrightarrow{e}\langle x',s'\rangle}{\langle x+ y,s\rangle\xrightarrow{e}\langle x',s'\rangle}$$
        $$\frac{\langle y,s\rangle\xrightarrow{e}\langle\surd,s'\rangle}{\langle x+ y,s\rangle\xrightarrow{e}\langle\surd,s'\rangle} \quad\frac{\langle y,s\rangle\xrightarrow{e}\langle y',s'\rangle}{\langle x+ y,s\rangle\xrightarrow{e}\langle y',s'\rangle}$$
        $$\frac{\langle x,s\rangle\xrightarrow{e}\langle\surd,s'\rangle}{\langle x\cdot y,s\rangle\xrightarrow{e} \langle y,s'\rangle} \quad\frac{\langle x,s\rangle\xrightarrow{e}\langle x',s'\rangle}{\langle x\cdot y,s\rangle\xrightarrow{e}\langle x'\cdot y,s'\rangle}$$
        \caption{Single event transition rules of $BATC_G$}
        \label{SETRForBATCG21}
    \end{table}
\end{center}

Note that, we replace the single atomic event $e\in\mathbb{E}$ by $X\subseteq\mathbb{E}$, we can obtain the pomset transition rules of $BATC_G$, and omit them.

\begin{theorem}[Congruence of $BATC_G$ with respect to truly concurrent bisimulation equivalences]
(1) Pomset bisimulation equivalence $\sim_{p}$ is a congruence with respect to $BATC_G$.

(2) Step bisimulation equivalence $\sim_{s}$ is a congruence with respect to $BATC_G$.

(3) Hp-bisimulation equivalence $\sim_{hp}$ is a congruence with respect to $BATC_G$.

(4) Hhp-bisimulation equivalence $\sim_{hhp}$ is a congruence with respect to $BATC_G$.
\end{theorem}

\begin{theorem}[Soundness of $BATC_G$ modulo truly concurrent bisimulation equivalences]
(1) Let $x$ and $y$ be $BATC_G$ terms. If $BATC\vdash x=y$, then $x\sim_{p} y$.

(2) Let $x$ and $y$ be $BATC_G$ terms. If $BATC\vdash x=y$, then $x\sim_{s} y$.

(3) Let $x$ and $y$ be $BATC_G$ terms. If $BATC\vdash x=y$, then $x\sim_{hp} y$.

(4) Let $x$ and $y$ be $BATC_G$ terms. If $BATC\vdash x=y$, then $x\sim_{hhp} y$.
\end{theorem}

\begin{theorem}[Completeness of $BATC_G$ modulo truly concurrent bisimulation equivalences]
(1) Let $p$ and $q$ be closed $BATC_G$ terms, if $p\sim_{p} q$ then $p=q$.

(2) Let $p$ and $q$ be closed $BATC_G$ terms, if $p\sim_{s} q$ then $p=q$.

(3) Let $p$ and $q$ be closed $BATC_G$ terms, if $p\sim_{hp} q$ then $p=q$.

(4) Let $p$ and $q$ be closed $BATC_G$ terms, if $p\sim_{hhp} q$ then $p=q$.
\end{theorem}

\subsubsection{$APTC$ with Guards}

In this subsection, we will extend $APTC$ with guards, which is abbreviated $APTC_G$. The set of basic guards $G$ with element $\phi,\psi,\cdots$, which is extended by the following formation rules:

$$\phi::=\delta|\epsilon|\neg\phi|\psi\in G_{at}|\phi+\psi|\phi\cdot\psi|\phi\parallel\psi$$

The set of axioms of $APTC_G$ including axioms of $BATC_G$ in Table \ref{AxiomsForBATCG21} and the axioms are shown in Table \ref{AxiomsForAPTCG21}.

\begin{center}
    \begin{table}
        \begin{tabular}{@{}ll@{}}
            \hline No. &Axiom\\
            $P1$ & $x\between y = x\parallel y + x\mid y$\\
            $P2$ & $x\parallel y = y \parallel x$\\
            $P3$ & $(x\parallel y)\parallel z = x\parallel (y\parallel z)$\\
            $P4$ & $x\parallel y = x\leftmerge y + y\leftmerge x$\\
            $P5$ & $(e_1\leq e_2)\quad e_1\leftmerge (e_2\cdot y) = (e_1\leftmerge e_2)\cdot y$\\
            $P6$ & $(e_1\leq e_2)\quad (e_1\cdot x)\leftmerge e_2 = (e_1\leftmerge e_2)\cdot x$\\
            $P7$ & $(e_1\leq e_2)\quad (e_1\cdot x)\leftmerge (e_2\cdot y) = (e_1\leftmerge e_2)\cdot (x\between y)$\\
            $P8$ & $(x+ y)\leftmerge z = (x\leftmerge z)+ (y\leftmerge z)$\\
            $P9$ & $\delta\leftmerge x = \delta$\\
            $P10$ & $\epsilon\leftmerge x = x$\\
            $P11$ & $x\leftmerge \epsilon = x$\\
            $C1$ & $e_1\mid e_2 = \gamma(e_1,e_2)$\\
            $C2$ & $e_1\mid (e_2\cdot y) = \gamma(e_1,e_2)\cdot y$\\
            $C3$ & $(e_1\cdot x)\mid e_2 = \gamma(e_1,e_2)\cdot x$\\
            $C4$ & $(e_1\cdot x)\mid (e_2\cdot y) = \gamma(e_1,e_2)\cdot (x\between y)$\\
            $C5$ & $(x+ y)\mid z = (x\mid z) + (y\mid z)$\\
            $C6$ & $x\mid (y+ z) = (x\mid y)+ (x\mid z)$\\
            $C7$ & $\delta\mid x = \delta$\\
            $C8$ & $x\mid\delta = \delta$\\
            $C9$ & $\epsilon\mid x = \delta$\\
            $C10$ & $x\mid\epsilon = \delta$\\
            $CE1$ & $\Theta(e) = e$\\
            $CE2$ & $\Theta(\delta) = \delta$\\
            $CE3$ & $\Theta(\epsilon) = \epsilon$\\
            $CE4$ & $\Theta(x+ y) = \Theta(x)\triangleleft y + \Theta(y)\triangleleft x$\\
            $CE5$ & $\Theta(x\cdot y)=\Theta(x)\cdot\Theta(y)$\\
            $CE6$ & $\Theta(x\leftmerge y) = ((\Theta(x)\triangleleft y)\leftmerge y)+ ((\Theta(y)\triangleleft x)\leftmerge x)$\\
            $CE7$ & $\Theta(x\mid y) = ((\Theta(x)\triangleleft y)\mid y)+ ((\Theta(y)\triangleleft x)\mid x)$\\
        \end{tabular}
        \caption{Axioms of $APTC_G$}
        \label{AxiomsForAPTCG21}
    \end{table}
\end{center}

\begin{center}
    \begin{table}
        \begin{tabular}{@{}ll@{}}
            \hline No. &Axiom\\
            $U1$ & $(\sharp(e_1,e_2))\quad e_1\triangleleft e_2 = \tau$\\
            $U2$ & $(\sharp(e_1,e_2),e_2\leq e_3)\quad e_1\triangleleft e_3 = e_1$\\
            $U3$ & $(\sharp(e_1,e_2),e_2\leq e_3)\quad e3\triangleleft e_1 = \tau$\\
            $U4$ & $e\triangleleft \delta = e$\\
            $U5$ & $\delta \triangleleft e = \delta$\\
            $U6$ & $e\triangleleft \epsilon = e$\\
            $U7$ & $\epsilon \triangleleft e = e$\\
            $U8$ & $(x+ y)\triangleleft z = (x\triangleleft z)+ (y\triangleleft z)$\\
            $U9$ & $(x\cdot y)\triangleleft z = (x\triangleleft z)\cdot (y\triangleleft z)$\\
            $U10$ & $(x\leftmerge y)\triangleleft z = (x\triangleleft z)\leftmerge (y\triangleleft z)$\\
            $U11$ & $(x\mid y)\triangleleft z = (x\triangleleft z)\mid (y\triangleleft z)$\\
            $U12$ & $x\triangleleft (y+ z) = (x\triangleleft y)\triangleleft z$\\
            $U13$ & $x\triangleleft (y\cdot z)=(x\triangleleft y)\triangleleft z$\\
            $U14$ & $x\triangleleft (y\leftmerge z) = (x\triangleleft y)\triangleleft z$\\
            $U15$ & $x\triangleleft (y\mid z) = (x\triangleleft y)\triangleleft z$\\
            $D1$ & $e\notin H\quad\partial_H(e) = e$\\
            $D2$ & $e\in H\quad \partial_H(e) = \delta$\\
            $D3$ & $\partial_H(\delta) = \delta$\\
            $D4$ & $\partial_H(x+ y) = \partial_H(x)+\partial_H(y)$\\
            $D5$ & $\partial_H(x\cdot y) = \partial_H(x)\cdot\partial_H(y)$\\
            $D6$ & $\partial_H(x\leftmerge y) = \partial_H(x)\leftmerge\partial_H(y)$\\
            $G12$ & $\phi(x\leftmerge y) =\phi x\leftmerge \phi y$\\
            $G13$ & $\phi(x\mid y) =\phi x\mid \phi y$\\
            $G14$ & $\delta\leftmerge \phi = \delta$\\
            $G15$ & $\phi\mid \delta = \delta$\\
            $G16$ & $\delta\mid \phi = \delta$\\
            $G17$ & $\phi\leftmerge \epsilon = \phi$\\
            $G18$ & $\epsilon\leftmerge \phi = \phi$\\
            $G19$ & $\phi\mid \epsilon = \delta$\\
            $G20$ & $\epsilon\mid \phi = \delta$\\
            $G21$ & $\phi\leftmerge\neg\phi = \delta$\\
            $G22$ & $\Theta(\phi) = \phi$\\
            $G23$ & $\partial_H(\phi) = \phi$\\
            $G24$ & $\phi_0\leftmerge\cdots\leftmerge\phi_n = \delta$ if $\forall s_0,\cdots,s_n\in S,\exists i\leq n.test(\neg\phi_i,s_0\cup\cdots\cup s_n)$\\
        \end{tabular}
        \caption{Axioms of $APTC_G$(continuing)}
        \label{AxiomsForAPTCG212}
    \end{table}
\end{center}

\begin{definition}[Basic terms of $APTC_G$]
The set of basic terms of $APTC_G$, $\mathcal{B}(APTC_G)$, is inductively defined as follows:
\begin{enumerate}
    \item $\mathbb{E}\subset\mathcal{B}(APTC_G)$;
    \item $G\subset\mathcal{B}(APTC_G)$;
    \item if $e\in \mathbb{E}, t\in\mathcal{B}(APTC_G)$ then $e\cdot t\in\mathcal{B}(APTC_G)$;
    \item if $\phi\in G, t\in\mathcal{B}(APTC_G)$ then $\phi\cdot t\in\mathcal{B}(APTC_G)$;
    \item if $t,s\in\mathcal{B}(APTC_G)$ then $t+ s\in\mathcal{B}(APTC_G)$.
    \item if $t,s\in\mathcal{B}(APTC_G)$ then $t\leftmerge s\in\mathcal{B}(APTC_G)$.
\end{enumerate}
\end{definition}

Based on the definition of basic terms for $APTC_G$ and axioms of $APTC_G$, we can prove the elimination theorem of $APTC_G$.

\begin{theorem}[Elimination theorem of $APTC_G$]
Let $p$ be a closed $APTC_G$ term. Then there is a basic $APTC_G$ term $q$ such that $APTC_G\vdash p=q$.
\end{theorem}

We will define a term-deduction system which gives the operational semantics of $APTC_G$. Two atomic events $e_1$ and $e_2$ are in race condition, which are denoted $e_1\% e_2$.

\begin{center}
    \begin{table}
        $$\frac{}{\langle e_1\parallel\cdots \parallel e_n,s\rangle\xrightarrow{\{e_1,\cdots,e_n\}}\langle\surd,s'\rangle}\textrm{ if }s'\in effect(e_1,s)\cup\cdots\cup effect(e_n,s)$$

        $$\frac{}{\langle\phi_1\parallel\cdots\parallel \phi_n,s\rangle\rightarrow\langle\surd,s\rangle}\textrm{ if }test(\phi_1,s),\cdots,test(\phi_n,s)$$

        $$\frac{\langle x,s\rangle\xrightarrow{e_1}\langle\surd,s'\rangle\quad \langle y,s\rangle\xrightarrow{e_2}\langle\surd,s''\rangle}{\langle x\parallel y,s\rangle\xrightarrow{\{e_1,e_2\}}\langle\surd,s'\cup s''\rangle} \quad\frac{\langle x,s\rangle\xrightarrow{e_1}\langle x',s'\rangle\quad \langle y,s\rangle\xrightarrow{e_2}\langle\surd,s''\rangle}{\langle x\parallel y,s\rangle\xrightarrow{\{e_1,e_2\}}\langle x',s'\cup s''\rangle}$$

        $$\frac{\langle x,s\rangle\xrightarrow{e_1}\langle\surd,s'\rangle\quad \langle y,s\rangle\xrightarrow{e_2}\langle y',s''\rangle}{\langle x\parallel y,s\rangle\xrightarrow{\{e_1,e_2\}}\langle y',s'\cup s''\rangle} \quad\frac{\langle x,s\rangle\xrightarrow{e_1}\langle x',s'\rangle\quad \langle y,s\rangle\xrightarrow{e_2}\langle y',s''\rangle}{\langle x\parallel y,s\rangle\xrightarrow{\{e_1,e_2\}}\langle x'\between y',s'\cup s''\rangle}$$

        $$\frac{\langle x,s\rangle\xrightarrow{e_1}\langle\surd,s'\rangle\quad \langle y,s\rangle\xnrightarrow{e_2}\quad(e_1\%e_2)}{\langle x\parallel y,s\rangle\xrightarrow{e_1}\langle y,s'\rangle} \quad\frac{\langle x,s\rangle\xrightarrow{e_1}\langle x',s'\rangle\quad \langle y,s\rangle\xnrightarrow{e_2}\quad(e_1\%e_2)}{\langle x\parallel y,s\rangle\xrightarrow{e_1}\langle x'\between y,s'\rangle}$$

        $$\frac{\langle x,s\rangle\xnrightarrow{e_1}\quad \langle y,s\rangle\xrightarrow{e_2}\langle\surd,s''\rangle\quad(e_1\%e_2)}{\langle x\parallel y,s\rangle\xrightarrow{e_2}\langle x,s''\rangle} \quad\frac{\langle x,s\rangle\xnrightarrow{e_1}\quad \langle y,s\rangle\xrightarrow{e_2}\langle y',s''\rangle\quad(e_1\%e_2)}{\langle x\parallel y,s\rangle\xrightarrow{e_2}\langle x\between y',s''\rangle}$$

        $$\frac{\langle x,s\rangle\xrightarrow{e_1}\langle\surd,s'\rangle\quad \langle y,s\rangle\xrightarrow{e_2}\langle\surd,s''\rangle \quad(e_1\leq e_2)}{\langle x\leftmerge y,s\rangle\xrightarrow{\{e_1,e_2\}}\langle \surd,s'\cup s''\rangle} \quad\frac{\langle x,s\rangle\xrightarrow{e_1}\langle x',s'\rangle\quad \langle y,s\rangle\xrightarrow{e_2}\langle\surd,s''\rangle \quad(e_1\leq e_2)}{\langle x\leftmerge y,s\rangle\xrightarrow{\{e_1,e_2\}}\langle x',s'\cup s''\rangle}$$

        $$\frac{\langle x,s\rangle\xrightarrow{e_1}\langle\surd,s'\rangle\quad \langle y,s\rangle\xrightarrow{e_2}\langle y',s''\rangle \quad(e_1\leq e_2)}{\langle x\leftmerge y,s\rangle\xrightarrow{\{e_1,e_2\}}\langle y',s'\cup s''\rangle} \quad\frac{\langle x,s\rangle\xrightarrow{e_1}\langle x',s'\rangle\quad \langle y,s\rangle\xrightarrow{e_2}\langle y',s''\rangle \quad(e_1\leq e_2)}{\langle x\leftmerge y,s\rangle\xrightarrow{\{e_1,e_2\}}\langle x'\between y',s'\cup s''\rangle}$$

        $$\frac{\langle x,s\rangle\xrightarrow{e_1}\langle\surd,s'\rangle\quad \langle y,s\rangle\xrightarrow{e_2}\langle\surd,s''\rangle}{\langle x\mid y,s\rangle\xrightarrow{\gamma(e_1,e_2)}\langle\surd,effect(\gamma(e_1,e_2),s)\rangle} \quad\frac{\langle x,s\rangle\xrightarrow{e_1}\langle x',s'\rangle\quad \langle y,s\rangle\xrightarrow{e_2}\langle\surd,s''\rangle}{\langle x\mid y,s\rangle\xrightarrow{\gamma(e_1,e_2)}\langle x',effect(\gamma(e_1,e_2),s)\rangle}$$

        $$\frac{\langle x,s\rangle\xrightarrow{e_1}\langle\surd,s'\rangle\quad \langle y,s\rangle\xrightarrow{e_2}\langle y',s''\rangle}{\langle x\mid y,s\rangle\xrightarrow{\gamma(e_1,e_2)}\langle y',effect(\gamma(e_1,e_2),s)\rangle} \quad\frac{\langle x,s\rangle\xrightarrow{e_1}\langle x',s'\rangle\quad \langle y,s\rangle\xrightarrow{e_2}\langle y',s''\rangle}{\langle x\mid y,s\rangle\xrightarrow{\gamma(e_1,e_2)}\langle x'\between y',effect(\gamma(e_1,e_2),s)\rangle}$$

        \caption{Transition rules of $APTC_G$}
        \label{TRForAPTCG21}
    \end{table}
\end{center}

\begin{center}
    \begin{table}

        $$\frac{\langle x,s\rangle\xrightarrow{e_1}\langle\surd,s'\rangle\quad (\sharp(e_1,e_2))}{\langle \Theta(x),s\rangle\xrightarrow{e_1}\langle\surd,s'\rangle} \quad\frac{\langle x,s\rangle\xrightarrow{e_2}\langle\surd,s''\rangle\quad (\sharp(e_1,e_2))}{\langle\Theta(x),s\rangle\xrightarrow{e_2}\langle\surd,s''\rangle}$$

        $$\frac{\langle x,s\rangle\xrightarrow{e_1}\langle x',s'\rangle\quad (\sharp(e_1,e_2))}{\langle\Theta(x),s\rangle\xrightarrow{e_1}\langle\Theta(x'),s'\rangle} \quad\frac{\langle x,s\rangle\xrightarrow{e_2}\langle x'',s''\rangle\quad (\sharp(e_1,e_2))}{\langle\Theta(x),s\rangle\xrightarrow{e_2}\langle\Theta(x''),s''\rangle}$$

        $$\frac{\langle x,s\rangle\xrightarrow{e_1}\langle\surd,s'\rangle \quad \langle y,s\rangle\nrightarrow^{e_2}\quad (\sharp(e_1,e_2))}{\langle x\triangleleft y,s\rangle\xrightarrow{\tau}\langle\surd,s'\rangle}
        \quad\frac{\langle x,s\rangle\xrightarrow{e_1}\langle x',s'\rangle \quad \langle y,s\rangle\nrightarrow^{e_2}\quad (\sharp(e_1,e_2))}{\langle x\triangleleft y,s\rangle\xrightarrow{\tau}\langle x',s'\rangle}$$

        $$\frac{\langle x,s\rangle\xrightarrow{e_1}\langle\surd,s\rangle \quad \langle y,s\rangle\nrightarrow^{e_3}\quad (\sharp(e_1,e_2),e_2\leq e_3)}{\langle x\triangleleft y,s\rangle\xrightarrow{e_1}\langle\surd,s'\rangle}
        \quad\frac{\langle x,s\rangle\xrightarrow{e_1}\langle x',s'\rangle \quad \langle y,s\rangle\nrightarrow^{e_3}\quad (\sharp(e_1,e_2),e_2\leq e_3)}{\langle x\triangleleft y,s\rangle\xrightarrow{e_1}\langle x',s'\rangle}$$

        $$\frac{\langle x,s\rangle\xrightarrow{e_3}\langle\surd,s'\rangle \quad \langle y,s\rangle\nrightarrow^{e_2}\quad (\sharp(e_1,e_2),e_1\leq e_3)}{\langle x\triangleleft y,s\rangle\xrightarrow{\tau}\langle\surd,s'\rangle}
        \quad\frac{\langle x,s\rangle\xrightarrow{e_3}\langle x',s'\rangle \quad \langle y,s\rangle\nrightarrow^{e_2}\quad (\sharp(e_1,e_2),e_1\leq e_3)}{\langle x\triangleleft y,s\rangle\xrightarrow{\tau}\langle x',s'\rangle}$$

        $$\frac{\langle x,s\rangle\xrightarrow{e}\langle\surd,s'\rangle}{\langle\partial_H(x),s\rangle\xrightarrow{e}\langle\surd,s'\rangle}\quad (e\notin H)\quad\frac{\langle x,s\rangle\xrightarrow{e}\langle x',s'\rangle}{\langle\partial_H(x),s\rangle\xrightarrow{e}\langle\partial_H(x'),s'\rangle}\quad(e\notin H)$$
        \caption{Transition rules of $APTC_G$ (continuing)}
        \label{TRForAPTCG212}
    \end{table}
\end{center}

\begin{theorem}[Generalization of $APTC_G$ with respect to $BATC_G$]
$APTC_G$ is a generalization of $BATC_G$.
\end{theorem}

\begin{theorem}[Congruence of $APTC_G$ with respect to truly concurrent bisimulation equivalences]
(1) Pomset bisimulation equivalence $\sim_{p}$ is a congruence with respect to $APTC_G$.

(2) Step bisimulation equivalence $\sim_{s}$ is a congruence with respect to $APTC_G$.

(3) Hp-bisimulation equivalence $\sim_{hp}$ is a congruence with respect to $APTC_G$.

(4) Hhp-bisimulation equivalence $\sim_{hhp}$ is a congruence with respect to $APTC_G$.
\end{theorem}

\begin{theorem}[Soundness of $APTC_G$ modulo truly concurrent bisimulation equivalences]
(1) Let $x$ and $y$ be $APTC_G$ terms. If $APTC\vdash x=y$, then $x\sim_{p} y$.

(2) Let $x$ and $y$ be $APTC_G$ terms. If $APTC\vdash x=y$, then $x\sim_{s} y$.

(3) Let $x$ and $y$ be $APTC_G$ terms. If $APTC\vdash x=y$, then $x\sim_{hp} y$.

(4) Let $x$ and $y$ be $APTC_G$ terms. If $APTC\vdash x=y$, then $x\sim_{hhp} y$.
\end{theorem}

\begin{theorem}[Completeness of $APTC_G$ modulo truly concurrent bisimulation equivalences]
(1) Let $p$ and $q$ be closed $APTC_G$ terms, if $p\sim_{p} q$ then $p=q$.

(2) Let $p$ and $q$ be closed $APTC_G$ terms, if $p\sim_{s} q$ then $p=q$.

(3) Let $p$ and $q$ be closed $APTC_G$ terms, if $p\sim_{hp} q$ then $p=q$.

(4) Let $p$ and $q$ be closed $APTC_G$ terms, if $p\sim_{hhp} q$ then $p=q$.
\end{theorem}

\subsubsection{Recursion}

In this subsection, we introduce recursion to capture infinite processes based on $APTC_G$. In the following, $E,F,G$ are recursion specifications, $X,Y,Z$ are recursive variables.

\begin{definition}[Guarded recursive specification]
A recursive specification

$$X_1=t_1(X_1,\cdots,X_n)$$
$$...$$
$$X_n=t_n(X_1,\cdots,X_n)$$

is guarded if the right-hand sides of its recursive equations can be adapted to the form by applications of the axioms in $APTC$ and replacing recursion variables by the right-hand sides of their recursive equations,

$$(a_{11}\leftmerge\cdots\leftmerge a_{1i_1})\cdot s_1(X_1,\cdots,X_n)+\cdots+(a_{k1}\leftmerge\cdots\leftmerge a_{ki_k})\cdot s_k(X_1,\cdots,X_n)+(b_{11}\leftmerge\cdots\leftmerge b_{1j_1})+\cdots+(b_{1j_1}\leftmerge\cdots\leftmerge b_{lj_l})$$

where $a_{11},\cdots,a_{1i_1},a_{k1},\cdots,a_{ki_k},b_{11},\cdots,b_{1j_1},b_{1j_1},\cdots,b_{lj_l}\in \mathbb{E}$, and the sum above is allowed to be empty, in which case it represents the deadlock $\delta$. And there does not exist an infinite sequence of $\epsilon$-transitions $\langle X|E\rangle\rightarrow\langle X'|E\rangle\rightarrow\langle X''|E\rangle\rightarrow\cdots$.
\end{definition}

\begin{center}
    \begin{table}
        $$\frac{\langle t_i(\langle X_1|E\rangle,\cdots,\langle X_n|E\rangle),s\rangle\xrightarrow{\{e_1,\cdots,e_k\}}\langle\surd,s'\rangle}{\langle\langle X_i|E\rangle,s\rangle\xrightarrow{\{e_1,\cdots,e_k\}}\langle\surd,s'\rangle}$$
        $$\frac{\langle t_i(\langle X_1|E\rangle,\cdots,\langle X_n|E\rangle),s\rangle\xrightarrow{\{e_1,\cdots,e_k\}} \langle y,s'\rangle}{\langle\langle X_i|E\rangle,s\rangle\xrightarrow{\{e_1,\cdots,e_k\}} \langle y,s'\rangle}$$
        \caption{Transition rules of guarded recursion}
        \label{TRForGRG21}
    \end{table}
\end{center}

The $RDP$ (Recursive Definition Principle) and the $RSP$ (Recursive Specification Principle) are shown in Table \ref{RDPRSP21}.

\begin{center}
\begin{table}
  \begin{tabular}{@{}ll@{}}
\hline No. &Axiom\\
  $RDP$ & $\langle X_i|E\rangle = t_i(\langle X_1|E,\cdots,X_n|E\rangle)\quad (i\in\{1,\cdots,n\})$\\
  $RSP$ & if $y_i=t_i(y_1,\cdots,y_n)$ for $i\in\{1,\cdots,n\}$, then $y_i=\langle X_i|E\rangle \quad(i\in\{1,\cdots,n\})$\\
\end{tabular}
\caption{Recursive definition and specification principle}
\label{RDPRSP21}
\end{table}
\end{center}

\begin{theorem}[Conservitivity of $APTC_G$ with guarded recursion]
$APTC_G$ with guarded recursion is a conservative extension of $APTC_G$.
\end{theorem}

\begin{theorem}[Congruence theorem of $APTC_G$ with guarded recursion]
Truly concurrent bisimulation equivalences $\sim_{p}$, $\sim_s$ and $\sim_{hp}$ are all congruences with respect to $APTC_G$ with guarded recursion.
\end{theorem}

\begin{theorem}[Elimination theorem of $APTC_G$ with linear recursion]
Each process term in $APTC_G$ with linear recursion is equal to a process term $\langle X_1|E\rangle$ with $E$ a linear recursive specification.
\end{theorem}

\begin{theorem}[Soundness of $APTC_G$ with guarded recursion]
Let $x$ and $y$ be $APTC_G$ with guarded recursion terms. If $APTC_G\textrm{ with guarded recursion}\vdash x=y$, then

(1) $x\sim_{s} y$.

(2) $x\sim_{p} y$.

(3) $x\sim_{hp} y$.

(4) $x\sim_{hhp} y$.
\end{theorem}

\begin{theorem}[Completeness of $APTC_G$ with linear recursion]
Let $p$ and $q$ be closed $APTC_G$ with linear recursion terms, then,

(1) if $p\sim_{s} q$ then $p=q$.

(2) if $p\sim_{p} q$ then $p=q$.

(3) if $p\sim_{hp} q$ then $p=q$.

(4) if $p\sim_{hhp} q$ then $p=q$.
\end{theorem}

\subsubsection{Abstraction}

To abstract away from the internal implementations of a program, and verify that the program exhibits the desired external behaviors, the silent step $\tau$ and abstraction operator
 $\tau_I$ are introduced, where $I\subseteq \mathbb{E}\cup G_{at}$ denotes the internal events or guards. The silent step $\tau$ represents the internal events or guards, when we
  consider the external behaviors of a process, $\tau$ steps can be removed, that is, $\tau$ steps must keep silent. The transition rule of $\tau$ is shown in Table \ref{TRForTauG21}. In the following, let the atomic event $e$ range over $\mathbb{E}\cup\{\epsilon\}\cup\{\delta\}\cup\{\tau\}$, and $\phi$ range over $G\cup \{\tau\}$, and let the communication function $\gamma:\mathbb{E}\cup\{\tau\}\times \mathbb{E}\cup\{\tau\}\rightarrow \mathbb{E}\cup\{\delta\}$, with each communication involved $\tau$ resulting in $\delta$. We use $\tau(s)$ to denote $effect(\tau,s)$, for the fact that $\tau$ only change the state of internal data environment, that is, for the external data environments, $s=\tau(s)$.

\begin{center}
    \begin{table}
        $$\frac{}{\langle\tau,s\rangle\rightarrow\langle\surd,s\rangle}\textrm{ if }test(\tau,s)$$
        $$\frac{}{\langle\tau,s\rangle\xrightarrow{\tau}\langle\surd,\tau(s)\rangle}$$
        \caption{Transition rule of the silent step}
        \label{TRForTauG21}
    \end{table}
\end{center}

\begin{definition}[Guarded linear recursive specification]\label{GLRSG}
A linear recursive specification $E$ is guarded if there does not exist an infinite sequence of $\tau$-transitions $\langle X|E\rangle\xrightarrow{\tau}\langle X'|E\rangle\xrightarrow{\tau}\langle X''|E\rangle\xrightarrow{\tau}\cdots$, and there does not exist an infinite sequence of $\epsilon$-transitions $\langle X|E\rangle\rightarrow\langle X'|E\rangle\rightarrow\langle X''|E\rangle\rightarrow\cdots$.
\end{definition}

\begin{theorem}[Conservitivity of $APTC_G$ with silent step and guarded linear recursion]
$APTC_G$ with silent step and guarded linear recursion is a conservative extension of $APTC_G$ with linear recursion.
\end{theorem}

\begin{theorem}[Congruence theorem of $APTC_G$ with silent step and guarded linear recursion]
Rooted branching truly concurrent bisimulation equivalences $\approx_{rbp}$, $\approx_{rbs}$ and $\approx_{rbhp}$ are all congruences with respect to $APTC_G$ with silent step and guarded linear recursion.
\end{theorem}

We design the axioms for the silent step $\tau$ in Table \ref{AxiomsForTauG21}.

\begin{center}
\begin{table}
  \begin{tabular}{@{}ll@{}}
\hline No. &Axiom\\
  $B1$ & $e\cdot\tau=e$\\
  $B2$ & $e\cdot(\tau\cdot(x+y)+x)=e\cdot(x+y)$\\
  $B3$ & $x\leftmerge\tau=x$\\
  $G26$ & $\phi\cdot\tau=\phi$\\
  $G27$ & $\phi\cdot(\tau\cdot(x+y)+x)=\phi\cdot(x+y)$\\
\end{tabular}
\caption{Axioms of silent step}
\label{AxiomsForTauG21}
\end{table}
\end{center}

\begin{theorem}[Elimination theorem of $APTC_G$ with silent step and guarded linear recursion]
Each process term in $APTC_G$ with silent step and guarded linear recursion is equal to a process term $\langle X_1|E\rangle$ with $E$ a guarded linear recursive specification.
\end{theorem}

\begin{theorem}[Soundness of $APTC_G$ with silent step and guarded linear recursion]
Let $x$ and $y$ be $APTC_G$ with silent step and guarded linear recursion terms. If $APTC_G$ with silent step and guarded linear recursion $\vdash x=y$, then

(1) $x\approx_{rbs} y$.

(2) $x\approx_{rbp} y$.

(3) $x\approx_{rbhp} y$.

(4) $x\approx_{rbhhp} y$.
\end{theorem}

\begin{theorem}[Completeness of $APTC_G$ with silent step and guarded linear recursion]
Let $p$ and $q$ be closed $APTC_G$ with silent step and guarded linear recursion terms, then,

(1) if $p\approx_{rbs} q$ then $p=q$.

(2) if $p\approx_{rbp} q$ then $p=q$.

(3) if $p\approx_{rbhp} q$ then $p=q$.

(4) if $p\approx_{rbhhp} q$ then $p=q$.
\end{theorem}

The unary abstraction operator $\tau_I$ ($I\subseteq \mathbb{E}\cup G_{at}$) renames all atomic events or atomic guards in $I$ into $\tau$. $APTC_G$ with silent step and abstraction 
operator is called $APTC_{G_{\tau}}$. The transition rules of operator $\tau_I$ are shown in Table \ref{TRForAbstractionG21}.

\begin{center}
    \begin{table}
        $$\frac{\langle x,s\rangle\xrightarrow{e}\langle\surd,s'\rangle}{\langle\tau_I(x),s\rangle\xrightarrow{e}\langle\surd,s'\rangle}\quad e\notin I
        \quad\quad\frac{\langle x,s\rangle\xrightarrow{e}\langle x',s'\rangle}{\langle\tau_I(x),s\rangle\xrightarrow{e}\langle\tau_I(x'),s'\rangle}\quad e\notin I$$

        $$\frac{\langle x,s\rangle\xrightarrow{e}\langle\surd,s'\rangle}{\langle\tau_I(x),s\rangle\xrightarrow{\tau}\langle\surd,\tau(s)\rangle}\quad e\in I
        \quad\quad\frac{\langle x,s\rangle\xrightarrow{e}\langle x',s'\rangle}{\langle\tau_I(x),s\rangle\xrightarrow{\tau}\langle\tau_I(x'),\tau(s)\rangle}\quad e\in I$$
        \caption{Transition rule of the abstraction operator}
        \label{TRForAbstractionG21}
    \end{table}
\end{center}

\begin{theorem}[Conservitivity of $APTC_{G_{\tau}}$ with guarded linear recursion]
$APTC_{G_{\tau}}$ with guarded linear recursion is a conservative extension of $APTC_G$ with silent step and guarded linear recursion.
\end{theorem}

\begin{theorem}[Congruence theorem of $APTC_{G_{\tau}}$ with guarded linear recursion]
Rooted branching truly concurrent bisimulation equivalences $\approx_{rbp}$, $\approx_{rbs}$ and $\approx_{rbhp}$ are all congruences with respect to $APTC_{G_{\tau}}$ with guarded linear recursion.
\end{theorem}

We design the axioms for the abstraction operator $\tau_I$ in Table \ref{AxiomsForAbstractionG21}.

\begin{center}
\begin{table}
  \begin{tabular}{@{}ll@{}}
\hline No. &Axiom\\
  $TI1$ & $e\notin I\quad \tau_I(e)=e$\\
  $TI2$ & $e\in I\quad \tau_I(e)=\tau$\\
  $TI3$ & $\tau_I(\delta)=\delta$\\
  $TI4$ & $\tau_I(x+y)=\tau_I(x)+\tau_I(y)$\\
  $TI5$ & $\tau_I(x\cdot y)=\tau_I(x)\cdot\tau_I(y)$\\
  $TI6$ & $\tau_I(x\leftmerge y)=\tau_I(x)\leftmerge\tau_I(y)$\\
  $G28$ & $\phi\notin I\quad \tau_I(\phi)=\phi$\\
  $G29$ & $\phi\in I\quad \tau_I(\phi)=\tau$\\
\end{tabular}
\caption{Axioms of abstraction operator}
\label{AxiomsForAbstractionG21}
\end{table}
\end{center}

\begin{theorem}[Soundness of $APTC_{G_{\tau}}$ with guarded linear recursion]
Let $x$ and $y$ be $APTC_{G_{\tau}}$ with guarded linear recursion terms. If $APTC_{G_{\tau}}$ with guarded linear recursion $\vdash x=y$, then

(1) $x\approx_{rbs} y$.

(2) $x\approx_{rbp} y$.

(3) $x\approx_{rbhp} y$.

(4) $x\approx_{rbhhp} y$.
\end{theorem}

\begin{definition}[Cluster]
Let $E$ be a guarded linear recursive specification, and $I\subseteq \mathbb{E}$. Two recursion variable $X$ and $Y$ in $E$ are in the same cluster for $I$ iff there exist sequences of
transitions $\langle X|E\rangle\xrightarrow{\{b_{11},\cdots, b_{1i}\}}\cdots\xrightarrow{\{b_{m1},\cdots, b_{mi}\}}\langle Y|E\rangle$ and $\langle Y|E\rangle\xrightarrow{\{c_{11},\cdots, c_{1j}\}}\cdots\xrightarrow{\{c_{n1},\cdots, c_{nj}\}}\langle X|E\rangle$, where $b_{11},\cdots,b_{mi},c_{11},\cdots,c_{nj}\in I\cup\{\tau\}$.

$a_1\leftmerge\cdots\leftmerge a_k$ or $(a_1\leftmerge\cdots\leftmerge a_k) X$ is an exit for the cluster $C$ iff: (1) $a_1\leftmerge\cdots\leftmerge a_k$ or
$(a_1\leftmerge\cdots\leftmerge a_k) X$ is a summand at the right-hand side of the recursive equation for a recursion variable in $C$, and (2) in the case of
$(a_1\leftmerge\cdots\leftmerge a_k) X$, either $a_l\notin I\cup\{\tau\}(l\in\{1,2,\cdots,k\})$ or $X\notin C$.
\end{definition}

\begin{center}
\begin{table}
  \begin{tabular}{@{}ll@{}}
\hline No. &Axiom\\
  $CFAR$ & If $X$ is in a cluster for $I$ with exits \\
           & $\{(a_{11}\leftmerge\cdots\leftmerge a_{1i})Y_1,\cdots,(a_{m1}\leftmerge\cdots\leftmerge a_{mi})Y_m, b_{11}\leftmerge\cdots\leftmerge b_{1j},\cdots,b_{n1}\leftmerge\cdots\leftmerge b_{nj}\}$, \\
           & then $\tau\cdot\tau_I(\langle X|E\rangle)=$\\
           & $\tau\cdot\tau_I((a_{11}\leftmerge\cdots\leftmerge a_{1i})\langle Y_1|E\rangle+\cdots+(a_{m1}\leftmerge\cdots\leftmerge a_{mi})\langle Y_m|E\rangle+b_{11}\leftmerge\cdots\leftmerge b_{1j}+\cdots+b_{n1}\leftmerge\cdots\leftmerge b_{nj})$\\
\end{tabular}
\caption{Cluster fair abstraction rule}
\label{CFARLeft21}
\end{table}
\end{center}

\begin{theorem}[Soundness of $CFAR$]
$CFAR$ is sound modulo rooted branching truly concurrent bisimulation equivalences $\approx_{rbs}$, $\approx_{rbp}$, $\approx_{rbhp}$ and $\approx_{rbhhp}$.
\end{theorem}

\begin{theorem}[Completeness of $APTC_{G_{\tau}}$ with guarded linear recursion and $CFAR$]
Let $p$ and $q$ be closed $APTC_{G_{\tau}}$ with guarded linear recursion and $CFAR$ terms, then,

(1) if $p\approx_{rbs} q$ then $p=q$.

(2) if $p\approx_{rbp} q$ then $p=q$.

(3) if $p\approx_{rbhp} q$ then $p=q$.

(4) if $p\approx_{rbhhp} q$ then $p=q$.
\end{theorem}

\subsection{APPTC with Guards -- $APPTC_G$}

\begin{definition}[Prime event structure with silent event and empty event]
Let $\Lambda$ be a fixed set of labels, ranged over $a,b,c,\cdots$ and $\tau,\epsilon$. A ($\Lambda$-labelled) prime event structure with silent event $\tau$ and empty event $\epsilon$
is a tuple $\mathcal{E}=\langle \mathbb{E}, \leq, \sharp, \sharp_{\pi} \lambda\rangle$, where $\mathbb{E}$ is a denumerable set of events, including the silent event $\tau$ and
empty event $\epsilon$. Let $\hat{\mathbb{E}}=\mathbb{E}\backslash\{\tau,\epsilon\}$, exactly excluding $\tau$ and $\epsilon$, it is obvious that $\hat{\tau^*}=\epsilon$. Let
$\lambda:\mathbb{E}\rightarrow\Lambda$ be a labelling function and let $\lambda(\tau)=\tau$ and $\lambda(\epsilon)=\epsilon$. And $\leq$, $\sharp$, $\sharp_{\pi}$ are binary relations
on $\mathbb{E}$, called causality, conflict and probabilistic conflict respectively, such that:

\begin{enumerate}
  \item $\leq$ is a partial order and $\lceil e \rceil = \{e'\in \mathbb{E}|e'\leq e\}$ is finite for all $e\in \mathbb{E}$. It is easy to see that
  $e\leq\tau^*\leq e'=e\leq\tau\leq\cdots\leq\tau\leq e'$, then $e\leq e'$.
  \item $\sharp$ is irreflexive, symmetric and hereditary with respect to $\leq$, that is, for all $e,e',e''\in \mathbb{E}$, if $e\sharp e'\leq e''$, then $e\sharp e''$;
  \item $\sharp_{\pi}$ is irreflexive, symmetric and hereditary with respect to $\leq$, that is, for all $e,e',e''\in \mathbb{E}$, if $e\sharp_{\pi} e'\leq e''$, then $e\sharp_{\pi} e''$.
\end{enumerate}

Then, the concepts of consistency and concurrency can be drawn from the above definition:

\begin{enumerate}
  \item $e,e'\in \mathbb{E}$ are consistent, denoted as $e\frown e'$, if $\neg(e\sharp e')$ and $\neg(e\sharp_{\pi} e')$. A subset $X\subseteq \mathbb{E}$ is called consistent, if
  $e\frown e'$ for all $e,e'\in X$.
  \item $e,e'\in \mathbb{E}$ are concurrent, denoted as $e\parallel e'$, if $\neg(e\leq e')$, $\neg(e'\leq e)$, and $\neg(e\sharp e')$ and $\neg(e\sharp_{\pi} e')$.
\end{enumerate}
\end{definition}

\begin{definition}[Configuration]
Let $\mathcal{E}$ be a PES. A (finite) configuration in $\mathcal{E}$ is a (finite) consistent subset of events $C\subseteq \mathcal{E}$, closed with respect to causality
(i.e. $\lceil C\rceil=C$), and a data state $s\in S$ with $S$ the set of all data states, denoted $\langle C, s\rangle$. The set of finite configurations of $\mathcal{E}$ is denoted by
$\langle\mathcal{C}(\mathcal{E}), S\rangle$. We let $\hat{C}=C\backslash\{\tau\}\cup\{\epsilon\}$.
\end{definition}

A consistent subset of $X\subseteq \mathbb{E}$ of events can be seen as a pomset. Given $X, Y\subseteq \mathbb{E}$, $\hat{X}\sim \hat{Y}$ if $\hat{X}$ and $\hat{Y}$ are isomorphic as
pomsets. In the following of the paper, we say $C_1\sim C_2$, we mean $\hat{C_1}\sim\hat{C_2}$.

\begin{definition}[Pomset transitions and step]
Let $\mathcal{E}$ be a PES and let $C\in\mathcal{C}(\mathcal{E})$, and $\emptyset\neq X\subseteq \mathbb{E}$, if $C\cap X=\emptyset$ and $C'=C\cup X\in\mathcal{C}(\mathcal{E})$, then
$\langle C,s\rangle\xrightarrow{X} \langle C',s'\rangle$ is called a pomset transition from $\langle C,s\rangle$ to $\langle C',s'\rangle$. When the events in $X$ are pairwise
concurrent, we say that $\langle C,s\rangle\xrightarrow{X}\langle C',s'\rangle$ is a step. It is obvious that $\rightarrow^*\xrightarrow{X}\rightarrow^*=\xrightarrow{X}$ and
$\rightarrow^*\xrightarrow{e}\rightarrow^*=\xrightarrow{e}$ for any $e\in\mathbb{E}$ and $X\subseteq\mathbb{E}$.
\end{definition}

\begin{definition}[Probabilistic transitions]
Let $\mathcal{E}$ be a PES and let $C\in\mathcal{C}(\mathcal{E})$, the transition $\langle C,s\rangle\xrsquigarrow{\pi} \langle C^{\pi},s\rangle$ is called a probabilistic transition
from $\langle C,s\rangle$ to $\langle C^{\pi},s\rangle$.
\end{definition}

\begin{definition}[Weak pomset transitions and weak step]
Let $\mathcal{E}$ be a PES and let $C\in\mathcal{C}(\mathcal{E})$, and $\emptyset\neq X\subseteq \hat{\mathbb{E}}$, if $C\cap X=\emptyset$ and
$\hat{C'}=\hat{C}\cup X\in\mathcal{C}(\mathcal{E})$, then $\langle C,s\rangle\xRightarrow{X} \langle C',s'\rangle$ is called a weak pomset transition from $\langle C,s\rangle$ to
$\langle C',s'\rangle$, where we define $\xRightarrow{e}\triangleq\xrightarrow{\tau^*}\xrightarrow{e}\xrightarrow{\tau^*}$. And
$\xRightarrow{X}\triangleq\xrightarrow{\tau^*}\xrightarrow{e}\xrightarrow{\tau^*}$, for every $e\in X$. When the events in $X$ are pairwise concurrent, we say that
$\langle C,s\rangle\xRightarrow{X}\langle C',s'\rangle$ is a weak step.
\end{definition}

We will also suppose that all the PESs in this chapter are image finite, that is, for any PES $\mathcal{E}$ and $C\in \mathcal{C}(\mathcal{E})$ and $a\in \Lambda$,
$\{\langle C,s\rangle\xrsquigarrow{\pi} \langle C^{\pi},s\rangle\}$,
$\{e\in \mathbb{E}|\langle C,s\rangle\xrightarrow{e} \langle C',s'\rangle\wedge \lambda(e)=a\}$ and
$\{e\in\hat{\mathbb{E}}|\langle C,s\rangle\xRightarrow{e} \langle C',s'\rangle\wedge \lambda(e)=a\}$ is finite.

\begin{definition}[Probabilistic pomset, step bisimulation]
Let $\mathcal{E}_1$, $\mathcal{E}_2$ be PESs. A probabilistic pomset bisimulation is a relation $R\subseteq\langle\mathcal{C}(\mathcal{E}_1),S\rangle\times\langle\mathcal{C}(\mathcal{E}_2),S\rangle$,
such that (1) if $(\langle C_1,s\rangle,\langle C_2,s\rangle)\in R$, and $\langle C_1,s\rangle\xrightarrow{X_1}\langle C_1',s'\rangle$ then
$\langle C_2,s\rangle\xrightarrow{X_2}\langle C_2',s'\rangle$, with $X_1\subseteq \mathbb{E}_1$, $X_2\subseteq \mathbb{E}_2$, $X_1\sim X_2$ and
$(\langle C_1',s'\rangle,\langle C_2',s'\rangle)\in R$ for all $s,s'\in S$, and vice-versa; (2) if $(\langle C_1,s\rangle,\langle C_2,s\rangle)\in R$, and $\langle C_1,s\rangle\xrsquigarrow{\pi}\langle C_1^{\pi},s\rangle$
then $\langle C_2,s\rangle\xrsquigarrow{\pi}\langle C_2^{\pi},s\rangle$ and $(\langle C_1^{\pi},s\rangle,\langle C_2^{\pi},s\rangle)\in R$, and vice-versa; (3) if $(\langle C_1,s\rangle,\langle C_2,s\rangle)\in R$,
then $\mu(C_1,C)=\mu(C_2,C)$ for each $C\in\mathcal{C}(\mathcal{E})/R$; (4) $[\surd]_R=\{\surd\}$. We say that $\mathcal{E}_1$, $\mathcal{E}_2$ are probabilistic pomset bisimilar, written
$\mathcal{E}_1\sim_{pp}\mathcal{E}_2$, if there exists a probabilistic pomset bisimulation $R$, such that $(\langle\emptyset,\emptyset\rangle,\langle\emptyset,\emptyset\rangle)\in R$.
By replacing probabilistic pomset transitions with probabilistic steps, we can get the definition of probabilistic step bisimulation. When PESs $\mathcal{E}_1$ and $\mathcal{E}_2$ are
probabilistic step bisimilar, we write $\mathcal{E}_1\sim_{ps}\mathcal{E}_2$.
\end{definition}

\begin{definition}[Weakly probabilistic pomset, step bisimulation]
Let $\mathcal{E}_1$, $\mathcal{E}_2$ be PESs. A weakly probabilistic pomset bisimulation is a relation $R\subseteq\langle\mathcal{C}(\mathcal{E}_1),S\rangle\times\langle\mathcal{C}(\mathcal{E}_2),S\rangle$,
such that (1) if $(\langle C_1,s\rangle,\langle C_2,s\rangle)\in R$, and $\langle C_1,s\rangle\xRightarrow{X_1}\langle C_1',s'\rangle$ then
$\langle C_2,s\rangle\xRightarrow{X_2}\langle C_2',s'\rangle$, with $X_1\subseteq \hat{\mathbb{E}_1}$, $X_2\subseteq \hat{\mathbb{E}_2}$, $X_1\sim X_2$ and
$(\langle C_1',s'\rangle,\langle C_2',s'\rangle)\in R$ for all $s,s'\in S$, and vice-versa; (2) if $(\langle C_1,s\rangle,\langle C_2,s\rangle)\in R$, and $\langle C_1,s\rangle\xrsquigarrow{\pi}\langle C_1^{\pi},s\rangle$
then $\langle C_2,s\rangle\xrsquigarrow{\pi}\langle C_2^{\pi},s\rangle$ and $(\langle C_1^{\pi},s\rangle,\langle C_2^{\pi},s\rangle)\in R$, and vice-versa; (3) if $(\langle C_1,s\rangle,\langle C_2,s\rangle)\in R$,
then $\mu(C_1,C)=\mu(C_2,C)$ for each $C\in\mathcal{C}(\mathcal{E})/R$; (4) $[\surd]_R=\{\surd\}$. We say that $\mathcal{E}_1$, $\mathcal{E}_2$ are weakly probabilistic pomset bisimilar,
written $\mathcal{E}_1\approx_{pp}\mathcal{E}_2$, if there exists a weakly probabilistic pomset bisimulation $R$, such that
$(\langle\emptyset,\emptyset\rangle,\langle\emptyset,\emptyset\rangle)\in R$. By replacing weakly probabilistic pomset transitions with weakly probabilistic steps, we can get the
definition of weakly probabilistic step bisimulation. When PESs $\mathcal{E}_1$ and $\mathcal{E}_2$ are weakly probabilistic step bisimilar, we write
$\mathcal{E}_1\approx_{ps}\mathcal{E}_2$.
\end{definition}

\begin{definition}[Posetal product]
Given two PESs $\mathcal{E}_1$, $\mathcal{E}_2$, the posetal product of their configurations, denoted
$\langle\mathcal{C}(\mathcal{E}_1),S\rangle\overline{\times}\langle\mathcal{C}(\mathcal{E}_2),S\rangle$, is defined as

$$\{(\langle C_1,s\rangle,f,\langle C_2,s\rangle)|C_1\in\mathcal{C}(\mathcal{E}_1),C_2\in\mathcal{C}(\mathcal{E}_2),f:C_1\rightarrow C_2 \textrm{ isomorphism}\}.$$

A subset $R\subseteq\langle\mathcal{C}(\mathcal{E}_1),S\rangle\overline{\times}\langle\mathcal{C}(\mathcal{E}_2),S\rangle$ is called a posetal relation. We say that $R$ is downward
closed when for any $(\langle C_1,s\rangle,f,\langle C_2,s\rangle),(\langle C_1',s'\rangle,f',\langle C_2',s'\rangle)\in \langle\mathcal{C}(\mathcal{E}_1),S\rangle\overline{\times}\langle\mathcal{C}(\mathcal{E}_2),S\rangle$,
if $(\langle C_1,s\rangle,f,\langle C_2,s\rangle)\subseteq (\langle C_1',s'\rangle,f',\langle C_2',s'\rangle)$ pointwise and
$(\langle C_1',s'\rangle,f',\langle C_2',s'\rangle)\in R$, then $(\langle C_1,s\rangle,f,\langle C_2,s\rangle)\in R$.

For $f:X_1\rightarrow X_2$, we define $f[x_1\mapsto x_2]:X_1\cup\{x_1\}\rightarrow X_2\cup\{x_2\}$, $z\in X_1\cup\{x_1\}$,(1)$f[x_1\mapsto x_2](z)=
x_2$,if $z=x_1$;(2)$f[x_1\mapsto x_2](z)=f(z)$, otherwise. Where $X_1\subseteq \mathbb{E}_1$, $X_2\subseteq \mathbb{E}_2$, $x_1\in \mathbb{E}_1$, $x_2\in \mathbb{E}_2$.
\end{definition}

\begin{definition}[Weakly posetal product]
Given two PESs $\mathcal{E}_1$, $\mathcal{E}_2$, the weakly posetal product of their configurations, denoted
$\langle\mathcal{C}(\mathcal{E}_1),S\rangle\overline{\times}\langle\mathcal{C}(\mathcal{E}_2),S\rangle$, is defined as

$$\{(\langle C_1,s\rangle,f,\langle C_2,s\rangle)|C_1\in\mathcal{C}(\mathcal{E}_1),C_2\in\mathcal{C}(\mathcal{E}_2),f:\hat{C_1}\rightarrow \hat{C_2} \textrm{ isomorphism}\}.$$

A subset $R\subseteq\langle\mathcal{C}(\mathcal{E}_1),S\rangle\overline{\times}\langle\mathcal{C}(\mathcal{E}_2),S\rangle$ is called a weakly posetal relation. We say that $R$ is
downward closed when for any $(\langle C_1,s\rangle,f,\langle C_2,s\rangle),(\langle C_1',s'\rangle,f,\langle C_2',s'\rangle)\in \langle\mathcal{C}(\mathcal{E}_1),S\rangle\overline{\times}\langle\mathcal{C}(\mathcal{E}_2),S\rangle$,
if $(\langle C_1,s\rangle,f,\langle C_2,s\rangle)\subseteq (\langle C_1',s'\rangle,f',\langle C_2',s'\rangle)$ pointwise and
$(\langle C_1',s'\rangle,f',\langle C_2',s'\rangle)\in R$, then $(\langle C_1,s\rangle,f,\langle C_2,s\rangle)\in R$.

For $f:X_1\rightarrow X_2$, we define $f[x_1\mapsto x_2]:X_1\cup\{x_1\}\rightarrow X_2\cup\{x_2\}$, $z\in X_1\cup\{x_1\}$,(1)$f[x_1\mapsto x_2](z)=
x_2$,if $z=x_1$;(2)$f[x_1\mapsto x_2](z)=f(z)$, otherwise. Where $X_1\subseteq \hat{\mathbb{E}_1}$, $X_2\subseteq \hat{\mathbb{E}_2}$, $x_1\in \hat{\mathbb{E}}_1$,
$x_2\in \hat{\mathbb{E}}_2$. Also, we define $f(\tau^*)=f(\tau^*)$.
\end{definition}

\begin{definition}[Probabilistic (hereditary) history-preserving bisimulation]
A probabilistic history-preserving (hp-) bisimulation is a posetal relation
$R\subseteq\langle\mathcal{C}(\mathcal{E}_1),S\rangle\overline{\times}\langle\mathcal{C}(\mathcal{E}_2),S\rangle$ such that (1) if $(\langle C_1,s\rangle,f,\langle C_2,s\rangle)\in R$,
and $\langle C_1,s\rangle\xrightarrow{e_1} \langle C_1',s'\rangle$, then $\langle C_2,s\rangle\xrightarrow{e_2} \langle C_2',s'\rangle$, with
$(\langle C_1',s'\rangle,f[e_1\mapsto e_2],\langle C_2',s'\rangle)\in R$ for all $s,s'\in S$, and vice-versa; (2) if $(\langle C_1,s\rangle,f,\langle C_2,s\rangle)\in R$, and
$\langle C_1,s\rangle\xrsquigarrow{\pi}\langle C_1^{\pi},s\rangle$ then $\langle C_2,s\rangle\xrsquigarrow{\pi}\langle C_2^{\pi},s\rangle$ and $(\langle C_1^{\pi},s\rangle,f,\langle C_2^{\pi},s\rangle)\in R$,
and vice-versa; (3) if $(C_1,f,C_2)\in R$, then $\mu(C_1,C)=\mu(C_2,C)$ for each $C\in\mathcal{C}(\mathcal{E})/R$; (4) $[\surd]_R=\{\surd\}$. $\mathcal{E}_1,\mathcal{E}_2$ are
probabilistic history-preserving (hp-)bisimilar and are written $\mathcal{E}_1\sim_{php}\mathcal{E}_2$ if there exists a probabilistic hp-bisimulation $R$ such that
$(\langle\emptyset,\emptyset\rangle,\emptyset,\langle\emptyset,\emptyset\rangle)\in R$.

A probabilistic hereditary history-preserving (hhp-)bisimulation is a downward closed probabilistic hp-bisimulation. $\mathcal{E}_1,\mathcal{E}_2$ are probabilistic hereditary
history-preserving (hhp-)bisimilar and are written $\mathcal{E}_1\sim_{phhp}\mathcal{E}_2$.
\end{definition}

\begin{definition}[Weakly probabilistic (hereditary) history-preserving bisimulation]
A weakly probabilistic history-preserving (hp-) bisimulation is a weakly posetal relation
$R\subseteq\langle\mathcal{C}(\mathcal{E}_1),S\rangle\overline{\times}\langle\mathcal{C}(\mathcal{E}_2),S\rangle$ such that (1) if $(\langle C_1,s\rangle,f,\langle C_2,s\rangle)\in R$,
and $\langle C_1,s\rangle\xRightarrow{e_1} \langle C_1',s'\rangle$, then $\langle C_2,s\rangle\xRightarrow{e_2} \langle C_2',s'\rangle$, with
$(\langle C_1',s'\rangle,f[e_1\mapsto e_2],\langle C_2',s'\rangle)\in R$ for all $s,s'\in S$, and vice-versa; (2) if $(\langle C_1,s\rangle,f,\langle C_2,s\rangle)\in R$, and
$\langle C_1,s\rangle\xrsquigarrow{\pi}\langle C_1^{\pi},s\rangle$ then $\langle C_2,s\rangle\xrsquigarrow{\pi}\langle C_2^{\pi},s\rangle$ and
$(\langle C_1^{\pi},s\rangle,f,\langle C_2^{\pi},s\rangle)\in R$, and vice-versa; (3) if $(C_1,f,C_2)\in R$, then $\mu(C_1,C)=\mu(C_2,C)$ for each $C\in\mathcal{C}(\mathcal{E})/R$;
(4) $[\surd]_R=\{\surd\}$. $\mathcal{E}_1,\mathcal{E}_2$ are weakly probabilistic history-preserving (hp-)bisimilar and are written $\mathcal{E}_1\approx_{php}\mathcal{E}_2$ if there
exists a weakly probabilistic hp-bisimulation $R$ such that $(\langle\emptyset,\emptyset\rangle,\emptyset,\langle\emptyset,\emptyset\rangle)\in R$.

A weakly probabilistic hereditary history-preserving (hhp-)bisimulation is a downward closed weakly probabilistic hp-bisimulation. $\mathcal{E}_1,\mathcal{E}_2$ are weakly
probabilistic hereditary history-preserving (hhp-)bisimilar and are written $\mathcal{E}_1\approx_{phhp}\mathcal{E}_2$.
\end{definition}

\begin{definition}[Probabilistic branching pomset, step bisimulation]
Assume a special termination predicate $\downarrow$, and let $\surd$ represent a state with $\surd\downarrow$. Let $\mathcal{E}_1$, $\mathcal{E}_2$ be PESs. A probabilistic branching
pomset bisimulation is a relation $R\subseteq\langle\mathcal{C}(\mathcal{E}_1),S\rangle\times\langle\mathcal{C}(\mathcal{E}_2),S\rangle$, such that:

 \begin{enumerate}
   \item if $(\langle C_1,s\rangle,\langle C_2,s\rangle)\in R$, and $\langle C_1,s\rangle\xrightarrow{X}\langle C_1',s'\rangle$ then
   \begin{itemize}
     \item either $X\equiv \tau^*$, and $(\langle C_1',s'\rangle,\langle C_2,s\rangle)\in R$ with $s'\in \tau(s)$;
     \item or there is a sequence of (zero or more) probabilistic transitions and $\tau$-transitions $\langle C_2,s\rangle\rightsquigarrow^*\xrightarrow{\tau^*} \langle C_2^0,s^0\rangle$, such that
     $(\langle C_1,s\rangle,\langle C_2^0,s^0\rangle)\in R$ and $\langle C_2^0,s^0\rangle\xRightarrow{X}\langle C_2',s'\rangle$ with
     $(\langle C_1',s'\rangle,\langle C_2',s'\rangle)\in R$;
   \end{itemize}
   \item if $(\langle C_1,s\rangle,\langle C_2,s\rangle)\in R$, and $\langle C_2,s\rangle\xrightarrow{X}\langle C_2',s'\rangle$ then
   \begin{itemize}
     \item either $X\equiv \tau^*$, and $(\langle C_1,s\rangle,\langle C_2',s'\rangle)\in R$;
     \item or there is a sequence of (zero or more) probabilistic transitions and $\tau$-transitions $\langle C_1,s\rangle\rightsquigarrow^*\xrightarrow{\tau^*} \langle C_1^0,s^0\rangle$, such that
     $(\langle C_1^0,s^0\rangle,\langle C_2,s\rangle)\in R$ and $\langle C_1^0,s^0\rangle\xRightarrow{X}\langle C_1',s'\rangle$ with
     $(\langle C_1',s'\rangle,\langle C_2',s'\rangle)\in R$;
   \end{itemize}
   \item if $(\langle C_1,s\rangle,\langle C_2,s\rangle)\in R$ and $\langle C_1,s\rangle\downarrow$, then there is a sequence of (zero or more) probabilistic transitions and $\tau$-transitions
   $\langle C_2,s\rangle\rightsquigarrow^*\xrightarrow{\tau^*}\langle C_2^0,s^0\rangle$ such that $(\langle C_1,s\rangle,\langle C_2^0,s^0\rangle)\in R$ and
   $\langle C_2^0,s^0\rangle\downarrow$;
   \item if $(\langle C_1,s\rangle,\langle C_2,s\rangle)\in R$ and $\langle C_2,s\rangle\downarrow$, then there is a sequence of (zero or more) probabilistic transitions and $\tau$-transitions
   $\langle C_1,s\rangle\rightsquigarrow^*\xrightarrow{\tau^*}\langle C_1^0,s^0\rangle$ such that $(\langle C_1^0,s^0\rangle,\langle C_2,s\rangle)\in R$ and
   $\langle C_1^0,s^0\rangle\downarrow$;
   \item if $(C_1,C_2)\in R$,then $\mu(C_1,C)=\mu(C_2,C)$ for each $C\in\mathcal{C}(\mathcal{E})/R$;
   \item $[\surd]_R=\{\surd\}$.
 \end{enumerate}

We say that $\mathcal{E}_1$, $\mathcal{E}_2$ are probabilistic branching pomset bisimilar, written $\mathcal{E}_1\approx_{pbp}\mathcal{E}_2$, if there exists a probabilistic branching
pomset bisimulation $R$, such that $(\langle\emptyset,\emptyset\rangle,\langle\emptyset,\emptyset\rangle)\in R$.

By replacing probabilistic pomset transitions with steps, we can get the definition of probabilistic branching step bisimulation. When PESs $\mathcal{E}_1$ and $\mathcal{E}_2$ are
probabilistic branching step bisimilar, we write $\mathcal{E}_1\approx_{pbs}\mathcal{E}_2$.
\end{definition}

\begin{definition}[Probabilistic rooted branching pomset, step bisimulation]
Assume a special termination predicate $\downarrow$, and let $\surd$ represent a state with $\surd\downarrow$. Let $\mathcal{E}_1$, $\mathcal{E}_2$ be PESs. A probabilistic rooted
branching pomset bisimulation is a relation $R\subseteq\langle\mathcal{C}(\mathcal{E}_1),S\rangle\times\langle\mathcal{C}(\mathcal{E}_2),S\rangle$, such that:

 \begin{enumerate}
   \item if $(\langle C_1,s\rangle,\langle C_2,s\rangle)\in R$, and $\langle C_1,s\rangle\rightsquigarrow\xrightarrow{X}\langle C_1',s'\rangle$ then
   $\langle C_2,s\rangle\rightsquigarrow\xrightarrow{X}\langle C_2',s'\rangle$ with $\langle C_1',s'\rangle\approx_{pbp}\langle C_2',s'\rangle$;
   \item if $(\langle C_1,s\rangle,\langle C_2,s\rangle)\in R$, and $\langle C_2,s\rangle\rightsquigarrow\xrightarrow{X}\langle C_2',s'\rangle$ then
   $\langle C_1,s\rangle\rightsquigarrow\xrightarrow{X}\langle C_1',s'\rangle$ with $\langle C_1',s'\rangle\approx_{pbp}\langle C_2',s'\rangle$;
   \item if $(\langle C_1,s\rangle,\langle C_2,s\rangle)\in R$ and $\langle C_1,s\rangle\downarrow$, then $\langle C_2,s\rangle\downarrow$;
   \item if $(\langle C_1,s\rangle,\langle C_2,s\rangle)\in R$ and $\langle C_2,s\rangle\downarrow$, then $\langle C_1,s\rangle\downarrow$.
 \end{enumerate}

We say that $\mathcal{E}_1$, $\mathcal{E}_2$ are probabilistic rooted branching pomset bisimilar, written $\mathcal{E}_1\approx_{prbp}\mathcal{E}_2$, if there exists a probabilistic
rooted branching pomset bisimulation $R$, such that $(\langle\emptyset,\emptyset\rangle,\langle\emptyset,\emptyset\rangle)\in R$.

By replacing pomset transitions with steps, we can get the definition of probabilistic rooted branching step bisimulation. When PESs $\mathcal{E}_1$ and $\mathcal{E}_2$ are probabilistic
rooted branching step bisimilar, we write $\mathcal{E}_1\approx_{prbs}\mathcal{E}_2$.
\end{definition}

\begin{definition}[Probabilistic branching (hereditary) history-preserving bisimulation]
Assume a special termination predicate $\downarrow$, and let $\surd$ represent a state with $\surd\downarrow$. A probabilistic branching history-preserving (hp-) bisimulation is a
weakly posetal relation $R\subseteq\langle\mathcal{C}(\mathcal{E}_1),S\rangle\overline{\times}\langle\mathcal{C}(\mathcal{E}_2),S\rangle$ such that:

 \begin{enumerate}
   \item if $(\langle C_1,s\rangle,f,\langle C_2,s\rangle)\in R$, and $\langle C_1,s\rangle\xrightarrow{e_1}\langle C_1',s'\rangle$ then
   \begin{itemize}
     \item either $e_1\equiv \tau$, and $(\langle C_1',s'\rangle,f[e_1\mapsto \tau],\langle C_2,s\rangle)\in R$;
     \item or there is a sequence of (zero or more) probabilistic transitions and $\tau$-transitions $\langle C_2,s\rangle\rightsquigarrow^*\xrightarrow{\tau^*} \langle C_2^0,s^0\rangle$, such that
     $(\langle C_1,s\rangle,f,\langle C_2^0,s^0\rangle)\in R$ and $\langle C_2^0,s^0\rangle\xrightarrow{e_2}\langle C_2',s'\rangle$ with
     $(\langle C_1',s'\rangle,f[e_1\mapsto e_2],\langle C_2',s'\rangle)\in R$;
   \end{itemize}
   \item if $(\langle C_1,s\rangle,f,\langle C_2,s\rangle)\in R$, and $\langle C_2,s\rangle\xrightarrow{e_2}\langle C_2',s'\rangle$ then
   \begin{itemize}
     \item either $e_2\equiv \tau$, and $(\langle C_1,s\rangle,f[e_2\mapsto \tau],\langle C_2',s'\rangle)\in R$;
     \item or there is a sequence of (zero or more) probabilistic transitions and $\tau$-transitions $\langle C_1,s\rangle\rightsquigarrow^*\xrightarrow{\tau^*} \langle C_1^0,s^0\rangle$, such that
     $(\langle C_1^0,s^0\rangle,f,\langle C_2,s\rangle)\in R$ and $\langle C_1^0,s^0\rangle\xrightarrow{e_1}\langle C_1',s'\rangle$ with
     $(\langle C_1',s'\rangle,f[e_2\mapsto e_1],\langle C_2',s'\rangle)\in R$;
   \end{itemize}
   \item if $(\langle C_1,s\rangle,f,\langle C_2,s\rangle)\in R$ and $\langle C_1,s\rangle\downarrow$, then there is a sequence of (zero or more) probabilistic transitions and $\tau$-transitions
   $\langle C_2,s\rangle\rightsquigarrow^*\xrightarrow{\tau^*}\langle C_2^0,s^0\rangle$ such that $(\langle C_1,s\rangle,f,\langle C_2^0,s^0\rangle)\in R$ and
   $\langle C_2^0,s^0\rangle\downarrow$;
   \item if $(\langle C_1,s\rangle,f,\langle C_2,s\rangle)\in R$ and $\langle C_2,s\rangle\downarrow$, then there is a sequence of (zero or more) probabilistic transitions and $\tau$-transitions
   $\langle C_1,s\rangle\rightsquigarrow^*\xrightarrow{\tau^*}\langle C_1^0,s^0\rangle$ such that $(\langle C_1^0,s^0\rangle,f,\langle C_2,s\rangle)\in R$ and
   $\langle C_1^0,s^0\rangle\downarrow$;
   \item if $(C_1,C_2)\in R$,then $\mu(C_1,C)=\mu(C_2,C)$ for each $C\in\mathcal{C}(\mathcal{E})/R$;
   \item $[\surd]_R=\{\surd\}$.
 \end{enumerate}

$\mathcal{E}_1,\mathcal{E}_2$ are probabilistic branching history-preserving (hp-)bisimilar and are written $\mathcal{E}_1\approx_{pbhp}\mathcal{E}_2$ if there exists a probabilistic
branching hp-bisimulation $R$ such that $(\langle\emptyset,\emptyset\rangle,\emptyset,\langle\emptyset,\emptyset\rangle)\in R$.

A probabilistic branching hereditary history-preserving (hhp-)bisimulation is a downward closed probabilistic branching hp-bisimulation. $\mathcal{E}_1,\mathcal{E}_2$ are probabilistic
branching hereditary history-preserving (hhp-)bisimilar and are written $\mathcal{E}_1\approx_{pbhhp}\mathcal{E}_2$.
\end{definition}

\begin{definition}[Probabilistic rooted branching (hereditary) history-preserving bisimulation]
Assume a special termination predicate $\downarrow$, and let $\surd$ represent a state with $\surd\downarrow$. A probabilistic rooted branching history-preserving (hp-) bisimulation is
a weakly posetal relation $R\subseteq\langle\mathcal{C}(\mathcal{E}_1),S\rangle\overline{\times}\langle\mathcal{C}(\mathcal{E}_2),S\rangle$ such that:

 \begin{enumerate}
   \item if $(\langle C_1,s\rangle,f,\langle C_2,s\rangle)\in R$, and $\langle C_1,s\rangle\rightsquigarrow\xrightarrow{e_1}\langle C_1',s'\rangle$, then
   $\langle C_2,s\rangle\rightsquigarrow\xrightarrow{e_2}\langle C_2',s'\rangle$ with $\langle C_1',s'\rangle\approx_{pbhp}\langle C_2',s'\rangle$;
   \item if $(\langle C_1,s\rangle,f,\langle C_2,s\rangle)\in R$, and $\langle C_2,s\rangle\rightsquigarrow\xrightarrow{e_2}\langle C_2',s'\rangle$, then
   $\langle C_1,s\rangle\rightsquigarrow\xrightarrow{e_1}\langle C_1',s'\rangle$ with $\langle C_1',s'\rangle\approx_{pbhp}\langle C_2',s'\rangle$;
   \item if $(\langle C_1,s\rangle,f,\langle C_2,s\rangle)\in R$ and $\langle C_1,s\rangle\downarrow$, then $\langle C_2,s\rangle\downarrow$;
   \item if $(\langle C_1,s\rangle,f,\langle C_2,s\rangle)\in R$ and $\langle C_2,s\rangle\downarrow$, then $\langle C_1,s\rangle\downarrow$.
 \end{enumerate}

$\mathcal{E}_1,\mathcal{E}_2$ are probabilistic rooted branching history-preserving (hp-)bisimilar and are written $\mathcal{E}_1\approx_{prbhp}\mathcal{E}_2$ if there exists a probabilistic
rooted branching hp-bisimulation $R$ such that $(\langle\emptyset,\emptyset\rangle,\emptyset,\langle\emptyset,\emptyset\rangle)\in R$.

A probabilistic rooted branching hereditary history-preserving (hhp-)bisimulation is a downward closed probabilistic rooted branching hp-bisimulation. $\mathcal{E}_1,\mathcal{E}_2$ are
probabilistic rooted branching hereditary history-preserving (hhp-)bisimilar and are written $\mathcal{E}_1\approx_{prbhhp}\mathcal{E}_2$.
\end{definition}

\subsubsection{$BAPTC$ with Guards}

In this subsection, we will discuss the guards for $BAPTC$, which is denoted as $BAPTC_G$. Let $\mathbb{E}$ be the set of atomic events (actions), $G_{at}$ be the set of atomic guards,
$\delta$ be the deadlock constant, and $\epsilon$ be the empty event. We extend $G_{at}$ to the set of basic guards $G$ with element $\phi,\psi,\cdots$, which is generated by the
following formation rules:

$$\phi::=\delta|\epsilon|\neg\phi|\psi\in G_{at}|\phi+\psi|\phi\boxplus_{\pi}\psi|\phi\cdot\psi$$

In the following, let $e_1, e_2, e_1', e_2'\in \mathbb{E}$, $\phi,\psi\in G$ and let variables $x,y,z$ range over the set of terms for true concurrency, $p,q,s$ range over the set of
closed terms. The predicate $test(\phi,s)$ represents that $\phi$ holds in the state $s$, and $test(\epsilon,s)$ holds and $test(\delta,s)$ does not hold. $effect(e,s)\in S$ denotes
$s'$ in $s\xrightarrow{e}s'$. The predicate weakest precondition $wp(e,\phi)$ denotes that $\forall s,s'\in S, test(\phi,effect(e,s))$ holds.

The set of axioms of $BAPTC_G$ consists of the laws given in Table \ref{AxiomsForBATCG22}.

\begin{center}
    \begin{table}
        \begin{tabular}{@{}ll@{}}
            \hline No. &Axiom\\
            $A1$ & $x+ y = y+ x$\\
            $A2$ & $(x+ y)+ z = x+ (y+ z)$\\
            $A3$ & $e+ e = e$\\
            $A4$ & $(x+ y)\cdot z = x\cdot z + y\cdot z$\\
            $A5$ & $(x\cdot y)\cdot z = x\cdot(y\cdot z)$\\
            $A6$ & $x+\delta = x$\\
            $A7$ & $\delta\cdot x = \delta$\\
            $A8$ & $\epsilon\cdot x = x$\\
            $A9$ & $x\cdot\epsilon = x$\\
            $PA1$ & $x\boxplus_{\pi} y=y\boxplus_{1-\pi} x$\\
            $PA2$ & $x\boxplus_{\pi}(y\boxplus_{\rho} z)=(x\boxplus_{\frac{\pi}{\pi+\rho-\pi\rho}}y)\boxplus_{\pi+\rho-\pi\rho} z$\\
            $PA3$ & $x\boxplus_{\pi}x=x$\\
            $PA4$ & $(x\boxplus_{\pi}y)\cdot z=x\cdot z\boxplus_{\pi}y\cdot z$\\
            $PA5$ & $(x\boxplus_{\pi}y)+z=(x+z)\boxplus_{\pi}(y+z)$\\
            $G1$ & $\phi\cdot\neg\phi = \delta$\\
            $G2$ & $\phi+\neg\phi = \epsilon$\\
            $PG1$ & $\phi\boxplus_{\pi}\neg\phi = \epsilon$\\
            $G3$ & $\phi\delta = \delta$\\
            $G4$ & $\phi(x+y)=\phi x+\phi y$\\
            $PG2$ & $\phi(x\boxplus_{\pi}y)=\phi x\boxplus_{\pi}\phi y$\\
            $G5$ & $\phi(x\cdot y)= \phi x\cdot y$\\
            $G6$ & $(\phi+\psi)x = \phi x + \psi x$\\
            $PG3$ & $(\phi\boxplus_{\pi}\psi)x = \phi x \boxplus_{\pi} \psi x$\\
            $G7$ & $(\phi\cdot \psi)\cdot x = \phi\cdot(\psi\cdot x)$\\
            $G8$ & $\phi=\epsilon$ if $\forall s\in S.test(\phi,s)$\\
            $G9$ & $\phi_0\cdot\cdots\cdot\phi_n = \delta$ if $\forall s\in S,\exists i\leq n.test(\neg\phi_i,s)$\\
            $G10$ & $wp(e,\phi)e\phi=wp(e,\phi)e$\\
            $G11$ & $\neg wp(e,\phi)e\neg\phi=\neg wp(e,\phi)e$\\
        \end{tabular}
        \caption{Axioms of $BAPTC_G$}
        \label{AxiomsForBATCG22}
    \end{table}
\end{center}

Note that, by eliminating atomic event from the process terms, the axioms in Table \ref{AxiomsForBATCG22} will lead to a Boolean Algebra. And $G8$ and $G9$ are preconditions of $e$ and
$\phi$, $G10$ is the weakest precondition of $e$ and $\phi$. A data environment with $effect$ function is sufficiently deterministic, and it is obvious that if the weakest precondition
is expressible and $G10$, $G11$ are sound, then the related data environment is sufficiently deterministic.

\begin{definition}[Basic terms of $BAPTC_G$]
The set of basic terms of $BAPTC_G$, $\mathcal{B}(BAPTC_G)$, is inductively defined as follows:

\begin{enumerate}
  \item $\mathbb{E}\subset\mathcal{B}(BAPTC_G)$;
  \item $G\subset\mathcal{B}(BAPTC_G)$;
  \item if $e\in \mathbb{E}, t\in\mathcal{B}(BAPTC_G)$ then $e\cdot t\in\mathcal{B}(BAPTC_G)$;
  \item if $\phi\in G, t\in\mathcal{B}(BAPTC_G)$ then $\phi\cdot t\in\mathcal{B}(BAPTC_G)$;
  \item if $t,s\in\mathcal{B}(BAPTC_G)$ then $t+ s\in\mathcal{B}(BAPTC_G)$;
  \item if $t,s\in\mathcal{B}(BAPTC_G)$ then $t\boxplus_{\pi} s\in\mathcal{B}(BAPTC_G)$.
\end{enumerate}
\end{definition}

\begin{theorem}[Elimination theorem of $BAPTC_G$]
Let $p$ be a closed $BAPTC_G$ term. Then there is a basic $BAPTC_G$ term $q$ such that $BAPTC_G\vdash p=q$.
\end{theorem}

In this subsection, we will define a term-deduction system which gives the operational semantics of $BAPTC$. Like the way in \cite{PPA}, we also introduce the counterpart $\breve{e}$
of the event $e$, and also the set $\breve{\mathbb{E}}=\{\breve{e}|e\in\mathbb{E}\}$.

We give the definition of PDFs of $BAPTC$ in Table \ref{PDFBAPTC22}.

\begin{center}
    \begin{table}
        $$\mu(e,\breve{e})=1$$
        $$\mu(x\cdot y, x'\cdot y)=\mu(x,x')$$
        $$\mu(x+y,x'+y')=\mu(x,x')\cdot \mu(y,y')$$
        $$\mu(x\boxplus_{\pi}y,z)=\pi\mu(x,z)+(1-\pi)\mu(y,z)$$
        $$\mu(x,y)=0,\textrm{otherwise}$$
        \caption{PDF definitions of $BAPTC$}
        \label{PDFBAPTC22}
    \end{table}
\end{center}

We will define a term-deduction system which gives the operational semantics of $BAPTC_G$. We give the operational transition rules for $\epsilon$, atomic guard $\phi\in G_{at}$,
atomic event $e\in\mathbb{E}$, operators $\cdot$ and $+$ as Table \ref{SETRForBATCG22} shows. And the predicate $\xrightarrow{e}\surd$ represents successful termination after execution
of the event $e$.

\begin{center}
    \begin{table}
        $$\frac{}{\langle\epsilon,s\rangle\rightsquigarrow\langle\breve{\epsilon},s\rangle}$$
        $$\frac{}{\langle e,s\rangle\rightsquigarrow\langle\breve{e},s\rangle}$$
        $$\frac{}{\langle\phi,s\rangle\rightsquigarrow\langle\breve{\phi},s\rangle}$$
        $$\frac{\langle x,s\rangle\rightsquigarrow \langle x',s\rangle}{\langle x\cdot y,s\rangle\rightsquigarrow \langle x'\cdot y,s\rangle}$$
        $$\frac{\langle x,s\rangle\rightsquigarrow \langle x',s\rangle\quad \langle y,s\rangle\rightsquigarrow \langle y',s\rangle}{\langle x+y,s\rangle\rightsquigarrow \langle x'+y',s\rangle}$$
        $$\frac{\langle x,s\rangle\rightsquigarrow \langle x',s\rangle}{\langle x\boxplus_{\pi}y,s\rangle\rightsquigarrow \langle x',s\rangle}\quad \frac{\langle y,s\rangle\rightsquigarrow \langle y',s\rangle}{\langle x\boxplus_{\pi}y,s\rangle\rightsquigarrow \langle y',s\rangle}$$

        $$\frac{}{\langle\breve{\epsilon},s\rangle\rightarrow\langle\surd,s\rangle}$$
        $$\frac{}{\langle \breve{e},s\rangle\xrightarrow{e}\langle\surd,s'\rangle}\textrm{ if }s'\in effect(e,s)$$
        $$\frac{}{\langle\breve{\phi},s\rangle\rightarrow\langle\surd,s\rangle}\textrm{ if }test(\phi,s)$$
        $$\frac{\langle x,s\rangle\xrightarrow{e}\langle\surd,s'\rangle}{\langle x+ y,s\rangle\xrightarrow{e}\langle\surd,s'\rangle} \quad\frac{\langle x,s\rangle\xrightarrow{e}\langle x',s'\rangle}{\langle x+ y,s\rangle\xrightarrow{e}\langle x',s'\rangle}$$
        $$\frac{\langle y,s\rangle\xrightarrow{e}\langle\surd,s'\rangle}{\langle x+ y,s\rangle\xrightarrow{e}\langle\surd,s'\rangle} \quad\frac{\langle y,s\rangle\xrightarrow{e}\langle y',s'\rangle}{\langle x+ y,s\rangle\xrightarrow{e}\langle y',s'\rangle}$$
        $$\frac{\langle x,s\rangle\xrightarrow{e}\langle\surd,s'\rangle}{\langle x\cdot y,s\rangle\xrightarrow{e} \langle y,s'\rangle} \quad\frac{\langle x,s\rangle\xrightarrow{e}\langle x',s'\rangle}{\langle x\cdot y,s\rangle\xrightarrow{e}\langle x'\cdot y,s'\rangle}$$
        \caption{Single event transition rules of $BAPTC_G$}
        \label{SETRForBATCG22}
    \end{table}
\end{center}

Note that, we replace the single atomic event $e\in\mathbb{E}$ by $X\subseteq\mathbb{E}$, we can obtain the pomset transition rules of $BAPTC_G$, and omit them.

\begin{theorem}[Congruence of $BAPTC_G$ with respect to probabilistic truly concurrent bisimulation equivalences]
(1) Probabilistic pomset bisimulation equivalence $\sim_{pp}$ is a congruence with respect to $BAPTC_G$.

(2) Probabilistic step bisimulation equivalence $\sim_{ps}$ is a congruence with respect to $BAPTC_G$.

(3) Probabilistic hp-bisimulation equivalence $\sim_{php}$ is a congruence with respect to $BAPTC_G$.

(4) Probabilistic hhp-bisimulation equivalence $\sim_{phhp}$ is a congruence with respect to $BAPTC_G$.
\end{theorem}

\begin{theorem}[Soundness of $BAPTC_G$ modulo probabilistic truly concurrent bisimulation equivalences]
(1) Let $x$ and $y$ be $BAPTC_G$ terms. If $BATC\vdash x=y$, then $x\sim_{pp} y$.

(2) Let $x$ and $y$ be $BAPTC_G$ terms. If $BATC\vdash x=y$, then $x\sim_{ps} y$.

(3) Let $x$ and $y$ be $BAPTC_G$ terms. If $BATC\vdash x=y$, then $x\sim_{php} y$.

(4) Let $x$ and $y$ be $BAPTC_G$ terms. If $BATC\vdash x=y$, then $x\sim_{phhp} y$.
\end{theorem}

\begin{theorem}[Completeness of $BAPTC_G$ modulo probabilistic truly concurrent bisimulation equivalences]
(1) Let $p$ and $q$ be closed $BAPTC_G$ terms, if $p\sim_{pp} q$ then $p=q$.

(2) Let $p$ and $q$ be closed $BAPTC_G$ terms, if $p\sim_{ps} q$ then $p=q$.

(3) Let $p$ and $q$ be closed $BAPTC_G$ terms, if $p\sim_{php} q$ then $p=q$.

(4) Let $p$ and $q$ be closed $BAPTC_G$ terms, if $p\sim_{phhp} q$ then $p=q$.
\end{theorem}

\subsubsection{$APPTC$ with Guards}

In this subsection, we will extend $APPTC$ with guards, which is abbreviated $APPTC_G$. The set of basic guards $G$ with element $\phi,\psi,\cdots$, which is extended by the following
formation rules:

$$\phi::=\delta|\epsilon|\neg\phi|\psi\in G_{at}|\phi+\psi|\phi\boxplus_{\pi}\psi|\phi\cdot\psi|\phi\leftmerge\psi$$

The set of axioms of $APPTC_G$ including axioms of $BATC_G$ in Table \ref{AxiomsForBATCG22} and the axioms are shown in Table \ref{AxiomsForAPTCG22}.

\begin{center}
    \begin{table}
        \begin{tabular}{@{}ll@{}}
            \hline No. &Axiom\\
            $P1$ & $(x+x=x,y+y=y)\quad x\between y = x\parallel y + x\mid y$\\
            $P2$ & $x\parallel y = y \parallel x$\\
            $P3$ & $(x\parallel y)\parallel z = x\parallel (y\parallel z)$\\
            $P4$ & $(x+x=x,y+y=y)\quad x\parallel y = x\leftmerge y + y\leftmerge x$\\
            $P5$ & $(e_1\leq e_2)\quad e_1\leftmerge (e_2\cdot y) = (e_1\leftmerge e_2)\cdot y$\\
            $P6$ & $(e_1\leq e_2)\quad (e_1\cdot x)\leftmerge e_2 = (e_1\leftmerge e_2)\cdot x$\\
            $P7$ & $(e_1\leq e_2)\quad (e_1\cdot x)\leftmerge (e_2\cdot y) = (e_1\leftmerge e_2)\cdot (x\between y)$\\
            $P8$ & $(x+ y)\leftmerge z = (x\leftmerge z)+ (y\leftmerge z)$\\
            $P9$ & $\delta\leftmerge x = \delta$\\
            $P10$ & $\epsilon\leftmerge x = x$\\
            $P11$ & $x\leftmerge \epsilon = x$\\
            $C1$ & $e_1\mid e_2 = \gamma(e_1,e_2)$\\
            $C2$ & $e_1\mid (e_2\cdot y) = \gamma(e_1,e_2)\cdot y$\\
            $C3$ & $(e_1\cdot x)\mid e_2 = \gamma(e_1,e_2)\cdot x$\\
            $C4$ & $(e_1\cdot x)\mid (e_2\cdot y) = \gamma(e_1,e_2)\cdot (x\between y)$\\
            $C5$ & $(x+ y)\mid z = (x\mid z) + (y\mid z)$\\
            $C6$ & $x\mid (y+ z) = (x\mid y)+ (x\mid z)$\\
            $C7$ & $\delta\mid x = \delta$\\
            $C8$ & $x\mid\delta = \delta$\\
            $C9$ & $\epsilon\mid x = \delta$\\
            $C10$ & $x\mid\epsilon = \delta$\\
            $PM1$ & $x\parallel (y\boxplus_{\pi} z)=(x\parallel y)\boxplus_{\pi}(x\parallel z)$\\
            $PM2$ & $(x\boxplus_{\pi} y)\parallel z=(x\parallel z)\boxplus_{\pi}(y\parallel z)$\\
            $PM3$ & $x\mid (y\boxplus_{\pi} z)=(x\mid y)\boxplus_{\pi}(x\mid z)$\\
            $PM4$ & $(x\boxplus_{\pi} y)\mid z=(x\mid z)\boxplus_{\pi}(y\mid z)$\\
            $CE1$ & $\Theta(e) = e$\\
            $CE2$ & $\Theta(\delta) = \delta$\\
            $CE3$ & $\Theta(\epsilon) = \epsilon$\\
            $CE4$ & $\Theta(x+ y) = \Theta(x)\triangleleft y + \Theta(y)\triangleleft x$\\
            $PCE1$ & $\Theta(x\boxplus_{\pi} y) = \Theta(x)\triangleleft y \boxplus_{\pi} \Theta(y)\triangleleft x$\\
            $CE5$ & $\Theta(x\cdot y)=\Theta(x)\cdot\Theta(y)$\\
            $CE6$ & $\Theta(x\leftmerge y) = ((\Theta(x)\triangleleft y)\leftmerge y)+ ((\Theta(y)\triangleleft x)\leftmerge x)$\\
            $CE7$ & $\Theta(x\mid y) = ((\Theta(x)\triangleleft y)\mid y)+ ((\Theta(y)\triangleleft x)\mid x)$\\
        \end{tabular}
        \caption{Axioms of $APPTC_G$}
        \label{AxiomsForAPTCG22}
    \end{table}
\end{center}

\begin{center}
    \begin{table}
        \begin{tabular}{@{}ll@{}}
            \hline No. &Axiom\\
            $U1$ & $(\sharp(e_1,e_2))\quad e_1\triangleleft e_2 = \tau$\\
            $U2$ & $(\sharp(e_1,e_2),e_2\leq e_3)\quad e_1\triangleleft e_3 = e_1$\\
            $U3$ & $(\sharp(e_1,e_2),e_2\leq e_3)\quad e3\triangleleft e_1 = \tau$\\
            $PU1$ & $(\sharp_{\pi}(e_1,e_2))\quad e_1\triangleleft e_2 = \tau$\\
            $PU2$ & $(\sharp_{\pi}(e_1,e_2),e_2\leq e_3)\quad e_1\triangleleft e_3 = e_1$\\
            $PU3$ & $(\sharp_{\pi}(e_1,e_2),e_2\leq e_3)\quad e_3\triangleleft e_1 = \tau$\\
            $U4$ & $e\triangleleft \delta = e$\\
            $U5$ & $\delta \triangleleft e = \delta$\\
            $U6$ & $e\triangleleft \epsilon = e$\\
            $U7$ & $\epsilon \triangleleft e = e$\\
            $U8$ & $(x+ y)\triangleleft z = (x\triangleleft z)+ (y\triangleleft z)$\\
            $PU4$ & $(x\boxplus_{\pi} y)\triangleleft z = (x\triangleleft z)\boxplus_{\pi} (y\triangleleft z)$\\
            $U9$ & $(x\cdot y)\triangleleft z = (x\triangleleft z)\cdot (y\triangleleft z)$\\
            $U10$ & $(x\leftmerge y)\triangleleft z = (x\triangleleft z)\leftmerge (y\triangleleft z)$\\
            $U11$ & $(x\mid y)\triangleleft z = (x\triangleleft z)\mid (y\triangleleft z)$\\
            $U12$ & $x\triangleleft (y+ z) = (x\triangleleft y)\triangleleft z$\\
            $PU5$ & $x\triangleleft (y\boxplus_{\pi} z) = (x\triangleleft y)\triangleleft z$\\
            $U13$ & $x\triangleleft (y\cdot z)=(x\triangleleft y)\triangleleft z$\\
            $U14$ & $x\triangleleft (y\leftmerge z) = (x\triangleleft y)\triangleleft z$\\
            $U15$ & $x\triangleleft (y\mid z) = (x\triangleleft y)\triangleleft z$\\
            $D1$ & $e\notin H\quad\partial_H(e) = e$\\
            $D2$ & $e\in H\quad \partial_H(e) = \delta$\\
            $D3$ & $\partial_H(\delta) = \delta$\\
            $D4$ & $\partial_H(x+ y) = \partial_H(x)+\partial_H(y)$\\
            $D5$ & $\partial_H(x\cdot y) = \partial_H(x)\cdot\partial_H(y)$\\
            $D6$ & $\partial_H(x\leftmerge y) = \partial_H(x)\leftmerge\partial_H(y)$\\
            $PD1$ & $\partial_H(x\boxplus_{\pi}y)=\partial_H(x)\boxplus_{\pi}\partial_H(y)$\\
            $G12$ & $\phi(x\leftmerge y) =\phi x\leftmerge \phi y$\\
            $G13$ & $\phi(x\mid y) =\phi x\mid \phi y$\\
            $G14$ & $\delta\leftmerge \phi = \delta$\\
            $G15$ & $\phi\mid \delta = \delta$\\
            $G16$ & $\delta\mid \phi = \delta$\\
            $G17$ & $\phi\leftmerge \epsilon = \phi$\\
            $G18$ & $\epsilon\leftmerge \phi = \phi$\\
            $G19$ & $\phi\mid \epsilon = \delta$\\
            $G20$ & $\epsilon\mid \phi = \delta$\\
            $G21$ & $\phi\leftmerge\neg\phi = \delta$\\
            $G22$ & $\Theta(\phi) = \phi$\\
            $G23$ & $\partial_H(\phi) = \phi$\\
            $G24$ & $\phi_0\leftmerge\cdots\leftmerge\phi_n = \delta$ if $\forall s_0,\cdots,s_n\in S,\exists i\leq n.test(\neg\phi_i,s_0\cup\cdots\cup s_n)$\\        \end{tabular}
        \caption{Axioms of $APPTC_G$ (continuing)}
        \label{AxiomsForAPTCG222}
    \end{table}
\end{center}

\begin{definition}[Basic terms of $APPTC_G$]
The set of basic terms of $APPTC_G$, $\mathcal{B}(APPTC_G)$, is inductively defined as follows:

\begin{enumerate}
    \item $\mathbb{E}\subset\mathcal{B}(APPTC_G)$;
    \item $G\subset\mathcal{B}(APPTC_G)$;
    \item if $e\in \mathbb{E}, t\in\mathcal{B}(APPTC_G)$ then $e\cdot t\in\mathcal{B}(APPTC_G)$;
    \item if $\phi\in G, t\in\mathcal{B}(APPTC_G)$ then $\phi\cdot t\in\mathcal{B}(APPTC_G)$;
    \item if $t,s\in\mathcal{B}(APPTC_G)$ then $t+ s\in\mathcal{B}(APPTC_G)$;
    \item if $t,s\in\mathcal{B}(APPTC_G)$ then $t\boxplus_{\pi} s\in\mathcal{B}(APPTC_G)$
    \item if $t,s\in\mathcal{B}(APPTC_G)$ then $t\leftmerge s\in\mathcal{B}(APPTC_G)$.
\end{enumerate}
\end{definition}

Based on the definition of basic terms for $APPTC_G$ and axioms of $APPTC_G$, we can prove the elimination theorem of $APPTC_G$.

\begin{theorem}[Elimination theorem of $APPTC_G$]
Let $p$ be a closed $APPTC_G$ term. Then there is a basic $APPTC_G$ term $q$ such that $APPTC_G\vdash p=q$.
\end{theorem}

We give the definition of PDFs of $APPTC$ in Table \ref{PDFAPPTC22}.

\begin{center}
    \begin{table}
        $$\mu(\delta,\breve{\delta})=1$$
        $$\mu(x\between y,x'\parallel y'+x'\mid y')=\mu(x,x')\cdot\mu(y,y')$$
        $$\mu(x\parallel y,x'\leftmerge y+y'\leftmerge x)=\mu(x,x')\cdot \mu(y,y')$$
        $$\mu(x\leftmerge y, x'\leftmerge y)=\mu(x,x')$$
        $$\mu(x\mid y,x'\mid y')=\mu(x,x')\cdot \mu(y,y')$$
        $$\mu(\Theta(x),\Theta(x'))=\mu(x,x')$$
        $$\mu(x\triangleleft y, x'\triangleleft y)=\mu(x,x')$$
        $$\mu(x,y)=0,\textrm{otherwise}$$
        \caption{PDF definitions of $APPTC$}
        \label{PDFAPPTC22}
    \end{table}
\end{center}

We will define a term-deduction system which gives the operational semantics of $APPTC_G$. Two atomic events $e_1$ and $e_2$ are in race condition, which are denoted $e_1\% e_2$.

\begin{center}
    \begin{table}
        $$\frac{x\rightsquigarrow x'\quad y\rightsquigarrow y'}{x\between y\rightsquigarrow x'\parallel y'+x'\mid y'}$$
        $$\frac{x\rightsquigarrow x'\quad y\rightsquigarrow y'}{x\parallel y\rightsquigarrow x'\leftmerge y+y'\leftmerge x}$$
        $$\frac{x\rightsquigarrow x'}{x\leftmerge y\rightsquigarrow x'\leftmerge y}$$
        $$\frac{x\rightsquigarrow x'\quad y\rightsquigarrow y'}{x\mid y\rightsquigarrow x'\mid y'}$$
        $$\frac{x\rightsquigarrow x'}{\Theta(x)\rightsquigarrow \Theta(x')}$$
        $$\frac{x\rightsquigarrow x'}{x\triangleleft y\rightsquigarrow x'\triangleleft y}$$
        \caption{Probabilistic transition rules of $APPTC_G$}
        \label{TRForAPPTCG122}
    \end{table}
\end{center}

\begin{center}
    \begin{table}
        $$\frac{}{\langle \breve{e_1}\parallel\cdots \parallel \breve{e_n},s\rangle\xrightarrow{\{e_1,\cdots,e_n\}}\langle\surd,s'\rangle}\textrm{ if }s'\in effect(e_1,s)\cup\cdots\cup effect(e_n,s)$$

        $$\frac{}{\langle\breve{\phi_1}\parallel\cdots\parallel \breve{\phi_n},s\rangle\rightarrow\langle\surd,s\rangle}\textrm{ if }test(\phi_1,s),\cdots,test(\phi_n,s)$$

        $$\frac{\langle x,s\rangle\xrightarrow{e_1}\langle\surd,s'\rangle\quad \langle y,s\rangle\xrightarrow{e_2}\langle\surd,s''\rangle}{\langle x\parallel y,s\rangle\xrightarrow{\{e_1,e_2\}}\langle\surd,s'\cup s''\rangle} \quad\frac{\langle x,s\rangle\xrightarrow{e_1}\langle x',s'\rangle\quad \langle y,s\rangle\xrightarrow{e_2}\langle\surd,s''\rangle}{\langle x\parallel y,s\rangle\xrightarrow{\{e_1,e_2\}}\langle x',s'\cup s''\rangle}$$

        $$\frac{\langle x,s\rangle\xrightarrow{e_1}\langle\surd,s'\rangle\quad \langle y,s\rangle\xrightarrow{e_2}\langle y',s''\rangle}{\langle x\parallel y,s\rangle\xrightarrow{\{e_1,e_2\}}\langle y',s'\cup s''\rangle} \quad\frac{\langle x,s\rangle\xrightarrow{e_1}\langle x',s'\rangle\quad \langle y,s\rangle\xrightarrow{e_2}\langle y',s''\rangle}{\langle x\parallel y,s\rangle\xrightarrow{\{e_1,e_2\}}\langle x'\between y',s'\cup s''\rangle}$$

        $$\frac{\langle x,s\rangle\xrightarrow{e_1}\langle\surd,s'\rangle\quad \langle y,s\rangle\xnrightarrow{e_2}\quad(e_1\%e_2)}{\langle x\parallel y,s\rangle\xrightarrow{e_1}\langle y,s'\rangle} \quad\frac{\langle x,s\rangle\xrightarrow{e_1}\langle x',s'\rangle\quad \langle y,s\rangle\xnrightarrow{e_2}\quad(e_1\%e_2)}{\langle x\parallel y,s\rangle\xrightarrow{e_1}\langle x'\between y,s'\rangle}$$

        $$\frac{\langle x,s\rangle\xnrightarrow{e_1}\quad \langle y,s\rangle\xrightarrow{e_2}\langle\surd,s''\rangle\quad(e_1\%e_2)}{\langle x\parallel y,s\rangle\xrightarrow{e_2}\langle x,s''\rangle} \quad\frac{\langle x,s\rangle\xnrightarrow{e_1}\quad \langle y,s\rangle\xrightarrow{e_2}\langle y',s''\rangle\quad(e_1\%e_2)}{\langle x\parallel y,s\rangle\xrightarrow{e_2}\langle x\between y',s''\rangle}$$

        $$\frac{\langle x,s\rangle\xrightarrow{e_1}\langle\surd,s'\rangle\quad \langle y,s\rangle\xrightarrow{e_2}\langle\surd,s''\rangle \quad(e_1\leq e_2)}{\langle x\leftmerge y,s\rangle\xrightarrow{\{e_1,e_2\}}\langle \surd,s'\cup s''\rangle} \quad\frac{\langle x,s\rangle\xrightarrow{e_1}\langle x',s'\rangle\quad \langle y,s\rangle\xrightarrow{e_2}\langle\surd,s''\rangle \quad(e_1\leq e_2)}{\langle x\leftmerge y,s\rangle\xrightarrow{\{e_1,e_2\}}\langle x',s'\cup s''\rangle}$$

        $$\frac{\langle x,s\rangle\xrightarrow{e_1}\langle\surd,s'\rangle\quad \langle y,s\rangle\xrightarrow{e_2}\langle y',s''\rangle \quad(e_1\leq e_2)}{\langle x\leftmerge y,s\rangle\xrightarrow{\{e_1,e_2\}}\langle y',s'\cup s''\rangle} \quad\frac{\langle x,s\rangle\xrightarrow{e_1}\langle x',s'\rangle\quad \langle y,s\rangle\xrightarrow{e_2}\langle y',s''\rangle \quad(e_1\leq e_2)}{\langle x\leftmerge y,s\rangle\xrightarrow{\{e_1,e_2\}}\langle x'\between y',s'\cup s''\rangle}$$

        $$\frac{\langle x,s\rangle\xrightarrow{e_1}\langle\surd,s'\rangle\quad \langle y,s\rangle\xrightarrow{e_2}\langle\surd,s''\rangle}{\langle x\mid y,s\rangle\xrightarrow{\gamma(e_1,e_2)}\langle\surd,effect(\gamma(e_1,e_2),s)\rangle} \quad\frac{\langle x,s\rangle\xrightarrow{e_1}\langle x',s'\rangle\quad \langle y,s\rangle\xrightarrow{e_2}\langle\surd,s''\rangle}{\langle x\mid y,s\rangle\xrightarrow{\gamma(e_1,e_2)}\langle x',effect(\gamma(e_1,e_2),s)\rangle}$$

        $$\frac{\langle x,s\rangle\xrightarrow{e_1}\langle\surd,s'\rangle\quad \langle y,s\rangle\xrightarrow{e_2}\langle y',s''\rangle}{\langle x\mid y,s\rangle\xrightarrow{\gamma(e_1,e_2)}\langle y',effect(\gamma(e_1,e_2),s)\rangle} \quad\frac{\langle x,s\rangle\xrightarrow{e_1}\langle x',s'\rangle\quad \langle y,s\rangle\xrightarrow{e_2}\langle y',s''\rangle}{\langle x\mid y,s\rangle\xrightarrow{\gamma(e_1,e_2)}\langle x'\between y',effect(\gamma(e_1,e_2),s)\rangle}$$

        \caption{Action transition rules of $APPTC_G$}
        \label{TRForAPTCG22}
    \end{table}
\end{center}

\begin{center}
    \begin{table}
        $$\frac{\langle x,s\rangle\xrightarrow{e_1}\langle\surd,s'\rangle\quad (\sharp(e_1,e_2))}{\langle \Theta(x),s\rangle\xrightarrow{e_1}\langle\surd,s'\rangle} \quad\frac{\langle x,s\rangle\xrightarrow{e_2}\langle\surd,s''\rangle\quad (\sharp(e_1,e_2))}{\langle\Theta(x),s\rangle\xrightarrow{e_2}\langle\surd,s''\rangle}$$

        $$\frac{\langle x,s\rangle\xrightarrow{e_1}\langle x',s'\rangle\quad (\sharp(e_1,e_2))}{\langle\Theta(x),s\rangle\xrightarrow{e_1}\langle\Theta(x'),s'\rangle} \quad\frac{\langle x,s\rangle\xrightarrow{e_2}\langle x'',s''\rangle\quad (\sharp(e_1,e_2))}{\langle\Theta(x),s\rangle\xrightarrow{e_2}\langle\Theta(x''),s''\rangle}$$

        $$\frac{\langle x,s\rangle\xrightarrow{e_1}\langle\surd,s'\rangle \quad \langle y,s\rangle\nrightarrow^{e_2}\quad (\sharp(e_1,e_2))}{\langle x\triangleleft y,s\rangle\xrightarrow{\tau}\langle\surd,s'\rangle}
        \quad\frac{\langle x,s\rangle\xrightarrow{e_1}\langle x',s'\rangle \quad \langle y,s\rangle\nrightarrow^{e_2}\quad (\sharp(e_1,e_2))}{\langle x\triangleleft y,s\rangle\xrightarrow{\tau}\langle x',s'\rangle}$$

        $$\frac{\langle x,s\rangle\xrightarrow{e_1}\langle\surd,s\rangle \quad \langle y,s\rangle\nrightarrow^{e_3}\quad (\sharp(e_1,e_2),e_2\leq e_3)}{\langle x\triangleleft y,s\rangle\xrightarrow{e_1}\langle\surd,s'\rangle}
        \quad\frac{\langle x,s\rangle\xrightarrow{e_1}\langle x',s'\rangle \quad \langle y,s\rangle\nrightarrow^{e_3}\quad (\sharp(e_1,e_2),e_2\leq e_3)}{\langle x\triangleleft y,s\rangle\xrightarrow{e_1}\langle x',s'\rangle}$$

        $$\frac{\langle x,s\rangle\xrightarrow{e_3}\langle\surd,s'\rangle \quad \langle y,s\rangle\nrightarrow^{e_2}\quad (\sharp(e_1,e_2),e_1\leq e_3)}{\langle x\triangleleft y,s\rangle\xrightarrow{\tau}\langle\surd,s'\rangle}
        \quad\frac{\langle x,s\rangle\xrightarrow{e_3}\langle x',s'\rangle \quad \langle y,s\rangle\nrightarrow^{e_2}\quad (\sharp(e_1,e_2),e_1\leq e_3)}{\langle x\triangleleft y,s\rangle\xrightarrow{\tau}\langle x',s'\rangle}$$

        $$\frac{\langle x,s\rangle\xrightarrow{e_1}\langle\surd,s'\rangle\quad (\sharp_{\pi}(e_1,e_2))}{\langle \Theta(x),s\rangle\xrightarrow{e_1}\langle\surd,s'\rangle} \quad\frac{\langle x,s\rangle\xrightarrow{e_2}\langle\surd,s''\rangle\quad (\sharp_{\pi}(e_1,e_2))}{\langle\Theta(x),s\rangle\xrightarrow{e_2}\langle\surd,s''\rangle}$$

        $$\frac{\langle x,s\rangle\xrightarrow{e_1}\langle x',s'\rangle\quad (\sharp_{\pi}(e_1,e_2))}{\langle\Theta(x),s\rangle\xrightarrow{e_1}\langle\Theta(x'),s'\rangle} \quad\frac{\langle x,s\rangle\xrightarrow{e_2}\langle x'',s''\rangle\quad (\sharp_{\pi}(e_1,e_2))}{\langle\Theta(x),s\rangle\xrightarrow{e_2}\langle\Theta(x''),s''\rangle}$$

        $$\frac{\langle x,s\rangle\xrightarrow{e_1}\langle\surd,s'\rangle \quad \langle y,s\rangle\nrightarrow^{e_2}\quad (\sharp_{\pi}(e_1,e_2))}{\langle x\triangleleft y,s\rangle\xrightarrow{\tau}\langle\surd,s'\rangle}
        \quad\frac{\langle x,s\rangle\xrightarrow{e_1}\langle x',s'\rangle \quad \langle y,s\rangle\nrightarrow^{e_2}\quad (\sharp_{\pi}(e_1,e_2))}{\langle x\triangleleft y,s\rangle\xrightarrow{\tau}\langle x',s'\rangle}$$

        $$\frac{\langle x,s\rangle\xrightarrow{e_1}\langle\surd,s\rangle \quad \langle y,s\rangle\nrightarrow^{e_3}\quad (\sharp_{\pi}(e_1,e_2),e_2\leq e_3)}{\langle x\triangleleft y,s\rangle\xrightarrow{e_1}\langle\surd,s'\rangle}
        \quad\frac{\langle x,s\rangle\xrightarrow{e_1}\langle x',s'\rangle \quad \langle y,s\rangle\nrightarrow^{e_3}\quad (\sharp_{\pi}(e_1,e_2),e_2\leq e_3)}{\langle x\triangleleft y,s\rangle\xrightarrow{e_1}\langle x',s'\rangle}$$

        $$\frac{\langle x,s\rangle\xrightarrow{e_3}\langle\surd,s'\rangle \quad \langle y,s\rangle\nrightarrow^{e_2}\quad (\sharp_{\pi}(e_1,e_2),e_1\leq e_3)}{\langle x\triangleleft y,s\rangle\xrightarrow{\tau}\langle\surd,s'\rangle}
        \quad\frac{\langle x,s\rangle\xrightarrow{e_3}\langle x',s'\rangle \quad \langle y,s\rangle\nrightarrow^{e_2}\quad (\sharp_{\pi}(e_1,e_2),e_1\leq e_3)}{\langle x\triangleleft y,s\rangle\xrightarrow{\tau}\langle x',s'\rangle}$$

        $$\frac{\langle x,s\rangle\xrightarrow{e}\langle\surd,s'\rangle}{\langle\partial_H(x),s\rangle\xrightarrow{e}\langle\surd,s'\rangle}\quad (e\notin H)\quad\frac{\langle x,s\rangle\xrightarrow{e}\langle x',s'\rangle}{\langle\partial_H(x),s\rangle\xrightarrow{e}\langle\partial_H(x'),s'\rangle}\quad(e\notin H)$$
        \caption{Action transition rules of $APPTC_G$ (continuing)}
        \label{TRForAPTCG222}
    \end{table}
\end{center}

\begin{theorem}[Generalization of $APPTC_G$ with respect to $BAPTC_G$]
$APPTC_G$ is a generalization of $BAPTC_G$.
\end{theorem}

\begin{theorem}[Congruence of $APPTC_G$ with respect to probabilistic truly concurrent bisimulation equivalences]
(1) Probabilistic pomset bisimulation equivalence $\sim_{pp}$ is a congruence with respect to $APPTC_G$.

(2) Probabilistic step bisimulation equivalence $\sim_{ps}$ is a congruence with respect to $APPTC_G$.

(3) Probabilistic hp-bisimulation equivalence $\sim_{php}$ is a congruence with respect to $APPTC_G$.

(4) Probabilistic hhp-bisimulation equivalence $\sim_{phhp}$ is a congruence with respect to $APPTC_G$.
\end{theorem}

\begin{theorem}[Soundness of $APPTC_G$ modulo probabilistic truly concurrent bisimulation equivalences]
(1) Let $x$ and $y$ be $APPTC_G$ terms. If $APTC\vdash x=y$, then $x\sim_{pp} y$.

(2) Let $x$ and $y$ be $APPTC_G$ terms. If $APTC\vdash x=y$, then $x\sim_{ps} y$.

(3) Let $x$ and $y$ be $APPTC_G$ terms. If $APTC\vdash x=y$, then $x\sim_{php} y$;

(3) Let $x$ and $y$ be $APPTC_G$ terms. If $APTC\vdash x=y$, then $x\sim_{phhp} y$.
\end{theorem}

\begin{theorem}[Completeness of $APPTC_G$ modulo probabilistic truly concurrent bisimulation equivalences]
(1) Let $p$ and $q$ be closed $APPTC_G$ terms, if $p\sim_{pp} q$ then $p=q$.

(2) Let $p$ and $q$ be closed $APPTC_G$ terms, if $p\sim_{ps} q$ then $p=q$.

(3) Let $p$ and $q$ be closed $APPTC_G$ terms, if $p\sim_{php} q$ then $p=q$.

(3) Let $p$ and $q$ be closed $APPTC_G$ terms, if $p\sim_{phhp} q$ then $p=q$.
\end{theorem}

\subsubsection{Recursion}

In this subsection, we introduce recursion to capture infinite processes based on $APPTC_G$. In the following, $E,F,G$ are recursion specifications, $X,Y,Z$ are recursive variables.

\begin{definition}[Guarded recursive specification]
A recursive specification

$$X_1=t_1(X_1,\cdots,X_n)$$
$$...$$
$$X_n=t_n(X_1,\cdots,X_n)$$

is guarded if the right-hand sides of its recursive equations can be adapted to the form by applications of the axioms in $APTC$ and replacing recursion variables by the right-hand
sides of their recursive equations,

$((a_{111}\leftmerge\cdots\leftmerge a_{11i_1})\cdot s_1(X_1,\cdots,X_n)+\cdots+(a_{1k1}\leftmerge\cdots\leftmerge a_{1ki_k})\cdot s_k(X_1,\cdots,X_n)+(b_{111}\leftmerge\cdots\leftmerge
b_{11j_1})+\cdots+(b_{11j_1}\leftmerge\cdots\leftmerge b_{1lj_l}))\boxplus_{\pi_1}\cdots\boxplus_{\pi_{m-1}}((a_{m11}\leftmerge\cdots\leftmerge a_{m1i_1})\cdot s_1(X_1,\cdots,X_n)+
\cdots+(a_{mk1}\leftmerge\cdots\leftmerge a_{mki_k})\cdot s_k(X_1,\cdots,X_n)+(b_{m11}\leftmerge\cdots\leftmerge b_{m1j_1})+\cdots+(b_{m1j_1}\leftmerge\cdots\leftmerge b_{mlj_l}))$

where $a_{111},\cdots,a_{11i_1},a_{1k1},\cdots,a_{1ki_k},b_{111},\cdots,b_{11j_1},b_{11j_1},\cdots,b_{1lj_l},\cdots, a_{m11},\cdots,a_{m1i_1},a_{1k1},\cdots,a_{mki_k},\\b_{111},\cdots,
b_{m1j_1},b_{m1j_1},\cdots,b_{mlj_l}\in \mathbb{E}$, and the sum above is allowed to be empty, in which case it represents the deadlock $\delta$. And there does not exist an infinite
sequence of $\epsilon$-transitions $\langle X|E\rangle\rightarrow\langle X'|E\rangle\rightarrow\langle X''|E\rangle\rightarrow\cdots$.
\end{definition}

\begin{center}
    \begin{table}
        $$\frac{\langle t_i(\langle X_1|E\rangle,\cdots,\langle X_n|E\rangle),s\rangle\rightsquigarrow \langle y,s\rangle}{\langle\langle X_i|E\rangle,s\rangle\rightsquigarrow \langle y,s\rangle}$$
        $$\frac{\langle t_i(\langle X_1|E\rangle,\cdots,\langle X_n|E\rangle),s\rangle\xrightarrow{\{e_1,\cdots,e_k\}}\langle\surd,s'\rangle}{\langle\langle X_i|E\rangle,s\rangle\xrightarrow{\{e_1,\cdots,e_k\}}\langle\surd,s'\rangle}$$
        $$\frac{\langle t_i(\langle X_1|E\rangle,\cdots,\langle X_n|E\rangle),s\rangle\xrightarrow{\{e_1,\cdots,e_k\}} \langle y,s'\rangle}{\langle\langle X_i|E\rangle,s\rangle\xrightarrow{\{e_1,\cdots,e_k\}} \langle y,s'\rangle}$$
        \caption{Transition rules of guarded recursion}
        \label{TRForGRG22}
    \end{table}
\end{center}

\begin{theorem}[Conservitivity of $APPTC_G$ with guarded recursion]
$APPTC_G$ with guarded recursion is a conservative extension of $APPTC_G$.
\end{theorem}

\begin{theorem}[Congruence theorem of $APPTC_G$ with guarded recursion]
Probabilistic truly concurrent bisimulation equivalences $\sim_{pp}$, $\sim_{p}$, $\sim_{php}$ and $\sim_{phhp}$ are all congruences with respect to $APPTC_G$ with guarded recursion.
\end{theorem}

\begin{theorem}[Elimination theorem of $APPTC_G$ with linear recursion]
Each process term in $APPTC_G$ with linear recursion is equal to a process term $\langle X_1|E\rangle$ with $E$ a linear recursive specification.
\end{theorem}

\begin{theorem}[Soundness of $APPTC_G$ with guarded recursion]
Let $x$ and $y$ be $APPTC_G$ with guarded recursion terms. If $APPTC_G\textrm{ with guarded recursion}\vdash x=y$, then

(1) $x\sim_{ps} y$.

(2) $x\sim_{pp} y$.

(3) $x\sim_{php} y$.

(4) $x\sim_{phhp} y$.
\end{theorem}

\begin{theorem}[Completeness of $APPTC_G$ with linear recursion]
Let $p$ and $q$ be closed $APPTC_G$ with linear recursion terms, then,

(1) if $p\sim_{ps} q$ then $p=q$.

(2) if $p\sim_{pp} q$ then $p=q$.

(3) if $p\sim_{php} q$ then $p=q$.

(4) if $p\sim_{phhp} q$ then $p=q$.
\end{theorem}

\subsubsection{Abstraction}

To abstract away from the internal implementations of a program, and verify that the program exhibits the desired external behaviors, the silent step $\tau$ and abstraction operator
$\tau_I$ are introduced, where $I\subseteq \mathbb{E}\cup G_{at}$ denotes the internal events or guards. The silent step $\tau$ represents the internal events or guards, when we
consider the external behaviors of a process, $\tau$ steps can be removed, that is, $\tau$ steps must keep silent. The transition rule of $\tau$ is shown in Table \ref{TRForTauG22}. In
the following, let the atomic event $e$ range over $\mathbb{E}\cup\{\epsilon\}\cup\{\delta\}\cup\{\tau\}$, and $\phi$ range over $G\cup \{\tau\}$, and let the communication function
$\gamma:\mathbb{E}\cup\{\tau\}\times \mathbb{E}\cup\{\tau\}\rightarrow \mathbb{E}\cup\{\delta\}$, with each communication involved $\tau$ resulting in $\delta$. We use $\tau(s)$ to
denote $effect(\tau,s)$, for the fact that $\tau$ only change the state of internal data environment, that is, for the external data environments, $s=\tau(s)$.

\begin{center}
    \begin{table}
        $$\frac{}{\tau\rightsquigarrow\breve{\tau}}$$
        $$\frac{}{\langle\tau,s\rangle\rightarrow\langle\surd,s\rangle}\textrm{ if }test(\tau,s)$$
        $$\frac{}{\langle\tau,s\rangle\xrightarrow{\tau}\langle\surd,\tau(s)\rangle}$$
        \caption{Transition rule of the silent step}
        \label{TRForTauG22}
    \end{table}
\end{center}

\begin{definition}[Guarded linear recursive specification]\label{GLRSG}
A linear recursive specification $E$ is guarded if there does not exist an infinite sequence of $\tau$-transitions
$\langle X|E\rangle\xrightarrow{\tau}\langle X'|E\rangle\xrightarrow{\tau}\langle X''|E\rangle\xrightarrow{\tau}\cdots$, and there does not exist an infinite sequence of
$\epsilon$-transitions $\langle X|E\rangle\rightarrow\langle X'|E\rangle\rightarrow\langle X''|E\rangle\rightarrow\cdots$.
\end{definition}

\begin{theorem}[Conservitivity of $APPTC_G$ with silent step and guarded linear recursion]
$APPTC_G$ with silent step and guarded linear recursion is a conservative extension of $APPTC_G$ with linear recursion.
\end{theorem}

\begin{theorem}[Congruence theorem of $APPTC_G$ with silent step and guarded linear recursion]
Probabilistic rooted branching truly concurrent bisimulation equivalences $\approx_{prbp}$, $\approx_{prbs}$, $\approx_{prbhp}$ and $\approx_{rbhhp}$ are all congruences with respect
to $APPTC_G$ with silent step and guarded linear recursion.
\end{theorem}

We design the axioms for the silent step $\tau$ in Table \ref{AxiomsForTauG22}.

\begin{center}
\begin{table}
  \begin{tabular}{@{}ll@{}}
  \hline No. &Axiom\\
  $B1$ & $(y=y+y,z=z+z)\quad x\cdot((y+\tau\cdot(y+z))\boxplus_{\pi}w)=x\cdot((y+z)\boxplus_{\pi}w)$\\
  $B2$ & $(y=y+y,z=z+z)\quad x\leftmerge((y+\tau\leftmerge(y+z))\boxplus_{\pi}w)=x\leftmerge((y+z)\boxplus_{\pi}w)$\\
\end{tabular}
\caption{Axioms of silent step}
\label{AxiomsForTauG22}
\end{table}
\end{center}

\begin{theorem}[Elimination theorem of $APPTC_G$ with silent step and guarded linear recursion]
Each process term in $APPTC_G$ with silent step and guarded linear recursion is equal to a process term $\langle X_1|E\rangle$ with $E$ a guarded linear recursive specification.
\end{theorem}

\begin{theorem}[Soundness of $APPTC_G$ with silent step and guarded linear recursion]
Let $x$ and $y$ be $APPTC_G$ with silent step and guarded linear recursion terms. If $APPTC_G$ with silent step and guarded linear recursion $\vdash x=y$, then

(1) $x\approx_{prbs} y$.

(2) $x\approx_{prbp} y$.

(3) $x\approx_{prbhp} y$.

(4) $x\approx_{prbhhp} y$.
\end{theorem}

\begin{theorem}[Completeness of $APPTC_G$ with silent step and guarded linear recursion]
Let $p$ and $q$ be closed $APPTC_G$ with silent step and guarded linear recursion terms, then,

(1) if $p\approx_{prbs} q$ then $p=q$.

(2) if $p\approx_{prbp} q$ then $p=q$.

(3) if $p\approx_{prbhp} q$ then $p=q$.

(3) if $p\approx_{prbhhp} q$ then $p=q$.
\end{theorem}

The unary abstraction operator $\tau_I$ ($I\subseteq \mathbb{E}\cup G_{at}$) renames all atomic events or atomic guards in $I$ into $\tau$. $APPTC_G$ with silent step and abstraction
operator is called $APPTC_{G_{\tau}}$. The transition rules of operator $\tau_I$ are shown in Table \ref{TRForAbstractionG22}.

\begin{center}
    \begin{table}
        $$\frac{\langle x,s\rangle\rightsquigarrow \langle x',s\rangle}{\langle \tau_I(x),s\rangle\rightsquigarrow\langle\tau_I(x'),s\rangle}$$
        $$\frac{\langle x,s\rangle\xrightarrow{e}\langle\surd,s'\rangle}{\langle\tau_I(x),s\rangle\xrightarrow{e}\langle\surd,s'\rangle}\quad e\notin I
        \quad\quad\frac{\langle x,s\rangle\xrightarrow{e}\langle x',s'\rangle}{\langle\tau_I(x),s\rangle\xrightarrow{e}\langle\tau_I(x'),s'\rangle}\quad e\notin I$$

        $$\frac{\langle x,s\rangle\xrightarrow{e}\langle\surd,s'\rangle}{\langle\tau_I(x),s\rangle\xrightarrow{\tau}\langle\surd,\tau(s)\rangle}\quad e\in I
        \quad\quad\frac{\langle x,s\rangle\xrightarrow{e}\langle x',s'\rangle}{\langle\tau_I(x),s\rangle\xrightarrow{\tau}\langle\tau_I(x'),\tau(s)\rangle}\quad e\in I$$
        \caption{Transition rule of the abstraction operator}
        \label{TRForAbstractionG22}
    \end{table}
\end{center}

\begin{theorem}[Conservitivity of $APPTC_{G_{\tau}}$ with guarded linear recursion]
$APPTC_{G_{\tau}}$ with guarded linear recursion is a conservative extension of $APPTC_G$ with silent step and guarded linear recursion.
\end{theorem}

\begin{theorem}[Congruence theorem of $APPTC_{G_{\tau}}$ with guarded linear recursion]
Probabilistic rooted branching truly concurrent bisimulation equivalences $\approx_{prbp}$, $\approx_{prbs}$, $\approx_{prbhp}$ and $\approx_{prbhhp}$ are all congruences with respect
to $APPTC_{G_{\tau}}$ with guarded linear recursion.
\end{theorem}

We design the axioms for the abstraction operator $\tau_I$ in Table \ref{AxiomsForAbstractionG22}.

\begin{center}
\begin{table}
  \begin{tabular}{@{}ll@{}}
\hline No. &Axiom\\
  $TI1$ & $e\notin I\quad \tau_I(e)=e$\\
  $TI2$ & $e\in I\quad \tau_I(e)=\tau$\\
  $TI3$ & $\tau_I(\delta)=\delta$\\
  $TI4$ & $\tau_I(x+y)=\tau_I(x)+\tau_I(y)$\\
  $PTI1$ & $\tau_I(x\boxplus_{\pi}y)=\tau_I(x)\boxplus_{\pi}\tau_I(y)$\\
  $TI5$ & $\tau_I(x\cdot y)=\tau_I(x)\cdot\tau_I(y)$\\
  $TI6$ & $\tau_I(x\leftmerge y)=\tau_I(x)\leftmerge\tau_I(y)$\\
  $G28$ & $\phi\notin I\quad \tau_I(\phi)=\phi$\\
  $G29$ & $\phi\in I\quad \tau_I(\phi)=\tau$\\
\end{tabular}
\caption{Axioms of abstraction operator}
\label{AxiomsForAbstractionG22}
\end{table}
\end{center}

\begin{theorem}[Soundness of $APPTC_{G_{\tau}}$ with guarded linear recursion]
Let $x$ and $y$ be $APPTC_{G_{\tau}}$ with guarded linear recursion terms. If $APPTC_{G_{\tau}}$ with guarded linear recursion $\vdash x=y$, then

(1) $x\approx_{prbs} y$.

(2) $x\approx_{prbp} y$.

(3) $x\approx_{prbhp} y$.

(4) $x\approx_{prbhhp} y$.
\end{theorem}

Though $\tau$-loops are prohibited in guarded linear recursive specifications in a specifiable way, they can be constructed using the abstraction operator, for example, there exist
$\tau$-loops in the process term $\tau_{\{a\}}(\langle X|X=aX\rangle)$. To avoid $\tau$-loops caused by $\tau_I$ and ensure fairness, we introduce the following recursive verification
rules as Table \ref{RVR22} shows, note that $i_1,\cdots, i_m,j_1,\cdots,j_n\in I\subseteq \mathbb{E}\setminus\{\tau\}$.

\begin{center}
\begin{table}
    $$VR_1\quad \frac{x=y+(i_1\leftmerge\cdots\leftmerge i_m)\cdot x, y=y+y}{\tau\cdot\tau_I(x)=\tau\cdot \tau_I(y)}$$
    $$VR_2\quad \frac{x=z\boxplus_{\pi}(u+(i_1\leftmerge\cdots\leftmerge i_m)\cdot x),z=z+u,z=z+z}{\tau\cdot\tau_I(x)=\tau\cdot\tau_I(z)}$$
    $$VR_3\quad \frac{x=z+(i_1\leftmerge\cdots\leftmerge i_m)\cdot y,y=z\boxplus_{\pi}(u+(j_1\leftmerge\cdots\leftmerge j_n)\cdot x), z=z+u,z=z+z}{\tau\cdot\tau_I(x)=\tau\cdot\tau_I(y')\textrm{ for }y'=z\boxplus_{\pi}(u+(i_1\leftmerge\cdots\leftmerge i_m)\cdot y')}$$
\caption{Recursive verification rules}
\label{RVR22}
\end{table}
\end{center}

\begin{theorem}[Soundness of $VR_1,VR_2,VR_3$]
$VR_1$, $VR_2$ and $VR_3$ are sound modulo probabilistic rooted branching truly concurrent bisimulation equivalences $\approx_{prbp}$, $\approx_{prbs}$, $\approx_{prbhp}$ and $\approx_{prbhhp}$.
\end{theorem}

\subsection{Operational Semantics for Quantum Computing}

In quantum processes, to avoid the abuse of quantum information which may violate the no-cloning theorem, a quantum configuration $\langle C,s,\varrho\rangle$
\cite{PSQP} \cite{QPA} \cite{QPA2} \cite{CQP} \cite{CQP2} \cite{qCCS} \cite{BQP} \cite{PSQP} \cite{SBQP} is usually consisted of a traditional configuration $C$, traditional state information $s$ and state information $\varrho$ of
all (public) quantum information variables. Though quantum information variables are not explicitly defined and are hidden behind quantum operations or unitary operators, more importantly, the
state information $\varrho$ is the effects of execution of a series of quantum operations or unitary operators on involved quantum systems, the execution of a series of quantum operations
or unitary operators should not only obey the restrictions of the structure of the process terms, but also those of quantum mechanics principles. Through the state information
$\varrho$, we can check and observe the functions of quantum mechanics principles, such as quantum entanglement, quantum measurement, etc.

So, the operational semantics of quantum processes should be defined based on quantum process configuration $\langle C,s,\varrho\rangle$, in which $\varrho=\varsigma$ of two state
information $\varrho$ and $\varsigma$ means equality under the framework of quantum information and quantum computing, that is, these two quantum processes are in the same quantum
state.

\begin{definition}[Pomset transitions and step]
Let $\mathcal{E}$ be a PES and let $C\in\mathcal{C}(\mathcal{E})$, and $\emptyset\neq X\subseteq \mathbb{E}$, if $C\cap X=\emptyset$ and $C'=C\cup X\in\mathcal{C}(\mathcal{E})$, then
$\langle C,s,\varrho\rangle\xrightarrow{X} \langle C',s,\varrho'\rangle$ is called a pomset transition from $\langle C,s,\varrho\rangle$ to $\langle C',s,\varrho'\rangle$;
if $C\cap X=\emptyset$ and $C'=C\cup X\in\mathcal{C}(\mathcal{E})$, then
$\langle C,s,\varrho\rangle\xrightarrow{X} \langle C',s',\varrho\rangle$ is also called a pomset transition from $\langle C,s,\varrho\rangle$ to $\langle C',s',\varrho\rangle$. When the events in $X$ are pairwise
concurrent, we say that $\langle C,s,\varrho\rangle\xrightarrow{X}\langle C',s,\varrho'\rangle$ or $\langle C,s,\varrho\rangle\xrightarrow{X}\langle C',s',\varrho\rangle$ is a step. It is obvious that $\rightarrow^*\xrightarrow{X}\rightarrow^*=\xrightarrow{X}$ and
$\rightarrow^*\xrightarrow{e}\rightarrow^*=\xrightarrow{e}$ for any $e\in\mathbb{E}$ and $X\subseteq\mathbb{E}$.
\end{definition}

\begin{definition}[Weak pomset transitions and weak step]
Let $\mathcal{E}$ be a PES and let $C\in\mathcal{C}(\mathcal{E})$, and $\emptyset\neq X\subseteq \hat{\mathbb{E}}$, if $C\cap X=\emptyset$ and
$\hat{C'}=\hat{C}\cup X\in\mathcal{C}(\mathcal{E})$, then $\langle C,s,\varrho\rangle\xRightarrow{X} \langle C',s,\varrho'\rangle$ or $\langle C,s,\varrho\rangle\xRightarrow{X} \langle C',s',\varrho\rangle$
is called a weak pomset transition from $\langle C,s,\varrho\rangle$ to
$\langle C',s,\varrho'\rangle$ or $\langle C,s,\varrho\rangle$ to $\langle C',s',\varrho\rangle$, where we define $\xRightarrow{e}\triangleq\xrightarrow{\tau^*}\xrightarrow{e}\xrightarrow{\tau^*}$. And
$\xRightarrow{X}\triangleq\xrightarrow{\tau^*}\xrightarrow{e}\xrightarrow{\tau^*}$, for every $e\in X$. When the events in $X$ are pairwise concurrent, we say that
$\langle C,s,\varrho\rangle\xRightarrow{X}\langle C',s,\varrho'\rangle$ is a weak step.
\end{definition}

\begin{definition}[Probabilistic transitions]
Let $\mathcal{E}$ be a PES and let $C\in\mathcal{C}(\mathcal{E})$, the transition $\langle C,s,\varrho\rangle\xrsquigarrow{\pi} \langle C^{\pi},s,\varrho\rangle$ is called a probabilistic
transition
from $\langle C,s,\varrho\rangle$ to $\langle C^{\pi},s,\varrho\rangle$.
\end{definition}

We will also suppose that all the PESs in this chapter are image finite, that is, for any PES $\mathcal{E}$ and $C\in \mathcal{C}(\mathcal{E})$ and $a\in \Lambda$,
$\{\langle C,s,\varrho\rangle\xrsquigarrow{\pi} \langle C^{\pi},s,\varrho\rangle\}$,
$\{e\in \mathbb{E}|\langle C,s,\varrho\rangle\xrightarrow{e} \langle C',s,\varrho'\rangle\wedge \lambda(e)=a\}$, $\{e\in \mathbb{E}|\langle C,s,\varrho\rangle\xrightarrow{e} \langle C',s',\varrho\rangle\wedge \lambda(e)=a\}$ and
$\{e\in\hat{\mathbb{E}}|\langle C,s,\varrho\rangle\xRightarrow{e} \langle C',s,\varrho'\rangle\wedge \lambda(e)=a\}$, $\{e\in\hat{\mathbb{E}}|\langle C,s,\varrho\rangle\xRightarrow{e} \langle C',s',\varrho\rangle\wedge \lambda(e)=a\}$ is finite.

\begin{definition}[Branching pomset, step bisimulation]\label{BPSBG}
Assume a special termination predicate $\downarrow$, and let $\surd$ represent a state with $\surd\downarrow$. Let $\mathcal{E}_1$, $\mathcal{E}_2$ be PESs. A branching pomset
bisimulation is a relation $R\subseteq\langle\mathcal{C}(\mathcal{E}_1),S,\varrho\rangle\times\langle\mathcal{C}(\mathcal{E}_2),S,\varrho\rangle$, such that:
 \begin{enumerate}
   \item if $(\langle C_1,s,\varrho\rangle,\langle C_2,s,\varrho\rangle)\in R$, and $\langle C_1,s,\varrho\rangle\xrightarrow{X}\langle C_1',s,\varrho'\rangle$ then
   \begin{itemize}
     \item either $X\equiv \tau^*$, and $(\langle C_1',s,\varrho'\rangle,\langle C_2,s,\varrho\rangle)\in R$ with $\varrho'\in \tau(\varrho)$;
     \item or there is a sequence of (zero or more) $\tau$-transitions $\langle C_2,s,\varrho\rangle\xrightarrow{\tau^*} \langle C_2^0,\varrho^0\rangle$, such that
     $(\langle C_1,s,\varrho\rangle,\langle C_2^0,\varrho^0\rangle)\in R$ and $\langle C_2^0,\varrho^0\rangle\xRightarrow{X}\langle C_2',s,\varrho'\rangle$ with
     $(\langle C_1',s,\varrho'\rangle,\langle C_2',s,\varrho'\rangle)\in R$;
   \end{itemize}
   \item if $(\langle C_1,s,\varrho\rangle,\langle C_2,s,\varrho\rangle)\in R$, and $\langle C_1,s,\varrho\rangle\xrightarrow{X}\langle C_1',s',\varrho\rangle$ then
   \begin{itemize}
     \item either $X\equiv \tau^*$, and $(\langle C_1',s',\varrho\rangle,\langle C_2,s,\varrho\rangle)\in R$ with $s'\in \tau(s)$;
     \item or there is a sequence of (zero or more) $\tau$-transitions $\langle C_2,s,\varrho\rangle\xrightarrow{\tau^*} \langle C_2^0,\varrho^0\rangle$, such that
     $(\langle C_1,s,\varrho\rangle,\langle C_2^0,\varrho^0\rangle)\in R$ and $\langle C_2^0,\varrho^0\rangle\xRightarrow{X}\langle C_2',s',\varrho\rangle$ with
     $(\langle C_1',s',\varrho\rangle,\langle C_2',s',\varrho\rangle)\in R$;
   \end{itemize}
   \item if $(\langle C_1,s,\varrho\rangle,\langle C_2,s,\varrho\rangle)\in R$, and $\langle C_2,s,\varrho\rangle\xrightarrow{X}\langle C_2',s,\varrho'\rangle$ then
   \begin{itemize}
     \item either $X\equiv \tau^*$, and $(\langle C_1,s,\varrho\rangle,\langle C_2',s,\varrho'\rangle)\in R$;
     \item or there is a sequence of (zero or more) $\tau$-transitions $\langle C_1,s,\varrho\rangle\xrightarrow{\tau^*} \langle C_1^0,\varrho^0\rangle$, such that
     $(\langle C_1^0,\varrho^0\rangle,\langle C_2,s,\varrho\rangle)\in R$ and $\langle C_1^0,\varrho^0\rangle\xRightarrow{X}\langle C_1',s,\varrho'\rangle$ with $(\langle C_1',s,\varrho'\rangle,\langle C_2',s,\varrho'\rangle)\in R$;
   \end{itemize}
   \item if $(\langle C_1,s,\varrho\rangle,\langle C_2,s,\varrho\rangle)\in R$, and $\langle C_2,s,\varrho\rangle\xrightarrow{X}\langle C_2',s',\varrho\rangle$ then
   \begin{itemize}
     \item either $X\equiv \tau^*$, and $(\langle C_1,s,\varrho\rangle,\langle C_2',s',\varrho\rangle)\in R$;
     \item or there is a sequence of (zero or more) $\tau$-transitions $\langle C_1,s,\varrho\rangle\xrightarrow{\tau^*} \langle C_1^0,\varrho^0\rangle$, such that
     $(\langle C_1^0,\varrho^0\rangle,\langle C_2,s,\varrho\rangle)\in R$ and $\langle C_1^0,\varrho^0\rangle\xRightarrow{X}\langle C_1',s',\varrho\rangle$ with $(\langle C_1',s',\varrho\rangle,\langle C_2',s',\varrho\rangle)\in R$;
   \end{itemize}
   \item if $(\langle C_1,s,\varrho\rangle,\langle C_2,s,\varrho\rangle)\in R$ and $\langle C_1,s,\varrho\rangle\downarrow$, then there is a sequence of (zero or more) $\tau$-transitions
   $\langle C_2,s,\varrho\rangle\xrightarrow{\tau^*}\langle C_2^0,\varrho^0\rangle$ such that $(\langle C_1,s,\varrho\rangle,\langle C_2^0,\varrho^0\rangle)\in R$ and
   $\langle C_2^0,\varrho^0\rangle\downarrow$;
   \item if $(\langle C_1,s,\varrho\rangle,\langle C_2,s,\varrho\rangle)\in R$ and $\langle C_2,s,\varrho\rangle\downarrow$, then there is a sequence of (zero or more) $\tau$-transitions
   $\langle C_1,s,\varrho\rangle\xrightarrow{\tau^*}\langle C_1^0,\varrho^0\rangle$ such that $(\langle C_1^0,\varrho^0\rangle,\langle C_2,s,\varrho\rangle)\in R$ and
   $\langle C_1^0,\varrho^0\rangle\downarrow$.
 \end{enumerate}

We say that $\mathcal{E}_1$, $\mathcal{E}_2$ are branching pomset bisimilar, written $\mathcal{E}_1\approx_{bp}\mathcal{E}_2$, if there exists a branching pomset bisimulation $R$, such
that $(\langle\emptyset,\emptyset\rangle,\langle\emptyset,\emptyset\rangle)\in R$.

By replacing pomset transitions with steps, we can get the definition of branching step bisimulation. When PESs $\mathcal{E}_1$ and $\mathcal{E}_2$ are branching step bisimilar, we
write $\mathcal{E}_1\approx_{bs}\mathcal{E}_2$.
\end{definition}

\begin{definition}[Rooted branching pomset, step bisimulation]\label{RBPSBG}
Assume a special termination predicate $\downarrow$, and let $\surd$ represent a state with $\surd\downarrow$. Let $\mathcal{E}_1$, $\mathcal{E}_2$ be PESs. A rooted branching pomset bisimulation is a relation $R\subseteq\langle\mathcal{C}(\mathcal{E}_1),S\rangle\times\langle\mathcal{C}(\mathcal{E}_2),S\rangle$, such that:
 \begin{enumerate}
   \item if $(\langle C_1,s,\varrho\rangle,\langle C_2,s,\varrho\rangle)\in R$, and $\langle C_1,s,\varrho\rangle\xrightarrow{X}\langle C_1',s,\varrho'\rangle$ then
   $\langle C_2,s,\varrho\rangle\xrightarrow{X}\langle C_2',s,\varrho'\rangle$ with $\langle C_1',s,\varrho'\rangle\approx_{bp}\langle C_2',s,\varrho'\rangle$;
   \item if $(\langle C_1,s,\varrho\rangle,\langle C_2,s,\varrho\rangle)\in R$, and $\langle C_1,s,\varrho\rangle\xrightarrow{X}\langle C_1',s',\varrho\rangle$ then
   $\langle C_2,s,\varrho\rangle\xrightarrow{X}\langle C_2',s',\varrho\rangle$ with $\langle C_1',s',\varrho\rangle\approx_{bp}\langle C_2',s',\varrho\rangle$;
   \item if $(\langle C_1,s,\varrho\rangle,\langle C_2,s,\varrho\rangle)\in R$, and $\langle C_2,s,\varrho\rangle\xrightarrow{X}\langle C_2',s,\varrho'\rangle$ then
   $\langle C_1,s,\varrho\rangle\xrightarrow{X}\langle C_1',s,\varrho'\rangle$ with $\langle C_1',s,\varrho'\rangle\approx_{bp}\langle C_2',s,\varrho'\rangle$;
   \item if $(\langle C_1,s,\varrho\rangle,\langle C_2,s,\varrho\rangle)\in R$, and $\langle C_2,s,\varrho\rangle\xrightarrow{X}\langle C_2',s',\varrho\rangle$ then
   $\langle C_1,s,\varrho\rangle\xrightarrow{X}\langle C_1',s',\varrho\rangle$ with $\langle C_1',s',\varrho\rangle\approx_{bp}\langle C_2',s',\varrho\rangle$;
   \item if $(\langle C_1,s,\varrho\rangle,\langle C_2,s,\varrho\rangle)\in R$ and $\langle C_1,s,\varrho\rangle\downarrow$, then $\langle C_2,s,\varrho\rangle\downarrow$;
   \item if $(\langle C_1,s,\varrho\rangle,\langle C_2,s,\varrho\rangle)\in R$ and $\langle C_2,s,\varrho\rangle\downarrow$, then $\langle C_1,s,\varrho\rangle\downarrow$.
 \end{enumerate}

We say that $\mathcal{E}_1$, $\mathcal{E}_2$ are rooted branching pomset bisimilar, written $\mathcal{E}_1\approx_{rbp}\mathcal{E}_2$, if there exists a rooted branching pomset
bisimulation $R$, such that $(\langle\emptyset,\emptyset\rangle,\langle\emptyset,\emptyset\rangle)\in R$.

By replacing pomset transitions with steps, we can get the definition of rooted branching step bisimulation. When PESs $\mathcal{E}_1$ and $\mathcal{E}_2$ are rooted branching step
bisimilar, we write $\mathcal{E}_1\approx_{rbs}\mathcal{E}_2$.
\end{definition}

\begin{definition}[Branching (hereditary) history-preserving bisimulation]\label{BHHPBG}
Assume a special termination predicate $\downarrow$, and let $\surd$ represent a state with $\surd\downarrow$. A branching history-preserving (hp-) bisimulation is a weakly posetal
relation $R\subseteq\langle\mathcal{C}(\mathcal{E}_1),S\rangle\overline{\times}\langle\mathcal{C}(\mathcal{E}_2),S\rangle$ such that:

 \begin{enumerate}
   \item if $(\langle C_1,s,\varrho\rangle,f,\langle C_2,s,\varrho\rangle)\in R$, and $\langle C_1,s,\varrho\rangle\xrightarrow{e_1}\langle C_1',s,\varrho'\rangle$ then
   \begin{itemize}
     \item either $e_1\equiv \tau$, and $(\langle C_1',s,\varrho'\rangle,f[e_1\mapsto \tau^{e_1}],\langle C_2,s,\varrho\rangle)\in R$;
     \item or there is a sequence of (zero or more) $\tau$-transitions $\langle C_2,s,\varrho\rangle\xrightarrow{\tau^*} \langle C_2^0,\varrho^0\rangle$, such that
     $(\langle C_1,s,\varrho\rangle,f,\langle C_2^0,\varrho^0\rangle)\in R$ and $\langle C_2^0,\varrho^0\rangle\xrightarrow{e_2}\langle C_2',s,\varrho'\rangle$ with
     $(\langle C_1',s,\varrho'\rangle,f[e_1\mapsto e_2],\langle C_2',s,\varrho'\rangle)\in R$;
   \end{itemize}
   \item if $(\langle C_1,s,\varrho\rangle,f,\langle C_2,s,\varrho\rangle)\in R$, and $\langle C_1,s,\varrho\rangle\xrightarrow{e_1}\langle C_1',s',\varrho\rangle$ then
   \begin{itemize}
     \item either $e_1\equiv \tau$, and $(\langle C_1',s',\varrho\rangle,f[e_1\mapsto \tau^{e_1}],\langle C_2,s,\varrho\rangle)\in R$;
     \item or there is a sequence of (zero or more) $\tau$-transitions $\langle C_2,s,\varrho\rangle\xrightarrow{\tau^*} \langle C_2^0,\varrho^0\rangle$, such that
     $(\langle C_1,s,\varrho\rangle,f,\langle C_2^0,\varrho^0\rangle)\in R$ and $\langle C_2^0,\varrho^0\rangle\xrightarrow{e_2}\langle C_2',s',\varrho\rangle$ with
     $(\langle C_1',s',\varrho\rangle,f[e_1\mapsto e_2],\langle C_2',s',\varrho\rangle)\in R$;
   \end{itemize}
   \item if $(\langle C_1,s,\varrho\rangle,f,\langle C_2,s,\varrho\rangle)\in R$, and $\langle C_2,s,\varrho\rangle\xrightarrow{e_2}\langle C_2',s,\varrho'\rangle$ then
   \begin{itemize}
     \item either $e_2\equiv \tau$, and $(\langle C_1,s,\varrho\rangle,f[e_2\mapsto \tau^{e_2}],\langle C_2',s,\varrho'\rangle)\in R$;
     \item or there is a sequence of (zero or more) $\tau$-transitions $\langle C_1,s,\varrho\rangle\xrightarrow{\tau^*} \langle C_1^0,\varrho^0\rangle$, such that
     $(\langle C_1^0,\varrho^0\rangle,f,\langle C_2,s,\varrho\rangle)\in R$ and $\langle C_1^0,\varrho^0\rangle\xrightarrow{e_1}\langle C_1',s,\varrho'\rangle$ with
     $(\langle C_1',s,\varrho'\rangle,f[e_2\mapsto e_1],\langle C_2',s,\varrho'\rangle)\in R$;
   \end{itemize}
   \begin{itemize}
     \item either $e_2\equiv \tau$, and $(\langle C_1,s,\varrho\rangle,f[e_2\mapsto \tau^{e_2}],\langle C_2',s',\varrho\rangle)\in R$;
     \item or there is a sequence of (zero or more) $\tau$-transitions $\langle C_1,s,\varrho\rangle\xrightarrow{\tau^*} \langle C_1^0,\varrho^0\rangle$, such that
     $(\langle C_1^0,\varrho^0\rangle,f,\langle C_2,s,\varrho\rangle)\in R$ and $\langle C_1^0,\varrho^0\rangle\xrightarrow{e_1}\langle C_1',s',\varrho\rangle$ with
     $(\langle C_1',s',\varrho\rangle,f[e_2\mapsto e_1],\langle C_2',s',\varrho\rangle)\in R$;
   \end{itemize}
   \item if $(\langle C_1,s,\varrho\rangle,f,\langle C_2,s,\varrho\rangle)\in R$ and $\langle C_1,s,\varrho\rangle\downarrow$, then there is a sequence of (zero or more)
   $\tau$-transitions $\langle C_2,s,\varrho\rangle\xrightarrow{\tau^*}\langle C_2^0,\varrho^0\rangle$ such that $(\langle C_1,s,\varrho\rangle,f,\langle C_2^0,\varrho^0\rangle)\in R$
   and    $\langle C_2^0,\varrho^0\rangle\downarrow$;
   \item if $(\langle C_1,s,\varrho\rangle,f,\langle C_2,s,\varrho\rangle)\in R$ and $\langle C_2,s,\varrho\rangle\downarrow$, then there is a sequence of (zero or more) $\tau$-transitions
   $\langle C_1,s,\varrho\rangle\xrightarrow{\tau^*}\langle C_1^0,\varrho^0\rangle$ such that $(\langle C_1^0,\varrho^0\rangle,f,\langle C_2,s,\varrho\rangle)\in R$ and
   $\langle C_1^0,\varrho^0\rangle\downarrow$.
 \end{enumerate}

$\mathcal{E}_1,\mathcal{E}_2$ are branching history-preserving (hp-)bisimilar and are written $\mathcal{E}_1\approx_{bhp}\mathcal{E}_2$ if there exists a branching hp-bisimulation $R$
such that $(\langle\emptyset,\emptyset\rangle,\emptyset,\langle\emptyset,\emptyset\rangle)\in R$.

A branching hereditary history-preserving (hhp-)bisimulation is a downward closed branching hp-bisimulation. $\mathcal{E}_1,\mathcal{E}_2$ are branching hereditary history-preserving
(hhp-)bisimilar and are written $\mathcal{E}_1\approx_{bhhp}\mathcal{E}_2$.
\end{definition}

\begin{definition}[Rooted branching (hereditary) history-preserving bisimulation]\label{RBHHPBG}
Assume a special termination predicate $\downarrow$, and let $\surd$ represent a state with $\surd\downarrow$. A rooted branching history-preserving (hp-) bisimulation is a weakly
posetal relation $R\subseteq\langle\mathcal{C}(\mathcal{E}_1),S\rangle\overline{\times}\langle\mathcal{C}(\mathcal{E}_2),S\rangle$ such that:

 \begin{enumerate}
   \item if $(\langle C_1,s,\varrho\rangle,f,\langle C_2,s,\varrho\rangle)\in R$, and $\langle C_1,s,\varrho\rangle\xrightarrow{e_1}\langle C_1',s,\varrho'\rangle$, then
   $\langle C_2,s,\varrho\rangle\xrightarrow{e_2}\langle C_2',s,\varrho'\rangle$ with $\langle C_1',s,\varrho'\rangle\approx_{bhp}\langle C_2',s,\varrho'\rangle$;
   \item if $(\langle C_1,s,\varrho\rangle,f,\langle C_2,s,\varrho\rangle)\in R$, and $\langle C_1,s,\varrho\rangle\xrightarrow{e_1}\langle C_1',s',\varrho\rangle$, then
   $\langle C_2,s,\varrho\rangle\xrightarrow{e_2}\langle C_2',s',\varrho\rangle$ with $\langle C_1',s',\varrho\rangle\approx_{bhp}\langle C_2',s',\varrho\rangle$;
   \item if $(\langle C_1,s,\varrho\rangle,f,\langle C_2,s,\varrho\rangle)\in R$, and $\langle C_2,s,\varrho\rangle\xrightarrow{e_2}\langle C_2',s,\varrho'\rangle$, then
   $\langle C_1,s,\varrho\rangle\xrightarrow{e_1}\langle C_1',s,\varrho'\rangle$ with $\langle C_1',s,\varrho'\rangle\approx_{bhp}\langle C_2',s,\varrho'\rangle$;
   \item if $(\langle C_1,s,\varrho\rangle,f,\langle C_2,s,\varrho\rangle)\in R$, and $\langle C_2,s,\varrho\rangle\xrightarrow{e_2}\langle C_2',s',\varrho\rangle$, then
   $\langle C_1,s,\varrho\rangle\xrightarrow{e_1}\langle C_1',s',\varrho\rangle$ with $\langle C_1',s',\varrho\rangle\approx_{bhp}\langle C_2',s',\varrho\rangle$;
   \item if $(\langle C_1,s,\varrho\rangle,f,\langle C_2,s,\varrho\rangle)\in R$ and $\langle C_1,s,\varrho\rangle\downarrow$, then $\langle C_2,s,\varrho\rangle\downarrow$;
   \item if $(\langle C_1,s,\varrho\rangle,f,\langle C_2,s,\varrho\rangle)\in R$ and $\langle C_2,s,\varrho\rangle\downarrow$, then $\langle C_1,s,\varrho\rangle\downarrow$.
 \end{enumerate}

$\mathcal{E}_1,\mathcal{E}_2$ are rooted branching history-preserving (hp-)bisimilar and are written $\mathcal{E}_1\approx_{rbhp}\mathcal{E}_2$ if there exists a rooted branching
hp-bisimulation $R$ such that $(\langle\emptyset,\emptyset\rangle,\emptyset,\langle\emptyset,\emptyset\rangle)\in R$.

A rooted branching hereditary history-preserving (hhp-)bisimulation is a downward closed rooted branching hp-bisimulation. $\mathcal{E}_1,\mathcal{E}_2$ are rooted branching hereditary
history-preserving (hhp-)bisimilar and are written $\mathcal{E}_1\approx_{rbhhp}\mathcal{E}_2$.
\end{definition}

\begin{definition}[Probabilistic pomset, step bisimulation]\label{PSBG}
Let $\mathcal{E}_1$, $\mathcal{E}_2$ be PESs. A probabilistic pomset bisimulation is a relation $R\subseteq\langle\mathcal{C}(\mathcal{E}_1),S\rangle\times\langle\mathcal{C}(\mathcal{E}_2),S\rangle$,
such that (1) if $(\langle C_1,s,\varrho\rangle,\langle C_2,s,\varrho\rangle)\in R$, and $\langle C_1,s,\varrho\rangle\xrightarrow{X_1}\langle C_1',s,\varrho'\rangle$ then
$\langle C_2,s,\varrho\rangle\xrightarrow{X_2}\langle C_2',s,\varrho'\rangle$, with $X_1\subseteq \mathbb{E}_1$, $X_2\subseteq \mathbb{E}_2$, $X_1\sim X_2$ and
$(\langle C_1',s,\varrho'\rangle,\langle C_2',s,\varrho'\rangle)\in R$ for all $\varrho,\varrho'\in S$, and vice-versa; (2) if $(\langle C_1,s,\varrho\rangle,\langle C_2,s,\varrho\rangle)\in R$, and $\langle C_1,s,\varrho\rangle\xrightarrow{X_1}\langle C_1',s',\varrho\rangle$ then
$\langle C_2,s,\varrho\rangle\xrightarrow{X_2}\langle C_2',s',\varrho\rangle$, with $X_1\subseteq \mathbb{E}_1$, $X_2\subseteq \mathbb{E}_2$, $X_1\sim X_2$ and
$(\langle C_1',s',\varrho\rangle,\langle C_2',s',\varrho\rangle)\in R$ for all $s,s'\in S$, and vice-versa; (3) if $(\langle C_1,s,\varrho\rangle,\langle C_2,s,\varrho\rangle)\in R$, and $\langle C_1,s,\varrho\rangle\xrsquigarrow{\pi}\langle C_1^{\pi},s,\varrho\rangle$
then $\langle C_2,s,\varrho\rangle\xrsquigarrow{\pi}\langle C_2^{\pi},s,\varrho\rangle$ and $(\langle C_1^{\pi},s,\varrho\rangle,\langle C_2^{\pi},s,\varrho\rangle)\in R$, and vice-versa; (4) if $(\langle C_1,s,\varrho\rangle,\langle C_2,s,\varrho\rangle)\in R$,
then $\mu(C_1,C)=\mu(C_2,C)$ for each $C\in\mathcal{C}(\mathcal{E})/R$; (5) $[\surd]_R=\{\surd\}$. We say that $\mathcal{E}_1$, $\mathcal{E}_2$ are probabilistic pomset bisimilar, written
$\mathcal{E}_1\sim_{pp}\mathcal{E}_2$, if there exists a probabilistic pomset bisimulation $R$, such that $(\langle\emptyset,\emptyset\rangle,\langle\emptyset,\emptyset\rangle)\in R$.
By replacing probabilistic pomset transitions with probabilistic steps, we can get the definition of probabilistic step bisimulation. When PESs $\mathcal{E}_1$ and $\mathcal{E}_2$ are
probabilistic step bisimilar, we write $\mathcal{E}_1\sim_{ps}\mathcal{E}_2$.
\end{definition}

\begin{definition}[Weakly probabilistic pomset, step bisimulation]\label{WPSBG}
Let $\mathcal{E}_1$, $\mathcal{E}_2$ be PESs. A weakly probabilistic pomset bisimulation is a relation $R\subseteq\langle\mathcal{C}(\mathcal{E}_1),S\rangle\times\langle\mathcal{C}(\mathcal{E}_2),S\rangle$,
such that (1) if $(\langle C_1,s,\varrho\rangle,\langle C_2,s,\varrho\rangle)\in R$, and $\langle C_1,s,\varrho\rangle\xRightarrow{X_1}\langle C_1',s,\varrho'\rangle$ or
$\langle C_1,s,\varrho\rangle\xRightarrow{X_1}\langle C_1',s',\varrho\rangle$ then
$\langle C_2,s,\varrho\rangle\xRightarrow{X_2}\langle C_2',s,\varrho'\rangle$ or $\langle C_2,s,\varrho\rangle\xRightarrow{X_2}\langle C_2',s',\varrho\rangle$,
with $X_1\subseteq \hat{\mathbb{E}_1}$, $X_2\subseteq \hat{\mathbb{E}_2}$, $X_1\sim X_2$ and
$(\langle C_1',s,\varrho'\rangle,\langle C_2',s,\varrho'\rangle)\in R$ and $(\langle C_1',s,\varrho'\rangle,\langle C_2',s',\varrho\rangle)\in R$ for all $s,s'\in S$, and vice-versa; (2) if $(\langle C_1,s,\varrho\rangle,\langle C_2,s,\varrho\rangle)\in R$, and $\langle C_1,s,\varrho\rangle\xrsquigarrow{\pi}\langle C_1^{\pi},s,\varrho\rangle$
then $\langle C_2,s,\varrho\rangle\xrsquigarrow{\pi}\langle C_2^{\pi},s,\varrho\rangle$ and $(\langle C_1^{\pi},s,\varrho\rangle,\langle C_2^{\pi},s,\varrho\rangle)\in R$, and vice-versa; (3) if $(\langle C_1,s,\varrho\rangle,\langle C_2,s,\varrho\rangle)\in R$,
then $\mu(C_1,C)=\mu(C_2,C)$ for each $C\in\mathcal{C}(\mathcal{E})/R$; (4) $[\surd]_R=\{\surd\}$. We say that $\mathcal{E}_1$, $\mathcal{E}_2$ are weakly probabilistic pomset bisimilar,
written $\mathcal{E}_1\approx_{pp}\mathcal{E}_2$, if there exists a weakly probabilistic pomset bisimulation $R$, such that
$(\langle\emptyset,\emptyset\rangle,\langle\emptyset,\emptyset\rangle)\in R$. By replacing weakly probabilistic pomset transitions with weakly probabilistic steps, we can get the
definition of weakly probabilistic step bisimulation. When PESs $\mathcal{E}_1$ and $\mathcal{E}_2$ are weakly probabilistic step bisimilar, we write
$\mathcal{E}_1\approx_{ps}\mathcal{E}_2$.
\end{definition}

\begin{definition}[Posetal product]
Given two PESs $\mathcal{E}_1$, $\mathcal{E}_2$, the posetal product of their configurations, denoted
$\langle\mathcal{C}(\mathcal{E}_1),S\rangle\overline{\times}\langle\mathcal{C}(\mathcal{E}_2),S\rangle$, is defined as

$$\{(\langle C_1,s,\varrho\rangle,f,\langle C_2,s,\varrho\rangle)|C_1\in\mathcal{C}(\mathcal{E}_1),C_2\in\mathcal{C}(\mathcal{E}_2),f:C_1\rightarrow C_2 \textrm{ isomorphism}\}.$$

A subset $R\subseteq\langle\mathcal{C}(\mathcal{E}_1),S\rangle\overline{\times}\langle\mathcal{C}(\mathcal{E}_2),S\rangle$ is called a posetal relation. We say that $R$ is downward
closed when for any $(\langle C_1,s,\varrho\rangle,f,\langle C_2,s,\varrho\rangle),(\langle C_1',s,\varrho'\rangle,f',\langle C_2',s,\varrho'\rangle), (\langle C_1,s,\varrho\rangle,f,\langle C_2,s,\varrho\rangle),(\langle C_1',s',\varrho\rangle,f',\langle C_2',s',\varrho\rangle)\in \langle\mathcal{C}(\mathcal{E}_1),S\rangle\overline{\times}\langle\mathcal{C}(\mathcal{E}_2),S\rangle$,
if $(\langle C_1,s,\varrho\rangle,f,\langle C_2,s,\varrho\rangle)\subseteq (\langle C_1',s,\varrho'\rangle,f',\langle C_2',s,\varrho'\rangle)$ and
$(\langle C_1,s,\varrho\rangle,f,\langle C_2,s,\varrho\rangle)\subseteq (\langle C_1',s',\varrho\rangle,f',\langle C_2',s',\varrho\rangle)$ pointwise and
$(\langle C_1',s,\varrho'\rangle,f',\langle C_2',s,\varrho'\rangle),(\langle C_1',s',\varrho\rangle,f',\langle C_2',s',\varrho\rangle)\in R$, then $(\langle C_1,s,\varrho\rangle,f,\langle C_2,s,\varrho\rangle)\in R$.

For $f:X_1\rightarrow X_2$, we define $f[x_1\mapsto x_2]:X_1\cup\{x_1\}\rightarrow X_2\cup\{x_2\}$, $z\in X_1\cup\{x_1\}$,(1)$f[x_1\mapsto x_2](z)=
x_2$,if $z=x_1$;(2)$f[x_1\mapsto x_2](z)=f(z)$, otherwise. Where $X_1\subseteq \mathbb{E}_1$, $X_2\subseteq \mathbb{E}_2$, $x_1\in \mathbb{E}_1$, $x_2\in \mathbb{E}_2$.
\end{definition}

\begin{definition}[Weakly posetal product]
Given two PESs $\mathcal{E}_1$, $\mathcal{E}_2$, the weakly posetal product of their configurations, denoted
$\langle\mathcal{C}(\mathcal{E}_1),S\rangle\overline{\times}\langle\mathcal{C}(\mathcal{E}_2),S\rangle$, is defined as

$$\{(\langle C_1,s,\varrho\rangle,f,\langle C_2,s,\varrho\rangle)|C_1\in\mathcal{C}(\mathcal{E}_1),C_2\in\mathcal{C}(\mathcal{E}_2),f:\hat{C_1}\rightarrow \hat{C_2} \textrm{ isomorphism}\}.$$

A subset $R\subseteq\langle\mathcal{C}(\mathcal{E}_1),S\rangle\overline{\times}\langle\mathcal{C}(\mathcal{E}_2),S\rangle$ is called a weakly posetal relation. We say that $R$ is
downward closed when for any $(\langle C_1,s,\varrho\rangle,f,\langle C_2,s,\varrho\rangle),(\langle C_1',s,\varrho'\rangle,f,\langle C_2',s,\varrho'\rangle)\in \langle\mathcal{C}(\mathcal{E}_1),S,\varrho\rangle\overline{\times}\langle\mathcal{C}(\mathcal{E}_2),S,\varrho\rangle$,
$(\langle C_1,s,\varrho\rangle,f,\langle C_2,s,\varrho\rangle),(\langle C_1',s',\varrho\rangle,f,\langle C_2',s',\varrho\rangle)\in \langle\mathcal{C}(\mathcal{E}_1),S,\varrho\rangle\overline{\times}\langle\mathcal{C}(\mathcal{E}_2),S,\varrho\rangle$,
if $(\langle C_1,s,\varrho\rangle,f,\langle C_2,s,\varrho\rangle)\subseteq (\langle C_1',s,\varrho'\rangle,f',\langle C_2',s,\varrho'\rangle)$,
$(\langle C_1,s,\varrho\rangle,f,\langle C_2,s,\varrho\rangle)\subseteq (\langle C_1',s',\varrho\rangle,f',\langle C_2',s',\varrho\rangle)$ pointwise and
$(\langle C_1',s,\varrho'\rangle,f',\langle C_2',s,\varrho'\rangle)\in R$ and
$(\langle C_1',s',\varrho\rangle,f',\langle C_2',s',\varrho\rangle)\in R$, then $(\langle C_1,s,\varrho\rangle,f,\langle C_2,s,\varrho\rangle)\in R$.

For $f:X_1\rightarrow X_2$, we define $f[x_1\mapsto x_2]:X_1\cup\{x_1\}\rightarrow X_2\cup\{x_2\}$, $z\in X_1\cup\{x_1\}$,(1)$f[x_1\mapsto x_2](z)=
x_2$,if $z=x_1$;(2)$f[x_1\mapsto x_2](z)=f(z)$, otherwise. Where $X_1\subseteq \hat{\mathbb{E}_1}$, $X_2\subseteq \hat{\mathbb{E}_2}$, $x_1\in \hat{\mathbb{E}}_1$,
$x_2\in \hat{\mathbb{E}}_2$. Also, we define $f(\tau^*)=f(\tau^*)$.
\end{definition}

\begin{definition}[Probabilistic (hereditary) history-preserving bisimulation]\label{HHPBG}
A probabilistic history-preserving (hp-) bisimulation is a posetal relation
$R\subseteq\langle\mathcal{C}(\mathcal{E}_1),S\rangle\overline{\times}\langle\mathcal{C}(\mathcal{E}_2),S\rangle$ such that (1) if $(\langle C_1,s,\varrho\rangle,f,\langle C_2,s,\varrho\rangle)\in R$,
and $\langle C_1,s,\varrho\rangle\xrightarrow{e_1} \langle C_1',s,\varrho'\rangle$, then $\langle C_2,s,\varrho\rangle\xrightarrow{e_2} \langle C_2',s,\varrho'\rangle$, with
$(\langle C_1',s,\varrho'\rangle,f[e_1\mapsto e_2],\langle C_2',s,\varrho'\rangle)\in R$ for all $\varrho,\varrho'\in S$, and vice-versa;
(2) if $(\langle C_1,s,\varrho\rangle,f,\langle C_2,s,\varrho\rangle)\in R$,
and $\langle C_1,s,\varrho\rangle\xrightarrow{e_1} \langle C_1',s',\varrho\rangle$, then $\langle C_2,s,\varrho\rangle\xrightarrow{e_2} \langle C_2',s',\varrho\rangle$, with
$(\langle C_1',s',\varrho\rangle,f[e_1\mapsto e_2],\langle C_2',s',\varrho\rangle)\in R$ for all $s,s'\in S$, and vice-versa;
(3) if $(\langle C_1,s,\varrho\rangle,f,\langle C_2,s,\varrho\rangle)\in R$, and
$\langle C_1,s,\varrho\rangle\xrsquigarrow{\pi}\langle C_1^{\pi},s,\varrho\rangle$ then $\langle C_2,s,\varrho\rangle\xrsquigarrow{\pi}\langle C_2^{\pi},s,\varrho\rangle$ and $(\langle C_1^{\pi},s,\varrho\rangle,f,\langle C_2^{\pi},s,\varrho\rangle)\in R$,
and vice-versa; (4) if $(C_1,f,C_2)\in R$, then $\mu(C_1,C)=\mu(C_2,C)$ for each $C\in\mathcal{C}(\mathcal{E})/R$; (5) $[\surd]_R=\{\surd\}$. $\mathcal{E}_1,\mathcal{E}_2$ are
probabilistic history-preserving (hp-)bisimilar and are written $\mathcal{E}_1\sim_{php}\mathcal{E}_2$ if there exists a probabilistic hp-bisimulation $R$ such that
$(\langle\emptyset,\emptyset\rangle,\emptyset,\langle\emptyset,\emptyset\rangle)\in R$.

A probabilistic hereditary history-preserving (hhp-)bisimulation is a downward closed probabilistic hp-bisimulation. $\mathcal{E}_1,\mathcal{E}_2$ are probabilistic hereditary
history-preserving (hhp-)bisimilar and are written $\mathcal{E}_1\sim_{phhp}\mathcal{E}_2$.
\end{definition}

\begin{definition}[Weakly probabilistic (hereditary) history-preserving bisimulation]\label{WHHPBG}
A weakly probabilistic history-preserving (hp-) bisimulation is a weakly posetal relation
$R\subseteq\langle\mathcal{C}(\mathcal{E}_1),S\rangle\overline{\times}\langle\mathcal{C}(\mathcal{E}_2),S\rangle$ such that (1) if $(\langle C_1,s,\varrho\rangle,f,\langle C_2,s,\varrho\rangle)\in R$,
and $\langle C_1,s,\varrho\rangle\xRightarrow{e_1} \langle C_1',s,\varrho'\rangle$, then $\langle C_2,s,\varrho\rangle\xRightarrow{e_2} \langle C_2',s,\varrho'\rangle$, with
$(\langle C_1',s,\varrho'\rangle,f[e_1\mapsto e_2],\langle C_2',s,\varrho'\rangle)\in R$ for all $\varrho,\varrho'\in S$, and vice-versa;
(2) if $(\langle C_1,s,\varrho\rangle,f,\langle C_2,s,\varrho\rangle)\in R$,
and $\langle C_1,s,\varrho\rangle\xRightarrow{e_1} \langle C_1',s',\varrho\rangle$, then $\langle C_2,s,\varrho\rangle\xRightarrow{e_2} \langle C_2',s',\varrho\rangle$, with
$(\langle C_1',s',\varrho\rangle,f[e_1\mapsto e_2],\langle C_2',s',\varrho\rangle)\in R$ for all $s,s'\in S$, and vice-versa;
(3) if $(\langle C_1,s,\varrho\rangle,f,\langle C_2,s,\varrho\rangle)\in R$, and
$\langle C_1,s,\varrho\rangle\xrsquigarrow{\pi}\langle C_1^{\pi},s,\varrho\rangle$ then $\langle C_2,s,\varrho\rangle\xrsquigarrow{\pi}\langle C_2^{\pi},s,\varrho\rangle$ and
$(\langle C_1^{\pi},s,\varrho\rangle,f,\langle C_2^{\pi},s,\varrho\rangle)\in R$, and vice-versa; (4) if $(C_1,f,C_2)\in R$, then $\mu(C_1,C)=\mu(C_2,C)$ for each $C\in\mathcal{C}(\mathcal{E})/R$;
(5) $[\surd]_R=\{\surd\}$. $\mathcal{E}_1,\mathcal{E}_2$ are weakly probabilistic history-preserving (hp-)bisimilar and are written $\mathcal{E}_1\approx_{php}\mathcal{E}_2$ if there
exists a weakly probabilistic hp-bisimulation $R$ such that $(\langle\emptyset,\emptyset\rangle,\emptyset,\langle\emptyset,\emptyset\rangle)\in R$.

A weakly probabilistic hereditary history-preserving (hhp-)bisimulation is a downward closed weakly probabilistic hp-bisimulation. $\mathcal{E}_1,\mathcal{E}_2$ are weakly
probabilistic hereditary history-preserving (hhp-)bisimilar and are written $\mathcal{E}_1\approx_{phhp}\mathcal{E}_2$.
\end{definition}

\begin{definition}[Probabilistic branching pomset, step bisimulation]\label{BPSBG}
Assume a special termination predicate $\downarrow$, and let $\surd$ represent a state with $\surd\downarrow$. Let $\mathcal{E}_1$, $\mathcal{E}_2$ be PESs. A probabilistic branching
pomset bisimulation is a relation $R\subseteq\langle\mathcal{C}(\mathcal{E}_1),S\rangle\times\langle\mathcal{C}(\mathcal{E}_2),S\rangle$, such that:

 \begin{enumerate}
   \item if $(\langle C_1,s,\varrho\rangle,\langle C_2,s,\varrho\rangle)\in R$, and $\langle C_1,s,\varrho\rangle\xrightarrow{X}\langle C_1',s,\varrho'\rangle$ then
   \begin{itemize}
     \item either $X\equiv \tau^*$, and $(\langle C_1',s,\varrho'\rangle,\langle C_2,s,\varrho\rangle)\in R$ with $\varrho'\in \tau(\varrho)$;
     \item or there is a sequence of (zero or more) probabilistic transitions and $\tau$-transitions $\langle C_2,s,\varrho\rangle\rightsquigarrow^*\xrightarrow{\tau^*} \langle C_2^0,\varrho^0\rangle$, such that
     $(\langle C_1,s,\varrho\rangle,\langle C_2^0,\varrho^0\rangle)\in R$ and $\langle C_2^0,\varrho^0\rangle\xRightarrow{X}\langle C_2',s,\varrho'\rangle$ with
     $(\langle C_1',s,\varrho'\rangle,\langle C_2',s,\varrho'\rangle)\in R$;
   \end{itemize}
   \item if $(\langle C_1,s,\varrho\rangle,\langle C_2,s,\varrho\rangle)\in R$, and $\langle C_1,s,\varrho\rangle\xrightarrow{X}\langle C_1',s',\varrho\rangle$ then
   \begin{itemize}
     \item either $X\equiv \tau^*$, and $(\langle C_1',s',\varrho\rangle,\langle C_2,s,\varrho\rangle)\in R$ with $s'\in \tau(s)$;
     \item or there is a sequence of (zero or more) probabilistic transitions and $\tau$-transitions $\langle C_2,s,\varrho\rangle\rightsquigarrow^*\xrightarrow{\tau^*} \langle C_2^0,\varrho^0\rangle$, such that
     $(\langle C_1,s,\varrho\rangle,\langle C_2^0,\varrho^0\rangle)\in R$ and $\langle C_2^0,\varrho^0\rangle\xRightarrow{X}\langle C_2',s',\varrho\rangle$ with
     $(\langle C_1',s',\varrho\rangle,\langle C_2',s',\varrho\rangle)\in R$;
   \end{itemize}
   \item if $(\langle C_1,s,\varrho\rangle,\langle C_2,s,\varrho\rangle)\in R$, and $\langle C_2,s,\varrho\rangle\xrightarrow{X}\langle C_2',s,\varrho'\rangle$ then
   \begin{itemize}
     \item either $X\equiv \tau^*$, and $(\langle C_1,s,\varrho\rangle,\langle C_2',s,\varrho'\rangle)\in R$;
     \item or there is a sequence of (zero or more) probabilistic transitions and $\tau$-transitions $\langle C_1,s,\varrho\rangle\rightsquigarrow^*\xrightarrow{\tau^*} \langle C_1^0,\varrho^0\rangle$, such that
     $(\langle C_1^0,\varrho^0\rangle,\langle C_2,s,\varrho\rangle)\in R$ and $\langle C_1^0,\varrho^0\rangle\xRightarrow{X}\langle C_1',s,\varrho'\rangle$ with
     $(\langle C_1',s,\varrho'\rangle,\langle C_2',s,\varrho'\rangle)\in R$;
   \end{itemize}
   \item if $(\langle C_1,s,\varrho\rangle,\langle C_2,s,\varrho\rangle)\in R$, and $\langle C_2,s,\varrho\rangle\xrightarrow{X}\langle C_2',s',\varrho\rangle$ then
   \begin{itemize}
     \item either $X\equiv \tau^*$, and $(\langle C_1,s,\varrho\rangle,\langle C_2',s',\varrho\rangle)\in R$;
     \item or there is a sequence of (zero or more) probabilistic transitions and $\tau$-transitions $\langle C_1,s,\varrho\rangle\rightsquigarrow^*\xrightarrow{\tau^*} \langle C_1^0,\varrho^0\rangle$, such that
     $(\langle C_1^0,\varrho^0\rangle,\langle C_2,s,\varrho\rangle)\in R$ and $\langle C_1^0,\varrho^0\rangle\xRightarrow{X}\langle C_1',s',\varrho\rangle$ with
     $(\langle C_1',s',\varrho\rangle,\langle C_2',s',\varrho\rangle)\in R$;
   \end{itemize}
   \item if $(\langle C_1,s,\varrho\rangle,\langle C_2,s,\varrho\rangle)\in R$ and $\langle C_1,s,\varrho\rangle\downarrow$, then there is a sequence of (zero or more) probabilistic transitions and $\tau$-transitions
   $\langle C_2,s,\varrho\rangle\rightsquigarrow^*\xrightarrow{\tau^*}\langle C_2^0,\varrho^0\rangle$ such that $(\langle C_1,s,\varrho\rangle,\langle C_2^0,\varrho^0\rangle)\in R$ and
   $\langle C_2^0,\varrho^0\rangle\downarrow$;
   \item if $(\langle C_1,s,\varrho\rangle,\langle C_2,s,\varrho\rangle)\in R$ and $\langle C_2,s,\varrho\rangle\downarrow$, then there is a sequence of (zero or more) probabilistic transitions and $\tau$-transitions
   $\langle C_1,s,\varrho\rangle\rightsquigarrow^*\xrightarrow{\tau^*}\langle C_1^0,\varrho^0\rangle$ such that $(\langle C_1^0,\varrho^0\rangle,\langle C_2,s,\varrho\rangle)\in R$ and
   $\langle C_1^0,\varrho^0\rangle\downarrow$;
   \item if $(C_1,C_2)\in R$,then $\mu(C_1,C)=\mu(C_2,C)$ for each $C\in\mathcal{C}(\mathcal{E})/R$;
   \item $[\surd]_R=\{\surd\}$.
 \end{enumerate}

We say that $\mathcal{E}_1$, $\mathcal{E}_2$ are probabilistic branching pomset bisimilar, written $\mathcal{E}_1\approx_{pbp}\mathcal{E}_2$, if there exists a probabilistic branching
pomset bisimulation $R$, such that $(\langle\emptyset,\emptyset\rangle,\langle\emptyset,\emptyset\rangle)\in R$.

By replacing probabilistic pomset transitions with steps, we can get the definition of probabilistic branching step bisimulation. When PESs $\mathcal{E}_1$ and $\mathcal{E}_2$ are
probabilistic branching step bisimilar, we write $\mathcal{E}_1\approx_{pbs}\mathcal{E}_2$.
\end{definition}

\begin{definition}[Probabilistic rooted branching pomset, step bisimulation]\label{RBPSBG}
Assume a special termination predicate $\downarrow$, and let $\surd$ represent a state with $\surd\downarrow$. Let $\mathcal{E}_1$, $\mathcal{E}_2$ be PESs. A probabilistic rooted
branching pomset bisimulation is a relation $R\subseteq\langle\mathcal{C}(\mathcal{E}_1),S\rangle\times\langle\mathcal{C}(\mathcal{E}_2),S\rangle$, such that:

 \begin{enumerate}
   \item if $(\langle C_1,s,\varrho\rangle,\langle C_2,s,\varrho\rangle)\in R$, and $\langle C_1,s,\varrho\rangle\rightsquigarrow\xrightarrow{X}\langle C_1',s,\varrho'\rangle$ then
   $\langle C_2,s,\varrho\rangle\rightsquigarrow\xrightarrow{X}\langle C_2',s,\varrho'\rangle$ with $\langle C_1',s,\varrho'\rangle\approx_{pbp}\langle C_2',s,\varrho'\rangle$;
   \item if $(\langle C_1,s,\varrho\rangle,\langle C_2,s,\varrho\rangle)\in R$, and $\langle C_1,s,\varrho\rangle\rightsquigarrow\xrightarrow{X}\langle C_1',s',\varrho\rangle$ then
   $\langle C_2,s,\varrho\rangle\rightsquigarrow\xrightarrow{X}\langle C_2',s',\varrho\rangle$ with $\langle C_1',s',\varrho\rangle\approx_{pbp}\langle C_2',s',\varrho\rangle$;
   \item if $(\langle C_1,s,\varrho\rangle,\langle C_2,s,\varrho\rangle)\in R$, and $\langle C_2,s,\varrho\rangle\rightsquigarrow\xrightarrow{X}\langle C_2',s,\varrho'\rangle$ then
   $\langle C_1,s,\varrho\rangle\rightsquigarrow\xrightarrow{X}\langle C_1',s,\varrho'\rangle$ with $\langle C_1',s,\varrho'\rangle\approx_{pbp}\langle C_2',s,\varrho'\rangle$;
   \item if $(\langle C_1,s,\varrho\rangle,\langle C_2,s,\varrho\rangle)\in R$, and $\langle C_2,s,\varrho\rangle\rightsquigarrow\xrightarrow{X}\langle C_2',s',\varrho\rangle$ then
   $\langle C_1,s,\varrho\rangle\rightsquigarrow\xrightarrow{X}\langle C_1',s',\varrho\rangle$ with $\langle C_1',s',\varrho\rangle\approx_{pbp}\langle C_2',s',\varrho\rangle$;
   \item if $(\langle C_1,s,\varrho\rangle,\langle C_2,s,\varrho\rangle)\in R$ and $\langle C_1,s,\varrho\rangle\downarrow$, then $\langle C_2,s,\varrho\rangle\downarrow$;
   \item if $(\langle C_1,s,\varrho\rangle,\langle C_2,s,\varrho\rangle)\in R$ and $\langle C_2,s,\varrho\rangle\downarrow$, then $\langle C_1,s,\varrho\rangle\downarrow$.
 \end{enumerate}

We say that $\mathcal{E}_1$, $\mathcal{E}_2$ are probabilistic rooted branching pomset bisimilar, written $\mathcal{E}_1\approx_{prbp}\mathcal{E}_2$, if there exists a probabilistic
rooted branching pomset bisimulation $R$, such that $(\langle\emptyset,\emptyset\rangle,\langle\emptyset,\emptyset\rangle)\in R$.

By replacing pomset transitions with steps, we can get the definition of probabilistic rooted branching step bisimulation. When PESs $\mathcal{E}_1$ and $\mathcal{E}_2$ are probabilistic
rooted branching step bisimilar, we write $\mathcal{E}_1\approx_{prbs}\mathcal{E}_2$.
\end{definition}

\begin{definition}[Probabilistic branching (hereditary) history-preserving bisimulation]\label{BHHPBG}
Assume a special termination predicate $\downarrow$, and let $\surd$ represent a state with $\surd\downarrow$. A probabilistic branching history-preserving (hp-) bisimulation is a
weakly posetal relation $R\subseteq\langle\mathcal{C}(\mathcal{E}_1),S\rangle\overline{\times}\langle\mathcal{C}(\mathcal{E}_2),S\rangle$ such that:

 \begin{enumerate}
   \item if $(\langle C_1,s,\varrho\rangle,f,\langle C_2,s,\varrho\rangle)\in R$, and $\langle C_1,s,\varrho\rangle\xrightarrow{e_1}\langle C_1',s,\varrho'\rangle$ then
   \begin{itemize}
     \item either $e_1\equiv \tau$, and $(\langle C_1',s,\varrho'\rangle,f[e_1\mapsto \tau],\langle C_2,s,\varrho\rangle)\in R$;
     \item or there is a sequence of (zero or more) probabilistic transitions and $\tau$-transitions $\langle C_2,s,\varrho\rangle\rightsquigarrow^*\xrightarrow{\tau^*} \langle C_2^0,\varrho^0\rangle$, such that
     $(\langle C_1,s,\varrho\rangle,f,\langle C_2^0,\varrho^0\rangle)\in R$ and $\langle C_2^0,\varrho^0\rangle\xrightarrow{e_2}\langle C_2',s,\varrho'\rangle$ with
     $(\langle C_1',s,\varrho'\rangle,f[e_1\mapsto e_2],\langle C_2',s,\varrho'\rangle)\in R$;
   \end{itemize}
   \item if $(\langle C_1,s,\varrho\rangle,f,\langle C_2,s,\varrho\rangle)\in R$, and $\langle C_1,s,\varrho\rangle\xrightarrow{e_1}\langle C_1',s',\varrho\rangle$ then
   \begin{itemize}
     \item either $e_1\equiv \tau$, and $(\langle C_1',s',\varrho\rangle,f[e_1\mapsto \tau],\langle C_2,s,\varrho\rangle)\in R$;
     \item or there is a sequence of (zero or more) probabilistic transitions and $\tau$-transitions $\langle C_2,s,\varrho\rangle\rightsquigarrow^*\xrightarrow{\tau^*} \langle C_2^0,\varrho^0\rangle$, such that
     $(\langle C_1,s,\varrho\rangle,f,\langle C_2^0,\varrho^0\rangle)\in R$ and $\langle C_2^0,\varrho^0\rangle\xrightarrow{e_2}\langle C_2',s',\varrho\rangle$ with
     $(\langle C_1',s',\varrho\rangle,f[e_1\mapsto e_2],\langle C_2',s',\varrho\rangle)\in R$;
   \end{itemize}
   \item if $(\langle C_1,s,\varrho\rangle,f,\langle C_2,s,\varrho\rangle)\in R$, and $\langle C_2,s,\varrho\rangle\xrightarrow{e_2}\langle C_2',s,\varrho'\rangle$ then
   \begin{itemize}
     \item either $e_2\equiv \tau$, and $(\langle C_1,s,\varrho\rangle,f[e_2\mapsto \tau],\langle C_2',s,\varrho'\rangle)\in R$;
     \item or there is a sequence of (zero or more) probabilistic transitions and $\tau$-transitions $\langle C_1,s,\varrho\rangle\rightsquigarrow^*\xrightarrow{\tau^*} \langle C_1^0,\varrho^0\rangle$, such that
     $(\langle C_1^0,\varrho^0\rangle,f,\langle C_2,s,\varrho\rangle)\in R$ and $\langle C_1^0,\varrho^0\rangle\xrightarrow{e_1}\langle C_1',s,\varrho'\rangle$ with
     $(\langle C_1',s,\varrho'\rangle,f[e_2\mapsto e_1],\langle C_2',s,\varrho'\rangle)\in R$;
   \end{itemize}
   \item if $(\langle C_1,s,\varrho\rangle,f,\langle C_2,s,\varrho\rangle)\in R$, and $\langle C_2,s,\varrho\rangle\xrightarrow{e_2}\langle C_2',s',\varrho\rangle$ then
   \begin{itemize}
     \item either $e_2\equiv \tau$, and $(\langle C_1,s,\varrho\rangle,f[e_2\mapsto \tau],\langle C_2',s',\varrho\rangle)\in R$;
     \item or there is a sequence of (zero or more) probabilistic transitions and $\tau$-transitions $\langle C_1,s,\varrho\rangle\rightsquigarrow^*\xrightarrow{\tau^*} \langle C_1^0,\varrho^0\rangle$, such that
     $(\langle C_1^0,\varrho^0\rangle,f,\langle C_2,s,\varrho\rangle)\in R$ and $\langle C_1^0,\varrho^0\rangle\xrightarrow{e_1}\langle C_1',s',\varrho\rangle$ with
     $(\langle C_1',s',\varrho\rangle,f[e_2\mapsto e_1],\langle C_2',s',\varrho\rangle)\in R$;
   \end{itemize}
   \item if $(\langle C_1,s,\varrho\rangle,f,\langle C_2,s,\varrho\rangle)\in R$ and $\langle C_1,s,\varrho\rangle\downarrow$, then there is a sequence of (zero or more) probabilistic transitions and $\tau$-transitions
   $\langle C_2,s,\varrho\rangle\rightsquigarrow^*\xrightarrow{\tau^*}\langle C_2^0,\varrho^0\rangle$ such that $(\langle C_1,s,\varrho\rangle,f,\langle C_2^0,\varrho^0\rangle)\in R$ and
   $\langle C_2^0,\varrho^0\rangle\downarrow$;
   \item if $(\langle C_1,s,\varrho\rangle,f,\langle C_2,s,\varrho\rangle)\in R$ and $\langle C_2,s,\varrho\rangle\downarrow$, then there is a sequence of (zero or more) probabilistic transitions and $\tau$-transitions
   $\langle C_1,s,\varrho\rangle\rightsquigarrow^*\xrightarrow{\tau^*}\langle C_1^0,\varrho^0\rangle$ such that $(\langle C_1^0,\varrho^0\rangle,f,\langle C_2,s,\varrho\rangle)\in R$ and
   $\langle C_1^0,\varrho^0\rangle\downarrow$;
   \item if $(C_1,C_2)\in R$,then $\mu(C_1,C)=\mu(C_2,C)$ for each $C\in\mathcal{C}(\mathcal{E})/R$;
   \item $[\surd]_R=\{\surd\}$.
 \end{enumerate}

$\mathcal{E}_1,\mathcal{E}_2$ are probabilistic branching history-preserving (hp-)bisimilar and are written $\mathcal{E}_1\approx_{pbhp}\mathcal{E}_2$ if there exists a probabilistic
branching hp-bisimulation $R$ such that $(\langle\emptyset,\emptyset\rangle,\emptyset,\langle\emptyset,\emptyset\rangle)\in R$.

A probabilistic branching hereditary history-preserving (hhp-)bisimulation is a downward closed probabilistic branching hp-bisimulation. $\mathcal{E}_1,\mathcal{E}_2$ are probabilistic
branching hereditary history-preserving (hhp-)bisimilar and are written $\mathcal{E}_1\approx_{pbhhp}\mathcal{E}_2$.
\end{definition}

\begin{definition}[Probabilistic rooted branching (hereditary) history-preserving bisimulation]\label{RBHHPBG}
Assume a special termination predicate $\downarrow$, and let $\surd$ represent a state with $\surd\downarrow$. A probabilistic rooted branching history-preserving (hp-) bisimulation is
a weakly posetal relation $R\subseteq\langle\mathcal{C}(\mathcal{E}_1),S\rangle\overline{\times}\langle\mathcal{C}(\mathcal{E}_2),S\rangle$ such that:

 \begin{enumerate}
   \item if $(\langle C_1,s,\varrho\rangle,f,\langle C_2,s,\varrho\rangle)\in R$, and $\langle C_1,s,\varrho\rangle\rightsquigarrow\xrightarrow{e_1}\langle C_1',s,\varrho'\rangle$, then
   $\langle C_2,s,\varrho\rangle\rightsquigarrow\xrightarrow{e_2}\langle C_2',s,\varrho'\rangle$ with $\langle C_1',s,\varrho'\rangle\approx_{pbhp}\langle C_2',s,\varrho'\rangle$;
   \item if $(\langle C_1,s,\varrho\rangle,f,\langle C_2,s,\varrho\rangle)\in R$, and $\langle C_1,s,\varrho\rangle\rightsquigarrow\xrightarrow{e_1}\langle C_1',s',\varrho\rangle$, then
   $\langle C_2,s,\varrho\rangle\rightsquigarrow\xrightarrow{e_2}\langle C_2',s',\varrho\rangle$ with $\langle C_1',s',\varrho\rangle\approx_{pbhp}\langle C_2',s',\varrho\rangle$;
   \item if $(\langle C_1,s,\varrho\rangle,f,\langle C_2,s,\varrho\rangle)\in R$, and $\langle C_2,s,\varrho\rangle\rightsquigarrow\xrightarrow{e_2}\langle C_2',s,\varrho'\rangle$, then
   $\langle C_1,s,\varrho\rangle\rightsquigarrow\xrightarrow{e_1}\langle C_1',s,\varrho'\rangle$ with $\langle C_1',s,\varrho'\rangle\approx_{pbhp}\langle C_2',s,\varrho'\rangle$;
   \item if $(\langle C_1,s,\varrho\rangle,f,\langle C_2,s,\varrho\rangle)\in R$, and $\langle C_2,s,\varrho\rangle\rightsquigarrow\xrightarrow{e_2}\langle C_2',s',\varrho\rangle$, then
   $\langle C_1,s,\varrho\rangle\rightsquigarrow\xrightarrow{e_1}\langle C_1',s',\varrho\rangle$ with $\langle C_1',s',\varrho\rangle\approx_{pbhp}\langle C_2',s',\varrho\rangle$;
   \item if $(\langle C_1,s,\varrho\rangle,f,\langle C_2,s,\varrho\rangle)\in R$ and $\langle C_1,s,\varrho\rangle\downarrow$, then $\langle C_2,s,\varrho\rangle\downarrow$;
   \item if $(\langle C_1,s,\varrho\rangle,f,\langle C_2,s,\varrho\rangle)\in R$ and $\langle C_2,s,\varrho\rangle\downarrow$, then $\langle C_1,s,\varrho\rangle\downarrow$.
 \end{enumerate}

$\mathcal{E}_1,\mathcal{E}_2$ are probabilistic rooted branching history-preserving (hp-)bisimilar and are written $\mathcal{E}_1\approx_{prbhp}\mathcal{E}_2$ if there exists a probabilistic
rooted branching hp-bisimulation $R$ such that $(\langle\emptyset,\emptyset\rangle,\emptyset,\langle\emptyset,\emptyset\rangle)\in R$.

A probabilistic rooted branching hereditary history-preserving (hhp-)bisimulation is a downward closed probabilistic rooted branching hp-bisimulation. $\mathcal{E}_1,\mathcal{E}_2$ are
probabilistic rooted branching hereditary history-preserving (hhp-)bisimilar and are written $\mathcal{E}_1\approx_{prbhhp}\mathcal{E}_2$.
\end{definition}

\newpage\section{$APTC_G$ for Open Quantum Systems}\label{qaptcg}

In this chapter, we introduce $APTC_G$ for open quantum systems, including $BATC_G$ for open quantum systems abbreviated $qBATC_G$ in section \ref{qbatcg}, $APTC_G$ for open quantum systems
abbreviated $qAPTC_G$ in section \ref{qaptcg2}, recursion in section \ref{qorecg}, abstraction in section \ref{qoabsg}, quantum entanglement in section \ref{qe1} and unification of quantum
and classical computing for open quantum systems in section \ref{uni1}.

Note that, in open quantum systems, quantum operations denoted $\mathbb{E}$ are the atomic actions (events), and a quantum operation $e\in\mathbb{E}$.

\subsection{$BATC_G$ for Open Quantum Systems}\label{qbatcg}

In this subsection, we will discuss $qBATC_G$. Let $\mathbb{E}$ be the set of atomic events (actions), $G_{at}$ be the set of atomic guards, $\delta$ be the deadlock constant, and
$\epsilon$ be the empty event. We extend $G_{at}$ to the set of basic guards $G$ with element $\phi,\psi,\cdots$, which is generated by the following formation rules:

$$\phi::=\delta|\epsilon|\neg\phi|\psi\in G_{at}|\phi+\psi|\phi\cdot\psi$$

In the following, let $e_1, e_2, e_1', e_2'\in \mathbb{E}$, $\phi,\psi\in G$ and let variables $x,y,z$ range over the set of terms for true concurrency, $p,q,s$ range over the set of
closed terms. The predicate $test(\phi,s,\varrho)$ represents that $\phi$ holds in the state $s,\varrho$, and $test(\epsilon,s,\varrho)$ holds and $test(\delta,s,\varrho)$ does not hold.
$effect(e,s,\varrho)\in S$ denotes $\varrho'$ in $\varrho\xrightarrow{e}\varrho'$. The predicate weakest precondition $wp(e,\phi)$ denotes that $\forall \varrho,\varrho'\in S, test(\phi,effect(e,s,\varrho))$
holds.

The set of axioms of $qBATC_G$ consists of the laws given in Table \ref{AxiomsForqBATCG}.

\begin{center}
    \begin{table}
        \begin{tabular}{@{}ll@{}}
            \hline No. &Axiom\\
            $A1$ & $x+ y = y+ x$\\
            $A2$ & $(x+ y)+ z = x+ (y+ z)$\\
            $A3$ & $x+ x = x$\\
            $A4$ & $(x+ y)\cdot z = x\cdot z + y\cdot z$\\
            $A5$ & $(x\cdot y)\cdot z = x\cdot(y\cdot z)$\\
            $A6$ & $x+\delta = x$\\
            $A7$ & $\delta\cdot x = \delta$\\
            $A8$ & $\epsilon\cdot x = x$\\
            $A9$ & $x\cdot\epsilon = x$\\
            $G1$ & $\phi\cdot\neg\phi = \delta$\\
            $G2$ & $\phi+\neg\phi = \epsilon$\\
            $G3$ & $\phi\delta = \delta$\\
            $G4$ & $\phi(x+y)=\phi x+\phi y$\\
            $G5$ & $\phi(x\cdot y)= \phi x\cdot y$\\
            $G6$ & $(\phi+\psi)x = \phi x + \psi x$\\
            $G7$ & $(\phi\cdot \psi)\cdot x = \phi\cdot(\psi\cdot x)$\\
            $G8$ & $\phi=\epsilon$ if $\forall s,\varrho\in S.test(\phi,s,\varrho)$\\
            $G9$ & $\phi_0\cdot\cdots\cdot\phi_n = \delta$ if $\forall s,\varrho\in S,\exists i\leq n.test(\neg\phi_i,s,\varrho)$\\
            $G10$ & $wp(e,\phi)e\phi=wp(e,\phi)e$\\
            $G11$ & $\neg wp(e,\phi)e\neg\phi=\neg wp(e,\phi)e$\\
        \end{tabular}
        \caption{Axioms of $qBATC_G$}
        \label{AxiomsForqBATCG}
    \end{table}
\end{center}

Note that, by eliminating atomic event from the process terms, the axioms in Table \ref{AxiomsForqBATCG} will lead to a Boolean Algebra. And $G9$ is a precondition of $e$ and $\phi$,
$G10$ is the weakest precondition of $e$ and $\phi$. A data environment with $effect$ function is sufficiently deterministic, and it is obvious that if the weakest precondition is
expressible and $G9$, $G10$ are sound, then the related data environment is sufficiently deterministic.

\begin{definition}[Basic terms of $qBATC_G$]\label{BTBATCG}
The set of basic terms of $qBATC_G$, $\mathcal{B}(qBATC_G)$, is inductively defined as follows:
\begin{enumerate}
  \item $\mathbb{E}\subset\mathcal{B}(qBATC_G)$;
  \item $G\subset\mathcal{B}(qBATC_G)$;
  \item if $e\in \mathbb{E}, t\in\mathcal{B}(qBATC_G)$ then $e\cdot t\in\mathcal{B}(qBATC_G)$;
  \item if $\phi\in G, t\in\mathcal{B}(qBATC_G)$ then $\phi\cdot t\in\mathcal{B}(qBATC_G)$;
  \item if $t,s\in\mathcal{B}(qBATC_G)$ then $t+ s\in\mathcal{B}(qBATC_G)$.
\end{enumerate}
\end{definition}

\begin{theorem}[Elimination theorem of $qBATC_G$]\label{ETBATCG}
Let $p$ be a closed $qBATC_G$ term. Then there is a basic $qBATC_G$ term $q$ such that $qBATC_G\vdash p=q$.
\end{theorem}

\begin{proof}
The same as that of $BATC_G$, we omit the proof, please refer to \cite{ATC} for details.
\end{proof}

We will define a term-deduction system which gives the operational semantics of $qBATC_G$. We give the operational transition rules for $\epsilon$, atomic guard $\phi\in G_{at}$,
atomic event $e\in\mathbb{E}$, operators $\cdot$ and $+$ as Table \ref{SETRForqBATCG} shows. And the predicate $\xrightarrow{e}\surd$ represents successful termination after execution
of the event $e$.

\begin{center}
    \begin{table}
        $$\frac{}{\langle\epsilon,s,\varrho\rangle\rightarrow\langle\surd,s,\varrho\rangle}$$
        $$\frac{}{\langle e,s,\varrho\rangle\xrightarrow{e}\langle\surd,s,\varrho'\rangle}\textrm{ if }s'\in effect(e,s)$$
        $$\frac{}{\langle\phi,s,\varrho\rangle\rightarrow\langle\surd,s,\varrho\rangle}\textrm{ if }test(\phi,s)$$
        $$\frac{\langle x,s,\varrho\rangle\xrightarrow{e}\langle\surd,s,\varrho'\rangle}{\langle x+ y,s,\varrho\rangle\xrightarrow{e}\langle\surd,s,\varrho'\rangle} \quad\frac{\langle x,s,\varrho\rangle\xrightarrow{e}\langle x',s,\varrho'\rangle}{\langle x+ y,s,\varrho\rangle\xrightarrow{e}\langle x',s,\varrho'\rangle}$$
        $$\frac{\langle y,s,\varrho\rangle\xrightarrow{e}\langle\surd,s,\varrho'\rangle}{\langle x+ y,s,\varrho\rangle\xrightarrow{e}\langle\surd,s,\varrho'\rangle} \quad\frac{\langle y,s,\varrho\rangle\xrightarrow{e}\langle y',s,\varrho'\rangle}{\langle x+ y,s,\varrho\rangle\xrightarrow{e}\langle y',s,\varrho'\rangle}$$
        $$\frac{\langle x,s,\varrho\rangle\xrightarrow{e}\langle\surd,s,\varrho'\rangle}{\langle x\cdot y,s,\varrho\rangle\xrightarrow{e} \langle y,s,\varrho'\rangle} \quad\frac{\langle x,s,\varrho\rangle\xrightarrow{e}\langle x',s,\varrho'\rangle}{\langle x\cdot y,s,\varrho\rangle\xrightarrow{e}\langle x'\cdot y,s,\varrho'\rangle}$$
        \caption{Single event transition rules of $qBATC_G$}
        \label{SETRForqBATCG}
    \end{table}
\end{center}

Note that, we replace the single atomic event $e\in\mathbb{E}$ by $X\subseteq\mathbb{E}$, we can obtain the pomset transition rules of $qBATC_G$, and omit them.

\begin{theorem}[Congruence of $qBATC_G$ with respect to truly concurrent bisimulation equivalences]\label{CBATCG}
(1) Pomset bisimulation equivalence $\sim_{p}$ is a congruence with respect to $qBATC_G$.

(2) Step bisimulation equivalence $\sim_{s}$ is a congruence with respect to $qBATC_G$.

(3) Hp-bisimulation equivalence $\sim_{hp}$ is a congruence with respect to $qBATC_G$.

(4) Hhp-bisimulation equivalence $\sim_{hhp}$ is a congruence with respect to $qBATC_G$.
\end{theorem}

\begin{proof}
It is obvious that truly concurrent bisimulations $\sim_{p}$, $\sim_s$, $\sim_{hp}$ and $\sim_{hhp}$ are all equivalent relations with respect to $qBATC_G$. So, it is sufficient to prove
that truly concurrent bisimulations $\sim_{p}$, $\sim_s$, $\sim_{hp}$ and $\sim_{hhp}$ are preserved for $\cdot$ and $+$ according to the transition rules in Table \ref{SETRForqBATCG},
that is, if $x\sim_{p}x'$ and $y\sim_{p}y'$, then $x+ y\sim_{p}x'+ y'$ and $x\cdot y\sim_{p}x'\cdot y'$; if $x\sim_{s}x'$ and $y\sim_{s}y'$, then $x+ y\sim_{s}x'+ y'$ and $x\cdot y\sim_{s}x'\cdot y'$;
if $x\sim_{hp}x'$ and $y\sim_{hp}y'$, then$x+ y\sim_{hp}x'+ y'$ and  $x\cdot y\sim_{hp}x'\cdot y'$; and if $x\sim_{hhp}x'$ and $y\sim_{hhp}y'$, then $x+ y\sim_{hhp}x'+ y'$ and $x\cdot y\sim_{hhp}x'\cdot y'$.
The proof is quit trivial, and we leave the proof as an exercise for the readers.
\end{proof}

\begin{theorem}[Soundness of $qBATC_G$ modulo truly concurrent bisimulation equivalences]\label{SBATCG}
(1) Let $x$ and $y$ be $qBATC_G$ terms. If $qBATC\vdash x=y$, then $x\sim_{p} y$.

(2) Let $x$ and $y$ be $qBATC_G$ terms. If $qBATC\vdash x=y$, then $x\sim_{s} y$.

(3) Let $x$ and $y$ be $qBATC_G$ terms. If $qBATC\vdash x=y$, then $x\sim_{hp} y$.

(4) Let $x$ and $y$ be $qBATC_G$ terms. If $qBATC\vdash x=y$, then $x\sim_{hhp} y$.
\end{theorem}

\begin{proof}
(1) Since pomset bisimulation $\sim_{p}$ is both an equivalent and a congruent relation, we only need to check if each axiom in Table \ref{AxiomsForqBATCG} is sound
modulo pomset bisimulation equivalence. We leave the proof as an exercise for the readers.

(2) Since  step bisimulation $\sim_{s}$ is both an equivalent and a congruent relation, we only need to check if each axiom in Table \ref{AxiomsForqBATCG} is sound modulo
step bisimulation equivalence. We leave the proof as an exercise for the readers.

(3) Since hp-bisimulation $\sim_{hp}$ is both an equivalent and a congruent relation, we only need to check if each axiom in Table \ref{AxiomsForqBATCG} is sound modulo
hp-bisimulation equivalence. We leave the proof as an exercise for the readers.

(4) Since hhp-bisimulation $\sim_{hhp}$ is both an equivalent and a congruent relation, we only need to check if each axiom in Table \ref{AxiomsForqBATCG} is sound modulo
hhp-bisimulation equivalence. We leave the proof as an exercise for the readers.
\end{proof}

\begin{theorem}[Completeness of $qBATC_G$ modulo truly concurrent bisimulation equivalences]\label{CBATCG}
(1) Let $p$ and $q$ be closed $qBATC_G$ terms, if $p\sim_{p} q$ then $p=q$.

(2) Let $p$ and $q$ be closed $qBATC_G$ terms, if $p\sim_{s} q$ then $p=q$.

(3) Let $p$ and $q$ be closed $qBATC_G$ terms, if $p\sim_{hp} q$ then $p=q$.

(4) Let $p$ and $q$ be closed $qBATC_G$ terms, if $p\sim_{hhp} q$ then $p=q$.
\end{theorem}

\begin{proof}
According to the definition of truly concurrent bisimulation equivalences $\sim_{p}$, $\sim_{s}$, $\sim_{hp}$ and $\sim_{hhp}$, $p\sim_{p}q$, $p\sim_{s}q$, $p\sim_{hp}q$ and $p\sim_{hhp}q$ implies
both the bisimilarities between $p$ and $q$, and also the in the same quantum states. According to the completeness of $BATC_G$ (please refer to \cite{ATC} for details), we can get the
completeness of $qBATC_G$.
\end{proof}

\subsection{$APTC_G$ for Open Quantum Systems}\label{qaptcg2}

In this subsection, we will introduce $qAPTC_G$. The set of basic guards $G$ with element $\phi,\psi,\cdots$, which is extended by the following formation rules:

$$\phi::=\delta|\epsilon|\neg\phi|\psi\in G_{at}|\phi+\psi|\phi\cdot\psi|\phi\parallel\psi$$

The set of axioms of $qAPTC_G$ including axioms of $qBATC_G$ in Table \ref{AxiomsForqBATCG} and the axioms are shown in Table \ref{AxiomsForqAPTCG}.

\begin{center}
    \begin{table}
        \begin{tabular}{@{}ll@{}}
            \hline No. &Axiom\\
            $P1$ & $x\between y = x\parallel y + x\mid y$\\
            $P2$ & $x\parallel y = y \parallel x$\\
            $P3$ & $(x\parallel y)\parallel z = x\parallel (y\parallel z)$\\
            $P4$ & $x\parallel y = x\leftmerge y + y\leftmerge x$\\
            $P5$ & $(e_1\leq e_2)\quad e_1\leftmerge (e_2\cdot y) = (e_1\leftmerge e_2)\cdot y$\\
            $P6$ & $(e_1\leq e_2)\quad (e_1\cdot x)\leftmerge e_2 = (e_1\leftmerge e_2)\cdot x$\\
            $P7$ & $(e_1\leq e_2)\quad (e_1\cdot x)\leftmerge (e_2\cdot y) = (e_1\leftmerge e_2)\cdot (x\between y)$\\
            $P8$ & $(x+ y)\leftmerge z = (x\leftmerge z)+ (y\leftmerge z)$\\
            $P9$ & $\delta\leftmerge x = \delta$\\
            $P10$ & $\epsilon\leftmerge x = x$\\
            $P11$ & $x\leftmerge \epsilon = x$\\
            $C1$ & $e_1\mid e_2 = \gamma(e_1,e_2)$\\
            $C2$ & $e_1\mid (e_2\cdot y) = \gamma(e_1,e_2)\cdot y$\\
            $C3$ & $(e_1\cdot x)\mid e_2 = \gamma(e_1,e_2)\cdot x$\\
            $C4$ & $(e_1\cdot x)\mid (e_2\cdot y) = \gamma(e_1,e_2)\cdot (x\between y)$\\
            $C5$ & $(x+ y)\mid z = (x\mid z) + (y\mid z)$\\
            $C6$ & $x\mid (y+ z) = (x\mid y)+ (x\mid z)$\\
            $C7$ & $\delta\mid x = \delta$\\
            $C8$ & $x\mid\delta = \delta$\\
            $C9$ & $\epsilon\mid x = \delta$\\
            $C10$ & $x\mid\epsilon = \delta$\\
            $CE1$ & $\Theta(e) = e$\\
            $CE2$ & $\Theta(\delta) = \delta$\\
            $CE3$ & $\Theta(\epsilon) = \epsilon$\\
            $CE4$ & $\Theta(x+ y) = \Theta(x)\triangleleft y + \Theta(y)\triangleleft x$\\
            $CE5$ & $\Theta(x\cdot y)=\Theta(x)\cdot\Theta(y)$\\
            $CE6$ & $\Theta(x\leftmerge y) = ((\Theta(x)\triangleleft y)\leftmerge y)+ ((\Theta(y)\triangleleft x)\leftmerge x)$\\
            $CE7$ & $\Theta(x\mid y) = ((\Theta(x)\triangleleft y)\mid y)+ ((\Theta(y)\triangleleft x)\mid x)$\\
        \end{tabular}
        \caption{Axioms of $qAPTC_G$}
        \label{AxiomsForqAPTCG}
    \end{table}
\end{center}

\begin{center}
    \begin{table}
        \begin{tabular}{@{}ll@{}}
            \hline No. &Axiom\\
            $U1$ & $(\sharp(e_1,e_2))\quad e_1\triangleleft e_2 = \tau$\\
            $U2$ & $(\sharp(e_1,e_2),e_2\leq e_3)\quad e_1\triangleleft e_3 = e_1$\\
            $U3$ & $(\sharp(e_1,e_2),e_2\leq e_3)\quad e3\triangleleft e_1 = \tau$\\
            $U4$ & $e\triangleleft \delta = e$\\
            $U5$ & $\delta \triangleleft e = \delta$\\
            $U6$ & $e\triangleleft \epsilon = e$\\
            $U7$ & $\epsilon \triangleleft e = e$\\
            $U8$ & $(x+ y)\triangleleft z = (x\triangleleft z)+ (y\triangleleft z)$\\
            $U9$ & $(x\cdot y)\triangleleft z = (x\triangleleft z)\cdot (y\triangleleft z)$\\
            $U10$ & $(x\leftmerge y)\triangleleft z = (x\triangleleft z)\leftmerge (y\triangleleft z)$\\
            $U11$ & $(x\mid y)\triangleleft z = (x\triangleleft z)\mid (y\triangleleft z)$\\
            $U12$ & $x\triangleleft (y+ z) = (x\triangleleft y)\triangleleft z$\\
            $U13$ & $x\triangleleft (y\cdot z)=(x\triangleleft y)\triangleleft z$\\
            $U14$ & $x\triangleleft (y\leftmerge z) = (x\triangleleft y)\triangleleft z$\\
            $U15$ & $x\triangleleft (y\mid z) = (x\triangleleft y)\triangleleft z$\\
            $D1$ & $e\notin H\quad\partial_H(e) = e$\\
            $D2$ & $e\in H\quad \partial_H(e) = \delta$\\
            $D3$ & $\partial_H(\delta) = \delta$\\
            $D4$ & $\partial_H(x+ y) = \partial_H(x)+\partial_H(y)$\\
            $D5$ & $\partial_H(x\cdot y) = \partial_H(x)\cdot\partial_H(y)$\\
            $D6$ & $\partial_H(x\leftmerge y) = \partial_H(x)\leftmerge\partial_H(y)$\\
            $G12$ & $\phi(x\leftmerge y) =\phi x\leftmerge \phi y$\\
            $G13$ & $\phi(x\mid y) =\phi x\mid \phi y$\\
            $G14$ & $\delta\leftmerge \phi = \delta$\\
            $G15$ & $\phi\mid \delta = \delta$\\
            $G16$ & $\delta\mid \phi = \delta$\\
            $G17$ & $\phi\leftmerge \epsilon = \phi$\\
            $G18$ & $\epsilon\leftmerge \phi = \phi$\\
            $G19$ & $\phi\mid \epsilon = \delta$\\
            $G20$ & $\epsilon\mid \phi = \delta$\\
            $G21$ & $\phi\leftmerge\neg\phi = \delta$\\
            $G22$ & $\Theta(\phi) = \phi$\\
            $G23$ & $\partial_H(\phi) = \phi$\\
            $G24$ & $\phi_0\leftmerge\cdots\leftmerge\phi_n = \delta$ if $\forall s_0,\cdots,s_n,\varrho_1,\cdots,\varrho_n\in S,\exists i\leq n.test(\neg\phi_i,s_0\cup\cdots\cup s_n\cup \varrho_1\cup\cdots\cup\varrho_n)$\\
        \end{tabular}
        \caption{Axioms of $qAPTC_G$(continuing)}
        \label{AxiomsForqAPTCG2}
    \end{table}
\end{center}

\begin{definition}[Basic terms of $qAPTC_G$]\label{BTAPTCG}
The set of basic terms of $qAPTC_G$, $\mathcal{B}(qAPTC_G)$, is inductively defined as follows:
\begin{enumerate}
    \item $\mathbb{E}\subset\mathcal{B}(qAPTC_G)$;
    \item $G\subset\mathcal{B}(qAPTC_G)$;
    \item if $e\in \mathbb{E}, t\in\mathcal{B}(qAPTC_G)$ then $e\cdot t\in\mathcal{B}(qAPTC_G)$;
    \item if $\phi\in G, t\in\mathcal{B}(qAPTC_G)$ then $\phi\cdot t\in\mathcal{B}(qAPTC_G)$;
    \item if $t,s\in\mathcal{B}(qAPTC_G)$ then $t+ s\in\mathcal{B}(qAPTC_G)$.
    \item if $t,s\in\mathcal{B}(qAPTC_G)$ then $t\leftmerge s\in\mathcal{B}(qAPTC_G)$.
\end{enumerate}
\end{definition}

Based on the definition of basic terms for $qAPTC_G$ (see Definition \ref{BTAPTCG}) and axioms of $qAPTC_G$, we can prove the elimination theorem of $qAPTC_G$.

\begin{theorem}[Elimination theorem of $qAPTC_G$]\label{ETAPTCG}
Let $p$ be a closed $qAPTC_G$ term. Then there is a basic $qAPTC_G$ term $q$ such that $qAPTC_G\vdash p=q$.
\end{theorem}

\begin{proof}
The same as that of $qAPTC_G$, we omit the proof, please refer to \cite{ATC} for details.
\end{proof}

We will define a term-deduction system which gives the operational semantics of $qAPTC_G$. Two atomic events $e_1$ and $e_2$ are in race condition, which are denoted $e_1\% e_2$.

\begin{center}
    \begin{table}
        $$\frac{}{\langle e_1\parallel\cdots \parallel e_n,s,\varrho\rangle\xrightarrow{\{e_1,\cdots,e_n\}}\langle\surd,s,\varrho'\rangle}\textrm{ if }s'\in effect(e_1,s)\cup\cdots\cup effect(e_n,s)$$

        $$\frac{}{\langle\phi_1\parallel\cdots\parallel \phi_n,s,\varrho\rangle\rightarrow\langle\surd,s,\varrho\rangle}\textrm{ if }test(\phi_1,s),\cdots,test(\phi_n,s)$$

        $$\frac{\langle x,s,\varrho\rangle\xrightarrow{e_1}\langle\surd,s,\varrho'\rangle\quad \langle y,s,\varrho\rangle\xrightarrow{e_2}\langle\surd,s,\varrho''\rangle}{\langle x\parallel y,s,\varrho\rangle\xrightarrow{\{e_1,e_2\}}\langle\surd,s,\varrho'\cup\varrho''\rangle} \quad\frac{\langle x,s,\varrho\rangle\xrightarrow{e_1}\langle x',s,\varrho'\rangle\quad \langle y,s,\varrho\rangle\xrightarrow{e_2}\langle\surd,s,\varrho''\rangle}{\langle x\parallel y,s,\varrho\rangle\xrightarrow{\{e_1,e_2\}}\langle x',s,\varrho'\cup\varrho''\rangle}$$

        $$\frac{\langle x,s,\varrho\rangle\xrightarrow{e_1}\langle\surd,s,\varrho'\rangle\quad \langle y,s,\varrho\rangle\xrightarrow{e_2}\langle y',s,\varrho''\rangle}{\langle x\parallel y,s,\varrho\rangle\xrightarrow{\{e_1,e_2\}}\langle y',s,\varrho'\cup\varrho''\rangle} \quad\frac{\langle x,s,\varrho\rangle\xrightarrow{e_1}\langle x',s,\varrho'\rangle\quad \langle y,s,\varrho\rangle\xrightarrow{e_2}\langle y',s,\varrho''\rangle}{\langle x\parallel y,s,\varrho\rangle\xrightarrow{\{e_1,e_2\}}\langle x'\between y',s,\varrho'\cup\varrho''\rangle}$$

        $$\frac{\langle x,s,\varrho\rangle\xrightarrow{e_1}\langle\surd,s,\varrho'\rangle\quad \langle y,s,\varrho\rangle\xnrightarrow{e_2}\quad(e_1\%e_2)}{\langle x\parallel y,s,\varrho\rangle\xrightarrow{e_1}\langle y,s,\varrho'\rangle} \quad\frac{\langle x,s,\varrho\rangle\xrightarrow{e_1}\langle x',s,\varrho'\rangle\quad \langle y,s,\varrho\rangle\xnrightarrow{e_2}\quad(e_1\%e_2)}{\langle x\parallel y,s,\varrho\rangle\xrightarrow{e_1}\langle x'\between y,s,\varrho'\rangle}$$

        $$\frac{\langle x,s,\varrho\rangle\xnrightarrow{e_1}\quad \langle y,s,\varrho\rangle\xrightarrow{e_2}\langle\surd,s,\varrho''\rangle\quad(e_1\%e_2)}{\langle x\parallel y,s,\varrho\rangle\xrightarrow{e_2}\langle x,s,\varrho''\rangle} \quad\frac{\langle x,s,\varrho\rangle\xnrightarrow{e_1}\quad \langle y,s,\varrho\rangle\xrightarrow{e_2}\langle y',s,\varrho''\rangle\quad(e_1\%e_2)}{\langle x\parallel y,s,\varrho\rangle\xrightarrow{e_2}\langle x\between y',s,\varrho''\rangle}$$

        $$\frac{\langle x,s,\varrho\rangle\xrightarrow{e_1}\langle\surd,s,\varrho'\rangle\quad \langle y,s,\varrho\rangle\xrightarrow{e_2}\langle\surd,s,\varrho''\rangle \quad(e_1\leq e_2)}{\langle x\leftmerge y,s,\varrho\rangle\xrightarrow{\{e_1,e_2\}}\langle \surd,s,\varrho'\cup\varrho''\rangle}$$
        
        $$\frac{\langle x,s,\varrho\rangle\xrightarrow{e_1}\langle x',s,\varrho'\rangle\quad \langle y,s,\varrho\rangle\xrightarrow{e_2}\langle\surd,s,\varrho''\rangle \quad(e_1\leq e_2)}{\langle x\leftmerge y,s,\varrho\rangle\xrightarrow{\{e_1,e_2\}}\langle x',s,\varrho'\cup\varrho''\rangle}$$

        $$\frac{\langle x,s,\varrho\rangle\xrightarrow{e_1}\langle\surd,s,\varrho'\rangle\quad \langle y,s,\varrho\rangle\xrightarrow{e_2}\langle y',s,\varrho''\rangle \quad(e_1\leq e_2)}{\langle x\leftmerge y,s,\varrho\rangle\xrightarrow{\{e_1,e_2\}}\langle y',s,\varrho'\cup\varrho''\rangle}$$
        
        $$\frac{\langle x,s,\varrho\rangle\xrightarrow{e_1}\langle x',s,\varrho'\rangle\quad \langle y,s,\varrho\rangle\xrightarrow{e_2}\langle y',s,\varrho''\rangle \quad(e_1\leq e_2)}{\langle x\leftmerge y,s,\varrho\rangle\xrightarrow{\{e_1,e_2\}}\langle x'\between y',s,\varrho'\cup\varrho''\rangle}$$

        $$\frac{\langle x,s,\varrho\rangle\xrightarrow{e_1}\langle\surd,s,\varrho'\rangle\quad \langle y,s,\varrho\rangle\xrightarrow{e_2}\langle\surd,s,\varrho''\rangle}{\langle x\mid y,s,\varrho\rangle\xrightarrow{\gamma(e_1,e_2)}\langle\surd,effect(\gamma(e_1,e_2),s),\varrho\rangle} \quad\frac{\langle x,s,\varrho\rangle\xrightarrow{e_1}\langle x',s,\varrho'\rangle\quad \langle y,s,\varrho\rangle\xrightarrow{e_2}\langle\surd,s,\varrho''\rangle}{\langle x\mid y,s,\varrho\rangle\xrightarrow{\gamma(e_1,e_2)}\langle x',s,effect(\gamma(e_1,e_2),),\varrho\rangle}$$

        $$\frac{\langle x,s,\varrho\rangle\xrightarrow{e_1}\langle\surd,s,\varrho'\rangle\quad \langle y,s,\varrho\rangle\xrightarrow{e_2}\langle y',s,\varrho''\rangle}{\langle x\mid y,s,\varrho\rangle\xrightarrow{\gamma(e_1,e_2)}\langle y',effect(\gamma(e_1,e_2),s),\varrho\rangle} \quad\frac{\langle x,s,\varrho\rangle\xrightarrow{e_1}\langle x',s,\varrho'\rangle\quad \langle y,s,\varrho\rangle\xrightarrow{e_2}\langle y',s,\varrho''\rangle}{\langle x\mid y,s,\varrho\rangle\xrightarrow{\gamma(e_1,e_2)}\langle x'\between y',effect(\gamma(e_1,e_2),s),\varrho\rangle}$$
        \caption{Transition rules of $qAPTC_G$}
        \label{TRForqqAPTCG}
    \end{table}
\end{center}

\begin{center}
    \begin{table}

        $$\frac{\langle x,s,\varrho\rangle\xrightarrow{e_1}\langle\surd,s,\varrho'\rangle\quad (\sharp(e_1,e_2))}{\langle \Theta(x),s,\varrho\rangle\xrightarrow{e_1}\langle\surd,s,\varrho'\rangle} \quad\frac{\langle x,s,\varrho\rangle\xrightarrow{e_2}\langle\surd,s,\varrho''\rangle\quad (\sharp(e_1,e_2))}{\langle\Theta(x),s,\varrho\rangle\xrightarrow{e_2}\langle\surd,s,\varrho''\rangle}$$

        $$\frac{\langle x,s,\varrho\rangle\xrightarrow{e_1}\langle x',s,\varrho'\rangle\quad (\sharp(e_1,e_2))}{\langle\Theta(x),s,\varrho\rangle\xrightarrow{e_1}\langle\Theta(x'),s,\varrho'\rangle} \quad\frac{\langle x,s,\varrho\rangle\xrightarrow{e_2}\langle x'',s,\varrho''\rangle\quad (\sharp(e_1,e_2))}{\langle\Theta(x),s,\varrho\rangle\xrightarrow{e_2}\langle\Theta(x''),s,\varrho''\rangle}$$

        $$\frac{\langle x,s,\varrho\rangle\xrightarrow{e_1}\langle\surd,s,\varrho'\rangle \quad \langle y,s,\varrho\rangle\nrightarrow^{e_2}\quad (\sharp(e_1,e_2))}{\langle x\triangleleft y,s,\varrho\rangle\xrightarrow{\tau}\langle\surd,s,\varrho'\rangle}
        \quad\frac{\langle x,s,\varrho\rangle\xrightarrow{e_1}\langle x',s,\varrho'\rangle \quad \langle y,s,\varrho\rangle\nrightarrow^{e_2}\quad (\sharp(e_1,e_2))}{\langle x\triangleleft y,s,\varrho\rangle\xrightarrow{\tau}\langle x',s,\varrho'\rangle}$$

        $$\frac{\langle x,s,\varrho\rangle\xrightarrow{e_1}\langle\surd,s,\varrho\rangle \quad \langle y,s,\varrho\rangle\nrightarrow^{e_3}\quad (\sharp(e_1,e_2),e_2\leq e_3)}{\langle x\triangleleft y,s,\varrho\rangle\xrightarrow{e_1}\langle\surd,s,\varrho'\rangle}$$
        
        $$\frac{\langle x,s,\varrho\rangle\xrightarrow{e_1}\langle x',s,\varrho'\rangle \quad \langle y,s,\varrho\rangle\nrightarrow^{e_3}\quad (\sharp(e_1,e_2),e_2\leq e_3)}{\langle x\triangleleft y,s,\varrho\rangle\xrightarrow{e_1}\langle x',s,\varrho'\rangle}$$

        $$\frac{\langle x,s,\varrho\rangle\xrightarrow{e_3}\langle\surd,s,\varrho'\rangle \quad \langle y,s,\varrho\rangle\nrightarrow^{e_2}\quad (\sharp(e_1,e_2),e_1\leq e_3)}{\langle x\triangleleft y,s,\varrho\rangle\xrightarrow{\tau}\langle\surd,s,\varrho'\rangle}$$
        
        $$\frac{\langle x,s,\varrho\rangle\xrightarrow{e_3}\langle x',s,\varrho'\rangle \quad \langle y,s,\varrho\rangle\nrightarrow^{e_2}\quad (\sharp(e_1,e_2),e_1\leq e_3)}{\langle x\triangleleft y,s,\varrho\rangle\xrightarrow{\tau}\langle x',s,\varrho'\rangle}$$

        $$\frac{\langle x,s,\varrho\rangle\xrightarrow{e}\langle\surd,s,\varrho'\rangle}{\langle\partial_H(x),s,\varrho\rangle\xrightarrow{e}\langle\surd,s,\varrho'\rangle}\quad (e\notin H)\quad\frac{\langle x,s,\varrho\rangle\xrightarrow{e}\langle x',s,\varrho'\rangle}{\langle\partial_H(x),s,\varrho\rangle\xrightarrow{e}\langle\partial_H(x'),s,\varrho'\rangle}\quad(e\notin H)$$
        \caption{Transition rules of $qAPTC_G$ (continuing)}
        \label{TRForqqAPTCG2}
    \end{table}
\end{center}

\begin{theorem}[Generalization of $qAPTC_G$ with respect to $qBATC_G$]
$qAPTC_G$ is a generalization of $qBATC_G$.
\end{theorem}

\begin{proof}
It follows from the following three facts.

\begin{enumerate}
  \item The transition rules of $qBATC_G$ in are all source-dependent;
  \item The sources of the transition rules $qAPTC_G$ contain an occurrence of $\between$, or $\parallel$, or $\leftmerge$, or $\mid$, or $\Theta$, or $\triangleleft$, or $\partial_H$;
  \item The transition rules of $qAPTC_G$ are all source-dependent.
\end{enumerate}

So, $qAPTC_G$ is a generalization of $qBATC_G$, that is, $qBATC_G$ is an embedding of $qAPTC_G$, as desired.
\end{proof}

\begin{theorem}[Congruence of $qAPTC_G$ with respect to truly concurrent bisimulation equivalences]\label{CAPTCG}
(1) Pomset bisimulation equivalence $\sim_{p}$ is a congruence with respect to $qAPTC_G$.

(2) Step bisimulation equivalence $\sim_{s}$ is a congruence with respect to $qAPTC_G$.

(3) Hp-bisimulation equivalence $\sim_{hp}$ is a congruence with respect to $qAPTC_G$.

(4) Hhp-bisimulation equivalence $\sim_{hhp}$ is a congruence with respect to $qAPTC_G$.
\end{theorem}

\begin{proof}
It is obvious that truly concurrent bisimulations $\sim_{p}$, $\sim_s$, $\sim_{hp}$ and $\sim_{hhp}$ are all equivalent relations with respect to $qAPTC$. So, it is sufficient to prove
that truly concurrent bisimulations $\sim_{p}$, $\sim_s$, $\sim_{hp}$ and $\sim_{hhp}$ are preserved for $\between$, $\parallel$, $\leftmerge$, $\mid$, $\Theta$, $\triangleleft$ and $\partial_H$
according to the transition rules in Table \ref{TRForqqAPTCG}, that is, if $x\sim_{p}x'$ and $y\sim_{p}y'$, then $x\between y\sim_{p}x'\between y'$, $x\parallel y\sim_{p}x'\parallel y'$,
$x\leftmerge y\sim_{p}x'\leftmerge y'$, $x\mid y\sim_{p}x'\mid y'$, $\Theta(x)\sim_{p}\Theta(x')$, $x\triangleleft y\sim_{p}x'\triangleleft y'$, and $\partial_H(x)\sim_{p}\partial_H(x')$; if $x\sim_{s}x'$ and $y\sim_{s}y'$,
then $x\between y\sim_{s}x'\between y'$, $x\parallel y\sim_{s}x'\parallel y'$,
$x\leftmerge y\sim_{s}x'\leftmerge y'$, $x\mid y\sim_{s}x'\mid y'$, $\Theta(x)\sim_{s}\Theta(x')$, $x\triangleleft y\sim_{s}x'\triangleleft y'$, and $\partial_H(x)\sim_{s}\partial_H(x')$;
if $x\sim_{hp}x'$ and $y\sim_{hp}y'$, then $x\between y\sim_{hp}x'\between y'$, $x\parallel y\sim_{hp}x'\parallel y'$,
$x\leftmerge y\sim_{hp}x'\leftmerge y'$, $x\mid y\sim_{hp}x'\mid y'$, $\Theta(x)\sim_{hp}\Theta(x')$, $x\triangleleft y\sim_{hp}x'\triangleleft y'$, and $\partial_H(x)\sim_{hp}\partial_H(x')$; and if $x\sim_{hhp}x'$ and $y\sim_{hhp}y'$,
then $x\between y\sim_{hhp}x'\between y'$, $x\parallel y\sim_{hhp}x'\parallel y'$,
$x\leftmerge y\sim_{hhp}x'\leftmerge y'$, $x\mid y\sim_{hhp}x'\mid y'$, $\Theta(x)\sim_{hhp}\Theta(x')$, $x\triangleleft y\sim_{hhp}x'\triangleleft y'$ and $\partial_H(x)\sim_{hhp}\partial_H(x')$.
The proof is quit trivial, and we leave the proof as an exercise for the readers.
\end{proof}

\begin{theorem}[Soundness of $qAPTC_G$ modulo truly concurrent bisimulation equivalences]\label{SAPTCG}
(1) Let $x$ and $y$ be $qAPTC_G$ terms. If $qAPTC_G\vdash x=y$, then $x\sim_{p} y$.

(2) Let $x$ and $y$ be $qAPTC_G$ terms. If $qAPTC_G\vdash x=y$, then $x\sim_{s} y$.

(3) Let $x$ and $y$ be $qAPTC_G$ terms. If $qAPTC_G\vdash x=y$, then $x\sim_{hp} y$.

(4) Let $x$ and $y$ be $qAPTC_G$ terms. If $qAPTC_G\vdash x=y$, then $x\sim_{hhp} y$.
\end{theorem}

\begin{proof}
(1) Since pomset bisimulation $\sim_{p}$ is both an equivalent and a congruent relation, we only need to check if each axiom in Table \ref{AxiomsForqAPTCG} is sound
modulo pomset bisimulation equivalence. We leave the proof as an exercise for the readers.

(2) Since  step bisimulation $\sim_{s}$ is both an equivalent and a congruent relation, we only need to check if each axiom in Table \ref{AxiomsForqAPTCG} is sound modulo
step bisimulation equivalence. We leave the proof as an exercise for the readers.

(3) Since hp-bisimulation $\sim_{hp}$ is both an equivalent and a congruent relation, we only need to check if each axiom in Table \ref{AxiomsForqAPTCG} is sound modulo
hp-bisimulation equivalence. We leave the proof as an exercise for the readers.

(4) Since hhp-bisimulation $\sim_{hhp}$ is both an equivalent and a congruent relation, we only need to check if each axiom in Table \ref{AxiomsForqAPTCG} is sound modulo
hhp-bisimulation equivalence. We leave the proof as an exercise for the readers.
\end{proof}

\begin{theorem}[Completeness of $qAPTC_G$ modulo truly concurrent bisimulation equivalences]\label{CAPTCG}
(1) Let $p$ and $q$ be closed $qAPTC_G$ terms, if $p\sim_{p} q$ then $p=q$.

(2) Let $p$ and $q$ be closed $qAPTC_G$ terms, if $p\sim_{s} q$ then $p=q$.

(3) Let $p$ and $q$ be closed $qAPTC_G$ terms, if $p\sim_{hp} q$ then $p=q$.

(4) Let $p$ and $q$ be closed $qAPTC_G$ terms, if $p\sim_{hhp} q$ then $p=q$.
\end{theorem}

\begin{proof}
According to the definition of truly concurrent bisimulation equivalences $\sim_{p}$, $\sim_{s}$, $\sim_{hp}$ and $\sim_{hhp}$, $p\sim_{p}q$, $p\sim_{s}q$, $p\sim_{hp}q$ and $p\sim_{hhp}q$ implies
both the bisimilarities between $p$ and $q$, and also the in the same quantum states. According to the completeness of $APTC_G$ (please refer to \cite{ATC} for details), we can get the
completeness of $qAPTC_G$.
\end{proof}

\subsection{Recursion}\label{qorecg}

In this subsection, we introduce recursion to capture infinite processes based on $qAPTC_G$. In the following, $E,F,G$ are recursion specifications, $X,Y,Z$ are recursive variables.

\begin{definition}[Guarded recursive specification]
A recursive specification

$$X_1=t_1(X_1,\cdots,X_n)$$
$$...$$
$$X_n=t_n(X_1,\cdots,X_n)$$

is guarded if the right-hand sides of its recursive equations can be adapted to the form by applications of the axioms in $qAPTC$ and replacing recursion variables by the right-hand sides of their recursive equations,

$$(a_{11}\leftmerge\cdots\leftmerge a_{1i_1})\cdot s_1(X_1,\cdots,X_n)+\cdots+(a_{k1}\leftmerge\cdots\leftmerge a_{ki_k})\cdot s_k(X_1,\cdots,X_n)+(b_{11}\leftmerge\cdots\leftmerge b_{1j_1})+\cdots+(b_{1j_1}\leftmerge\cdots\leftmerge b_{lj_l})$$

where $a_{11},\cdots,a_{1i_1},a_{k1},\cdots,a_{ki_k},b_{11},\cdots,b_{1j_1},b_{1j_1},\cdots,b_{lj_l}\in \mathbb{E}$, and the sum above is allowed to be empty, in which case it represents the deadlock $\delta$. And there does not exist an infinite sequence of $\epsilon$-transitions $\langle X|E\rangle\rightarrow\langle X'|E\rangle\rightarrow\langle X''|E\rangle\rightarrow\cdots$.
\end{definition}

\begin{center}
    \begin{table}
        $$\frac{\langle t_i(\langle X_1|E\rangle,\cdots,\langle X_n|E\rangle),s,\varrho\rangle\xrightarrow{\{e_1,\cdots,e_k\}}\langle\surd,s,\varrho'\rangle}{\langle\langle X_i|E\rangle,s,\varrho\rangle\xrightarrow{\{e_1,\cdots,e_k\}}\langle\surd,s,\varrho'\rangle}$$
        $$\frac{\langle t_i(\langle X_1|E\rangle,\cdots,\langle X_n|E\rangle),s,\varrho\rangle\xrightarrow{\{e_1,\cdots,e_k\}} \langle y,s,\varrho'\rangle}{\langle\langle X_i|E\rangle,s,\varrho\rangle\xrightarrow{\{e_1,\cdots,e_k\}} \langle y,s,\varrho'\rangle}$$
        \caption{Transition rules of guarded recursion}
        \label{TRForGRG}
    \end{table}
\end{center}

The $RDP$ (Recursive Definition Principle) and the $RSP$ (Recursive Specification Principle) are shown in Table \ref{RDPRSP}.

\begin{center}
\begin{table}
  \begin{tabular}{@{}ll@{}}
\hline No. &Axiom\\
  $RDP$ & $\langle X_i|E\rangle = t_i(\langle X_1|E,\cdots,X_n|E\rangle)\quad (i\in\{1,\cdots,n\})$\\
  $RSP$ & if $y_i=t_i(y_1,\cdots,y_n)$ for $i\in\{1,\cdots,n\}$, then $y_i=\langle X_i|E\rangle \quad(i\in\{1,\cdots,n\})$\\
\end{tabular}
\caption{Recursive definition and specification principle}
\label{RDPRSP}
\end{table}
\end{center}

\begin{theorem}[Conservitivity of $qAPTC_G$ with guarded recursion]
$qAPTC_G$ with guarded recursion is a conservative extension of $qAPTC_G$.
\end{theorem}

\begin{proof}
It follows from the following three facts.

\begin{enumerate}
  \item The transition rules of $qAPTC_G$ in are all source-dependent;
  \item The sources of the transition rules $qAPTC_G$ with guarded recursion contain only one constant;
  \item The transition rules of $qAPTC_G$ with guarded recursion are all source-dependent.
\end{enumerate}

So, $qAPTC$ with guarded recursion is a conservative extension of $qAPTC_G$, as desired.
\end{proof}

\begin{theorem}[Congruence theorem of $qAPTC_G$ with guarded recursion]
Truly concurrent bisimulation equivalences $\sim_{p}$, $\sim_s$, $\sim_{hp}$ and $\sim_{hhp}$ are all congruences with respect to $qAPTC_G$ with guarded recursion.
\end{theorem}

\begin{proof}
It follows the following two facts:
\begin{enumerate}
  \item in a guarded recursive specification, right-hand sides of its recursive equations can be adapted to the form by applications of the axioms in $qAPTC_G$ and replacing recursion
  variables by the right-hand sides of their recursive equations;
  \item truly concurrent bisimulation equivalences $\sim_{p}$, $\sim_{s}$, $\sim_{hp}$ and $\sim_{hhp}$ are all congruences with respect to all operators of $qAPTC$.
\end{enumerate}
\end{proof}

\begin{theorem}[Elimination theorem of $qAPTC_G$ with linear recursion]\label{ETRecursionG}
Each process term in $qAPTC_G$ with linear recursion is equal to a process term $\langle X_1|E\rangle$ with $E$ a linear recursive specification.
\end{theorem}

\begin{proof}
The same as that of $qAPTC_G$ with linear recursion, we omit the proof, please refer to \cite{ATC} for details.
\end{proof}

\begin{theorem}[Soundness of $qAPTC_G$ with guarded recursion]\label{SAPTC_GRG}
Let $x$ and $y$ be $qAPTC_G$ with guarded recursion terms. If $qAPTC_G\textrm{ with guarded recursion}\vdash x=y$, then

(1) $x\sim_{s} y$.

(2) $x\sim_{p} y$.

(3) $x\sim_{hp} y$.

(4) $x\sim_{hhp} y$.
\end{theorem}

\begin{proof}
(1) Since pomset bisimulation $\sim_{p}$ is both an equivalent and a congruent relation, we only need to check if each axiom in Table \ref{RDPRSP} is sound
modulo pomset bisimulation equivalence. We leave the proof as an exercise for the readers.

(2) Since  step bisimulation $\sim_{s}$ is both an equivalent and a congruent relation, we only need to check if each axiom in Table \ref{RDPRSP} is sound modulo
step bisimulation equivalence. We leave the proof as an exercise for the readers.

(3) Since hp-bisimulation $\sim_{hp}$ is both an equivalent and a congruent relation, we only need to check if each axiom in Table \ref{RDPRSP} is sound modulo
hp-bisimulation equivalence. We leave the proof as an exercise for the readers.

(4) Since hhp-bisimulation $\sim_{hhp}$ is both an equivalent and a congruent relation, we only need to check if each axiom in Table \ref{RDPRSP} is sound modulo
hhp-bisimulation equivalence. We leave the proof as an exercise for the readers.
\end{proof}

\begin{theorem}[Completeness of $qAPTC_G$ with linear recursion]\label{CAPTC_GRG}
Let $p$ and $q$ be closed $qAPTC_G$ with linear recursion terms, then,

(1) if $p\sim_{s} q$ then $p=q$.

(2) if $p\sim_{p} q$ then $p=q$.

(3) if $p\sim_{hp} q$ then $p=q$.

(4) if $p\sim_{hhp} q$ then $p=q$.
\end{theorem}

\begin{proof}
According to the definition of truly concurrent bisimulation equivalences $\sim_{p}$, $\sim_{s}$, $\sim_{hp}$ and $\sim_{hhp}$, $p\sim_{p}q$, $p\sim_{s}q$, $p\sim_{hp}q$ and $p\sim_{hhp}q$ implies
both the bisimilarities between $p$ and $q$, and also the in the same quantum states. According to the completeness of $APTC_G$ with linear recursion (please refer to \cite{ATC} for details), we can get the
completeness of $qAPTC_G$ with linear recursion.
\end{proof}

\subsection{Abstraction}\label{qoabsg}

To abstract away from the internal implementations of a program, and verify that the program exhibits the desired external behaviors, the silent step $\tau$ and abstraction operator 
$\tau_I$ are introduced, where $I\subseteq \mathbb{E}\cup G_{at}$ denotes the internal events or guards. The silent step $\tau$ represents the internal events or guards, when we 
consider the external behaviors of a process, $\tau$ steps can be removed, that is, $\tau$ steps must keep silent. The transition rule of $\tau$ is shown in Table \ref{TRForqTauG}. 
In the following, let the atomic event $e$ range over $\mathbb{E}\cup\{\epsilon\}\cup\{\delta\}\cup\{\tau\}$, and $\phi$ range over $G\cup \{\tau\}$, and let the communication 
function $\gamma:\mathbb{E}\cup\{\tau\}\times \mathbb{E}\cup\{\tau\}\rightarrow \mathbb{E}\cup\{\delta\}$, with each communication involved $\tau$ resulting in $\delta$. We use 
$\tau(s)$ to denote $effect(\tau,s)$, for the fact that $\tau$ only change the state of internal data environment, that is, for the external data environments, $s=\tau(s)$.

\begin{center}
    \begin{table}
        $$\frac{}{\langle\tau,s,\varrho\rangle\rightarrow\langle\surd,s,\varrho\rangle}\textrm{ if }test(\tau,s)$$
        $$\frac{}{\langle\tau,s,\varrho\rangle\xrightarrow{\tau}\langle\surd,\tau(s)\rangle}$$
        \caption{Transition rule of the silent step}
        \label{TRForqTauG}
    \end{table}
\end{center}

\begin{definition}[Guarded linear recursive specification]\label{GLRSG}
A linear recursive specification $E$ is guarded if there does not exist an infinite sequence of $\tau$-transitions $\langle X|E\rangle\xrightarrow{\tau}\langle X'|E\rangle\xrightarrow{\tau}\langle X''|E\rangle\xrightarrow{\tau}\cdots$, and there does not exist an infinite sequence of $\epsilon$-transitions $\langle X|E\rangle\rightarrow\langle X'|E\rangle\rightarrow\langle X''|E\rangle\rightarrow\cdots$.
\end{definition}

\begin{theorem}[Conservitivity of $qAPTC_G$ with silent step and guarded linear recursion]
$qAPTC_G$ with silent step and guarded linear recursion is a conservative extension of $qAPTC_G$ with linear recursion.
\end{theorem}

\begin{proof}
Since the transition rules of $qAPTC_G$ with silent step and guarded linear recursion are source-dependent, and the transition rules for abstraction operator in Table
\ref{TRForqTauG} contain only a fresh constant $\tau$ in their source, so the transition rules of $qAPTC_G$ with silent step and guarded linear recursion is a conservative extension
of those of $qAPTC_G$ with guarded linear recursion.
\end{proof}

\begin{theorem}[Congruence theorem of $qAPTC_G$ with silent step and guarded linear recursion]
Rooted branching truly concurrent bisimulation equivalences $\approx_{rbp}$, $\approx_{rbs}$, $\approx_{rbhp}^{fr}$ and $\approx_{rbhhp}$ are all congruences with respect to $qAPTC_G$ with silent step and guarded linear recursion.
\end{theorem}

\begin{proof}
It follows the following three facts:
\begin{enumerate}
  \item in a guarded linear recursive specification, right-hand sides of its recursive equations can be adapted to the form by applications of the axioms in $qAPTC_G$ and replacing
  recursion variables by the right-hand sides of their recursive equations;
  \item truly concurrent bisimulation equivalences $\sim_{p}$, $\sim_{s}$, $\sim_{hp}$ and $\sim_{hhp}$ are all congruences with respect to all operators of
  $qAPTC_G$, while truly concurrent bisimulation equivalences $\sim_{p}$, $\sim_{s}$, $\sim_{hp}$ and $\sim_{hhp}$ imply the corresponding rooted
  branching truly concurrent bisimulations $\approx_{rbp}$, $\approx_{rbs}$, $\approx_{rbhp}$ and $\approx_{rbhhp}$, so rooted branching truly concurrent
  bisimulations $\approx_{rbp}$, $\approx_{rbs}$, $\approx_{rbhp}$ and $\approx_{rbhhp}$ are all congruences with respect to all operators of $qAPTC_G$;
  \item While $\mathbb{E}$ is extended to $\mathbb{E}\cup\{\tau\}$, it can be proved that rooted branching truly concurrent
  bisimulations $\approx_{rbp}$, $\approx_{rbs}$, $\approx_{rbhp}$ and $\approx_{rbhhp}$ are all congruences with respect to all operators of $qAPTC_G$, we omit it.
\end{enumerate}
\end{proof}

We design the axioms for the silent step $\tau$ in Table \ref{AxiomsForqTauG}.

\begin{center}
\begin{table}
  \begin{tabular}{@{}ll@{}}
\hline No. &Axiom\\
  $B1$ & $e\cdot\tau=e$\\
  $B2$ & $e\cdot(\tau\cdot(x+y)+x)=e\cdot(x+y)$\\
  $B3$ & $x\leftmerge\tau=x$\\
  $G26$ & $\phi\cdot\tau=\phi$\\
  $G27$ & $\phi\cdot(\tau\cdot(x+y)+x)=\phi\cdot(x+y)$\\
\end{tabular}
\caption{Axioms of silent step}
\label{AxiomsForqTauG}
\end{table}
\end{center}

\begin{theorem}[Elimination theorem of $qAPTC_G$ with silent step and guarded linear recursion]\label{ETTauG}
Each process term in $qAPTC_G$ with silent step and guarded linear recursion is equal to a process term $\langle X_1|E\rangle$ with $E$ a guarded linear recursive specification.
\end{theorem}

\begin{proof}
The same as that of $APTC_G$ with silent step and guarded linear recursion, we omit the proof, please refer to \cite{ATC} for details.
\end{proof}

\begin{theorem}[Soundness of $qAPTC_G$ with silent step and guarded linear recursion]\label{SAPTC_GTAUG}
Let $x$ and $y$ be $qAPTC_G$ with silent step and guarded linear recursion terms. If $qAPTC_G$ with silent step and guarded linear recursion $\vdash x=y$, then

(1) $x\approx_{rbs} y$.

(2) $x\approx_{rbp} y$.

(3) $x\approx_{rbhp} y$.

(4) $x\approx_{rbhhp} y$.
\end{theorem}

\begin{proof}
(1) Since rooted branching pomset bisimulation $\approx_{rbp}$ is both an equivalent and a congruent relation with respect to $qAPTC$ with silent step and guarded
linear recursion, we only need to check if each axiom in Table \ref{AxiomsForqTauG} is sound modulo rooted branching pomset bisimulation $\approx_{rbp}$. We leave them as
exercises to the readers.

(2) Since rooted branching step bisimulation $\approx_{rbs}$ is both an equivalent and a congruent relation with respect to $qAPTC$ with silent step and guarded
linear recursion, we only need to check if each axiom in Table \ref{AxiomsForqTauG} is sound modulo rooted branching step bisimulation $\approx_{rbs}$. We leave them
as exercises to the readers.

(3) Since rooted branching hp-bisimulation $\approx_{rbhp}$ is both an equivalent and a congruent relation with respect to $qAPTC$ with silent step and guarded linear
recursion, we only need to check if each axiom in Table \ref{AxiomsForqTauG} is sound modulo rooted branching hp-bisimulation $\approx_{rbhp}$. We leave them as exercises
to the readers.

(4) Since rooted branching hhp-bisimulation $\approx_{rbhhp}$ is both an equivalent and a congruent relation with respect to $APPTC_G$ with silent step and guarded linear
recursion, we only need to check if each axiom in Table \ref{AxiomsForqTauG} is sound modulo rooted branching hhp-bisimulation $\approx_{rbhhp}$. We leave them as exercises
to the readers.
\end{proof}

\begin{theorem}[Completeness of $qAPTC_G$ with silent step and guarded linear recursion]\label{CAPTC_GTAUG}
Let $p$ and $q$ be closed $qAPTC_G$ with silent step and guarded linear recursion terms, then,

(1) if $p\approx_{rbs} q$ then $p=q$.

(2) if $p\approx_{rbp} q$ then $p=q$.

(3) if $p\approx_{rbhp} q$ then $p=q$.

(4) if $p\approx_{rbhhp} q$ then $p=q$.
\end{theorem}

\begin{proof}
According to the definition of truly concurrent rooted branching bisimulation equivalences $\approx_{rbp}$, $\approx_{rbs}$, $\approx_{rbhp}$ and $\approx_{rbhhp}$,
$p\approx_{rbp}q$, $p\approx_{rbs}q$, $p\approx_{rbhp}q$ and $p\approx_{rbhhp}q$ implies
both the rooted branching bisimilarities between $p$ and $q$, and also the in the same quantum states. According to the completeness of $APTC_G$ with silent step and guarded linear recursion (please refer to \cite{ATC} for details), we can get the
completeness of $qAPTC_G$ with silent step and guarded linear recursion.
\end{proof}

The unary abstraction operator $\tau_I$ ($I\subseteq \mathbb{E}\cup G_{at}$) renames all atomic events or atomic guards in $I$ into $\tau$. $qAPTC_G$ with silent step and abstraction
operator is called $qAPTC_{G_{\tau}}$. The transition rules of operator $\tau_I$ are shown in Table \ref{TRForqAbstractionG}.

\begin{center}
    \begin{table}
        $$\frac{\langle x,s,\varrho\rangle\xrightarrow{e}\langle\surd,s,\varrho'\rangle}{\langle\tau_I(x),s,\varrho\rangle\xrightarrow{e}\langle\surd,s,\varrho'\rangle}\quad e\notin I
        \quad\quad\frac{\langle x,s,\varrho\rangle\xrightarrow{e}\langle x',s,\varrho'\rangle}{\langle\tau_I(x),s,\varrho\rangle\xrightarrow{e}\langle\tau_I(x'),s,\varrho'\rangle}\quad e\notin I$$

        $$\frac{\langle x,s,\varrho\rangle\xrightarrow{e}\langle\surd,s,\varrho'\rangle}{\langle\tau_I(x),s,\varrho\rangle\xrightarrow{\tau}\langle\surd,\tau(s)\rangle}\quad e\in I
        \quad\quad\frac{\langle x,s,\varrho\rangle\xrightarrow{e}\langle x',s,\varrho'\rangle}{\langle\tau_I(x),s,\varrho\rangle\xrightarrow{\tau}\langle\tau_I(x'),\tau(s)\rangle}\quad e\in I$$
        \caption{Transition rule of the abstraction operator}
        \label{TRForqAbstractionG}
    \end{table}
\end{center}

\begin{theorem}[Conservitivity of $qAPTC_{G_{\tau}}$ with guarded linear recursion]
$qAPTC_{G_{\tau}}$ with guarded linear recursion is a conservative extension of $qAPTC_G$ with silent step and guarded linear recursion.
\end{theorem}

\begin{proof}
Since the transition rules of $qAPTC_G$ with silent step and guarded linear recursion are source-dependent, and the transition rules for abstraction operator in Table
\ref{TRForqAbstractionG} contain only a fresh operator $\tau_I$ in their source, so the transition rules of $qAPTC_{G_{\tau}}$ with guarded linear recursion is a conservative extension
of those of $qAPTC_G$ with silent step and guarded linear recursion.
\end{proof}

\begin{theorem}[Congruence theorem of $qAPTC_{G_{\tau}}$ with guarded linear recursion]
Rooted branching truly concurrent bisimulation equivalences $\approx_{rbp}$, $\approx_{rbs}$ $\approx_{rbhp}^{fr}$, and $\approx_{rbhhp}$ are all congruences with respect to $qAPTC_{G_{\tau}}$ with guarded linear recursion.
\end{theorem}

\begin{proof}
(1) It is easy to see that rooted branching pomset bisimulation is an equivalent relation on $qAPTC_{G_{\tau}}$ with guarded linear recursion terms, we only need to
prove that $\approx_{rbp}$ is preserved by the operator $\tau_I$. It is trivial and we leave the proof as an exercise for the readers.

(2) It is easy to see that rooted branching step bisimulation is an equivalent relation on $qAPTC_{G_{\tau}}$ with guarded linear recursion terms, we only need to
prove that $\approx_{rbs}$ is preserved by the operator $\tau_I$. It is trivial and we leave the proof as an exercise for the readers.

(3) It is easy to see that rooted branching hp-bisimulation is an equivalent relation on $qAPTC_{G_{\tau}}$ with guarded linear recursion terms, we only need to
prove that $\approx_{rbhp}$ is preserved by the operator $\tau_I$. It is trivial and we leave the proof as an exercise for the readers.

(4) It is easy to see that rooted branching hhp-bisimulation is an equivalent relation on $qAPTC_{G_{\tau}}$ with guarded linear recursion terms, we only need to
prove that $\approx_{rbhhp}$ is preserved by the operator $\tau_I$. It is trivial and we leave the proof as an exercise for the readers.
\end{proof}

We design the axioms for the abstraction operator $\tau_I$ in Table \ref{AxiomsForqAbstractionG}.

\begin{center}
\begin{table}
  \begin{tabular}{@{}ll@{}}
\hline No. &Axiom\\
  $TI1$ & $e\notin I\quad \tau_I(e)=e$\\
  $TI2$ & $e\in I\quad \tau_I(e)=\tau$\\
  $TI3$ & $\tau_I(\delta)=\delta$\\
  $TI4$ & $\tau_I(x+y)=\tau_I(x)+\tau_I(y)$\\
  $TI5$ & $\tau_I(x\cdot y)=\tau_I(x)\cdot\tau_I(y)$\\
  $TI6$ & $\tau_I(x\leftmerge y)=\tau_I(x)\leftmerge\tau_I(y)$\\
  $G28$ & $\phi\notin I\quad \tau_I(\phi)=\phi$\\
  $G29$ & $\phi\in I\quad \tau_I(\phi)=\tau$\\
\end{tabular}
\caption{Axioms of abstraction operator}
\label{AxiomsForqAbstractionG}
\end{table}
\end{center}

\begin{theorem}[Soundness of $qAPTC_{G_{\tau}}$ with guarded linear recursion]\label{SAPTC_GABSG}
Let $x$ and $y$ be $qAPTC_{G_{\tau}}$ with guarded linear recursion terms. If $qAPTC_{G_{\tau}}$ with guarded linear recursion $\vdash x=y$, then

(1) $x\approx_{rbs} y$.

(2) $x\approx_{rbp} y$.

(3) $x\approx_{rbhp} y$.

(4) $x\approx_{rbhhp} y$.
\end{theorem}

\begin{proof}
(1) Since rooted branching step bisimulation $\approx_{rbs}$ is both an equivalent and a congruent relation with respect to $qAPTC_{G_{\tau}}$ with guarded linear
recursion, we only need to check if each axiom in Table \ref{AxiomsForqAbstractionG} is sound modulo rooted branching step bisimulation $\approx_{rbs}$. We leave them as
exercises to the readers.

(2) Since rooted branching pomset bisimulation $\approx_{rbp}$ is both an equivalent and a congruent relation with respect to $qAPTC_{G_{\tau}}$ with guarded linear
recursion, we only need to check if each axiom in Table \ref{AxiomsForqAbstractionG} is sound modulo rooted branching pomset bisimulation $\approx_{rbp}$. We leave them
as exercises to the readers.

(3) Since rooted branching hp-bisimulation $\approx_{rbhp}$ is both an equivalent and a congruent relation with respect to $qAPTC_{G_{\tau}}$ with guarded linear
recursion, we only need to check if each axiom in Table \ref{AxiomsForqAbstractionG} is sound modulo rooted branching hp-bisimulation $\approx_{rbhp}$. We leave them as
exercises to the readers.

(4) Since rooted branching hhp-bisimulation $\approx_{rbhhp}$ is both an equivalent and a congruent relation with respect to $qAPTC_{G_{\tau}}$ with guarded linear
recursion, we only need to check if each axiom in Table \ref{AxiomsForqAbstractionG} is sound modulo rooted branching hhp-bisimulation $\approx_{rbhhp}$. We leave them as
exercises to the readers.
\end{proof}

\begin{definition}[Cluster]
Let $E$ be a guarded linear recursive specification, and $I\subseteq \mathbb{E}$. Two recursion variable $X$ and $Y$ in $E$ are in the same cluster for $I$ iff there exist sequences of
transitions $\langle X|E\rangle\xrightarrow{\{b_{11},\cdots, b_{1i}\}}\cdots\xrightarrow{\{b_{m1},\cdots, b_{mi}\}}\langle Y|E\rangle$ and $\langle Y|E\rangle\xrightarrow{\{c_{11},\cdots, c_{1j}\}}\cdots\xrightarrow{\{c_{n1},\cdots, c_{nj}\}}\langle X|E\rangle$, where $b_{11},\cdots,b_{mi},c_{11},\cdots,c_{nj}\in I\cup\{\tau\}$.

$a_1\leftmerge\cdots\leftmerge a_k$ or $(a_1\leftmerge\cdots\leftmerge a_k) X$ is an exit for the cluster $C$ iff: (1) $a_1\leftmerge\cdots\leftmerge a_k$ or
$(a_1\leftmerge\cdots\leftmerge a_k) X$ is a summand at the right-hand side of the recursive equation for a recursion variable in $C$, and (2) in the case of
$(a_1\leftmerge\cdots\leftmerge a_k) X$, either $a_l\notin I\cup\{\tau\}(l\in\{1,2,\cdots,k\})$ or $X\notin C$.
\end{definition}

\begin{center}
\begin{table}
  \begin{tabular}{@{}ll@{}}
\hline No. &Axiom\\
  $CFAR$ & If $X$ is in a cluster for $I$ with exits \\
           & $\{(a_{11}\leftmerge\cdots\leftmerge a_{1i})Y_1,\cdots,(a_{m1}\leftmerge\cdots\leftmerge a_{mi})Y_m, b_{11}\leftmerge\cdots\leftmerge b_{1j},\cdots,b_{n1}\leftmerge\cdots\leftmerge b_{nj}\}$, \\
           & then $\tau\cdot\tau_I(\langle X|E\rangle)=$\\
           & $\tau\cdot\tau_I((a_{11}\leftmerge\cdots\leftmerge a_{1i})\langle Y_1|E\rangle+\cdots+(a_{m1}\leftmerge\cdots\leftmerge a_{mi})\langle Y_m|E\rangle+b_{11}\leftmerge\cdots\leftmerge b_{1j}+\cdots+b_{n1}\leftmerge\cdots\leftmerge b_{nj})$\\
\end{tabular}
\caption{Cluster fair abstraction rule}
\label{qCFARLeft}
\end{table}
\end{center}

\begin{theorem}[Soundness of $CFAR$]
$CFAR$ is sound modulo rooted branching truly concurrent bisimulation equivalences $\approx_{rbs}$, $\approx_{rbp}$, $\approx_{rbhp}$ and $\approx_{rbhhp}$.
\end{theorem}

\begin{proof}
(1) Since rooted branching step bisimulation $\approx_{rbs}$ is both an equivalent and a congruent relation with respect to $qAPTC_{G_{\tau}}$ with guarded linear
recursion, we only need to check if each axiom in Table \ref{qCFARLeft} is sound modulo rooted branching step bisimulation $\approx_{rbs}$. We leave them as
exercises to the readers.

(2) Since rooted branching pomset bisimulation $\approx_{rbp}$ is both an equivalent and a congruent relation with respect to $qAPTC_{G_{\tau}}$ with guarded linear
recursion, we only need to check if each axiom in Table \ref{qCFARLeft} is sound modulo rooted branching pomset bisimulation $\approx_{rbp}$. We leave them
as exercises to the readers.

(3) Since rooted branching hp-bisimulation $\approx_{rbhp}$ is both an equivalent and a congruent relation with respect to $qAPTC_{G_{\tau}}$ with guarded linear
recursion, we only need to check if each axiom in Table \ref{qCFARLeft} is sound modulo rooted branching hp-bisimulation $\approx_{rbhp}$. We leave them as
exercises to the readers.

(4) Since rooted branching hhp-bisimulation $\approx_{rbhhp}$ is both an equivalent and a congruent relation with respect to $qAPTC_{G_{\tau}}$ with guarded linear
recursion, we only need to check if each axiom in Table \ref{qCFARLeft} is sound modulo rooted branching hhp-bisimulation $\approx_{rbhhp}$. We leave them as
exercises to the readers.
\end{proof}

\begin{theorem}[Completeness of $qAPTC_{G_{\tau}}$ with guarded linear recursion and $CFAR$]\label{CCFARG}
Let $p$ and $q$ be closed $qAPTC_{G_{\tau}}$ with guarded linear recursion and $CFAR$ terms, then,

(1) if $p\approx_{rbs} q$ then $p=q$.

(2) if $p\approx_{rbp} q$ then $p=q$.

(3) if $p\approx_{rbhp} q$ then $p=q$.

(4) if $p\approx_{rbhhp} q$ then $p=q$.
\end{theorem}

\begin{proof}
According to the definition of truly concurrent rooted branching bisimulation equivalences $\approx_{rbp}$, $\approx_{rbs}$, $\approx_{rbhp}$ and $\approx_{rbhhp}$,
$p\approx_{rbp}q$, $p\approx_{rbs}q$, $p\approx_{rbhp}q$ and $p\approx_{rbhhp}q$ implies
both the rooted branching bisimilarities between $p$ and $q$, and also the in the same quantum states. According to the completeness of $APTC_{G_{\tau}}$ guarded linear recursion (please refer to \cite{ATC} for details), we can get the
completeness of $qAPTC_{G_{\tau}}$ with guarded linear recursion.
\end{proof}

\subsection{Quantum Entanglement}\label{qe1}

If two quantum variables are entangled, then a quantum operation performed on one variable, then state of the other quantum variable is also changed. So, the entangled states must be
all the inner variables or all the public variables. We will introduced a mechanism to explicitly define quantum entanglement in open quantum systems.
A new constant called shadow constant denoted $\circledS^e_i$ corresponding to a specific quantum operation.
If there are $n$ quantum variables entangled, they maybe be distributed in different quantum systems, with a quantum operation performed on one variable, there should be one
$\circledS^e_i$ ($1\leq i\leq n-1$) executed on each variable in the other $n-1$ variables. Thus, distributed variables are all hidden behind actions.
In the following, we let $\circledS\in \mathbb{E}$.

The axiom system of the shadow constant $\circledS$ is shown in Table \ref{AxiomsForQE1}.

\begin{center}
\begin{table}
  \begin{tabular}{@{}ll@{}}
\hline No. &Axiom\\
  $SC1$ & $\circledS\cdot x = x$ \\
  $SC2$ & $x\cdot\circledS = x$\\
  $SC3$ & $e\leftmerge\circledS^e=e$\\
  $SC4$ & $\circledS^e\leftmerge e=e$\\
  $SC5$ & $e\leftmerge(\circledS^e\cdot y) = e\cdot y$\\
  $SC6$ & $\circledS^e\leftmerge(e\cdot y) = e\cdot y$\\
  $SC7$ & $(e\cdot x)\leftmerge\circledS^e = e\cdot x$\\
  $SC8$ & $(\circledS^e\cdot x)\leftmerge e = e\cdot x$\\
  $SC9$ & $(e\cdot x)\leftmerge(\circledS^e\cdot y) = e\cdot (x\between y)$\\
  $SC10$ & $(\circledS^e\cdot x)\leftmerge(e\cdot y) = e\cdot (x\between y)$\\
\end{tabular}
\caption{Axioms of quantum entanglement}
\label{AxiomsForQE1}
\end{table}
\end{center}

The transition rules of constant $\circledS$ are as Table \ref{TRForENT1} shows.

\begin{center}
    \begin{table}
        $$\frac{}{\langle\circledS,s,\varrho\rangle\rightarrow\langle\surd,s,\varrho\rangle}$$
        $$\frac{\langle x, s,\varrho\rangle\xrightarrow{e}\langle x',s,\varrho'\rangle\quad \langle y, s,\varrho'\rangle\xrightarrow{\circledS^e}\langle y',s,\varrho'\rangle}{\langle x\leftmerge y,s,\varrho\rangle\xrightarrow{e}\langle x'\between y', s,\varrho'\rangle}$$
        $$\frac{\langle x, s,\varrho\rangle\xrightarrow{e}\langle\surd,s,\varrho'\rangle\quad \langle y, s,\varrho'\rangle\xrightarrow{\circledS^e}\langle y',s,\varrho'\rangle}{\langle x\leftmerge y,s,\varrho\rangle\xrightarrow{e}\langle y', s,\varrho'\rangle}$$
        $$\frac{\langle x, s,\varrho'\rangle\xrightarrow{\circledS^e}\langle\surd,s,\varrho'\rangle\quad \langle y, s,\varrho\rangle\xrightarrow{e}\langle y',s,\varrho'\rangle}{\langle x\leftmerge y,s,\varrho\rangle\xrightarrow{e}\langle y', s,\varrho'\rangle}$$
        $$\frac{\langle x, s,\varrho\rangle\xrightarrow{e}\langle\surd,s,\varrho'\rangle\quad \langle y, s,\varrho'\rangle\xrightarrow{\circledS^e}\langle\surd,s,\varrho'\rangle}{\langle x\leftmerge y,s,\varrho\rangle\xrightarrow{e}\langle \surd, s,\varrho'\rangle}$$
        \caption{Transition rules of constant $\circledS$}
        \label{TRForENT1}
    \end{table}
\end{center}

\begin{theorem}[Elimination theorem of $qAPTC_{G_{\tau}}$ with guarded linear recursion and shadow constant]
Let $p$ be a closed $qAPTC_{G_{\tau}}$ with guarded linear recursion and shadow constant term. Then there is a closed $qAPTC$ term such that $qAPTC_{G_{\tau}}$ with guarded linear
recursion and shadow constant$\vdash p=q$.
\end{theorem}

\begin{proof}
We leave the proof to the readers as an excise.
\end{proof}

\begin{theorem}[Conservitivity of $qAPTC_{G_{\tau}}$ with guarded linear recursion and shadow constant]
$qAPTC_{G_{\tau}}$ with guarded linear recursion and shadow constant is a conservative extension of $qAPTC_{G_{\tau}}$ with guarded linear recursion.
\end{theorem}

\begin{proof}
We leave the proof to the readers as an excise.
\end{proof}

\begin{theorem}[Congruence theorem of $qAPTC_{G_{\tau}}$ with guarded linear recursion and shadow constant]
Rooted branching truly concurrent bisimulation equivalences $\approx_{rbp}$, $\approx_{rbs}$, $\approx_{rbhp}$ and $\approx_{rbhhp}$ are all congruences with respect to $qAPTC_{G_{\tau}}$
with guarded linear recursion and shadow constant.
\end{theorem}

\begin{proof}
We leave the proof to the readers as an excise.
\end{proof}

\begin{theorem}[Soundness of $qAPTC_{G_{\tau}}$ with guarded linear recursion and shadow constant]
Let $p$ and $q$ be closed $qAPTC_{G_{\tau}}$ with guarded linear recursion and shadow constant terms. If $qAPTC_{G_{\tau}}$ with guarded linear recursion and shadow constant$\vdash x=y$, then

\begin{enumerate}
  \item $x\approx_{rbs} y$;
  \item $x\approx_{rbp} y$;
  \item $x\approx_{rbhp} y$;
  \item $x\approx_{rbhhp} y$.
\end{enumerate}
\end{theorem}

\begin{proof}
We leave the proof to the readers as an excise.
\end{proof}

\begin{theorem}[Completeness of $qAPTC_{G_{\tau}}$ with guarded linear recursion and shadow constant]
Let $p$ and $q$ are closed $qAPTC_{G_{\tau}}$ with guarded linear recursion and shadow constant terms, then,

\begin{enumerate}
  \item if $p\approx_{rbs} q$ then $p=q$;
  \item if $p\approx_{rbp} q$ then $p=q$;
  \item if $p\approx_{rbhp} q$ then $p=q$;
  \item if $p\approx_{rbhhp} q$ then $p=q$.
\end{enumerate}
\end{theorem}

\begin{proof}
We leave the proof to the readers as an excise.
\end{proof}

\subsection{Unification of Quantum and Classical Computing for Open Quantum Systems}\label{uni1}

We give the transition rules under quantum configuration for traditional atomic actions (events) $e'\in\mathbb{E}$ as Table \ref{TRForBPA3} shows.

\begin{center}
    \begin{table}
        $$\frac{}{\langle e',s,\varrho\rangle\xrightarrow{e'}\langle\surd,s',\varrho\rangle}$$
        $$\frac{\langle x,s,\varrho\rangle\xrightarrow{e'}\langle\surd,s',\varrho\rangle}{\langle x+y,s,\varrho\rangle\xrightarrow{e'}\langle\surd,s',\varrho\rangle}$$
        $$\frac{\langle x,s,\varrho\rangle\xrightarrow{e'}\langle x',s',\varrho\rangle}{\langle x+y,s,\varrho\rangle\xrightarrow{e'}\langle x',s',\varrho\rangle}$$
        $$\frac{\langle y,s,\varrho\rangle\xrightarrow{e'}\langle\surd,s',\varrho\rangle}{\langle x+y,s,\varrho\rangle\xrightarrow{e'}\langle\surd,s',\varrho\rangle}$$
        $$\frac{\langle y,s,\varrho\rangle\xrightarrow{e'}\langle y',s',\varrho\rangle}{\langle x+y,s,\varrho\rangle\xrightarrow{e'}\langle y',s',\varrho\rangle}$$
        $$\frac{\langle x,s,\varrho\rangle\xrightarrow{e'}\langle\surd,s',\varrho\rangle}{\langle x\cdot y,s,\varrho\rangle\xrightarrow{e'}\langle y,s',\varrho\rangle}$$
        $$\frac{\langle x,s,\varrho\rangle\xrightarrow{e'}\langle x',s',\varrho\rangle}{\langle x\cdot y,s,\varrho\rangle\xrightarrow{e'}\langle x'\cdot y,s',\varrho\rangle}$$
        \caption{Transition rules of $BATC_G$ under quantum configuration}
        \label{TRForBPA3}
    \end{table}
\end{center}

And the axioms for traditional actions are the same as those of $qBATC_G$. And it is natural can be extended to $qAPTC_G$, recursion and abstraction. So, quantum and classical computing
are unified under the framework of $qAPTC_G$ for open quantum systems.

\newpage\section{Applications of $qAPTC_G$}\label{aqaptcg}

Quantum and classical computing in open systems are unified with $qAPTC_G$, which have the same equational logic and the same quantum configuration based operational semantics.
The unification can be used widely in verification for the behaviors of quantum and classical computing mixed systems. In this chapter, we show its usage in verification of the
quantum communication protocols.

\subsection{Verification of BB84 Protocol}\label{VBB844}

\begin{figure}
  \centering
  \includegraphics{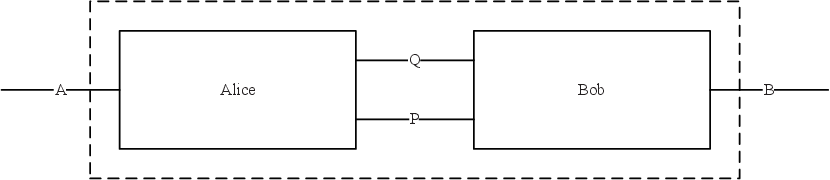}
  \caption{The BB84 protocol.}
  \label{BB844}
\end{figure}

The BB84 protocol is used to create a private key between two parities, Alice and Bob. Firstly, we introduce the basic BB84 protocol briefly, which is illustrated in Figure \ref{BB844}.

\begin{enumerate}
  \item Alice create two string of bits with size $n$ randomly, denoted as $B_a$ and $K_a$.
  \item Alice generates a string of qubits $q$ with size $n$, and the $i$th qubit in $q$ is $|x_y\rangle$, where $x$ is the $i$th bit of $B_a$ and $y$ is the $i$th bit of $K_a$.
  \item Alice sends $q$ to Bob through a quantum channel $Q$ between Alice and Bob.
  \item Bob receives $q$ and randomly generates a string of bits $B_b$ with size $n$.
  \item Bob measures each qubit of $q$ according to a basis by bits of $B_b$. And the measurement results would be $K_b$, which is also with size $n$.
  \item Bob sends his measurement bases $B_b$ to Alice through a public channel $P$.
  \item Once receiving $B_b$, Alice sends her bases $B_a$ to Bob through channel $P$, and Bob receives $B_a$.
  \item Alice and Bob determine that at which position the bit strings $B_a$ and $B_b$ are equal, and they discard the mismatched bits of $B_a$ and $B_b$. Then the remaining bits of 
  $K_a$ and $K_b$, denoted as $K_a'$ and $K_b'$ with $K_{a,b}=K_a'=K_b'$.
\end{enumerate}

We re-introduce the basic BB84 protocol in an abstract way with more technical details as Figure \ref{BB844} illustrates.

Now, we assume a special measurement operation $Rand[q;B_a]$ which create a string of $n$ random bits $B_a$ from the $q$ quantum system, and the same as $Rand[q;K_a]$, $Rand[q';B_b]$. $M[q;K_b]$ denotes the Bob's measurement operation of $q$. The generation of $n$ qubits $q$ through two quantum operations $Set_{K_a}[q]$ and $H_{B_a}[q]$. Alice sends $q$ to Bob through the quantum channel $Q$ by quantum communicating action $send_{Q}(q)$ and Bob receives $q$ through $Q$ by quantum communicating action $receive_{Q}(q)$. Bob sends $B_b$ to Alice through the public channel $P$ by classical communicating action $send_{P}(B_b)$ and Alice receives $B_b$ through channel $P$ by classical communicating action $receive_{P}(B_b)$, and the same as $send_{P}(B_a)$ and $receive_{P}(B_a)$. Alice and Bob generate the private key $K_{a,b}$ by a classical comparison action $cmp(K_{a,b},K_a,K_b,B_a,B_b)$. Let Alice and Bob be a system $AB$ and let interactions between Alice and Bob be internal actions. $AB$ receives external input $D_i$ through channel $A$ by communicating action $receive_A(D_i)$ and sends results $D_o$ through channel $B$ by communicating action $send_B(D_o)$.

Then the state transition of Alice can be described by qACP as follows.

\begin{eqnarray}
&&A=\sum_{D_i\in \Delta_i}receive_A(D_i)\cdot A_1\nonumber\\
&&A_1=Rand[q;B_a]\cdot A_2\nonumber\\
&&A_2=Rand[q;K_a]\cdot A_3\nonumber\\
&&A_3=Set_{K_a}[q]\cdot A_4\nonumber\\
&&A_4=H_{B_a}[q]\cdot A_5\nonumber\\
&&A_5=send_Q(q)\cdot A_6\nonumber\\
&&A_6=receive_P(B_b)\cdot A_7\nonumber\\
&&A_7=send_P(B_a)\cdot A_8\nonumber\\
&&A_8=cmp(K_{a,b},K_a,K_b,B_a,B_b)\cdot A_9\nonumber\\
&&A_9=\{B_{a_i}=B_{b_i}\}\cdot generate(K_a)\cdot A+\{B_{a_i}\neq B_{b_i}\}\cdot discard\cdot A\nonumber
\end{eqnarray}

where $\Delta_i$ is the collection of the input data.

And the state transition of Bob can be described by qACP as follows.

\begin{eqnarray}
&&B=receive_Q(q)\cdot B_1\nonumber\\
&&B_1=Rand[q';B_b]\cdot B_2\nonumber\\
&&B_2=M[q;K_b]\cdot B_3\nonumber\\
&&B_3=send_P(B_b)\cdot B_4\nonumber\\
&&B_4=receive_P(B_a)\cdot B_5\nonumber\\
&&B_5=cmp(K_{a,b},K_a,K_b,B_a,B_b)\cdot B_6\nonumber\\
&&B_6=\{B_{a_i}=B_{b_i}\}\cdot generate(K_b)\cdot B_7+\{B_{a_i}\neq B_{b_i}\}\cdot discard\cdot B_7\nonumber\\
&&B_7=\sum_{D_o\in\Delta_o}send_B(D_o)\cdot B\nonumber
\end{eqnarray}

where $\Delta_o$ is the collection of the output data.

The send action and receive action of the same data through the same channel can communicate each other, otherwise, a deadlock $\delta$ will be caused. We define the following communication functions.

\begin{eqnarray}
&&\gamma(send_Q(q),receive_Q(q))\triangleq c_Q(q)\nonumber\\
&&\gamma(send_P(B_b),receive_P(B_b))\triangleq c_P(B_b)\nonumber\\
&&\gamma(send_P(B_a),receive_P(B_a))\triangleq c_P(B_a)\nonumber
\end{eqnarray}

Let $A$ and $B$ in parallel, then the system $AB$ can be represented by the following process term.

$$\tau_I(\partial_H(\Theta(A\between B)))$$

where $H=\{send_Q(q),receive_Q(q),send_P(B_b),receive_P(B_b),send_P(B_a),receive_P(B_a)\}$ and 
$I=\{Rand[q;B_a], Rand[q;K_a], Set_{K_a}[q], H_{B_a}[q], Rand[q';B_b], M[q;K_b], c_Q(q), c_P(B_b),\\ c_P(B_a), cmp(K_{a,b},K_a,K_b,B_a,B_b),\{B_{a_i}=B_{b_i}\},\{B_{a_i}\neq B_{b_i}\},\\
generate(K_a),generate(K_b),discard\}$.

Then we get the following conclusion.

\begin{theorem}
The basic BB84 protocol $\tau_I(\partial_H(\Theta(A\between B)))$ exhibits desired external behaviors.
\end{theorem}

\begin{proof}
We can get $\tau_I(\partial_H(\Theta(A\between B)))=\sum_{D_i\in \Delta_i}\sum_{D_o\in\Delta_o}receive_A(D_i)\leftmerge send_B(D_o)\leftmerge \tau_I(\partial_H(\Theta(A\between B)))$. So, the basic
BB84 protocol $\tau_I(\partial_H(\Theta(A\between B)))$ exhibits desired external behaviors.
\end{proof}

\subsection{Verification of E91 Protocol}\label{VE914}

With support of Entanglement merge $\between$, now, qACP can be used to verify quantum protocols utilizing entanglement. The E91 protocol\cite{E91} is the first quantum protocol which utilizes entanglement and mixes quantum and classical information. In this section, we take an example of verification for the E91 protocol.

\begin{figure}
  \centering
  \includegraphics{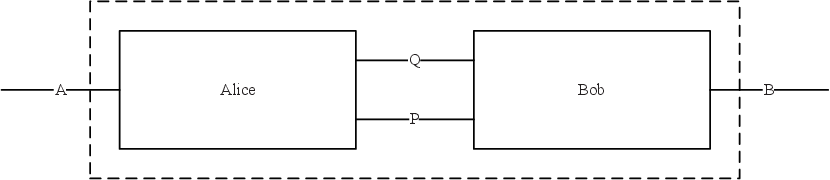}
  \caption{The E91 protocol.}
  \label{E914}
\end{figure}

The E91 protocol is used to create a private key between two parities, Alice and Bob. Firstly, we introduce the basic E91 protocol briefly, which is illustrated in Figure \ref{E914}.

\begin{enumerate}
  \item Alice generates a string of EPR pairs $q$ with size $n$, i.e., $2n$ particles, and sends a string of qubits $q_b$ from each EPR pair with $n$ to Bob through a quantum channel $Q$, remains the other string of qubits $q_a$ from each pair with size $n$.
  \item Alice create two string of bits with size $n$ randomly, denoted as $B_a$ and $K_a$.
  \item Bob receives $q_b$ and randomly generates a string of bits $B_b$ with size $n$.
  \item Alice measures each qubit of $q_a$ according to a basis by bits of $B_a$. And the measurement results would be $K_a$, which is also with size $n$.
  \item Bob measures each qubit of $q_b$ according to a basis by bits of $B_b$. And the measurement results would be $K_b$, which is also with size $n$.
  \item Bob sends his measurement bases $B_b$ to Alice through a public channel $P$.
  \item Once receiving $B_b$, Alice sends her bases $B_a$ to Bob through channel $P$, and Bob receives $B_a$.
  \item Alice and Bob determine that at which position the bit strings $B_a$ and $B_b$ are equal, and they discard the mismatched bits of $B_a$ and $B_b$. Then the remaining bits of $K_a$ and $K_b$, denoted as $K_a'$ and $K_b'$ with $K_{a,b}=K_a'=K_b'$.
\end{enumerate}

We re-introduce the basic E91 protocol in an abstract way with more technical details as Figure \ref{E914} illustrates.

Now, $M[q_a;K_a]$ denotes the Alice's measurement operation of $q_a$, and $\circledS_{M[q_a;K_a]}$ denotes the responding shadow constant; $M[q_b;K_b]$ denotes the Bob's measurement operation of $q_b$, and $\circledS_{M[q_b;K_b]}$ denotes the responding shadow constant. Alice sends $q_b$ to Bob through the quantum channel $Q$ by quantum communicating action $send_{Q}(q_b)$ and Bob receives $q_b$ through $Q$ by quantum communicating action $receive_{Q}(q_b)$. Bob sends $B_b$ to Alice through the public channel $P$ by classical communicating action $send_{P}(B_b)$ and Alice receives $B_b$ through channel $P$ by classical communicating action $receive_{P}(B_b)$, and the same as $send_{P}(B_a)$ and $receive_{P}(B_a)$. Alice and Bob generate the private key $K_{a,b}$ by a classical comparison action $cmp(K_{a,b},K_a,K_b,B_a,B_b)$. Let Alice and Bob be a system $AB$ and let interactions between Alice and Bob be internal actions. $AB$ receives external input $D_i$ through channel $A$ by communicating action $receive_A(D_i)$ and sends results $D_o$ through channel $B$ by communicating action $send_B(D_o)$.

Then the state transition of Alice can be described by qACP as follows.

\begin{eqnarray}
&&A=\sum_{D_i\in \Delta_i}receive_A(D_i)\cdot A_1\nonumber\\
&&A_1=send_Q(q_b)\cdot A_2\nonumber\\
&&A_2=M[q_a;K_a]\cdot A_3\nonumber\\
&&A_3=\circledS_{M[q_b;K_b]}\cdot A_4\nonumber\\
&&A_4=receive_P(B_b)\cdot A_5\nonumber\\
&&A_5=send_P(B_a)\cdot A_6\nonumber\\
&&A_6=cmp(K_{a,b},K_a,K_b,B_a,B_b)\cdot A_7\nonumber\\
&&A_7=\{B_{a_i}=B_{b_i}\}\cdot generate(K_a)\cdot A+\{B_{a_i}\neq B_{b_i}\}\cdot discard\cdot A\nonumber
\end{eqnarray}

where $\Delta_i$ is the collection of the input data.

And the state transition of Bob can be described by qACP as follows.

\begin{eqnarray}
&&B=receive_Q(q_b)\cdot B_1\nonumber\\
&&B_1=\circledS_{M[q_a;K_a]}\cdot B_2\nonumber\\
&&B_2=M[q_b;K_b]\cdot B_3\nonumber\\
&&B_3=send_P(B_b)\cdot B_4\nonumber\\
&&B_4=receive_P(B_a)\cdot B_5\nonumber\\
&&B_5=cmp(K_{a,b},K_a,K_b,B_a,B_b)\cdot B_6\nonumber\\
&&B_6=\{B_{a_i}=B_{b_i}\}\cdot generate(K_b)\cdot B_7+\{B_{a_i}\neq B_{b_i}\}\cdot discard\cdot B_7\nonumber\\
&&B_7=\sum_{D_o\in\Delta_o}send_B(D_o)\cdot B\nonumber
\end{eqnarray}

where $\Delta_o$ is the collection of the output data.

The send action and receive action of the same data through the same channel can communicate each other, otherwise, a deadlock $\delta$ will be caused. The quantum operation and its shadow constant pair will lead entanglement occur, otherwise, a deadlock $\delta$ will occur. We define the following communication functions.

\begin{eqnarray}
&&\gamma(send_Q(q_b),receive_Q(q_b))\triangleq c_Q(q_b)\nonumber\\
&&\gamma(send_P(B_b),receive_P(B_b))\triangleq c_P(B_b)\nonumber\\
&&\gamma(send_P(B_a),receive_P(B_a))\triangleq c_P(B_a)\nonumber
\end{eqnarray}

Let $A$ and $B$ in parallel, then the system $AB$ can be represented by the following process term.

$$\tau_I(\partial_H(\Theta(A\between B)))$$

where $H=\{send_Q(q_b),receive_Q(q_b),send_P(B_b),receive_P(B_b),send_P(B_a),receive_P(B_a),\\ M[q_a;K_a], \circledS_{M[q_a;K_a]}, M[q_b;K_b], \circledS_{M[q_b;K_b]}\}$ and
$I=\{c_Q(q_b), c_P(B_b), c_P(B_a), M[q_a;K_a], M[q_b;K_b],\\ cmp(K_{a,b},K_a,K_b,B_a,B_b),\{B_{a_i}=B_{b_i}\},\{B_{a_i}\neq B_{b_i}\},\\
generate(K_a),generate(K_b),discard\}$.

Then we get the following conclusion.

\begin{theorem}
The basic E91 protocol $\tau_I(\partial_H(A\parallel B))$ exhibits desired external behaviors.
\end{theorem}

\begin{proof}
We can get $\tau_I(\partial_H(\Theta(A\between B)))=\sum_{D_i\in \Delta_i}\sum_{D_o\in\Delta_o}receive_A(D_i)\leftmerge send_B(D_o)\leftmerge \tau_I(\partial_H(\Theta(A\between B)))$.
So, the basic E91 protocol $\tau_I(\partial_H(\Theta(A\between B)))$ exhibits desired external behaviors.
\end{proof}

\subsection{Verification of B92 Protocol}\label{VB924}

The famous B92 protocol\cite{B92} is a quantum key distribution protocol, in which quantum information and classical information are mixed. We take an example of the B92 protocol to illustrate the usage of qACP in verification of quantum protocols.

\begin{figure}
  \centering
  \includegraphics{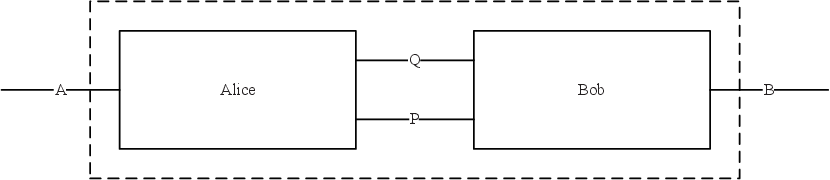}
  \caption{The B92 protocol.}
  \label{B924}
\end{figure}

The B92 protocol is used to create a private key between two parities, Alice and Bob. B92 is a protocol of quantum key distribution (QKD) which uses polarized photons as information carriers. Firstly, we introduce the basic B92 protocol briefly, which is illustrated in Figure \ref{B924}.

\begin{enumerate}
  \item Alice create a string of bits with size $n$ randomly, denoted as $A$.
  \item Alice generates a string of qubits $q$ with size $n$, carried by polarized photons. If $A_i=0$, the ith qubit is $|0\rangle$; else if $A_i=1$, the ith qubit is $|+\rangle$.
  \item Alice sends $q$ to Bob through a quantum channel $Q$ between Alice and Bob.
  \item Bob receives $q$ and randomly generates a string of bits $B$ with size $n$.
  \item If $B_i=0$, Bob chooses the basis $\oplus$; else if $B_i=1$, Bob chooses the basis $\otimes$. Bob measures each qubit of $q$ according to the above basses. And Bob builds a String of bits $T$, if the measurement produces $|0\rangle$ or $|+\rangle$, then $T_i=0$; else if the measurement produces $|1\rangle$ or $|-\rangle$, then $T_i=1$, which is also with size $n$.
  \item Bob sends $T$ to Alice through a public channel $P$.
  \item Alice and Bob determine that at which position the bit strings $A$ and $B$ are remained for which $T_i=1$. In absence of Eve, $A_i=1-B_i$, a shared raw key $K_{a,b}$ is formed by $A_i$.
\end{enumerate}

We re-introduce the basic B92 protocol in an abstract way with more technical details as Figure \ref{B924} illustrates.

Now, we assume a special measurement operation $Rand[q;A]$ which create a string of $n$ random bits $A$ from the $q$ quantum system, and the same as $Rand[q';B]$. $M[q;T]$ denotes the Bob's measurement operation of $q$. The generation of $n$ qubits $q$ through a quantum operation $Set_{A}[q]$. Alice sends $q$ to Bob through the quantum channel $Q$ by quantum communicating action $send_{Q}(q)$ and Bob receives $q$ through $Q$ by quantum communicating action $receive_{Q}(q)$. Bob sends $T$ to Alice through the public channel $P$ by classical communicating action $send_{P}(T)$ and Alice receives $T$ through channel $P$ by classical communicating action $receive_{P}(T)$. Alice and Bob generate the private key $K_{a,b}$ by a classical comparison action $cmp(K_{a,b},T,A,B)$. Let Alice and Bob be a system $AB$ and let interactions between Alice and Bob be internal actions. $AB$ receives external input $D_i$ through channel $A$ by communicating action $receive_A(D_i)$ and sends results $D_o$ through channel $B$ by communicating action $send_B(D_o)$.

Then the state transition of Alice can be described by qACP as follows.

\begin{eqnarray}
&&A=\sum_{D_i\in \Delta_i}receive_A(D_i)\cdot A_1\nonumber\\
&&A_1=Rand[q;A]\cdot A_2\nonumber\\
&&A_2=Set_{A}[q]\cdot A_3\nonumber\\
&&A_3=send_Q(q)\cdot A_4\nonumber\\
&&A_4=receive_P(T)\cdot A_5\nonumber\\
&&A_5=cmp(K_{a,b},T,A,B)\cdot A_6\nonumber\\
&&A_6=\{A_{i}=B_{i}\}\cdot generate(K_a)\cdot A+\{A_{i}\neq B_{i}\}\cdot discard\cdot A\nonumber
\end{eqnarray}

where $\Delta_i$ is the collection of the input data.

And the state transition of Bob can be described by qACP as follows.

\begin{eqnarray}
&&B=receive_Q(q)\cdot B_1\nonumber\\
&&B_1=Rand[q';B]\cdot B_2\nonumber\\
&&B_2=M[q;T]\cdot B_3\nonumber\\
&&B_3=send_P(T)\cdot B_4\nonumber\\
&&B_4=cmp(K_{a,b},T,A,B)\cdot B_5\nonumber\\
&&B_5=\{A_{i}=B_{b_i}\}\cdot generate(K_b)\cdot B_6+\{A_{i}\neq B_{i}\}\cdot discard\cdot B_6\nonumber\\
&&B_6=\sum_{D_o\in\Delta_o}send_B(D_o)\cdot B\nonumber
\end{eqnarray}

where $\Delta_o$ is the collection of the output data.

The send action and receive action of the same data through the same channel can communicate each other, otherwise, a deadlock $\delta$ will be caused. We define the following communication functions.

\begin{eqnarray}
&&\gamma(send_Q(q),receive_Q(q))\triangleq c_Q(q)\nonumber\\
&&\gamma(send_P(T),receive_P(T))\triangleq c_P(T)\nonumber
\end{eqnarray}

Let $A$ and $B$ in parallel, then the system $AB$ can be represented by the following process term.

$$\tau_I(\partial_H(\Theta(A\between B)))$$

where $H=\{send_Q(q),receive_Q(q),send_P(T),receive_P(T)\}$ and $I=\{Rand[q;A], Set_{A}[q], Rand[q';B], \\ M[q;T], c_Q(q), c_P(T), cmp(K_{a,b},T,A,B),\{A_{i}=B_{i}\},\{A_{i}\neq B_{i}\},\\
generate(K_a),generate(K_b),discard\}$.

Then we get the following conclusion.

\begin{theorem}
The basic B92 protocol $\tau_I(\partial_H(A\parallel B))$ exhibits desired external behaviors.
\end{theorem}

\begin{proof}
We can get $\tau_I(\partial_H(\Theta(A\between B)))=\sum_{D_i\in \Delta_i}\sum_{D_o\in\Delta_o}receive_A(D_i)\leftmerge send_B(D_o)\leftmerge \tau_I(\partial_H(\Theta(A\between B)))$.
So, the basic B92 protocol $\tau_I(\partial_H(\Theta(A\between B)))$ exhibits desired external behaviors.
\end{proof}

\subsection{Verification of DPS Protocol}\label{VDPS4}

The famous DPS protocol\cite{DPS} is a quantum key distribution protocol, in which quantum information and classical information are mixed. We take an example of the DPS protocol to illustrate the usage of qACP in verification of quantum protocols.

\begin{figure}
  \centering
  \includegraphics{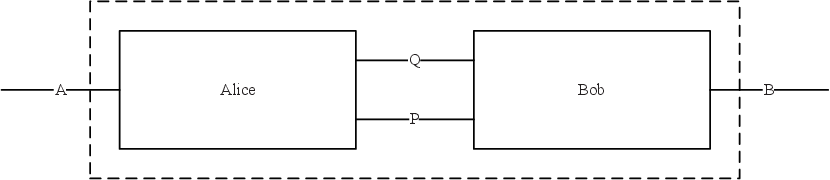}
  \caption{The DPS protocol.}
  \label{DPS4}
\end{figure}

The DPS protocol is used to create a private key between two parities, Alice and Bob. DPS is a protocol of quantum key distribution (QKD) which uses pulses of a photon which has nonorthogonal four states. Firstly, we introduce the basic DPS protocol briefly, which is illustrated in Figure \ref{DPS4}.

\begin{enumerate}
  \item Alice generates a string of qubits $q$ with size $n$, carried by a series of single photons possily at four time instances.
  \item Alice sends $q$ to Bob through a quantum channel $Q$ between Alice and Bob.
  \item Bob receives $q$ by detectors clicking at the second or third time instance, and records the time into $T$ with size $n$ and which detector clicks into $D$ with size $n$.
  \item Bob sends $T$ to Alice through a public channel $P$.
  \item Alice receives $T$. From $T$ and her modulation data, Alice knows which detector clicked in Bob's site, i.e. $D$.
  \item Alice and Bob have an identical bit string, provided that the first detector click represents "0" and the other detector represents "1", then a shared raw key $K_{a,b}$ is formed.
\end{enumerate}

We re-introduce the basic DPS protocol in an abstract way with more technical details as Figure \ref{DPS4} illustrates.

Now, we assume $M[q;T]$ denotes the Bob's measurement operation of $q$. The generation of $n$ qubits $q$ through a quantum operation $Set_{A}[q]$. Alice sends $q$ to Bob through the quantum channel $Q$ by quantum communicating action $send_{Q}(q)$ and Bob receives $q$ through $Q$ by quantum communicating action $receive_{Q}(q)$. Bob sends $T$ to Alice through the public channel $P$ by classical communicating action $send_{P}(T)$ and Alice receives $T$ through channel $P$ by classical communicating action $receive_{P}(T)$. Alice and Bob generate the private key $K_{a,b}$ by a classical comparison action $cmp(K_{a,b},D)$. Let Alice and Bob be a system $AB$ and let interactions between Alice and Bob be internal actions. $AB$ receives external input $D_i$ through channel $A$ by communicating action $receive_A(D_i)$ and sends results $D_o$ through channel $B$ by communicating action $send_B(D_o)$.

Then the state transition of Alice can be described by qACP as follows.

\begin{eqnarray}
&&A=\sum_{D_i\in \Delta_i}receive_A(D_i)\cdot A_1\nonumber\\
&&A_1=Set_{A}[q]\cdot A_2\nonumber\\
&&A_2=send_Q(q)\cdot A_3\nonumber\\
&&A_3=receive_P(T)\cdot A_4\nonumber\\
&&A_4=cmp(K_{a,b},D)\cdot A\nonumber
\end{eqnarray}

where $\Delta_i$ is the collection of the input data.

And the state transition of Bob can be described by qACP as follows.

\begin{eqnarray}
&&B=receive_Q(q)\cdot B_1\nonumber\\
&&B_1=M[q;T]\cdot B_2\nonumber\\
&&B_2=send_P(T)\cdot B_3\nonumber\\
&&B_3=cmp(K_{a,b},D)\cdot B_4\nonumber\\
&&B_4=\sum_{D_o\in\Delta_o}send_B(D_o)\cdot B\nonumber
\end{eqnarray}

where $\Delta_o$ is the collection of the output data.

The send action and receive action of the same data through the same channel can communicate each other, otherwise, a deadlock $\delta$ will be caused. We define the following communication functions.

\begin{eqnarray}
&&\gamma(send_Q(q),receive_Q(q))\triangleq c_Q(q)\nonumber\\
&&\gamma(send_P(T),receive_P(T))\triangleq c_P(T)\nonumber\\
\end{eqnarray}

Let $A$ and $B$ in parallel, then the system $AB$ can be represented by the following process term.

$$\tau_I(\partial_H(\Theta(A\between B)))$$

where $H=\{send_Q(q),receive_Q(q),send_P(T),receive_P(T)\}$

and $I=\{Set_{A}[q], M[q;T], c_Q(q), c_P(T), cmp(K_{a,b},D)\}$.

Then we get the following conclusion.

\begin{theorem}
The basic DPS protocol $\tau_I(\partial_H(\Theta(A\between B)))$ exhibits desired external behaviors.
\end{theorem}

\begin{proof}
We can get $\tau_I(\partial_H(\Theta(A\between B)))=\sum_{D_i\in \Delta_i}\sum_{D_o\in\Delta_o}receive_A(D_i)\leftmerge send_B(D_o)\leftmerge \tau_I(\partial_H(\Theta(A\between B)))$. 
So, the basic DPS protocol $\tau_I(\partial_H(\Theta(A\between B)))$ exhibits desired external behaviors.
\end{proof}

\subsection{Verification of BBM92 Protocol}\label{VBBM924}

The famous BBM92 protocol\cite{BBM92} is a quantum key distribution protocol, in which quantum information and classical information are mixed. We take an example of the BBM92 protocol to illustrate the usage of qACP in verification of quantum protocols.

\begin{figure}
  \centering
  \includegraphics{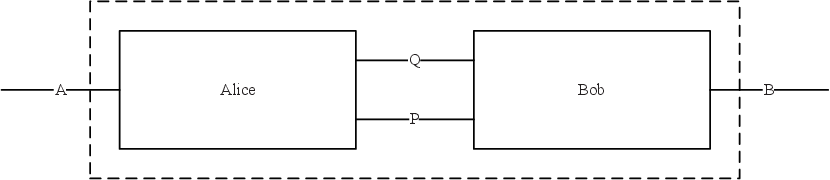}
  \caption{The BBM92 protocol.}
  \label{BBM924}
\end{figure}

The BBM92 protocol is used to create a private key between two parities, Alice and Bob. BBM92 is a protocol of quantum key distribution (QKD) which uses EPR pairs as information carriers. Firstly, we introduce the basic BBM92 protocol briefly, which is illustrated in Figure \ref{BBM924}.

\begin{enumerate}
  \item Alice generates a string of EPR pairs $q$ with size $n$, i.e., $2n$ particles, and sends a string of qubits $q_b$ from each EPR pair with $n$ to Bob through a quantum channel $Q$, remains the other string of qubits $q_a$ from each pair with size $n$.
  \item Alice create a string of bits with size $n$ randomly, denoted as $B_a$.
  \item Bob receives $q_b$ and randomly generates a string of bits $B_b$ with size $n$.
  \item Alice measures each qubit of $q_a$ according to bits of $B_a$, if $B_{a_i}=0$, then uses $x$ axis ($\rightarrow$); else if $B_{a_i}=1$, then uses $z$ axis ($\uparrow$).
  \item Bob measures each qubit of $q_b$ according to bits of $B_b$, if $B_{b_i}=0$, then uses $x$ axis ($\rightarrow$); else if $B_{b_i}=1$, then uses $z$ axis ($\uparrow$).
  \item Bob sends his measurement axis choices $B_b$ to Alice through a public channel $P$.
  \item Once receiving $B_b$, Alice sends her axis choices $B_a$ to Bob through channel $P$, and Bob receives $B_a$.
  \item Alice and Bob agree to discard all instances in which they happened to measure along different axes, as well as instances in which measurements fails because of imperfect quantum efficiency of the detectors. Then the remaining instances can be used to generate a private key $K_{a,b}$.
\end{enumerate}

We re-introduce the basic BBM92 protocol in an abstract way with more technical details as Figure \ref{BBM924} illustrates.

Now, $M[q_a;B_a]$ denotes the Alice's measurement operation of $q_a$, and $\circledS_{M[q_a;B_a]}$ denotes the responding shadow constant; $M[q_b;B_b]$ denotes the Bob's measurement operation of $q_b$, and $\circledS_{M[q_b;B_b]}$ denotes the responding shadow constant. Alice sends $q_b$ to Bob through the quantum channel $Q$ by quantum communicating action $send_{Q}(q_b)$ and Bob receives $q_b$ through $Q$ by quantum communicating action $receive_{Q}(q_b)$. Bob sends $B_b$ to Alice through the public channel $P$ by classical communicating action $send_{P}(B_b)$ and Alice receives $B_b$ through channel $P$ by classical communicating action $receive_{P}(B_b)$, and the same as $send_{P}(B_a)$ and $receive_{P}(B_a)$. Alice and Bob generate the private key $K_{a,b}$ by a classical comparison action $cmp(K_{a,b},B_a,B_b)$. Let Alice and Bob be a system $AB$ and let interactions between Alice and Bob be internal actions. $AB$ receives external input $D_i$ through channel $A$ by communicating action $receive_A(D_i)$ and sends results $D_o$ through channel $B$ by communicating action $send_B(D_o)$.

Then the state transition of Alice can be described by qACP as follows.

\begin{eqnarray}
&&A=\sum_{D_i\in \Delta_i}receive_A(D_i)\cdot A_1\nonumber\\
&&A_1=send_Q(q_b)\cdot A_2\nonumber\\
&&A_2=M[q_a;B_a]\cdot A_3\nonumber\\
&&A_3=\circledS_{M[q_b;B_b]}\cdot A_4\nonumber\\
&&A_4=receive_P(B_b)\cdot A_5\nonumber\\
&&A_5=send_P(B_a)\cdot A_6\nonumber\\
&&A_6=cmp(K_{a,b},B_a,B_b)\cdot A_7\nonumber\\
&&A_7=\{B_{a_i}=B_{b_i}\}\cdot generate(K_a)\cdot A+\{B_{a_i}\neq B_{b_i}\}\cdot discard\cdot A\nonumber
\end{eqnarray}

where $\Delta_i$ is the collection of the input data.

And the state transition of Bob can be described by qACP as follows.

\begin{eqnarray}
&&B=receive_Q(q_b)\cdot B_1\nonumber\\
&&B_1=\circledS_{M[q_a;B_a]}\cdot B_2\nonumber\\
&&B_2=M[q_b;B_b]\cdot B_3\nonumber\\
&&B_3=send_P(B_b)\cdot B_4\nonumber\\
&&B_4=receive_P(B_a)\cdot B_5\nonumber\\
&&B_5=cmp(K_{a,b},B_a,B_b)\cdot B_6\nonumber\\
&&B_6=\{B_{a_i}=B_{b_i}\}\cdot generate(K_b)\cdot B_7+\{B_{a_i}\neq B_{b_i}\}\cdot discard\cdot B_7\nonumber\\
&&B_7=\sum_{D_o\in\Delta_o}send_B(D_o)\cdot B\nonumber
\end{eqnarray}

where $\Delta_o$ is the collection of the output data.

The send action and receive action of the same data through the same channel can communicate each other, otherwise, a deadlock $\delta$ will be caused. The quantum operation and its shadow constant pair will lead entanglement occur, otherwise, a deadlock $\delta$ will occur. We define the following communication functions.

\begin{eqnarray}
&&\gamma(send_Q(q_b),receive_Q(q_b))\triangleq c_Q(q_b)\nonumber\\
&&\gamma(send_P(B_b),receive_P(B_b))\triangleq c_P(B_b)\nonumber\\
&&\gamma(send_P(B_a),receive_P(B_a))\triangleq c_P(B_a)\nonumber
\end{eqnarray}

Let $A$ and $B$ in parallel, then the system $AB$ can be represented by the following process term.

$$\tau_I(\partial_H(\Theta(A\between B)))$$

where $H=\{send_Q(q_b),receive_Q(q_b),send_P(B_b),receive_P(B_b),send_P(B_a),receive_P(B_a),\\ M[q_a;B_a], \circledS_{M[q_a;B_a]}, M[q_b;B_b], \circledS_{M[q_b;B_b]}\}$ and 
$I=\{c_Q(q_b), c_P(B_b), c_P(B_a), M[q_a;B_a], M[q_b;B_b],\\ cmp(K_{a,b},B_a,B_b),\{B_{a_i}=B_{b_i}\},\{B_{a_i}\neq B_{b_i}\},\\
generate(K_a),generate(K_b),discard\}$.
Then we get the following conclusion.

\begin{theorem}
The basic BBM92 protocol $\tau_I(\partial_H(\Theta(A\between B)))$ exhibits desired external behaviors.
\end{theorem}

\begin{proof}
We can get $\tau_I(\partial_H(\Theta(A\between B)))=\sum_{D_i\in \Delta_i}\sum_{D_o\in\Delta_o}receive_A(D_i)\leftmerge send_B(D_o)\leftmerge \tau_I(\partial_H(\Theta(A\between B)))$.
So, the basic BBM92 protocol $\tau_I(\partial_H(\Theta(A\between B)))$ exhibits desired external behaviors.
\end{proof}

\subsection{Verification of SARG04 Protocol}\label{VSARG044}

The famous SARG04 protocol\cite{SARG04} is a quantum key distribution protocol, in which quantum information and classical information are mixed. We take an example of the SARG04 protocol to illustrate the usage of qACP in verification of quantum protocols.

\begin{figure}
  \centering
  \includegraphics{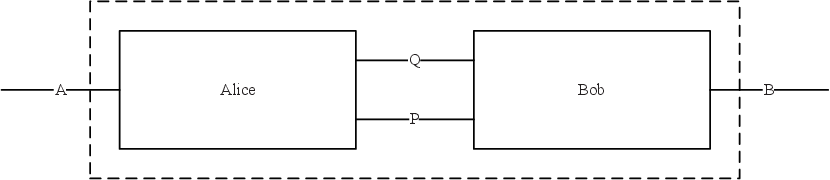}
  \caption{The SARG04 protocol.}
  \label{SARG044}
\end{figure}

The SARG04 protocol is used to create a private key between two parities, Alice and Bob. SARG04 is a protocol of quantum key distribution (QKD) which refines the BB84 protocol against PNS (Photon Number Splitting) attacks. The main innovations are encoding bits in nonorthogonal states and the classical sifting procedure. Firstly, we introduce the basic SARG04 protocol briefly, which is illustrated in Figure \ref{SARG044}.

\begin{enumerate}
  \item Alice create a string of bits with size $n$ randomly, denoted as $K_a$.
  \item Alice generates a string of qubits $q$ with size $n$, and the $i$th qubit of $q$ has four nonorthogonal states, it is $|\pm x\rangle$ if $K_a=0$; it is $|\pm z\rangle$ if $K_a=1$. And she records the corresponding one of the four pairs of nonorthogonal states into $B_a$ with size $2n$.
  \item Alice sends $q$ to Bob through a quantum channel $Q$ between Alice and Bob.
  \item Alice sends $B_a$ through a public channel $P$.
  \item Bob measures each qubit of $q$ $\sigma_x$ or $\sigma_z$. And he records the unambiguous discriminations into $K_b$ with a raw size $n/4$, and the unambiguous discrimination information into $B_b$ with size $n$.
  \item Bob sends $B_b$ to Alice through the public channel $P$.
  \item Alice and Bob determine that at which position the bit should be remained. Then the remaining bits of $K_a$ and $K_b$ is the private key $K_{a,b}$.
\end{enumerate}

We re-introduce the basic SARG04 protocol in an abstract way with more technical details as Figure \ref{SARG044} illustrates.

Now, we assume a special measurement operation $Rand[q;K_a]$ which create a string of $n$ random bits $K_a$ from the $q$ quantum system. $M[q;K_b]$ denotes the Bob's measurement operation of $q$. The generation of $n$ qubits $q$ through a quantum operation $Set_{K_a}[q]$. Alice sends $q$ to Bob through the quantum channel $Q$ by quantum communicating action $send_{Q}(q)$ and Bob receives $q$ through $Q$ by quantum communicating action $receive_{Q}(q)$. Bob sends $B_b$ to Alice through the public channel $P$ by classical communicating action $send_{P}(B_b)$ and Alice receives $B_b$ through channel $P$ by classical communicating action $receive_{P}(B_b)$, and the same as $send_{P}(B_a)$ and $receive_{P}(B_a)$. Alice and Bob generate the private key $K_{a,b}$ by a classical comparison action $cmp(K_{a,b},K_a,K_b,B_a,B_b)$. Let Alice and Bob be a system $AB$ and let interactions between Alice and Bob be internal actions. $AB$ receives external input $D_i$ through channel $A$ by communicating action $receive_A(D_i)$ and sends results $D_o$ through channel $B$ by communicating action $send_B(D_o)$.

Then the state transition of Alice can be described by qACP as follows.

\begin{eqnarray}
&&A=\sum_{D_i\in \Delta_i}receive_A(D_i)\cdot A_1\nonumber\\
&&A_1=Rand[q;K_a]\cdot A_2\nonumber\\
&&A_2=Set_{K_a}[q]\cdot A_3\nonumber\\
&&A_3=send_Q(q)\cdot A_4\nonumber\\
&&A_4=send_P(B_a)\cdot A_5\nonumber\\
&&A_5=receive_P(B_b)\cdot A_6\nonumber\\
&&A_6=cmp(K_{a,b},K_a,K_b,B_a,B_b)\cdot A_7\nonumber\\
&&A_7=\{B_{a_i}=B_{b_i}\}\cdot generate(K_a)\cdot A+\{B_{a_i}\neq B_{b_i}\}\cdot discard\cdot A\nonumber
\end{eqnarray}

where $\Delta_i$ is the collection of the input data.

And the state transition of Bob can be described by qACP as follows.

\begin{eqnarray}
&&B=receive_Q(q)\cdot B_1\nonumber\\
&&B_1=receive_P(B_a)\cdot B_2\nonumber\\
&&B_2=M[q;K_b]\cdot B_3\nonumber\\
&&B_3=send_P(B_b)\cdot B_4\nonumber\\
&&B_4=cmp(K_{a,b},K_a,K_b,B_a,B_b)\cdot B_5\nonumber\\
&&B_5=\{B_{a_i}=B_{b_i}\}\cdot generate(K_b)\cdot B_6+\{B_{a_i}\neq B_{b_i}\}\cdot discard\cdot B_6\nonumber\\
&&B_6=\sum_{D_o\in\Delta_o}send_B(D_o)\cdot B\nonumber
\end{eqnarray}

where $\Delta_o$ is the collection of the output data.

The send action and receive action of the same data through the same channel can communicate each other, otherwise, a deadlock $\delta$ will be caused. We define the following communication functions.

\begin{eqnarray}
&&\gamma(send_Q(q),receive_Q(q))\triangleq c_Q(q)\nonumber\\
&&\gamma(send_P(B_b),receive_P(B_b))\triangleq c_P(B_b)\nonumber\\
&&\gamma(send_P(B_a),receive_P(B_a))\triangleq c_P(B_a)\nonumber
\end{eqnarray}

Let $A$ and $B$ in parallel, then the system $AB$ can be represented by the following process term.

$$\tau_I(\partial_H(\Theta(A\between B)))$$

where $H=\{send_Q(q),receive_Q(q),send_P(B_b),receive_P(B_b),send_P(B_a),receive_P(B_a)\}$ and 
$I=\{Rand[q;K_a], Set_{K_a}[q], M[q;K_b], c_Q(q), c_P(B_b),\\ c_P(B_a), cmp(K_{a,b},K_a,K_b,B_a,B_b),\{B_{a_i}=B_{b_i}\},\{B_{a_i}\neq B_{b_i}\},\\
generate(K_a),generate(K_b),discard\}$.
Then we get the following conclusion.

\begin{theorem}
The basic SARG04 protocol $\tau_I(\partial_H(\Theta(A\between B)))$ exhibits desired external behaviors.
\end{theorem}

\begin{proof}
We can get $\tau_I(\partial_H(\Theta(A\between B)))=\sum_{D_i\in \Delta_i}\sum_{D_o\in\Delta_o}receive_A(D_i)\leftmerge send_B(D_o)\leftmerge \tau_I(\partial_H(\Theta(A\between B)))$.
So, the basic SARG04 protocol $\tau_I(\partial_H(\Theta(A\between B)))$ exhibits desired external behaviors.
\end{proof}

\subsection{Verification of COW Protocol}\label{VCOW4}

The famous COW protocol\cite{COW} is a quantum key distribution protocol, in which quantum information and classical information are mixed. We take an example of the COW protocol to illustrate the usage of qACP in verification of quantum protocols.

\begin{figure}
  \centering
  \includegraphics{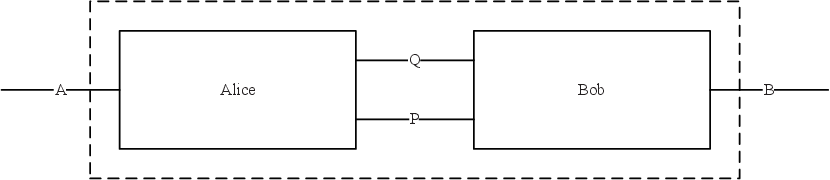}
  \caption{The COW protocol.}
  \label{COW4}
\end{figure}

The COW protocol is used to create a private key between two parities, Alice and Bob. COW is a protocol of quantum key distribution (QKD) which is practical. Firstly, we introduce the basic COW protocol briefly, which is illustrated in Figure \ref{COW4}.

\begin{enumerate}
  \item Alice generates a string of qubits $q$ with size $n$, and the $i$th qubit of $q$ is "0" with probability $\frac{1-f}{2}$, "1" with probability $\frac{1-f}{2}$ and the decoy sequence with probability $f$.
  \item Alice sends $q$ to Bob through a quantum channel $Q$ between Alice and Bob.
  \item Alice sends $A$ of the items corresponding to a decoy sequence through a public channel $P$.
  \item Bob removes all the detections at times $2A-1$ and $2A$ from his raw key and looks whether detector $D_{2M}$ has ever fired at time $2A$.
  \item Bob sends $B$ of the times $2A+1$ in which he had a detector in $D_{2M}$ to Alice through the public channel $P$.
  \item Alice receives $B$ and verifies if some of these items corresponding to a bit sequence "1,0".
  \item Bob sends $C$ of the items that he has detected through the public channel $P$.
  \item Alice and Bob run error correction and privacy amplification on these bits, and the private key $K_{a,b}$ is established.
\end{enumerate}

We re-introduce the basic COW protocol in an abstract way with more technical details as Figure \ref{COW4} illustrates.

Now, we assume The generation of $n$ qubits $q$ through a quantum operation $Set[q]$. $M[q]$ denotes the Bob's measurement operation of $q$.  Alice sends $q$ to Bob through the quantum channel $Q$ by quantum communicating action $send_{Q}(q)$ and Bob receives $q$ through $Q$ by quantum communicating action $receive_{Q}(q)$. Alice sends $A$ to Alice through the public channel $P$ by classical communicating action $send_{P}(A)$ and Alice receives $A$ through channel $P$ by classical communicating action $receive_{P}(A)$, and the same as $send_{P}(B)$ and $receive_{P}(B)$, and $send_{P}(C)$ and $receive_{P}(C)$. Alice and Bob generate the private key $K_{a,b}$ by a classical comparison action $cmp(K_{a,b})$. Let Alice and Bob be a system $AB$ and let interactions between Alice and Bob be internal actions. $AB$ receives external input $D_i$ through channel $A$ by communicating action $receive_A(D_i)$ and sends results $D_o$ through channel $B$ by communicating action $send_B(D_o)$.

Then the state transition of Alice can be described by qACP as follows.

\begin{eqnarray}
&&A=\sum_{D_i\in \Delta_i}receive_A(D_i)\cdot A_1\nonumber\\
&&A_1=Set[q]\cdot A_2\nonumber\\
&&A_2=send_Q(q)\cdot A_3\nonumber\\
&&A_3=send_P(A)\cdot A_4\nonumber\\
&&A_4=receive_P(B)\cdot A_5\nonumber\\
&&A_5=receive_P(C)\cdot A_6\nonumber\\
&&A_6=cmp(K_{a,b})\cdot A\nonumber
\end{eqnarray}

where $\Delta_i$ is the collection of the input data.

And the state transition of Bob can be described by qACP as follows.

\begin{eqnarray}
&&B=receive_Q(q)\cdot B_1\nonumber\\
&&B_1=receive_P(A)\cdot B_2\nonumber\\
&&B_2=M[q]\cdot B_3\nonumber\\
&&B_3=send_P(B)\cdot B_4\nonumber\\
&&B_4=send_P(C)\cdot B_5\nonumber\\
&&B_5=cmp(K_{a,b})\cdot B_6\nonumber\\
&&B_6=\sum_{D_o\in\Delta_o}send_B(D_o)\cdot B\nonumber
\end{eqnarray}

where $\Delta_o$ is the collection of the output data.

The send action and receive action of the same data through the same channel can communicate each other, otherwise, a deadlock $\delta$ will be caused. We define the following communication functions.

\begin{eqnarray}
&&\gamma(send_Q(q),receive_Q(q))\triangleq c_Q(q)\nonumber\\
&&\gamma(send_P(A),receive_P(A))\triangleq c_P(A)\nonumber\\
&&\gamma(send_P(B),receive_P(B))\triangleq c_P(B)\nonumber\\
&&\gamma(send_P(C),receive_P(C))\triangleq c_P(C)\nonumber
\end{eqnarray}

Let $A$ and $B$ in parallel, then the system $AB$ can be represented by the following process term.

$$\tau_I(\partial_H(\Theta(A\between B)))$$

where $H=\{send_Q(q),receive_Q(q),send_P(A),receive_P(A),send_P(B),receive_P(B),\\send_P(C),receive_P(C)\}$ and $I=\{Set[q], M[q], c_Q(q), c_P(A),\\ c_P(B),c_P(C), cmp(K_{a,b})\}$.

Then we get the following conclusion.

\begin{theorem}
The basic COW protocol $\tau_I(\partial_H(\Theta(A\between B)))$ exhibits desired external behaviors.
\end{theorem}

\begin{proof}
We can get $\tau_I(\partial_H(\Theta(A\between B)))=\sum_{D_i\in \Delta_i}\sum_{D_o\in\Delta_o}receive_A(D_i)\leftmerge send_B(D_o)\leftmerge \tau_I(\partial_H(\Theta(A\between B)))$. 
So, the basic COW protocol $\tau_I(\partial_H(\Theta(A\between B)))$ exhibits desired external behaviors.
\end{proof}

\subsection{Verification of SSP Protocol}\label{VSSP4}

The famous SSP protocol\cite{SSP} is a quantum key distribution protocol, in which quantum information and classical information are mixed. We take an example of the SSP protocol to illustrate the usage of qACP in verification of quantum protocols.

\begin{figure}
  \centering
  \includegraphics{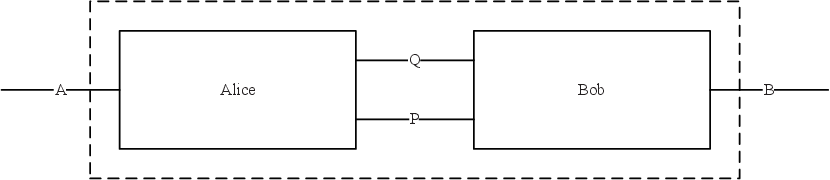}
  \caption{The SSP protocol.}
  \label{SSP4}
\end{figure}

The SSP protocol is used to create a private key between two parities, Alice and Bob. SSP is a protocol of quantum key distribution (QKD) which uses six states. Firstly, we introduce the basic SSP protocol briefly, which is illustrated in Figure \ref{SSP4}.

\begin{enumerate}
  \item Alice create two string of bits with size $n$ randomly, denoted as $B_a$ and $K_a$.
  \item Alice generates a string of qubits $q$ with size $n$, and the $i$th qubit in $q$ is one of the six states $\pm x$, $\pm y$ and $\pm z$.
  \item Alice sends $q$ to Bob through a quantum channel $Q$ between Alice and Bob.
  \item Bob receives $q$ and randomly generates a string of bits $B_b$ with size $n$.
  \item Bob measures each qubit of $q$ according to a basis by bits of $B_b$, i.e., $x$, $y$ or $z$ basis. And the measurement results would be $K_b$, which is also with size $n$.
  \item Bob sends his measurement bases $B_b$ to Alice through a public channel $P$.
  \item Once receiving $B_b$, Alice sends her bases $B_a$ to Bob through channel $P$, and Bob receives $B_a$.
  \item Alice and Bob determine that at which position the bit strings $B_a$ and $B_b$ are equal, and they discard the mismatched bits of $B_a$ and $B_b$. Then the remaining bits of $K_a$ and $K_b$, denoted as $K_a'$ and $K_b'$ with $K_{a,b}=K_a'=K_b'$.
\end{enumerate}

We re-introduce the basic SSP protocol in an abstract way with more technical details as Figure \ref{SSP4} illustrates.

Now, we assume a special measurement operation $Rand[q;B_a]$ which create a string of $n$ random bits $B_a$ from the $q$ quantum system, and the same as $Rand[q;K_a]$, $Rand[q';B_b]$. $M[q;K_b]$ denotes the Bob's measurement operation of $q$. The generation of $n$ qubits $q$ through two quantum operations $Set_{K_a}[q]$ and $H_{B_a}[q]$. Alice sends $q$ to Bob through the quantum channel $Q$ by quantum communicating action $send_{Q}(q)$ and Bob receives $q$ through $Q$ by quantum communicating action $receive_{Q}(q)$. Bob sends $B_b$ to Alice through the public channel $P$ by classical communicating action $send_{P}(B_b)$ and Alice receives $B_b$ through channel $P$ by classical communicating action $receive_{P}(B_b)$, and the same as $send_{P}(B_a)$ and $receive_{P}(B_a)$. Alice and Bob generate the private key $K_{a,b}$ by a classical comparison action $cmp(K_{a,b},K_a,K_b,B_a,B_b)$. Let Alice and Bob be a system $AB$ and let interactions between Alice and Bob be internal actions. $AB$ receives external input $D_i$ through channel $A$ by communicating action $receive_A(D_i)$ and sends results $D_o$ through channel $B$ by communicating action $send_B(D_o)$.

Then the state transition of Alice can be described by qACP as follows.

\begin{eqnarray}
&&A=\sum_{D_i\in \Delta_i}receive_A(D_i)\cdot A_1\nonumber\\
&&A_1=Rand[q;B_a]\cdot A_2\nonumber\\
&&A_2=Rand[q;K_a]\cdot A_3\nonumber\\
&&A_3=Set_{K_a}[q]\cdot A_4\nonumber\\
&&A_4=H_{B_a}[q]\cdot A_5\nonumber\\
&&A_5=send_Q(q)\cdot A_6\nonumber\\
&&A_6=receive_P(B_b)\cdot A_7\nonumber\\
&&A_7=send_P(B_a)\cdot A_8\nonumber\\
&&A_8=cmp(K_{a,b},K_a,K_b,B_a,B_b)\cdot A_9\nonumber\\
&&A_9=\{B_{a_i}=B_{b_i}\}\cdot generate(K_a)\cdot A+\{B_{a_i}\neq B_{b_i}\}\cdot discard\cdot A\nonumber
\end{eqnarray}

where $\Delta_i$ is the collection of the input data.

And the state transition of Bob can be described by qACP as follows.

\begin{eqnarray}
&&B=receive_Q(q)\cdot B_1\nonumber\\
&&B_1=Rand[q';B_b]\cdot B_2\nonumber\\
&&B_2=M[q;K_b]\cdot B_3\nonumber\\
&&B_3=send_P(B_b)\cdot B_4\nonumber\\
&&B_4=receive_P(B_a)\cdot B_5\nonumber\\
&&B_5=cmp(K_{a,b},K_a,K_b,B_a,B_b)\cdot B_6\nonumber\\
&&B_6=\{B_{a_i}=B_{b_i}\}\cdot generate(K_b)\cdot B_7+\{B_{a_i}\neq B_{b_i}\}\cdot discard\cdot B_7\nonumber\\
&&B_7=\sum_{D_o\in\Delta_o}send_B(D_o)\cdot B\nonumber
\end{eqnarray}

where $\Delta_o$ is the collection of the output data.

The send action and receive action of the same data through the same channel can communicate each other, otherwise, a deadlock $\delta$ will be caused. We define the following communication functions.

\begin{eqnarray}
&&\gamma(send_Q(q),receive_Q(q))\triangleq c_Q(q)\nonumber\\
&&\gamma(send_P(B_b),receive_P(B_b))\triangleq c_P(B_b)\nonumber\\
&&\gamma(send_P(B_a),receive_P(B_a))\triangleq c_P(B_a)\nonumber
\end{eqnarray}

Let $A$ and $B$ in parallel, then the system $AB$ can be represented by the following process term.

$$\tau_I(\partial_H(\Theta(A\between B)))$$

where $H=\{send_Q(q),receive_Q(q),send_P(B_b),receive_P(B_b),send_P(B_a),receive_P(B_a)\}$ and
$I=\{Rand[q;B_a], Rand[q;K_a], Set_{K_a}[q], H_{B_a}[q], Rand[q';B_b], M[q;K_b], c_Q(q), c_P(B_b),\\ c_P(B_a), cmp(K_{a,b},K_a,K_b,B_a,B_b),\{B_{a_i}=B_{b_i}\},\{B_{a_i}\neq B_{b_i}\},\\
generate(K_a),generate(K_b),discard\}$.

Then we get the following conclusion.

\begin{theorem}
The basic SSP protocol $\tau_I(\partial_H(\Theta(A\between B)))$ exhibits desired external behaviors.
\end{theorem}

\begin{proof}
We can get $\tau_I(\partial_H(\Theta(A\between B)))=\sum_{D_i\in \Delta_i}\sum_{D_o\in\Delta_o}receive_A(D_i)\leftmerge send_B(D_o)\leftmerge \tau_I(\partial_H(\Theta(A\between B)))$.
So, the basic SSP protocol $\tau_I(\partial_H(\Theta(A\between B)))$ exhibits desired external behaviors.
\end{proof}

\subsection{Verification of S09 Protocol}\label{VS094}

The famous S09 protocol\cite{S09} is a quantum key distribution protocol, in which quantum information and classical information are mixed. We take an example of the S09 protocol to illustrate the usage of qACP in verification of quantum protocols.

\begin{figure}
  \centering
  \includegraphics{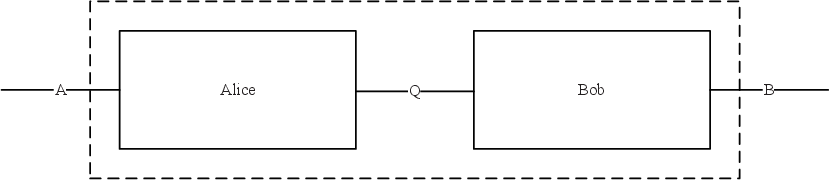}
  \caption{The S09 protocol.}
  \label{S094}
\end{figure}

The S09 protocol is used to create a private key between two parities, Alice and Bob, by use of pure quantum information. Firstly, we introduce the basic S09 protocol briefly, which is illustrated in Figure \ref{S094}.

\begin{enumerate}
  \item Alice create two string of bits with size $n$ randomly, denoted as $B_a$ and $K_a$.
  \item Alice generates a string of qubits $q$ with size $n$, and the $i$th qubit in $q$ is $|x_y\rangle$, where $x$ is the $i$th bit of $B_a$ and $y$ is the $i$th bit of $K_a$.
  \item Alice sends $q$ to Bob through a quantum channel $Q$ between Alice and Bob.
  \item Bob receives $q$ and randomly generates a string of bits $B_b$ with size $n$.
  \item Bob measures each qubit of $q$ according to a basis by bits of $B_b$. After the measurement, the state of $q$ evolves into $q'$.
  \item Bob sends $q'$ to Alice through the quantum channel $Q$.
  \item Alice measures each qubit of $q'$ to generate a string $C$.
  \item Alice sums $C_i\oplus B_{a_i}$ to get the private key $K_{a,b}=B_b$.
\end{enumerate}

We re-introduce the basic S09 protocol in an abstract way with more technical details as Figure \ref{S094} illustrates.

Now, we assume a special measurement operation $Rand[q;B_a]$ which create a string of $n$ random bits $B_a$ from the $q$ quantum system, and the same as $Rand[q;K_a]$, $Rand[q';B_b]$. $M[q;B_b]$ denotes the Bob's measurement operation of $q$, and the same as $M[q';C]$. The generation of $n$ qubits $q$ through two quantum operations $Set_{K_a}[q]$ and $H_{B_a}[q]$. Alice sends $q$ to Bob through the quantum channel $Q$ by quantum communicating action $send_{Q}(q)$ and Bob receives $q$ through $Q$ by quantum communicating action $receive_{Q}(q)$, and the same as $send_{Q}(q')$ and $receive_{Q}(q')$. Alice and Bob generate the private key $K_{a,b}$ by a classical comparison action $cmp(K_{a,b},B_b)$. We omit the sum classical $\oplus$ actions without of loss of generality. Let Alice and Bob be a system $AB$ and let interactions between Alice and Bob be internal actions. $AB$ receives external input $D_i$ through channel $A$ by communicating action $receive_A(D_i)$ and sends results $D_o$ through channel $B$ by communicating action $send_B(D_o)$.

Then the state transition of Alice can be described by qACP as follows.

\begin{eqnarray}
&&A=\sum_{D_i\in \Delta_i}receive_A(D_i)\cdot A_1\nonumber\\
&&A_1=Rand[q;B_a]\cdot A_2\nonumber\\
&&A_2=Rand[q;K_a]\cdot A_3\nonumber\\
&&A_3=Set_{K_a}[q]\cdot A_4\nonumber\\
&&A_4=H_{B_a}[q]\cdot A_5\nonumber\\
&&A_5=send_Q(q)\cdot A_6\nonumber\\
&&A_6=receive_Q(q')\cdot A_{7}\nonumber\\
&&A_7=M[q';C]\cdot A_8\nonumber\\
&&A_{8}=cmp(K_{a,b},B_b)\cdot A\nonumber
\end{eqnarray}

where $\Delta_i$ is the collection of the input data.

And the state transition of Bob can be described by qACP as follows.

\begin{eqnarray}
&&B=receive_Q(q)\cdot B_1\nonumber\\
&&B_1=Rand[q';B_b]\cdot B_2\nonumber\\
&&B_2=M[q;B_b]\cdot B_3\nonumber\\
&&B_3=send_Q(q')\cdot B_4\nonumber\\
&&B_4=cmp(K_{a,b},B_b)\cdot B_{5}\nonumber\\
&&B_{5}=\sum_{D_o\in\Delta_o}send_B(D_o)\cdot B\nonumber
\end{eqnarray}

where $\Delta_o$ is the collection of the output data.

The send action and receive action of the same data through the same channel can communicate each other, otherwise, a deadlock $\delta$ will be caused. We define the following communication functions.

\begin{eqnarray}
&&\gamma(send_Q(q),receive_Q(q))\triangleq c_Q(q)\nonumber\\
&&\gamma(send_Q(q'),receive_Q(q'))\triangleq c_Q(q')\nonumber
\end{eqnarray}

Let $A$ and $B$ in parallel, then the system $AB$ can be represented by the following process term.

$$\tau_I(\partial_H(\Theta(A\between B)))$$

where $H=\{send_Q(q),receive_Q(q),send_Q(q'),receive_Q(q')\}$ and $I=\{Rand[q;B_a], Rand[q;K_a], Set_{K_a}[q], \\ H_{B_a}[q], Rand[q';B_b], M[q;K_b], M[q';C], c_Q(q), c_Q(q'), cmp(K_{a,b},B_b)\}$.

Then we get the following conclusion.

\begin{theorem}
The basic S09 protocol $\tau_I(\partial_H(\Theta(A\between B)))$ exhibits desired external behaviors.
\end{theorem}

\begin{proof}
We can get $\tau_I(\partial_H(\Theta(A\between B)))=\sum_{D_i\in \Delta_i}\sum_{D_o\in\Delta_o}receive_A(D_i)\leftmerge send_B(D_o)\leftmerge \tau_I(\partial_H(\Theta(A\between B)))$. 
So, the basic S09 protocol $\tau_I(\partial_H(\Theta(A\between B)))$ exhibits desired external behaviors.
\end{proof}

\subsection{Verification of KMB09 Protocol}\label{VKMB094}

The famous KMB09 protocol\cite{KMB09} is a quantum key distribution protocol, in which quantum information and classical information are mixed. We take an example of the KMB09 protocol to illustrate the usage of qACP in verification of quantum protocols.

\begin{figure}
  \centering
  \includegraphics{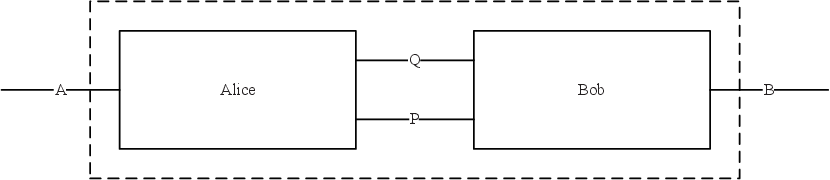}
  \caption{The KMB09 protocol.}
  \label{KMB094}
\end{figure}

The KMB09 protocol is used to create a private key between two parities, Alice and Bob. KMB09 is a protocol of quantum key distribution (QKD) which refines the BB84 protocol against PNS (Photon Number Splitting) attacks. The main innovations are encoding bits in nonorthogonal states and the classical sifting procedure. Firstly, we introduce the basic KMB09 protocol briefly, which is illustrated in Figure \ref{KMB094}.

\begin{enumerate}
  \item Alice create a string of bits with size $n$ randomly, denoted as $K_a$, and randomly assigns each bit value a random index $i=1,2,...,N$ into $B_a$.
  \item Alice generates a string of qubits $q$ with size $n$, accordingly either in $|e_i\rangle$ or $|f_i\rangle$.
  \item Alice sends $q$ to Bob through a quantum channel $Q$ between Alice and Bob.
  \item Alice sends $B_a$ through a public channel $P$.
  \item Bob measures each qubit of $q$ by randomly switching the measurement basis between $e$ and $f$. And he records the unambiguous discriminations into $K_b$, and the unambiguous discrimination information into $B_b$.
  \item Bob sends $B_b$ to Alice through the public channel $P$.
  \item Alice and Bob determine that at which position the bit should be remained. Then the remaining bits of $K_a$ and $K_b$ is the private key $K_{a,b}$.
\end{enumerate}

We re-introduce the basic KMB09 protocol in an abstract way with more technical details as Figure \ref{KMB094} illustrates.

Now, we assume a special measurement operation $Rand[q;K_a]$ which create a string of $n$ random bits $K_a$ from the $q$ quantum system. $M[q;K_b]$ denotes the Bob's measurement operation of $q$. The generation of $n$ qubits $q$ through a quantum operation $Set_{K_a}[q]$. Alice sends $q$ to Bob through the quantum channel $Q$ by quantum communicating action $send_{Q}(q)$ and Bob receives $q$ through $Q$ by quantum communicating action $receive_{Q}(q)$. Bob sends $B_b$ to Alice through the public channel $P$ by classical communicating action $send_{P}(B_b)$ and Alice receives $B_b$ through channel $P$ by classical communicating action $receive_{P}(B_b)$, and the same as $send_{P}(B_a)$ and $receive_{P}(B_a)$. Alice and Bob generate the private key $K_{a,b}$ by a classical comparison action $cmp(K_{a,b},K_a,K_b,B_a,B_b)$. Let Alice and Bob be a system $AB$ and let interactions between Alice and Bob be internal actions. $AB$ receives external input $D_i$ through channel $A$ by communicating action $receive_A(D_i)$ and sends results $D_o$ through channel $B$ by communicating action $send_B(D_o)$.

Then the state transition of Alice can be described by qACP as follows.

\begin{eqnarray}
&&A=\sum_{D_i\in \Delta_i}receive_A(D_i)\cdot A_1\nonumber\\
&&A_1=Rand[q;K_a]\cdot A_2\nonumber\\
&&A_2=Set_{K_a}[q]\cdot A_3\nonumber\\
&&A_3=send_Q(q)\cdot A_4\nonumber\\
&&A_4=send_P(B_a)\cdot A_5\nonumber\\
&&A_5=receive_P(B_b)\cdot A_6\nonumber\\
&&A_6=cmp(K_{a,b},K_a,K_b,B_a,B_b)\cdot A_7\nonumber\\
&&A_7=\{B_{a_i}=B_{b_i}\}\cdot generate(K_a)\cdot A+\{B_{a_i}\neq B_{b_i}\}\cdot discard\cdot A\nonumber
\end{eqnarray}

where $\Delta_i$ is the collection of the input data.

And the state transition of Bob can be described by qACP as follows.

\begin{eqnarray}
&&B=receive_Q(q)\cdot B_1\nonumber\\
&&B_1=receive_P(B_a)\cdot B_2\nonumber\\
&&B_2=M[q;K_b]\cdot B_3\nonumber\\
&&B_3=send_P(B_b)\cdot B_4\nonumber\\
&&B_4=cmp(K_{a,b},K_a,K_b,B_a,B_b)\cdot B_5\nonumber\\
&&B_5=\{B_{a_i}=B_{b_i}\}\cdot generate(K_b)\cdot B_6+\{B_{a_i}\neq B_{b_i}\}\cdot discard\cdot B_6\nonumber\\
&&B_6=\sum_{D_o\in\Delta_o}send_B(D_o)\cdot B\nonumber
\end{eqnarray}

where $\Delta_o$ is the collection of the output data.

The send action and receive action of the same data through the same channel can communicate each other, otherwise, a deadlock $\delta$ will be caused. We define the following communication functions.

\begin{eqnarray}
&&\gamma(send_Q(q),receive_Q(q))\triangleq c_Q(q)\nonumber\\
&&\gamma(send_P(B_b),receive_P(B_b))\triangleq c_P(B_b)\nonumber\\
&&\gamma(send_P(B_a),receive_P(B_a))\triangleq c_P(B_a)\nonumber
\end{eqnarray}

Let $A$ and $B$ in parallel, then the system $AB$ can be represented by the following process term.

$$\tau_I(\partial_H(\Theta(A\between B)))$$

where $H=\{send_Q(q),receive_Q(q),send_P(B_b),receive_P(B_b),send_P(B_a),receive_P(B_a)\}$ and
$I=\{Rand[q;K_a], Set_{K_a}[q], M[q;K_b], c_Q(q), c_P(B_b),\\ c_P(B_a), cmp(K_{a,b},K_a,K_b,B_a,B_b),\{B_{a_i}=B_{b_i}\},\{B_{a_i}\neq B_{b_i}\},\\
generate(K_a),generate(K_b),discard\}$.

Then we get the following conclusion.

\begin{theorem}
The basic KMB09 protocol $\tau_I(\partial_H(\Theta(A\between B)))$ exhibits desired external behaviors.
\end{theorem}

\begin{proof}
We can get $\tau_I(\partial_H(\Theta(A\between B)))=\sum_{D_i\in \Delta_i}\sum_{D_o\in\Delta_o}receive_A(D_i)\leftmerge send_B(D_o)\leftmerge \tau_I(\partial_H(\Theta(A\between B)))$.
So, the basic KMB09 protocol $\tau_I(\partial_H(\Theta(A\between B)))$ exhibits desired external behaviors.
\end{proof}

\subsection{Verification of S13 Protocol}\label{VS134}

The famous S13 protocol\cite{S13} is a quantum key distribution protocol, in which quantum information and classical information are mixed. We take an example of the S13 protocol to illustrate the usage of qACP in verification of quantum protocols.

\begin{figure}
  \centering
  \includegraphics{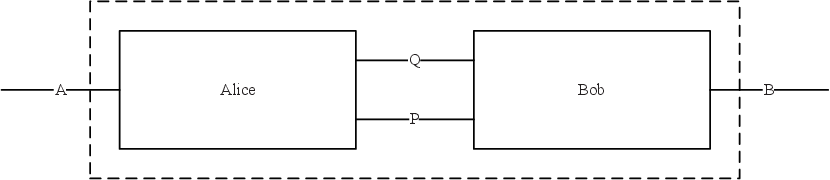}
  \caption{The S13 protocol.}
  \label{S134}
\end{figure}

The S13 protocol is used to create a private key between two parities, Alice and Bob. Firstly, we introduce the basic S13 protocol briefly, which is illustrated in Figure \ref{S134}.

\begin{enumerate}
  \item Alice create two string of bits with size $n$ randomly, denoted as $B_a$ and $K_a$.
  \item Alice generates a string of qubits $q$ with size $n$, and the $i$th qubit in $q$ is $|x_y\rangle$, where $x$ is the $i$th bit of $B_a$ and $y$ is the $i$th bit of $K_a$.
  \item Alice sends $q$ to Bob through a quantum channel $Q$ between Alice and Bob.
  \item Bob receives $q$ and randomly generates a string of bits $B_b$ with size $n$.
  \item Bob measures each qubit of $q$ according to a basis by bits of $B_b$. And the measurement results would be $K_b$, which is also with size $n$.
  \item Alice sends a random binary string $C$ to Bob through the public channel $P$.
  \item Alice sums $B_{a_i}\oplus C_i$ to obtain $T$ and generates other random string of binary values $J$. From the elements occupying a concrete position, $i$, of the preceding strings, Alice get the new states of $q'$, and sends it to Bob through the quantum channel $Q$.
  \item Bob sums $1\oplus B_{b_i}$ to obtain the string of binary basis $N$ and measures $q'$ according to these bases, and generating $D$.
  \item Alice sums $K_{a_i}\oplus J_i$ to obtain the binary string $Y$ and sends it to Bob through the public channel $P$.
  \item Bob encrypts $B_b$ to obtain $U$ and sends to Alice through the public channel $P$.
  \item Alice decrypts $U$ to obtain $B_b$. She sums $B_{a_i}\oplus B_{b_i}$ to obtain $L$ and sends $L$ to Bob through the public channel $P$.
  \item Bob sums $B_{b_i}\oplus L_i$ to get the private key $K_{a,b}$.
\end{enumerate}

We re-introduce the basic S13 protocol in an abstract way with more technical details as Figure \ref{S134} illustrates.

Now, we assume a special measurement operation $Rand[q;B_a]$ which create a string of $n$ random bits $B_a$ from the $q$ quantum system, and the same as $Rand[q;K_a]$, $Rand[q';B_b]$. $M[q;K_b]$ denotes the Bob's measurement operation of $q$, and the same as $M[q';D]$. The generation of $n$ qubits $q$ through two quantum operations $Set_{K_a}[q]$ and $H_{B_a}[q]$, and the same as $Set_{T}[q']$. Alice sends $q$ to Bob through the quantum channel $Q$ by quantum communicating action $send_{Q}(q)$ and Bob receives $q$ through $Q$ by quantum communicating action $receive_{Q}(q)$, and the same as $send_{Q}(q')$ and $receive_{Q}(q')$. Bob sends $B_b$ to Alice through the public channel $P$ by classical communicating action $send_{P}(B_b)$ and Alice receives $B_b$ through channel $P$ by classical communicating action $receive_{P}(B_b)$, and the same as $send_{P}(B_a)$ and $receive_{P}(B_a)$, $send_{P}(C)$ and $receive_{P}(C)$, $send_{P}(Y)$ and $receive_{P}(Y)$, $send_{P}(U)$ and $receive_{P}(U)$, $send_{P}(L)$ and $receive_{P}(L)$. Alice and Bob generate the private key $K_{a,b}$ by a classical comparison action $cmp(K_{a,b},K_a,K_b,B_a,B_b)$. We omit the sum classical $\oplus$ actions without of loss of generality. Let Alice and Bob be a system $AB$ and let interactions between Alice and Bob be internal actions. $AB$ receives external input $D_i$ through channel $A$ by communicating action $receive_A(D_i)$ and sends results $D_o$ through channel $B$ by communicating action $send_B(D_o)$.

Then the state transition of Alice can be described by qACP as follows.

\begin{eqnarray}
&&A=\sum_{D_i\in \Delta_i}receive_A(D_i)\cdot A_1\nonumber\\
&&A_1=Rand[q;B_a]\cdot A_2\nonumber\\
&&A_2=Rand[q;K_a]\cdot A_3\nonumber\\
&&A_3=Set_{K_a}[q]\cdot A_4\nonumber\\
&&A_4=H_{B_a}[q]\cdot A_5\nonumber\\
&&A_5=send_Q(q)\cdot A_6\nonumber\\
&&A_6=send_P(C)\cdot A_7\nonumber\\
&&A_7=send_Q(q')\cdot A_8\nonumber\\
&&A_8=send_P(Y)\cdot A_9\nonumber\\
&&A_9=receive_P(U)\cdot A_{10}\nonumber\\
&&A_{10}=send_P(L)\cdot A_{11}\nonumber\\
&&A_{11}=cmp(K_{a,b},K_a,K_b,B_a,B_b)\cdot A_{12}\nonumber\\
&&A_{12}=\{B_{a_i}=B_{b_i}\}\cdot generate(K_a)\cdot A+\{B_{a_i}\neq B_{b_i}\}\cdot discard\cdot A\nonumber
\end{eqnarray}

where $\Delta_i$ is the collection of the input data.

And the state transition of Bob can be described by qACP as follows.

\begin{eqnarray}
&&B=receive_Q(q)\cdot B_1\nonumber\\
&&B_1=Rand[q';B_b]\cdot B_2\nonumber\\
&&B_2=M[q;K_b]\cdot B_3\nonumber\\
&&B_3=receive_P(C)\cdot B_4\nonumber\\
&&B_4=receive_Q(q')\cdot B_5\nonumber\\
&&B_5=M[q';D]\cdot B_6\nonumber\\
&&B_6=receive_P(Y)\cdot B_7\nonumber\\
&&B_7=send_P(U)\cdot B_8\nonumber\\
&&B_8=receive_P(L)\cdot B_9\nonumber\\
&&B_9=cmp(K_{a,b},K_a,K_b,B_a,B_b)\cdot B_{10}\nonumber\\
&&B_{10}=\{B_{a_i}=B_{b_i}\}\cdot generate(K_b)\cdot B_{11}+\{B_{a_i}\neq B_{b_i}\}\cdot discard\cdot B_{11}\nonumber\\
&&B_{11}=\sum_{D_o\in\Delta_o}send_B(D_o)\cdot B\nonumber
\end{eqnarray}

where $\Delta_o$ is the collection of the output data.

The send action and receive action of the same data through the same channel can communicate each other, otherwise, a deadlock $\delta$ will be caused. We define the following communication functions.

\begin{eqnarray}
&&\gamma(send_Q(q),receive_Q(q))\triangleq c_Q(q)\nonumber\\
&&\gamma(send_Q(q'),receive_Q(q'))\triangleq c_Q(q')\nonumber\\
&&\gamma(send_P(C),receive_P(C))\triangleq c_P(C)\nonumber\\
&&\gamma(send_P(Y),receive_P(Y))\triangleq c_P(Y)\nonumber\\
&&\gamma(send_P(U),receive_P(U))\triangleq c_P(U)\nonumber\\
&&\gamma(send_P(L),receive_P(L))\triangleq c_P(L)\nonumber
\end{eqnarray}

Let $A$ and $B$ in parallel, then the system $AB$ can be represented by the following process term.

$$\tau_I(\partial_H(\Theta(A\between B)))$$

where $H=\{send_Q(q),receive_Q(q),send_Q(q'),receive_Q(q'),send_P(C),\\receive_P(C),send_P(Y),receive_P(Y),send_P(U),receive_P(U),send_P(L),receive_P(L)\}$ and 
$I=\{Rand[q;B_a], Rand[q;K_a],\\ Set_{K_a}[q], H_{B_a}[q], Rand[q';B_b], M[q;K_b], M[q';D], c_Q(q), c_P(C),\\c_Q(q'), c_P(Y), c_P(U), c_P(L), cmp(K_{a,b},K_a,K_b,B_a,B_b),\{B_{a_i}=B_{b_i}\},\{B_{a_i}\neq B_{b_i}\},\\
generate(K_a),generate(K_b),discard\}$.

Then we get the following conclusion.

\begin{theorem}
The basic S13 protocol $\tau_I(\partial_H(\Theta(A\between B)))$ exhibits desired external behaviors.
\end{theorem}

\begin{proof}
We can get $\tau_I(\partial_H(\Theta(A\between B)))=\sum_{D_i\in \Delta_i}\sum_{D_o\in\Delta_o}receive_A(D_i)\leftmerge send_B(D_o)\leftmerge \tau_I(\partial_H(\Theta(A\between B)))$.
So, the basic S13 protocol $\tau_I(\partial_H(\Theta(A\between B)))$ exhibits desired external behaviors.
\end{proof}

\newpage\section{$APPTC_G$ for Closed Quantum Systems}\label{qapptcg2}

The theory $APPTC_G$ for closed quantum systems abbreviated $qAPPTC_G$ has four modules: $qBAPTC_G$ , $qAPPTC_G$, recursion and abstraction.

This chapter is organized as follows. We introduce $qBAPTC_G$ in section \ref{qbaptcg}, $APPTC_G$ in section \ref{qapptcg}, recursion in section \ref{qcrecg}, and abstraction in section
\ref{qcabsg}. And we introduce quantum measurement in section \ref{qm}, quantum entanglement in section \ref{qe2}, and unification of quantum and classical computing in section \ref{uni2}.

Note that, for a closed quantum system, the unitary operators are the atomic actions (events) and let unitary operators into $\mathbb{E}$. And for the existence of quantum measurement,
the probabilism is unavoidable.

\subsection{$BAPTC_G$ for Closed Quantum Systems}{\label{qbaptcg}}

In this subsection, we will discuss $qBAPTC_G$. Let $\mathbb{E}$ be the set of atomic events (actions), $G_{at}$ be the set of atomic guards,
$\delta$ be the deadlock constant, and $\epsilon$ be the empty event. We extend $G_{at}$ to the set of basic guards $G$ with element $\phi,\psi,\cdots$, which is generated by the
following formation rules:

$$\phi::=\delta|\epsilon|\neg\phi|\psi\in G_{at}|\phi+\psi|\phi\boxplus_{\pi}\psi|\phi\cdot\psi$$

In the following, let $e_1, e_2, e_1', e_2'\in \mathbb{E}$, $\phi,\psi\in G$ and let variables $x,y,z$ range over the set of terms for true concurrency, $p,q,s$ range over the set of
closed terms. The predicate $test(\phi,s,\varrho)$ represents that $\phi$ holds in the state $s,\varrho$, and $test(\epsilon,s,\varrho)$ holds and $test(\delta,s,\varrho)$ does not hold.
$effect(e,s,\varrho)\in S$ denotes
$\varrho'$ in $\varrho\xrightarrow{e}\varrho'$. The predicate weakest precondition $wp(e,\phi)$ denotes that $\forall s,s',\varrho,\varrho'\in S, test(\phi,effect(e,s,\varrho))$ holds.

The set of axioms of $qBAPTC_G$ consists of the laws given in Table \ref{AxiomsForqBAPTCG}.

\begin{center}
    \begin{table}
        \begin{tabular}{@{}ll@{}}
            \hline No. &Axiom\\
            $A1$ & $x+ y = y+ x$\\
            $A2$ & $(x+ y)+ z = x+ (y+ z)$\\
            $A3$ & $e+ e = e$\\
            $A4$ & $(x+ y)\cdot z = x\cdot z + y\cdot z$\\
            $A5$ & $(x\cdot y)\cdot z = x\cdot(y\cdot z)$\\
            $A6$ & $x+\delta = x$\\
            $A7$ & $\delta\cdot x = \delta$\\
            $A8$ & $\epsilon\cdot x = x$\\
            $A9$ & $x\cdot\epsilon = x$\\
            $PA1$ & $x\boxplus_{\pi} y=y\boxplus_{1-\pi} x$\\
            $PA2$ & $x\boxplus_{\pi}(y\boxplus_{\rho} z)=(x\boxplus_{\frac{\pi}{\pi+\rho-\pi\rho}}y)\boxplus_{\pi+\rho-\pi\rho} z$\\
            $PA3$ & $x\boxplus_{\pi}x=x$\\
            $PA4$ & $(x\boxplus_{\pi}y)\cdot z=x\cdot z\boxplus_{\pi}y\cdot z$\\
            $PA5$ & $(x\boxplus_{\pi}y)+z=(x+z)\boxplus_{\pi}(y+z)$\\
            $G1$ & $\phi\cdot\neg\phi = \delta$\\
            $G2$ & $\phi+\neg\phi = \epsilon$\\
            $PG1$ & $\phi\boxplus_{\pi}\neg\phi = \epsilon$\\
            $G3$ & $\phi\delta = \delta$\\
            $G4$ & $\phi(x+y)=\phi x+\phi y$\\
            $PG2$ & $\phi(x\boxplus_{\pi}y)=\phi x\boxplus_{\pi}\phi y$\\
            $G5$ & $\phi(x\cdot y)= \phi x\cdot y$\\
            $G6$ & $(\phi+\psi)x = \phi x + \psi x$\\
            $PG3$ & $(\phi\boxplus_{\pi}\psi)x = \phi x \boxplus_{\pi} \psi x$\\
            $G7$ & $(\phi\cdot \psi)\cdot x = \phi\cdot(\psi\cdot x)$\\
            $G8$ & $\phi=\epsilon$ if $\forall s,\varrho\in S.test(\phi,s,\varrho)$\\
            $G9$ & $\phi_0\cdot\cdots\cdot\phi_n = \delta$ if $\forall s,\varrho\in S,\exists i\leq n.test(\neg\phi_i,s,\varrho)$\\
            $G10$ & $wp(e,\phi)e\phi=wp(e,\phi)e$\\
            $G11$ & $\neg wp(e,\phi)e\neg\phi=\neg wp(e,\phi)e$\\
        \end{tabular}
        \caption{Axioms of $qBAPTC_G$}
        \label{AxiomsForqBAPTCG}
    \end{table}
\end{center}

Note that, by eliminating atomic event from the process terms, the axioms in Table \ref{AxiomsForqBAPTCG} will lead to a Boolean Algebra. And $G8$ and $G9$ are preconditions of $e$ and
$\phi$, $G10$ is the weakest precondition of $e$ and $\phi$. A data environment with $effect$ function is sufficiently deterministic, and it is obvious that if the weakest precondition
is expressible and $G10$, $G11$ are sound, then the related data environment is sufficiently deterministic.

\begin{definition}[Basic terms of $qBAPTC_G$]\label{BTBAPTCG}
The set of basic terms of $qBAPTC_G$, $\mathcal{B}(qBAPTC_G)$, is inductively defined as follows:

\begin{enumerate}
  \item $\mathbb{E}\subset\mathcal{B}(qBAPTC_G)$;
  \item $G\subset\mathcal{B}(qBAPTC_G)$;
  \item if $e\in \mathbb{E}, t\in\mathcal{B}(qBAPTC_G)$ then $e\cdot t\in\mathcal{B}(qBAPTC_G)$;
  \item if $\phi\in G, t\in\mathcal{B}(qBAPTC_G)$ then $\phi\cdot t\in\mathcal{B}(qBAPTC_G)$;
  \item if $t,s\in\mathcal{B}(qBAPTC_G)$ then $t+ s\in\mathcal{B}(qBAPTC_G)$;
  \item if $t,s\in\mathcal{B}(qBAPTC_G)$ then $t\boxplus_{\pi} s\in\mathcal{B}(qBAPTC_G)$.
\end{enumerate}
\end{definition}

\begin{theorem}[Elimination theorem of $qBAPTC_G$]\label{ETBAPTCG}
Let $p$ be a closed $qBAPTC_G$ term. Then there is a basic $qBAPTC_G$ term $q$ such that $qBAPTC_G\vdash p=q$.
\end{theorem}

\begin{proof}
The same as that of $BAPTC_G$, we omit the proof, please refer to \cite{APPTC} for details.
\end{proof}

In this subsection, we will define a term-deduction system which gives the operational semantics of $qBAPTC$. Like the way in \cite{PPA}, we also introduce the counterpart $\breve{e}$
of the event $e$, and also the set $\breve{\mathbb{E}}=\{\breve{e}|e\in\mathbb{E}\}$.

We give the definition of PDFs of $qBAPTC$ in Table \ref{PDFBAPTC}.

\begin{center}
    \begin{table}
        $$\mu(e,\breve{e})=1$$
        $$\mu(x\cdot y, x'\cdot y)=\mu(x,x')$$
        $$\mu(x+y,x'+y')=\mu(x,x')\cdot \mu(y,y')$$
        $$\mu(x\boxplus_{\pi}y,z)=\pi\mu(x,z)+(1-\pi)\mu(y,z)$$
        $$\mu(x,y)=0,\textrm{otherwise}$$
        \caption{PDF definitions of $qBAPTC$}
        \label{PDFBAPTC}
    \end{table}
\end{center}

We will define a term-deduction system which gives the operational semantics of $qBAPTC_G$. We give the operational transition rules for $\epsilon$, atomic guard $\phi\in G_{at}$,
atomic event $e\in\mathbb{E}$, operators $\cdot$ and $+$ as Table \ref{SETRForqBAPTCG} shows. And the predicate $\xrightarrow{e}\surd$ represents successful termination after execution
of the event $e$.

\begin{center}
    \begin{table}
        $$\frac{}{\langle\epsilon,s,\varrho\rangle\rightsquigarrow\langle\breve{\epsilon},s,\varrho\rangle}$$
        $$\frac{}{\langle e,s,\varrho\rangle\rightsquigarrow\langle\breve{e},s,\varrho\rangle}$$
        $$\frac{}{\langle\phi,s,\varrho\rangle\rightsquigarrow\langle\breve{\phi},s,\varrho\rangle}$$
        $$\frac{\langle x,s,\varrho\rangle\rightsquigarrow \langle x',s,\varrho\rangle}{\langle x\cdot y,s,\varrho\rangle\rightsquigarrow \langle x'\cdot y,s,\varrho\rangle}$$
        $$\frac{\langle x,s,\varrho\rangle\rightsquigarrow \langle x',s,\varrho\rangle\quad \langle y,s,\varrho\rangle\rightsquigarrow \langle y',s,\varrho\rangle}{\langle x+y,s,\varrho\rangle\rightsquigarrow \langle x'+y',s,\varrho\rangle}$$
        $$\frac{\langle x,s,\varrho\rangle\rightsquigarrow \langle x',s,\varrho\rangle}{\langle x\boxplus_{\pi}y,s,\varrho\rangle\rightsquigarrow \langle x',s,\varrho\rangle}\quad \frac{\langle y,s,\varrho\rangle\rightsquigarrow \langle y',s,\varrho\rangle}{\langle x\boxplus_{\pi}y,s,\varrho\rangle\rightsquigarrow \langle y',s,\varrho\rangle}$$

        $$\frac{}{\langle\breve{\epsilon},s,\varrho\rangle\rightarrow\langle\surd,s,\varrho\rangle}$$
        $$\frac{}{\langle \breve{e},s,\varrho\rangle\xrightarrow{e}\langle\surd,s,\varrho'\rangle}\textrm{ if }s'\in effect(e,s)$$
        $$\frac{}{\langle\breve{\phi},s,\varrho\rangle\rightarrow\langle\surd,s,\varrho\rangle}\textrm{ if }test(\phi,s)$$
        $$\frac{\langle x,s,\varrho\rangle\xrightarrow{e}\langle\surd,s,\varrho'\rangle}{\langle x+ y,s,\varrho\rangle\xrightarrow{e}\langle\surd,s,\varrho'\rangle} \quad\frac{\langle x,s,\varrho\rangle\xrightarrow{e}\langle x',s,\varrho'\rangle}{\langle x+ y,s,\varrho\rangle\xrightarrow{e}\langle x',s,\varrho'\rangle}$$
        $$\frac{\langle y,s,\varrho\rangle\xrightarrow{e}\langle\surd,s,\varrho'\rangle}{\langle x+ y,s,\varrho\rangle\xrightarrow{e}\langle\surd,s,\varrho'\rangle} \quad\frac{\langle y,s,\varrho\rangle\xrightarrow{e}\langle y',s,\varrho'\rangle}{\langle x+ y,s,\varrho\rangle\xrightarrow{e}\langle y',s,\varrho'\rangle}$$
        $$\frac{\langle x,s,\varrho\rangle\xrightarrow{e}\langle\surd,s,\varrho'\rangle}{\langle x\cdot y,s,\varrho\rangle\xrightarrow{e} \langle y,s,\varrho'\rangle} \quad\frac{\langle x,s,\varrho\rangle\xrightarrow{e}\langle x',s,\varrho'\rangle}{\langle x\cdot y,s,\varrho\rangle\xrightarrow{e}\langle x'\cdot y,s,\varrho'\rangle}$$
        \caption{Single event transition rules of $qBAPTC_G$}
        \label{SETRForqBAPTCG}
    \end{table}
\end{center}

Note that, we replace the single atomic event $e\in\mathbb{E}$ by $X\subseteq\mathbb{E}$, we can obtain the pomset transition rules of $qBAPTC_G$, and omit them.

\begin{theorem}[Congruence of $qBAPTC_G$ with respect to probabilistic truly concurrent bisimulation equivalences]\label{CBAPTCG}
(1) Probabilistic pomset bisimulation equivalence $\sim_{pp}$ is a congruence with respect to $qBAPTC_G$.

(2) Probabilistic step bisimulation equivalence $\sim_{ps}$ is a congruence with respect to $qBAPTC_G$.

(3) Probabilistic hp-bisimulation equivalence $\sim_{php}$ is a congruence with respect to $qBAPTC_G$.

(4) Probabilistic hhp-bisimulation equivalence $\sim_{phhp}$ is a congruence with respect to $qBAPTC_G$.
\end{theorem}

\begin{proof}
(1) It is easy to see that probabilistic pomset bisimulation is an equivalent relation on $qBAPTC_G$ terms, we only need to prove that $\sim_{pp}$ is preserved by the operators $\cdot$
, $+$ and $\boxplus_{\pi}$. It is trivial and we leave the proof as an exercise for the readers.

(2) It is easy to see that probabilistic step bisimulation is an equivalent relation on $qBAPTC_G$ terms, we only need to prove that $\sim_{ps}$ is preserved by the operators $\cdot$,
$+$ and $\boxplus_{\pi}$. It is trivial and we leave the proof as an exercise for the readers.

(3) It is easy to see that probabilistic hp-bisimulation is an equivalent relation on $qBAPTC_G$ terms, we only need to prove that $\sim_{php}$ is preserved by the operators $\cdot$,
$+$, and $\boxplus_{\pi}$. It is trivial and we leave the proof as an exercise for the readers.

(4) It is easy to see that probabilistic hhp-bisimulation is an equivalent relation on $qBAPTC_G$ terms, we only need to prove that $\sim_{phhp}$ is preserved by the operators $\cdot$,
$+$, and $\boxplus_{\pi}$. It is trivial and we leave the proof as an exercise for the readers.
\end{proof}

\begin{theorem}[Soundness of $qBAPTC_G$ modulo probabilistic truly concurrent bisimulation equivalences]\label{SBAPTCG}
(1) Let $x$ and $y$ be $qBAPTC_G$ terms. If $qBAPTC\vdash x=y$, then $x\sim_{pp} y$.

(2) Let $x$ and $y$ be $qBAPTC_G$ terms. If $qBAPTC\vdash x=y$, then $x\sim_{ps} y$.

(3) Let $x$ and $y$ be $qBAPTC_G$ terms. If $qBAPTC\vdash x=y$, then $x\sim_{php} y$.

(4) Let $x$ and $y$ be $qBAPTC_G$ terms. If $qBAPTC\vdash x=y$, then $x\sim_{phhp} y$.
\end{theorem}

\begin{proof}
(1) Since probabilistic pomset bisimulation $\sim_{pp}$ is both an equivalent and a congruent relation, we only need to check if each axiom in Table \ref{AxiomsForqBAPTCG} is sound
modulo probabilistic pomset bisimulation equivalence. We leave the proof as an exercise for the readers.

(2) Since probabilistic step bisimulation $\sim_{ps}$ is both an equivalent and a congruent relation, we only need to check if each axiom in Table \ref{AxiomsForqBAPTCG} is sound modulo
probabilistic step bisimulation equivalence. We leave the proof as an exercise for the readers.

(3) Since probabilistic hp-bisimulation $\sim_{php}$ is both an equivalent and a congruent relation, we only need to check if each axiom in Table \ref{AxiomsForqBAPTCG} is sound modulo
probabilistic hp-bisimulation equivalence. We leave the proof as an exercise for the readers.

(4) Since probabilistic hhp-bisimulation $\sim_{phhp}$ is both an equivalent and a congruent relation, we only need to check if each axiom in Table \ref{AxiomsForqBAPTCG} is sound modulo
probabilistic hhp-bisimulation equivalence. We leave the proof as an exercise for the readers.
\end{proof}

\begin{theorem}[Completeness of $qBAPTC_G$ modulo probabilistic truly concurrent bisimulation equivalences]\label{CBAPTCG}
(1) Let $p$ and $q$ be closed $qBAPTC_G$ terms, if $p\sim_{pp} q$ then $p=q$.

(2) Let $p$ and $q$ be closed $qBAPTC_G$ terms, if $p\sim_{ps} q$ then $p=q$.

(3) Let $p$ and $q$ be closed $qBAPTC_G$ terms, if $p\sim_{php} q$ then $p=q$.

(4) Let $p$ and $q$ be closed $qBAPTC_G$ terms, if $p\sim_{phhp} q$ then $p=q$.
\end{theorem}

\begin{proof}
According to the definition of probabilistic truly concurrent bisimulation equivalences $\sim_{pp}$, $\sim_{ps}$, $\sim_{php}$ and $\sim_{phhp}$, $p\sim_{pp}q$, $p\sim_{ps}q$, $p\sim_{php}q$ and $p\sim_{phhp}q$ implies
both the bisimilarities between $p$ and $q$, and also the in the same quantum states. According to the completeness of $BAPTC_G$ (please refer to \cite{APPTC} for details), we can get the
completeness of $qBAPTC_G$.
\end{proof}

\subsection{$APPTC_G$ for Closed Quantum Systems}{\label{qapptcg}}

In this subsection, we will extend $qAPPTC$ with guards, which is abbreviated $qAPPTC_G$. The set of basic guards $G$ with element $\phi,\psi,\cdots$, which is extended by the
following formation rules:

$$\phi::=\delta|\epsilon|\neg\phi|\psi\in G_{at}|\phi+\psi|\phi\boxplus_{\pi}\psi|\phi\cdot\psi|\phi\leftmerge\psi$$

The set of axioms of $qAPPTC_G$ including axioms of $qBAPTC_G$ in Table \ref{AxiomsForqBAPTCG} and the axioms are shown in Table \ref{AxiomsForqAPPTCG}.

\begin{center}
    \begin{table}
        \begin{tabular}{@{}ll@{}}
            \hline No. &Axiom\\
            $P1$ & $(x+x=x,y+y=y)\quad x\between y = x\parallel y + x\mid y$\\
            $P2$ & $x\parallel y = y \parallel x$\\
            $P3$ & $(x\parallel y)\parallel z = x\parallel (y\parallel z)$\\
            $P4$ & $(x+x=x,y+y=y)\quad x\parallel y = x\leftmerge y + y\leftmerge x$\\
            $P5$ & $(e_1\leq e_2)\quad e_1\leftmerge (e_2\cdot y) = (e_1\leftmerge e_2)\cdot y$\\
            $P6$ & $(e_1\leq e_2)\quad (e_1\cdot x)\leftmerge e_2 = (e_1\leftmerge e_2)\cdot x$\\
            $P7$ & $(e_1\leq e_2)\quad (e_1\cdot x)\leftmerge (e_2\cdot y) = (e_1\leftmerge e_2)\cdot (x\between y)$\\
            $P8$ & $(x+ y)\leftmerge z = (x\leftmerge z)+ (y\leftmerge z)$\\
            $P9$ & $\delta\leftmerge x = \delta$\\
            $P10$ & $\epsilon\leftmerge x = x$\\
            $P11$ & $x\leftmerge \epsilon = x$\\
            $C1$ & $e_1\mid e_2 = \gamma(e_1,e_2)$\\
            $C2$ & $e_1\mid (e_2\cdot y) = \gamma(e_1,e_2)\cdot y$\\
            $C3$ & $(e_1\cdot x)\mid e_2 = \gamma(e_1,e_2)\cdot x$\\
            $C4$ & $(e_1\cdot x)\mid (e_2\cdot y) = \gamma(e_1,e_2)\cdot (x\between y)$\\
            $C5$ & $(x+ y)\mid z = (x\mid z) + (y\mid z)$\\
            $C6$ & $x\mid (y+ z) = (x\mid y)+ (x\mid z)$\\
            $C7$ & $\delta\mid x = \delta$\\
            $C8$ & $x\mid\delta = \delta$\\
            $C9$ & $\epsilon\mid x = \delta$\\
            $C10$ & $x\mid\epsilon = \delta$\\
            $PM1$ & $x\parallel (y\boxplus_{\pi} z)=(x\parallel y)\boxplus_{\pi}(x\parallel z)$\\
            $PM2$ & $(x\boxplus_{\pi} y)\parallel z=(x\parallel z)\boxplus_{\pi}(y\parallel z)$\\
            $PM3$ & $x\mid (y\boxplus_{\pi} z)=(x\mid y)\boxplus_{\pi}(x\mid z)$\\
            $PM4$ & $(x\boxplus_{\pi} y)\mid z=(x\mid z)\boxplus_{\pi}(y\mid z)$\\
            $CE1$ & $\Theta(e) = e$\\
            $CE2$ & $\Theta(\delta) = \delta$\\
            $CE3$ & $\Theta(\epsilon) = \epsilon$\\
            $CE4$ & $\Theta(x+ y) = \Theta(x)\triangleleft y + \Theta(y)\triangleleft x$\\
            $PCE1$ & $\Theta(x\boxplus_{\pi} y) = \Theta(x)\triangleleft y \boxplus_{\pi} \Theta(y)\triangleleft x$\\
            $CE5$ & $\Theta(x\cdot y)=\Theta(x)\cdot\Theta(y)$\\
            $CE6$ & $\Theta(x\leftmerge y) = ((\Theta(x)\triangleleft y)\leftmerge y)+ ((\Theta(y)\triangleleft x)\leftmerge x)$\\
            $CE7$ & $\Theta(x\mid y) = ((\Theta(x)\triangleleft y)\mid y)+ ((\Theta(y)\triangleleft x)\mid x)$\\
        \end{tabular}
        \caption{Axioms of $qAPPTC_G$}
        \label{AxiomsForqAPPTCG}
    \end{table}
\end{center}

\begin{center}
    \begin{table}
        \begin{tabular}{@{}ll@{}}
            \hline No. &Axiom\\
            $U1$ & $(\sharp(e_1,e_2))\quad e_1\triangleleft e_2 = \tau$\\
            $U2$ & $(\sharp(e_1,e_2),e_2\leq e_3)\quad e_1\triangleleft e_3 = e_1$\\
            $U3$ & $(\sharp(e_1,e_2),e_2\leq e_3)\quad e3\triangleleft e_1 = \tau$\\
            $PU1$ & $(\sharp_{\pi}(e_1,e_2))\quad e_1\triangleleft e_2 = \tau$\\
            $PU2$ & $(\sharp_{\pi}(e_1,e_2),e_2\leq e_3)\quad e_1\triangleleft e_3 = e_1$\\
            $PU3$ & $(\sharp_{\pi}(e_1,e_2),e_2\leq e_3)\quad e_3\triangleleft e_1 = \tau$\\
            $U4$ & $e\triangleleft \delta = e$\\
            $U5$ & $\delta \triangleleft e = \delta$\\
            $U6$ & $e\triangleleft \epsilon = e$\\
            $U7$ & $\epsilon \triangleleft e = e$\\
            $U8$ & $(x+ y)\triangleleft z = (x\triangleleft z)+ (y\triangleleft z)$\\
            $PU4$ & $(x\boxplus_{\pi} y)\triangleleft z = (x\triangleleft z)\boxplus_{\pi} (y\triangleleft z)$\\
            $U9$ & $(x\cdot y)\triangleleft z = (x\triangleleft z)\cdot (y\triangleleft z)$\\
            $U10$ & $(x\leftmerge y)\triangleleft z = (x\triangleleft z)\leftmerge (y\triangleleft z)$\\
            $U11$ & $(x\mid y)\triangleleft z = (x\triangleleft z)\mid (y\triangleleft z)$\\
            $U12$ & $x\triangleleft (y+ z) = (x\triangleleft y)\triangleleft z$\\
            $PU5$ & $x\triangleleft (y\boxplus_{\pi} z) = (x\triangleleft y)\triangleleft z$\\
            $U13$ & $x\triangleleft (y\cdot z)=(x\triangleleft y)\triangleleft z$\\
            $U14$ & $x\triangleleft (y\leftmerge z) = (x\triangleleft y)\triangleleft z$\\
            $U15$ & $x\triangleleft (y\mid z) = (x\triangleleft y)\triangleleft z$\\
            $D1$ & $e\notin H\quad\partial_H(e) = e$\\
            $D2$ & $e\in H\quad \partial_H(e) = \delta$\\
            $D3$ & $\partial_H(\delta) = \delta$\\
            $D4$ & $\partial_H(x+ y) = \partial_H(x)+\partial_H(y)$\\
            $D5$ & $\partial_H(x\cdot y) = \partial_H(x)\cdot\partial_H(y)$\\
            $D6$ & $\partial_H(x\leftmerge y) = \partial_H(x)\leftmerge\partial_H(y)$\\
            $PD1$ & $\partial_H(x\boxplus_{\pi}y)=\partial_H(x)\boxplus_{\pi}\partial_H(y)$\\
            $G12$ & $\phi(x\leftmerge y) =\phi x\leftmerge \phi y$\\
            $G13$ & $\phi(x\mid y) =\phi x\mid \phi y$\\
            $G14$ & $\delta\leftmerge \phi = \delta$\\
            $G15$ & $\phi\mid \delta = \delta$\\
            $G16$ & $\delta\mid \phi = \delta$\\
            $G17$ & $\phi\leftmerge \epsilon = \phi$\\
            $G18$ & $\epsilon\leftmerge \phi = \phi$\\
            $G19$ & $\phi\mid \epsilon = \delta$\\
            $G20$ & $\epsilon\mid \phi = \delta$\\
            $G21$ & $\phi\leftmerge\neg\phi = \delta$\\
            $G22$ & $\Theta(\phi) = \phi$\\
            $G23$ & $\partial_H(\phi) = \phi$\\
            $G24$ & $\phi_0\leftmerge\cdots\leftmerge\phi_n = \delta$ if $\forall s_0,\cdots,s_n,\varrho_1,\cdots,\varrho_n\in S,\exists i\leq n.test(\neg\phi_i,s_0\cup\cdots\cup s_n\cup\varrho_1\cup\cdots\cup\varrho_n)$\\        \end{tabular}
        \caption{Axioms of $qAPPTC_G$ (continuing)}
        \label{AxiomsForqAPPTCG2}
    \end{table}
\end{center}

\begin{definition}[Basic terms of $qAPPTC_G$]\label{BTAPPTCG}
The set of basic terms of $qAPPTC_G$, $\mathcal{B}(qAPPTC_G)$, is inductively defined as follows:

\begin{enumerate}
    \item $\mathbb{E}\subset\mathcal{B}(qAPPTC_G)$;
    \item $G\subset\mathcal{B}(qAPPTC_G)$;
    \item if $e\in \mathbb{E}, t\in\mathcal{B}(qAPPTC_G)$ then $e\cdot t\in\mathcal{B}(qAPPTC_G)$;
    \item if $\phi\in G, t\in\mathcal{B}(qAPPTC_G)$ then $\phi\cdot t\in\mathcal{B}(qAPPTC_G)$;
    \item if $t,s\in\mathcal{B}(qAPPTC_G)$ then $t+ s\in\mathcal{B}(qAPPTC_G)$;
    \item if $t,s\in\mathcal{B}(qAPPTC_G)$ then $t\boxplus_{\pi} s\in\mathcal{B}(qAPPTC_G)$
    \item if $t,s\in\mathcal{B}(qAPPTC_G)$ then $t\leftmerge s\in\mathcal{B}(qAPPTC_G)$.
\end{enumerate}
\end{definition}

Based on the definition of basic terms for $qAPPTC_G$ (see Definition \ref{BTAPPTCG}) and axioms of $qAPPTC_G$, we can prove the elimination theorem of $qAPPTC_G$.

\begin{theorem}[Elimination theorem of $qAPPTC_G$]\label{ETAPPTCG}
Let $p$ be a closed $qAPPTC_G$ term. Then there is a basic $qAPPTC_G$ term $q$ such that $qAPPTC_G\vdash p=q$.
\end{theorem}

\begin{proof}
The same as that of $APPTC_G$, we omit the proof, please refer to \cite{APPTC} for details.
\end{proof}

We give the definition of PDFs of $qAPPTC$ in Table \ref{PDFAPPTC}.

\begin{center}
    \begin{table}
        $$\mu(\delta,\breve{\delta})=1$$
        $$\mu(x\between y,x'\parallel y'+x'\mid y')=\mu(x,x')\cdot\mu(y,y')$$
        $$\mu(x\parallel y,x'\leftmerge y+y'\leftmerge x)=\mu(x,x')\cdot \mu(y,y')$$
        $$\mu(x\leftmerge y, x'\leftmerge y)=\mu(x,x')$$
        $$\mu(x\mid y,x'\mid y')=\mu(x,x')\cdot \mu(y,y')$$
        $$\mu(\Theta(x),\Theta(x'))=\mu(x,x')$$
        $$\mu(x\triangleleft y, x'\triangleleft y)=\mu(x,x')$$
        $$\mu(x,y)=0,\textrm{otherwise}$$
        \caption{PDF definitions of $qAPPTC$}
        \label{PDFAPPTC}
    \end{table}
\end{center}

We will define a term-deduction system which gives the operational semantics of $qAPPTC_G$. Two atomic events $e_1$ and $e_2$ are in race condition, which are denoted $e_1\% e_2$.

\begin{center}
    \begin{table}
        $$\frac{x\rightsquigarrow x'\quad y\rightsquigarrow y'}{x\between y\rightsquigarrow x'\parallel y'+x'\mid y'}$$
        $$\frac{x\rightsquigarrow x'\quad y\rightsquigarrow y'}{x\parallel y\rightsquigarrow x'\leftmerge y+y'\leftmerge x}$$
        $$\frac{x\rightsquigarrow x'}{x\leftmerge y\rightsquigarrow x'\leftmerge y}$$
        $$\frac{x\rightsquigarrow x'\quad y\rightsquigarrow y'}{x\mid y\rightsquigarrow x'\mid y'}$$
        $$\frac{x\rightsquigarrow x'}{\Theta(x)\rightsquigarrow \Theta(x')}$$
        $$\frac{x\rightsquigarrow x'}{x\triangleleft y\rightsquigarrow x'\triangleleft y}$$
        \caption{Probabilistic transition rules of $qAPPTC_G$}
        \label{TRForAPPTCG1}
    \end{table}
\end{center}

\begin{center}
    \begin{table}
        $$\frac{}{\langle \breve{e_1}\parallel\cdots \parallel \breve{e_n},s,\varrho\rangle\xrightarrow{\{e_1,\cdots,e_n\}}\langle\surd,s,\varrho'\rangle}\textrm{ if }s'\in effect(e_1,s)\cup\cdots\cup effect(e_n,s)$$

        $$\frac{}{\langle\breve{\phi_1}\parallel\cdots\parallel \breve{\phi_n},s,\varrho\rangle\rightarrow\langle\surd,s,\varrho\rangle}\textrm{ if }test(\phi_1,s),\cdots,test(\phi_n,s)$$

        $$\frac{\langle x,s,\varrho\rangle\xrightarrow{e_1}\langle\surd,s,\varrho'\rangle\quad \langle y,s,\varrho\rangle\xrightarrow{e_2}\langle\surd,s,\varrho''\rangle}{\langle x\parallel y,s,\varrho\rangle\xrightarrow{\{e_1,e_2\}}\langle\surd,s,\varrho'\cup\varrho''\rangle} \quad\frac{\langle x,s,\varrho\rangle\xrightarrow{e_1}\langle x',s,\varrho'\rangle\quad \langle y,s,\varrho\rangle\xrightarrow{e_2}\langle\surd,s,\varrho''\rangle}{\langle x\parallel y,s,\varrho\rangle\xrightarrow{\{e_1,e_2\}}\langle x',s'\cup s,\varrho''\rangle}$$

        $$\frac{\langle x,s,\varrho\rangle\xrightarrow{e_1}\langle\surd,s,\varrho'\rangle\quad \langle y,s,\varrho\rangle\xrightarrow{e_2}\langle y',s,\varrho''\rangle}{\langle x\parallel y,s,\varrho\rangle\xrightarrow{\{e_1,e_2\}}\langle y',s,\varrho'\cup\varrho''\rangle} \quad\frac{\langle x,s,\varrho\rangle\xrightarrow{e_1}\langle x',s,\varrho'\rangle\quad \langle y,s,\varrho\rangle\xrightarrow{e_2}\langle y',s,\varrho''\rangle}{\langle x\parallel y,s,\varrho\rangle\xrightarrow{\{e_1,e_2\}}\langle x'\between y',s'\cup s,\varrho''\rangle}$$

        $$\frac{\langle x,s,\varrho\rangle\xrightarrow{e_1}\langle\surd,s,\varrho'\rangle\quad \langle y,s,\varrho\rangle\xnrightarrow{e_2}\quad(e_1\%e_2)}{\langle x\parallel y,s,\varrho\rangle\xrightarrow{e_1}\langle y,s,\varrho'\rangle} \quad\frac{\langle x,s,\varrho\rangle\xrightarrow{e_1}\langle x',s,\varrho'\rangle\quad \langle y,s,\varrho\rangle\xnrightarrow{e_2}\quad(e_1\%e_2)}{\langle x\parallel y,s,\varrho\rangle\xrightarrow{e_1}\langle x'\between y,s,\varrho'\rangle}$$

        $$\frac{\langle x,s,\varrho\rangle\xnrightarrow{e_1}\quad \langle y,s,\varrho\rangle\xrightarrow{e_2}\langle\surd,s,\varrho''\rangle\quad(e_1\%e_2)}{\langle x\parallel y,s,\varrho\rangle\xrightarrow{e_2}\langle x,s,\varrho''\rangle} \quad\frac{\langle x,s,\varrho\rangle\xnrightarrow{e_1}\quad \langle y,s,\varrho\rangle\xrightarrow{e_2}\langle y',s,\varrho''\rangle\quad(e_1\%e_2)}{\langle x\parallel y,s,\varrho\rangle\xrightarrow{e_2}\langle x\between y',s,\varrho''\rangle}$$

        $$\frac{\langle x,s,\varrho\rangle\xrightarrow{e_1}\langle\surd,s,\varrho'\rangle\quad \langle y,s,\varrho\rangle\xrightarrow{e_2}\langle\surd,s,\varrho''\rangle \quad(e_1\leq e_2)}{\langle x\leftmerge y,s,\varrho\rangle\xrightarrow{\{e_1,e_2\}}\langle \surd,s,\varrho'\cup\varrho''\rangle}$$
        
        $$\frac{\langle x,s,\varrho\rangle\xrightarrow{e_1}\langle x',s,\varrho'\rangle\quad \langle y,s,\varrho\rangle\xrightarrow{e_2}\langle\surd,s,\varrho''\rangle \quad(e_1\leq e_2)}{\langle x\leftmerge y,s,\varrho\rangle\xrightarrow{\{e_1,e_2\}}\langle x',s,\varrho'\cup\varrho''\rangle}$$

        $$\frac{\langle x,s,\varrho\rangle\xrightarrow{e_1}\langle\surd,s,\varrho'\rangle\quad \langle y,s,\varrho\rangle\xrightarrow{e_2}\langle y',s,\varrho''\rangle \quad(e_1\leq e_2)}{\langle x\leftmerge y,s,\varrho\rangle\xrightarrow{\{e_1,e_2\}}\langle y',s,\varrho'\cup\varrho''\rangle}$$
        
        $$\frac{\langle x,s,\varrho\rangle\xrightarrow{e_1}\langle x',s,\varrho'\rangle\quad \langle y,s,\varrho\rangle\xrightarrow{e_2}\langle y',s,\varrho''\rangle \quad(e_1\leq e_2)}{\langle x\leftmerge y,s,\varrho\rangle\xrightarrow{\{e_1,e_2\}}\langle x'\between y',s,\varrho'\cup\varrho''\rangle}$$

        $$\frac{\langle x,s,\varrho\rangle\xrightarrow{e_1}\langle\surd,s,\varrho'\rangle\quad \langle y,s,\varrho\rangle\xrightarrow{e_2}\langle\surd,s,\varrho''\rangle}{\langle x\mid y,s,\varrho\rangle\xrightarrow{\gamma(e_1,e_2)}\langle\surd,effect(\gamma(e_1,e_2),s),\varrho\rangle} \quad\frac{\langle x,s,\varrho\rangle\xrightarrow{e_1}\langle x',s,\varrho'\rangle\quad \langle y,s,\varrho\rangle\xrightarrow{e_2}\langle\surd,s,\varrho''\rangle}{\langle x\mid y,s,\varrho\rangle\xrightarrow{\gamma(e_1,e_2)}\langle x',effect(\gamma(e_1,e_2),s,\varrho)\rangle}$$

        $$\frac{\langle x,s,\varrho\rangle\xrightarrow{e_1}\langle\surd,s,\varrho'\rangle\quad \langle y,s,\varrho\rangle\xrightarrow{e_2}\langle y',s,\varrho''\rangle}{\langle x\mid y,s,\varrho\rangle\xrightarrow{\gamma(e_1,e_2)}\langle y',effect(\gamma(e_1,e_2),s),\varrho\rangle} \quad\frac{\langle x,s,\varrho\rangle\xrightarrow{e_1}\langle x',s,\varrho'\rangle\quad \langle y,s,\varrho\rangle\xrightarrow{e_2}\langle y',s,\varrho''\rangle}{\langle x\mid y,s,\varrho\rangle\xrightarrow{\gamma(e_1,e_2)}\langle x'\between y',effect(\gamma(e_1,e_2),s),\varrho\rangle}$$

        \caption{Action transition rules of $qAPPTC_G$}
        \label{TRForAPPTCG}
    \end{table}
\end{center}

\begin{center}
    \begin{table}
        $$\frac{\langle x,s,\varrho\rangle\xrightarrow{e_1}\langle\surd,s,\varrho'\rangle\quad (\sharp(e_1,e_2))}{\langle \Theta(x),s,\varrho\rangle\xrightarrow{e_1}\langle\surd,s,\varrho'\rangle} \quad\frac{\langle x,s,\varrho\rangle\xrightarrow{e_2}\langle\surd,s,\varrho''\rangle\quad (\sharp(e_1,e_2))}{\langle\Theta(x),s,\varrho\rangle\xrightarrow{e_2}\langle\surd,s,\varrho''\rangle}$$

        $$\frac{\langle x,s,\varrho\rangle\xrightarrow{e_1}\langle x',s,\varrho'\rangle\quad (\sharp(e_1,e_2))}{\langle\Theta(x),s,\varrho\rangle\xrightarrow{e_1}\langle\Theta(x'),s,\varrho'\rangle} \quad\frac{\langle x,s,\varrho\rangle\xrightarrow{e_2}\langle x'',s,\varrho''\rangle\quad (\sharp(e_1,e_2))}{\langle\Theta(x),s,\varrho\rangle\xrightarrow{e_2}\langle\Theta(x''),s,\varrho''\rangle}$$

        $$\frac{\langle x,s,\varrho\rangle\xrightarrow{e_1}\langle\surd,s,\varrho'\rangle \quad \langle y,s,\varrho\rangle\nrightarrow^{e_2}\quad (\sharp(e_1,e_2))}{\langle x\triangleleft y,s,\varrho\rangle\xrightarrow{\tau}\langle\surd,s,\varrho'\rangle}
        \quad\frac{\langle x,s,\varrho\rangle\xrightarrow{e_1}\langle x',s,\varrho'\rangle \quad \langle y,s,\varrho\rangle\nrightarrow^{e_2}\quad (\sharp(e_1,e_2))}{\langle x\triangleleft y,s,\varrho\rangle\xrightarrow{\tau}\langle x',s,\varrho'\rangle}$$

        $$\frac{\langle x,s,\varrho\rangle\xrightarrow{e_1}\langle\surd,s,\varrho\rangle \quad \langle y,s,\varrho\rangle\nrightarrow^{e_3}\quad (\sharp(e_1,e_2),e_2\leq e_3)}{\langle x\triangleleft y,s,\varrho\rangle\xrightarrow{e_1}\langle\surd,s,\varrho'\rangle}$$
        $$\frac{\langle x,s,\varrho\rangle\xrightarrow{e_1}\langle x',s,\varrho'\rangle \quad \langle y,s,\varrho\rangle\nrightarrow^{e_3}\quad (\sharp(e_1,e_2),e_2\leq e_3)}{\langle x\triangleleft y,s,\varrho\rangle\xrightarrow{e_1}\langle x',s,\varrho'\rangle}$$

        $$\frac{\langle x,s,\varrho\rangle\xrightarrow{e_3}\langle\surd,s,\varrho'\rangle \quad \langle y,s,\varrho\rangle\nrightarrow^{e_2}\quad (\sharp(e_1,e_2),e_1\leq e_3)}{\langle x\triangleleft y,s,\varrho\rangle\xrightarrow{\tau}\langle\surd,s,\varrho'\rangle}$$
        $$\frac{\langle x,s,\varrho\rangle\xrightarrow{e_3}\langle x',s,\varrho'\rangle \quad \langle y,s,\varrho\rangle\nrightarrow^{e_2}\quad (\sharp(e_1,e_2),e_1\leq e_3)}{\langle x\triangleleft y,s,\varrho\rangle\xrightarrow{\tau}\langle x',s,\varrho'\rangle}$$

        $$\frac{\langle x,s,\varrho\rangle\xrightarrow{e_1}\langle\surd,s,\varrho'\rangle\quad (\sharp_{\pi}(e_1,e_2))}{\langle \Theta(x),s,\varrho\rangle\xrightarrow{e_1}\langle\surd,s,\varrho'\rangle} \quad\frac{\langle x,s,\varrho\rangle\xrightarrow{e_2}\langle\surd,s,\varrho''\rangle\quad (\sharp_{\pi}(e_1,e_2))}{\langle\Theta(x),s,\varrho\rangle\xrightarrow{e_2}\langle\surd,s,\varrho''\rangle}$$

        $$\frac{\langle x,s,\varrho\rangle\xrightarrow{e_1}\langle x',s,\varrho'\rangle\quad (\sharp_{\pi}(e_1,e_2))}{\langle\Theta(x),s,\varrho\rangle\xrightarrow{e_1}\langle\Theta(x'),s,\varrho'\rangle} \quad\frac{\langle x,s,\varrho\rangle\xrightarrow{e_2}\langle x'',s,\varrho''\rangle\quad (\sharp_{\pi}(e_1,e_2))}{\langle\Theta(x),s,\varrho\rangle\xrightarrow{e_2}\langle\Theta(x''),s,\varrho''\rangle}$$

        $$\frac{\langle x,s,\varrho\rangle\xrightarrow{e_1}\langle\surd,s,\varrho'\rangle \quad \langle y,s,\varrho\rangle\nrightarrow^{e_2}\quad (\sharp_{\pi}(e_1,e_2))}{\langle x\triangleleft y,s,\varrho\rangle\xrightarrow{\tau}\langle\surd,s,\varrho'\rangle}
        \quad\frac{\langle x,s,\varrho\rangle\xrightarrow{e_1}\langle x',s,\varrho'\rangle \quad \langle y,s,\varrho\rangle\nrightarrow^{e_2}\quad (\sharp_{\pi}(e_1,e_2))}{\langle x\triangleleft y,s,\varrho\rangle\xrightarrow{\tau}\langle x',s,\varrho'\rangle}$$

        $$\frac{\langle x,s,\varrho\rangle\xrightarrow{e_1}\langle\surd,s,\varrho\rangle \quad \langle y,s,\varrho\rangle\nrightarrow^{e_3}\quad (\sharp_{\pi}(e_1,e_2),e_2\leq e_3)}{\langle x\triangleleft y,s,\varrho\rangle\xrightarrow{e_1}\langle\surd,s,\varrho'\rangle}$$
        $$\frac{\langle x,s,\varrho\rangle\xrightarrow{e_1}\langle x',s,\varrho'\rangle \quad \langle y,s,\varrho\rangle\nrightarrow^{e_3}\quad (\sharp_{\pi}(e_1,e_2),e_2\leq e_3)}{\langle x\triangleleft y,s,\varrho\rangle\xrightarrow{e_1}\langle x',s,\varrho'\rangle}$$

        $$\frac{\langle x,s,\varrho\rangle\xrightarrow{e_3}\langle\surd,s,\varrho'\rangle \quad \langle y,s,\varrho\rangle\nrightarrow^{e_2}\quad (\sharp_{\pi}(e_1,e_2),e_1\leq e_3)}{\langle x\triangleleft y,s,\varrho\rangle\xrightarrow{\tau}\langle\surd,s,\varrho'\rangle}$$
        $$\frac{\langle x,s,\varrho\rangle\xrightarrow{e_3}\langle x',s,\varrho'\rangle \quad \langle y,s,\varrho\rangle\nrightarrow^{e_2}\quad (\sharp_{\pi}(e_1,e_2),e_1\leq e_3)}{\langle x\triangleleft y,s,\varrho\rangle\xrightarrow{\tau}\langle x',s,\varrho'\rangle}$$

        $$\frac{\langle x,s,\varrho\rangle\xrightarrow{e}\langle\surd,s,\varrho'\rangle}{\langle\partial_H(x),s,\varrho\rangle\xrightarrow{e}\langle\surd,s,\varrho'\rangle}\quad (e\notin H)\quad\frac{\langle x,s,\varrho\rangle\xrightarrow{e}\langle x',s,\varrho'\rangle}{\langle\partial_H(x),s,\varrho\rangle\xrightarrow{e}\langle\partial_H(x'),s,\varrho'\rangle}\quad(e\notin H)$$
        \caption{Action transition rules of $qAPPTC_G$ (continuing)}
        \label{TRForAPPTCG2}
    \end{table}
\end{center}

\begin{theorem}[Generalization of $qAPPTC_G$ with respect to $BAPTC_G$]
$qAPPTC_G$ is a generalization of $BAPTC_G$.
\end{theorem}

\begin{proof}
It follows from the following three facts.

\begin{enumerate}
  \item The transition rules of $qBAPTC_G$ in section \ref{qbaptcg} are all source-dependent;
  \item The sources of the transition rules $qAPPTC_G$ contain an occurrence of $\between$, or $\parallel$, or $\leftmerge$, or $\mid$, or $\Theta$, or $\triangleleft$;
  \item The transition rules of $qAPPTC_G$ are all source-dependent.
\end{enumerate}

So, $qAPPTC_G$ is a generalization of $qBAPTC_G$, that is, $qBAPTC_G$ is an embedding of $qAPPTC_G$, as desired.
\end{proof}

\begin{theorem}[Congruence of $qAPPTC_G$ with respect to probabilistic truly concurrent bisimulation equivalences]\label{CAPPTCG}
(1) Probabilistic pomset bisimulation equivalence $\sim_{pp}$ is a congruence with respect to $qAPPTC_G$.

(2) Probabilistic step bisimulation equivalence $\sim_{ps}$ is a congruence with respect to $qAPPTC_G$.

(3) Probabilistic hp-bisimulation equivalence $\sim_{php}$ is a congruence with respect to $qAPPTC_G$.

(4) Probabilistic hhp-bisimulation equivalence $\sim_{phhp}$ is a congruence with respect to $qAPPTC_G$.
\end{theorem}

\begin{proof}
(1) It is easy to see that probabilistic pomset bisimulation is an equivalent relation on $qAPPTC$ terms, we only need to prove that $\sim_{pp}$ is preserved by the operators
$\parallel$, $\leftmerge$, $\mid$, $\Theta$, $\triangleleft$, $\partial_H$. It is trivial and we leave the proof as an exercise for the readers.

(2) It is easy to see that probabilistic step bisimulation is an equivalent relation on $qAPPTC$ terms, we only need to prove that $\sim_{ps}$ is preserved by the operators
$\parallel$, $\leftmerge$, $\mid$, $\Theta$, $\triangleleft$, $\partial_H$. It is trivial and we leave the proof as an exercise for the readers.

(3) It is easy to see that probabilistic hp-bisimulation is an equivalent relation on $qAPPTC$ terms, we only need to prove that $\sim_{php}$ is preserved by the operators
$\parallel$, $\leftmerge$, $\mid$, $\Theta$, $\triangleleft$, $\partial_H$. It is trivial and we leave the proof as an exercise for the readers.

(4) It is easy to see that probabilistic hhp-bisimulation is an equivalent relation on $qAPPTC$ terms, we only need to prove that $\sim_{phhp}$ is preserved by the operators
$\parallel$, $\leftmerge$, $\mid$, $\Theta$, $\triangleleft$, $\partial_H$. It is trivial and we leave the proof as an exercise for the readers.
\end{proof}

\begin{theorem}[Soundness of $qAPPTC_G$ modulo probabilistic truly concurrent bisimulation equivalences]\label{SAPPTCG}
(1) Let $x$ and $y$ be $qAPPTC_G$ terms. If $qAPPTC\vdash x=y$, then $x\sim_{pp} y$.

(2) Let $x$ and $y$ be $qAPPTC_G$ terms. If $qAPPTC\vdash x=y$, then $x\sim_{ps} y$.

(3) Let $x$ and $y$ be $qAPPTC_G$ terms. If $qAPPTC\vdash x=y$, then $x\sim_{php} y$;

(3) Let $x$ and $y$ be $qAPPTC_G$ terms. If $qAPPTC\vdash x=y$, then $x\sim_{phhp} y$.
\end{theorem}

\begin{proof}
(1) Since probabilistic pomset bisimulation $\sim_{pp}$ is both an equivalent and a congruent relation, we only need to check if each axiom in Table \ref{AxiomsForqAPPTCG} is sound modulo
probabilistic pomset bisimulation equivalence. We leave the proof as an exercise for the readers.

(2) Since probabilistic step bisimulation $\sim_{ps}$ is both an equivalent and a congruent relation, we only need to check if each axiom in Table \ref{AxiomsForqAPPTCG} is sound modulo
probabilistic step bisimulation equivalence. We leave the proof as an exercise for the readers.

(3) Since probabilistic hp-bisimulation $\sim_{php}$ is both an equivalent and a congruent relation, we only need to check if each axiom in Table \ref{AxiomsForqAPPTCG} is sound modulo
probabilistic hp-bisimulation equivalence. We leave the proof as an exercise for the readers.

(4) Since probabilistic hhp-bisimulation $\sim_{phhp}$ is both an equivalent and a congruent relation, we only need to check if each axiom in Table \ref{AxiomsForqAPPTCG} is sound modulo
probabilistic hhp-bisimulation equivalence. We leave the proof as an exercise for the readers.
\end{proof}

\begin{theorem}[Completeness of $qAPPTC_G$ modulo probabilistic truly concurrent bisimulation equivalences]\label{CAPPTCG}
(1) Let $p$ and $q$ be closed $qAPPTC_G$ terms, if $p\sim_{pp} q$ then $p=q$.

(2) Let $p$ and $q$ be closed $qAPPTC_G$ terms, if $p\sim_{ps} q$ then $p=q$.

(3) Let $p$ and $q$ be closed $qAPPTC_G$ terms, if $p\sim_{php} q$ then $p=q$.

(3) Let $p$ and $q$ be closed $qAPPTC_G$ terms, if $p\sim_{phhp} q$ then $p=q$.
\end{theorem}

\begin{proof}
According to the definition of probabilistic truly concurrent bisimulation equivalences $\sim_{pp}$, $\sim_{ps}$, $\sim_{php}$ and $\sim_{phhp}$, $p\sim_{pp}q$, $p\sim_{ps}q$, $p\sim_{php}q$ and $p\sim_{phhp}q$ implies
both the bisimilarities between $p$ and $q$, and also the in the same quantum states. According to the completeness of $APPTC_G$ (please refer to \cite{APPTC} for details), we can get the
completeness of $qAPPTC_G$.
\end{proof}

\subsection{Recursion}{\label{qcrecg}}

In this subsection, we introduce recursion to capture infinite processes based on $qAPPTC_G$. In the following, $E,F,G$ are recursion specifications, $X,Y,Z$ are recursive variables.

\begin{definition}[Guarded recursive specification]
A recursive specification

$$X_1=t_1(X_1,\cdots,X_n)$$
$$...$$
$$X_n=t_n(X_1,\cdots,X_n)$$

is guarded if the right-hand sides of its recursive equations can be adapted to the form by applications of the axioms in $qAPPTC$ and replacing recursion variables by the right-hand
sides of their recursive equations,

$((a_{111}\leftmerge\cdots\leftmerge a_{11i_1})\cdot s_1(X_1,\cdots,X_n)+\cdots+(a_{1k1}\leftmerge\cdots\leftmerge a_{1ki_k})\cdot s_k(X_1,\cdots,X_n)+(b_{111}\leftmerge\cdots\leftmerge
b_{11j_1})+\cdots+(b_{11j_1}\leftmerge\cdots\leftmerge b_{1lj_l}))\boxplus_{\pi_1}\cdots\boxplus_{\pi_{m-1}}((a_{m11}\leftmerge\cdots\leftmerge a_{m1i_1})\cdot s_1(X_1,\cdots,X_n)+
\cdots+(a_{mk1}\leftmerge\cdots\leftmerge a_{mki_k})\cdot s_k(X_1,\cdots,X_n)+(b_{m11}\leftmerge\cdots\leftmerge b_{m1j_1})+\cdots+(b_{m1j_1}\leftmerge\cdots\leftmerge b_{mlj_l}))$

where $a_{111},\cdots,a_{11i_1},a_{1k1},\cdots,a_{1ki_k},b_{111},\cdots,b_{11j_1},b_{11j_1},\cdots,b_{1lj_l},\cdots, a_{m11},\cdots,a_{m1i_1},a_{1k1},\cdots,a_{mki_k},\\b_{111},\cdots,
b_{m1j_1},b_{m1j_1},\cdots,b_{mlj_l}\in \mathbb{E}$, and the sum above is allowed to be empty, in which case it represents the deadlock $\delta$. And there does not exist an infinite
sequence of $\epsilon$-transitions $\langle X|E\rangle\rightarrow\langle X'|E\rangle\rightarrow\langle X''|E\rangle\rightarrow\cdots$.
\end{definition}

\begin{center}
    \begin{table}
        $$\frac{\langle t_i(\langle X_1|E\rangle,\cdots,\langle X_n|E\rangle),s,\varrho\rangle\rightsquigarrow \langle y,s,\varrho\rangle}{\langle\langle X_i|E\rangle,s,\varrho\rangle\rightsquigarrow \langle y,s,\varrho\rangle}$$
        $$\frac{\langle t_i(\langle X_1|E\rangle,\cdots,\langle X_n|E\rangle),s,\varrho\rangle\xrightarrow{\{e_1,\cdots,e_k\}}\langle\surd,s,\varrho'\rangle}{\langle\langle X_i|E\rangle,s,\varrho\rangle\xrightarrow{\{e_1,\cdots,e_k\}}\langle\surd,s,\varrho'\rangle}$$
        $$\frac{\langle t_i(\langle X_1|E\rangle,\cdots,\langle X_n|E\rangle),s,\varrho\rangle\xrightarrow{\{e_1,\cdots,e_k\}} \langle y,s,\varrho'\rangle}{\langle\langle X_i|E\rangle,s,\varrho\rangle\xrightarrow{\{e_1,\cdots,e_k\}} \langle y,s,\varrho'\rangle}$$
        \caption{Transition rules of guarded recursion}
        \label{TRForGRG}
    \end{table}
\end{center}

\begin{theorem}[Conservitivity of $qAPPTC_G$ with guarded recursion]
$qAPPTC_G$ with guarded recursion is a conservative extension of $qAPPTC_G$.
\end{theorem}

\begin{proof}
Since the transition rules of $qAPPTC_G$ are source-dependent, and the transition rules for guarded recursion in Table \ref{TRForGRG} contain only a fresh constant in their source, so
the transition rules of $qAPPTC_G$ with guarded recursion are a conservative extension of those of $qAPPTC_G$.
\end{proof}

\begin{theorem}[Congruence theorem of $qAPPTC_G$ with guarded recursion]
Probabilistic truly concurrent bisimulation equivalences $\sim_{pp}$, $\sim_{p}$, $\sim_{php}$ and $\sim_{phhp}$ are all congruences with respect to $qAPPTC_G$ with guarded recursion.
\end{theorem}

\begin{proof}
It follows the following two facts:
\begin{enumerate}
  \item in a guarded recursive specification, right-hand sides of its recursive equations can be adapted to the form by applications of the axioms in $qAPPTC_G$ and replacing recursion
  variables by the right-hand sides of their recursive equations;
  \item probabilistic truly concurrent bisimulation equivalences $\sim_{pp}$, $\sim_{ps}$, $\sim_{php}$ and $\sim_{phhp}$ are all congruences with respect to all operators of $qAPPTC_G$.
\end{enumerate}
\end{proof}

\begin{theorem}[Elimination theorem of $qAPPTC_G$ with linear recursion]\label{ETRecursionG}
Each process term in $qAPPTC_G$ with linear recursion is equal to a process term $\langle X_1|E\rangle$ with $E$ a linear recursive specification.
\end{theorem}

\begin{proof}
The same as that of $APPTC_G$, we omit the proof, please refer to \cite{APPTC} for details.
\end{proof}

\begin{theorem}[Soundness of $qAPPTC_G$ with guarded recursion]\label{SAPPTC_GRG}
Let $x$ and $y$ be $qAPPTC_G$ with guarded recursion terms. If $qAPPTC_G\textrm{ with guarded recursion}\vdash x=y$, then

(1) $x\sim_{ps} y$.

(2) $x\sim_{pp} y$.

(3) $x\sim_{php} y$.

(4) $x\sim_{phhp} y$.
\end{theorem}

\begin{proof}
(1) Since probabilistic step bisimulation $\sim_{ps}$ is both an equivalent and a congruent relation with respect to $qAPPTC_G$ with guarded recursion, we only need to check if each
axiom in Table \ref{RDPRSP} is sound modulo probabilistic step bisimulation equivalence. We leave them as exercises to the readers.

(2) Since probabilistic pomset bisimulation $\sim_{pp}$ is both an equivalent and a congruent relation with respect to the guarded recursion, we only need to check if each axiom in
Table \ref{RDPRSP} is sound modulo probabilistic pomset bisimulation equivalence. We leave them as exercises to the readers.

(3) Since probabilistic hp-bisimulation $\sim_{php}$ is both an equivalent and a congruent relation with respect to guarded recursion, we only need to check if each axiom in Table
\ref{RDPRSP} is sound modulo probabilistic hp-bisimulation equivalence. We leave them as exercises to the readers.

(4) Since probabilistic hhp-bisimulation $\sim_{phhp}$ is both an equivalent and a congruent relation with respect to guarded recursion, we only need to check if each axiom in Table
\ref{RDPRSP} is sound modulo probabilistic hhp-bisimulation equivalence. We leave them as exercises to the readers.
\end{proof}

\begin{theorem}[Completeness of $qAPPTC_G$ with linear recursion]\label{CAPPTC_GRG}
Let $p$ and $q$ be closed $qAPPTC_G$ with linear recursion terms, then,

(1) if $p\sim_{ps} q$ then $p=q$.

(2) if $p\sim_{pp} q$ then $p=q$.

(3) if $p\sim_{php} q$ then $p=q$.

(4) if $p\sim_{phhp} q$ then $p=q$.
\end{theorem}

\begin{proof}
According to the definition of probabilistic truly concurrent bisimulation equivalences $\sim_{pp}$, $\sim_{ps}$, $\sim_{php}$ and $\sim_{phhp}$, $p\sim_{pp}q$, $p\sim_{ps}q$, $p\sim_{php}q$ and $p\sim_{phhp}q$ implies
both the bisimilarities between $p$ and $q$, and also the in the same quantum states. According to the completeness of $APPTC_G$ with linear recursion (please refer to \cite{APPTC} for details), we can get the
completeness of $qAPPTC_G$ with linear recursion.
\end{proof}

\subsection{Abstraction}{\label{qcabsg}}

To abstract away from the internal implementations of a program, and verify that the program exhibits the desired external behaviors, the silent step $\tau$ and abstraction operator
$\tau_I$ are introduced, where $I\subseteq \mathbb{E}\cup G_{at}$ denotes the internal events or guards. The silent step $\tau$ represents the internal events or guards, when we
consider the external behaviors of a process, $\tau$ steps can be removed, that is, $\tau$ steps must keep silent. The transition rule of $\tau$ is shown in Table \ref{TRForqTauG2}. In
the following, let the atomic event $e$ range over $\mathbb{E}\cup\{\epsilon\}\cup\{\delta\}\cup\{\tau\}$, and $\phi$ range over $G\cup \{\tau\}$, and let the communication function
$\gamma:\mathbb{E}\cup\{\tau\}\times \mathbb{E}\cup\{\tau\}\rightarrow \mathbb{E}\cup\{\delta\}$, with each communication involved $\tau$ resulting in $\delta$. We use $\tau(\varrho)$ to
denote $effect(\tau,\varrho)$, for the fact that $\tau$ only change the state of internal data environment, that is, for the external data environments, $\varrho=\tau(\varrho)$.

\begin{center}
    \begin{table}
        $$\frac{}{\tau\rightsquigarrow\breve{\tau}}$$
        $$\frac{}{\langle\tau,s,\varrho\rangle\rightarrow\langle\surd,s,\varrho\rangle}\textrm{ if }test(\tau,s)$$
        $$\frac{}{\langle\tau,s,\varrho\rangle\xrightarrow{\tau}\langle\surd,\tau(s)\rangle}$$
        \caption{Transition rule of the silent step}
        \label{TRForqTauG2}
    \end{table}
\end{center}

\begin{definition}[Guarded linear recursive specification]\label{GLRSG}
A linear recursive specification $E$ is guarded if there does not exist an infinite sequence of $\tau$-transitions
$\langle X|E\rangle\xrightarrow{\tau}\langle X'|E\rangle\xrightarrow{\tau}\langle X''|E\rangle\xrightarrow{\tau}\cdots$, and there does not exist an infinite sequence of
$\epsilon$-transitions $\langle X|E\rangle\rightarrow\langle X'|E\rangle\rightarrow\langle X''|E\rangle\rightarrow\cdots$.
\end{definition}

\begin{theorem}[Conservitivity of $qAPPTC_G$ with silent step and guarded linear recursion]
$qAPPTC_G$ with silent step and guarded linear recursion is a conservative extension of $qAPPTC_G$ with linear recursion.
\end{theorem}

\begin{proof}
Since the transition rules of $qAPPTC_G$ with linear recursion are source-dependent, and the transition rules for silent step in Table \ref{TRForqTauG2} contain only a fresh constant
$\tau$ in their source, so the transition rules of $qAPPTC_G$ with silent step and guarded linear recursion is a conservative extension of those of $qAPPTC_G$ with linear recursion.
\end{proof}

\begin{theorem}[Congruence theorem of $qAPPTC_G$ with silent step and guarded linear recursion]
Probabilistic rooted branching truly concurrent bisimulation equivalences $\approx_{prbp}$, $\approx_{prbs}$, $\approx_{prbhp}$ and $\approx_{rbhhp}$ are all congruences with respect
to $qAPPTC_G$ with silent step and guarded linear recursion.
\end{theorem}

\begin{proof}
It follows the following three facts:
\begin{enumerate}
  \item in a guarded linear recursive specification, right-hand sides of its recursive equations can be adapted to the form by applications of the axioms in $qAPPTC_G$ and replacing
  recursion variables by the right-hand sides of their recursive equations;
  \item probabilistic truly concurrent bisimulation equivalences $\sim_{pp}$, $\sim_{ps}$, $\sim_{php}$ and $\sim_{phhp}$ are all congruences with respect to all operators of
  $qAPPTC_G$, while probabilistic truly concurrent bisimulation equivalences $\sim_{pp}$, $\sim_{ps}$, $\sim_{php}$ and $\sim_{phhp}$ imply the corresponding probabilistic rooted
  branching truly concurrent bisimulations $\approx_{prbp}$, $\approx_{prbs}$, $\approx_{prbhp}$ and $\approx_{prbhhp}$, so probabilistic rooted branching truly concurrent
  bisimulations $\approx_{prbp}$, $\approx_{prbs}$, $\approx_{prbhp}$ and $\approx_{prbhhp}$ are all congruences with respect to all operators of $qAPPTC_G$;
  \item While $\mathbb{E}$ is extended to $\mathbb{E}\cup\{\tau\}$, and $G$ is extended to $G\cup\{\tau\}$, it can be proved that probabilistic rooted branching truly concurrent
  bisimulations $\approx_{prbp}$, $\approx_{prbs}$, $\approx_{prbhp}$ and $\approx_{prbhhp}$ are all congruences with respect to all operators of $qAPPTC_G$, we omit it.
\end{enumerate}
\end{proof}

We design the axioms for the silent step $\tau$ in Table \ref{AxiomsForqTauG2}.

\begin{center}
\begin{table}
  \begin{tabular}{@{}ll@{}}
  \hline No. &Axiom\\
  $B1$ & $(y=y+y,z=z+z)\quad x\cdot((y+\tau\cdot(y+z))\boxplus_{\pi}w)=x\cdot((y+z)\boxplus_{\pi}w)$\\
  $B2$ & $(y=y+y,z=z+z)\quad x\leftmerge((y+\tau\leftmerge(y+z))\boxplus_{\pi}w)=x\leftmerge((y+z)\boxplus_{\pi}w)$\\
\end{tabular}
\caption{Axioms of silent step}
\label{AxiomsForqTauG2}
\end{table}
\end{center}

\begin{theorem}[Elimination theorem of $qAPPTC_G$ with silent step and guarded linear recursion]\label{ETTauG}
Each process term in $qAPPTC_G$ with silent step and guarded linear recursion is equal to a process term $\langle X_1|E\rangle$ with $E$ a guarded linear recursive specification.
\end{theorem}

\begin{proof}
The same as that of $APPTC_G$, we omit the proof, please refer to \cite{APPTC} for details.
\end{proof}

\begin{theorem}[Soundness of $qAPPTC_G$ with silent step and guarded linear recursion]\label{SAPPTC_GTAUG}
Let $x$ and $y$ be $qAPPTC_G$ with silent step and guarded linear recursion terms. If $qAPPTC_G$ with silent step and guarded linear recursion $\vdash x=y$, then

(1) $x\approx_{prbs} y$.

(2) $x\approx_{prbp} y$.

(3) $x\approx_{prbhp} y$.

(4) $x\approx_{prbhhp} y$.
\end{theorem}

\begin{proof}
(1) Since probabilistic rooted branching step bisimulation $\approx_{prbs}$ is both an equivalent and a congruent relation with respect to $qAPPTC_G$ with silent step and guarded
linear recursion, we only need to check if each axiom in Table \ref{AxiomsForqTauG2} is sound modulo probabilistic rooted branching step bisimulation $\approx_{prbs}$. We leave them as
exercises to the readers.

(2) Since probabilistic rooted branching pomset bisimulation $\approx_{prbp}$ is both an equivalent and a congruent relation with respect to $qAPPTC_G$ with silent step and guarded
linear recursion, we only need to check if each axiom in Table \ref{AxiomsForqTauG2} is sound modulo probabilistic rooted branching pomset bisimulation $\approx_{prbp}$. We leave them
as exercises to the readers.

(3) Since probabilistic rooted branching hp-bisimulation $\approx_{prbhp}$ is both an equivalent and a congruent relation with respect to $qAPPTC_G$ with silent step and guarded linear
recursion, we only need to check if each axiom in Table \ref{AxiomsForqTauG2} is sound modulo probabilistic rooted branching hp-bisimulation $\approx_{prbhp}$. We leave them as exercises
to the readers.

(4) Since probabilistic rooted branching hhp-bisimulation $\approx_{prbhhp}$ is both an equivalent and a congruent relation with respect to $qAPPTC_G$ with silent step and guarded linear
recursion, we only need to check if each axiom in Table \ref{AxiomsForqTauG2} is sound modulo probabilistic rooted branching hhp-bisimulation $\approx_{prbhhp}$. We leave them as exercises
to the readers.
\end{proof}

\begin{theorem}[Completeness of $qAPPTC_G$ with silent step and guarded linear recursion]\label{CAPPTC_GTAUG}
Let $p$ and $q$ be closed $qAPPTC_G$ with silent step and guarded linear recursion terms, then,

(1) if $p\approx_{prbs} q$ then $p=q$.

(2) if $p\approx_{prbp} q$ then $p=q$.

(3) if $p\approx_{prbhp} q$ then $p=q$.

(3) if $p\approx_{prbhhp} q$ then $p=q$.
\end{theorem}

\begin{proof}
According to the definition of probabilistic rooted branching truly concurrent bisimulation equivalences $\approx_{prbp}$, $\approx_{prbs}$, $\approx_{prbhp}$ and $\approx_{prbhhp}$, and $\approx_{prbp}$, $\approx_{prbs}$, $\approx_{prbhp}$ and $\approx_{prbhhp}$ implies
both the bisimilarities between $p$ and $q$, and also the in the same quantum states. According to the completeness of $APPTC_G$ with silent step and guarded linear recursion (please refer to \cite{APPTC} for details), we can get the
completeness of $qAPPTC_G$ with silent step and guarded linear recursion.
\end{proof}

The unary abstraction operator $\tau_I$ ($I\subseteq \mathbb{E}\cup G_{at}$) renames all atomic events or atomic guards in $I$ into $\tau$. $qAPPTC_G$ with silent step and abstraction
operator is called $qAPPTC_{G_{\tau}}$. The transition rules of operator $\tau_I$ are shown in Table \ref{TRForqAbstractionG2}.

\begin{center}
    \begin{table}
        $$\frac{\langle x,s,\varrho\rangle\rightsquigarrow \langle x',s,\varrho\rangle}{\langle \tau_I(x),s,\varrho\rangle\rightsquigarrow\langle\tau_I(x'),s,\varrho\rangle}$$
        $$\frac{\langle x,s,\varrho\rangle\xrightarrow{e}\langle\surd,s,\varrho'\rangle}{\langle\tau_I(x),s,\varrho\rangle\xrightarrow{e}\langle\surd,s,\varrho'\rangle}\quad e\notin I
        \quad\quad\frac{\langle x,s,\varrho\rangle\xrightarrow{e}\langle x',s,\varrho'\rangle}{\langle\tau_I(x),s,\varrho\rangle\xrightarrow{e}\langle\tau_I(x'),s,\varrho'\rangle}\quad e\notin I$$

        $$\frac{\langle x,s,\varrho\rangle\xrightarrow{e}\langle\surd,s,\varrho'\rangle}{\langle\tau_I(x),s,\varrho\rangle\xrightarrow{\tau}\langle\surd,\tau(s)\rangle}\quad e\in I
        \quad\quad\frac{\langle x,s,\varrho\rangle\xrightarrow{e}\langle x',s,\varrho'\rangle}{\langle\tau_I(x),s,\varrho\rangle\xrightarrow{\tau}\langle\tau_I(x'),\tau(s)\rangle}\quad e\in I$$
        \caption{Transition rule of the abstraction operator}
        \label{TRForqAbstractionG2}
    \end{table}
\end{center}

\begin{theorem}[Conservitivity of $qAPPTC_{G_{\tau}}$ with guarded linear recursion]
$qAPPTC_{G_{\tau}}$ with guarded linear recursion is a conservative extension of $qAPPTC_G$ with silent step and guarded linear recursion.
\end{theorem}

\begin{proof}
Since the transition rules of $qAPPTC_G$ with silent step and guarded linear recursion are source-dependent, and the transition rules for abstraction operator in Table
\ref{TRForqAbstractionG2} contain only a fresh operator $\tau_I$ in their source, so the transition rules of $qAPPTC_{G_{\tau}}$ with guarded linear recursion is a conservative extension
of those of $qAPPTC_G$ with silent step and guarded linear recursion.
\end{proof}

\begin{theorem}[Congruence theorem of $qAPPTC_{G_{\tau}}$ with guarded linear recursion]
Probabilistic rooted branching truly concurrent bisimulation equivalences $\approx_{prbp}$, $\approx_{prbs}$, $\approx_{prbhp}$ and $\approx_{prbhhp}$ are all congruences with respect
to $qAPPTC_{G_{\tau}}$ with guarded linear recursion.
\end{theorem}

\begin{proof}
(1) It is easy to see that probabilistic rooted branching pomset bisimulation is an equivalent relation on $qAPPTC_{G_{\tau}}$ with guarded linear recursion terms, we only need to
prove that $\approx_{prbp}$ is preserved by the operators $\tau_I$. It is trivial and we leave the proof as an exercise for the readers.

(2) It is easy to see that probabilistic rooted branching step bisimulation is an equivalent relation on $qAPPTC_{G_{\tau}}$ with guarded linear recursion terms, we only need to
prove that $\approx_{prbs}$ is preserved by the operators $\tau_I$. It is trivial and we leave the proof as an exercise for the readers.

(3) It is easy to see that probabilistic rooted branching hp-bisimulation is an equivalent relation on $qAPPTC_{G_{\tau}}$ with guarded linear recursion terms, we only need to
prove that $\approx_{prbhp}$ is preserved by the operators $\tau_I$. It is trivial and we leave the proof as an exercise for the readers.

(4) It is easy to see that probabilistic rooted branching hhp-bisimulation is an equivalent relation on $qAPPTC_{G_{\tau}}$ with guarded linear recursion terms, we only need to
prove that $\approx_{prbhhp}$ is preserved by the operators $\tau_I$. It is trivial and we leave the proof as an exercise for the readers.
\end{proof}

We design the axioms for the abstraction operator $\tau_I$ in Table \ref{AxiomsForqAbstractionG2}.

\begin{center}
\begin{table}
  \begin{tabular}{@{}ll@{}}
\hline No. &Axiom\\
  $TI1$ & $e\notin I\quad \tau_I(e)=e$\\
  $TI2$ & $e\in I\quad \tau_I(e)=\tau$\\
  $TI3$ & $\tau_I(\delta)=\delta$\\
  $TI4$ & $\tau_I(x+y)=\tau_I(x)+\tau_I(y)$\\
  $PTI1$ & $\tau_I(x\boxplus_{\pi}y)=\tau_I(x)\boxplus_{\pi}\tau_I(y)$\\
  $TI5$ & $\tau_I(x\cdot y)=\tau_I(x)\cdot\tau_I(y)$\\
  $TI6$ & $\tau_I(x\leftmerge y)=\tau_I(x)\leftmerge\tau_I(y)$\\
  $G28$ & $\phi\notin I\quad \tau_I(\phi)=\phi$\\
  $G29$ & $\phi\in I\quad \tau_I(\phi)=\tau$\\
\end{tabular}
\caption{Axioms of abstraction operator}
\label{AxiomsForqAbstractionG2}
\end{table}
\end{center}

\begin{theorem}[Soundness of $qAPPTC_{G_{\tau}}$ with guarded linear recursion]\label{SAPPTC_GABSG}
Let $x$ and $y$ be $qAPPTC_{G_{\tau}}$ with guarded linear recursion terms. If $qAPPTC_{G_{\tau}}$ with guarded linear recursion $\vdash x=y$, then

(1) $x\approx_{prbs} y$.

(2) $x\approx_{prbp} y$.

(3) $x\approx_{prbhp} y$.

(4) $x\approx_{prbhhp} y$.
\end{theorem}

\begin{proof}
(1) Since probabilistic rooted branching step bisimulation $\approx_{prbs}$ is both an equivalent and a congruent relation with respect to $qAPPTC_{G_{\tau}}$ with guarded linear
recursion, we only need to check if each axiom in Table \ref{AxiomsForqAbstractionG2} is sound modulo probabilistic rooted branching step bisimulation $\approx_{prbs}$. We leave them as
exercises to the readers.

(2) Since probabilistic rooted branching pomset bisimulation $\approx_{prbp}$ is both an equivalent and a congruent relation with respect to $qAPPTC_{G_{\tau}}$ with guarded linear
recursion, we only need to check if each axiom in Table \ref{AxiomsForqAbstractionG2} is sound modulo probabilistic rooted branching pomset bisimulation $\approx_{prbp}$. We leave them
as exercises to the readers.

(3) Since probabilistic rooted branching hp-bisimulation $\approx_{prbhp}$ is both an equivalent and a congruent relation with respect to $qAPPTC_{G_{\tau}}$ with guarded linear
recursion, we only need to check if each axiom in Table \ref{AxiomsForqAbstractionG2} is sound modulo probabilistic rooted branching hp-bisimulation $\approx_{prbhp}$. We leave them as
exercises to the readers.

(4) Since probabilistic rooted branching hhp-bisimulation $\approx_{prbhhp}$ is both an equivalent and a congruent relation with respect to $qAPPTC_{G_{\tau}}$ with guarded linear
recursion, we only need to check if each axiom in Table \ref{AxiomsForqAbstractionG2} is sound modulo probabilistic rooted branching hhp-bisimulation $\approx_{prbhhp}$. We leave them as
exercises to the readers.
\end{proof}

Though $\tau$-loops are prohibited in guarded linear recursive specifications in a specifiable way, they can be constructed using the abstraction operator, for example, there exist
$\tau$-loops in the process term $\tau_{\{a\}}(\langle X|X=aX\rangle)$. To avoid $\tau$-loops caused by $\tau_I$ and ensure fairness, we introduce the following recursive verification
rules as Table \ref{RVR} shows, note that $i_1,\cdots, i_m,j_1,\cdots,j_n\in I\subseteq \mathbb{E}\setminus\{\tau\}$.

\begin{center}
\begin{table}
    $$VR_1\quad \frac{x=y+(i_1\leftmerge\cdots\leftmerge i_m)\cdot x, y=y+y}{\tau\cdot\tau_I(x)=\tau\cdot \tau_I(y)}$$
    $$VR_2\quad \frac{x=z\boxplus_{\pi}(u+(i_1\leftmerge\cdots\leftmerge i_m)\cdot x),z=z+u,z=z+z}{\tau\cdot\tau_I(x)=\tau\cdot\tau_I(z)}$$
    $$VR_3\quad \frac{x=z+(i_1\leftmerge\cdots\leftmerge i_m)\cdot y,y=z\boxplus_{\pi}(u+(j_1\leftmerge\cdots\leftmerge j_n)\cdot x), z=z+u,z=z+z}{\tau\cdot\tau_I(x)=\tau\cdot\tau_I(y')\textrm{ for }y'=z\boxplus_{\pi}(u+(i_1\leftmerge\cdots\leftmerge i_m)\cdot y')}$$
\caption{Recursive verification rules}
\label{RVR}
\end{table}
\end{center}

\begin{theorem}[Soundness of $VR_1,VR_2,VR_3$]
$VR_1$, $VR_2$ and $VR_3$ are sound modulo probabilistic rooted branching truly concurrent bisimulation equivalences $\approx_{prbp}$, $\approx_{prbs}$, $\approx_{prbhp}$ and $\approx_{prbhhp}$.
\end{theorem}

\subsection{Quantum Measurement}\label{qm}

In closed quantum systems, there is another basic quantum operation -- quantum measurement, besides the unitary operator. Quantum measurements have a probabilistic nature.

There is a concrete but non-trivial problem in modeling quantum measurement.

Let the following process term represent quantum measurement during modeling phase,

$$\beta_1\cdot t_1\boxplus_{\pi_1}\beta_2\cdot t_2\boxplus_{\pi_2}\cdots\boxplus_{\pi_{i-1}}\beta_i\cdot t_i$$

where $\sum_i \pi_i=1$, $t_i\in\mathcal{B}(qBAPTC)$, $\beta$ denotes a quantum measurement, and $\beta=\sum_i\lambda_i \beta_i$, $\beta_i$ denotes the projection performed on the quantum
system $\varrho$, $\pi_i=Tr(\beta_i\varrho)$, $\varrho_i=\beta_i\varrho \beta_i/\pi_i$.

The above term means that, firstly, we choose a projection $\beta_i$ in a quantum measurement $\beta=\sum_i\lambda_i\beta_i$ probabilistically, then, we execute (perform) the
projection $\beta_i$ on the closed quantum system. This also adheres to the intuition on quantum mechanics.

We define $B$ as the collection of all projections of all quantum measurements, and make the collection of atomic actions be $\mathbb{E}=\mathbb{E}\cup B$. We see that a
projection $\beta_i\in B$ has the almost same semantics as a unitary operator $\alpha\in A$. So, we add the following (probabilistic and action)
transition rules into those of $PQRA$.

$$\frac{}{\langle\beta_i,s,\varrho\rangle\rightsquigarrow\langle\breve{\beta_i},s,\varrho\rangle}$$

$$\frac{}{\langle\breve{\beta_i},s,\varrho\rangle\xrightarrow{\beta_i}\langle\surd,s,\varrho'\rangle}$$

Until now, $qAPPTC_G$ works again. The two main quantum operations in a closed quantum system -- the unitary operator and the quantum measurement, are fully modeled in probabilistic
process algebra.

\subsection{Quantum Entanglement}\label{qe2}

As in section \ref{qe1}, The axiom system of the shadow constant $\circledS$ is shown in Table \ref{AxiomsForQE2}.

\begin{center}
\begin{table}
  \begin{tabular}{@{}ll@{}}
\hline No. &Axiom\\
  $SC1$ & $\circledS\cdot x = x$ \\
  $SC2$ & $x\cdot\circledS = x$\\
  $SC3$ & $e\leftmerge\circledS^e=e$\\
  $SC4$ & $\circledS^e\leftmerge e=e$\\
  $SC5$ & $e\leftmerge(\circledS^e\cdot y) = e\cdot y$\\
  $SC6$ & $\circledS^e\leftmerge(e\cdot y) = e\cdot y$\\
  $SC7$ & $(e\cdot x)\leftmerge\circledS^e = e\cdot x$\\
  $SC8$ & $(\circledS^e\cdot x)\leftmerge e = e\cdot x$\\
  $SC9$ & $(e\cdot x)\leftmerge(\circledS^e\cdot y) = e\cdot (x\between y)$\\
  $SC10$ & $(\circledS^e\cdot x)\leftmerge(e\cdot y) = e\cdot (x\between y)$\\
\end{tabular}
\caption{Axioms of quantum entanglement}
\label{AxiomsForQE2}
\end{table}
\end{center}

The transition rules of constant $\circledS$ are as Table \ref{TRForENT2} shows.

\begin{center}
    \begin{table}
        $$\frac{}{\langle\circledS,s,\varrho\rangle\rightsquigarrow\langle\breve{\circledS},s,\varrho\rangle}$$
        $$\frac{}{\langle\breve{\circledS},s,\varrho\rangle\rightarrow\langle\surd,s,\varrho\rangle}$$
        $$\frac{\langle x, s,\varrho\rangle\xrightarrow{e}\langle x',s,\varrho'\rangle\quad \langle y, s,\varrho'\rangle\xrightarrow{\circledS^e}\langle y',s,\varrho'\rangle}{\langle x\leftmerge y,s,\varrho\rangle\xrightarrow{e}\langle x'\between y', s,\varrho'\rangle}$$
        $$\frac{\langle x, s,\varrho\rangle\xrightarrow{e}\langle\surd,s,\varrho'\rangle\quad \langle y, s,\varrho'\rangle\xrightarrow{\circledS^e}\langle y',s,\varrho'\rangle}{\langle x\leftmerge y,s,\varrho\rangle\xrightarrow{e}\langle y', s,\varrho'\rangle}$$
        $$\frac{\langle x, s,\varrho'\rangle\xrightarrow{\circledS^e}\langle\surd,s,\varrho'\rangle\quad \langle y, s,\varrho\rangle\xrightarrow{e}\langle y',s,\varrho'\rangle}{\langle x\leftmerge y,s,\varrho\rangle\xrightarrow{e}\langle y', s,\varrho'\rangle}$$
        $$\frac{\langle x, s,\varrho\rangle\xrightarrow{e}\langle\surd,s,\varrho'\rangle\quad \langle y, s,\varrho'\rangle\xrightarrow{\circledS^e}\langle\surd,s,\varrho'\rangle}{\langle x\leftmerge y,s,\varrho\rangle\xrightarrow{e}\langle \surd, s,\varrho'\rangle}$$
        \caption{Transition rules of constant $\circledS$}
        \label{TRForENT2}
    \end{table}
\end{center}

\begin{theorem}[Elimination theorem of $qAPPTC_{G_{\tau}}$ with guarded linear recursion and shadow constant]
Let $p$ be a closed $qAPPTC_{G_{\tau}}$ with guarded linear recursion and shadow constant term. Then there is a closed $qAPPTC$ term such that $qAPPTC_{G_{\tau}}$ with guarded linear 
recursion and shadow constant$\vdash p=q$.
\end{theorem}

\begin{proof}
We leave the proof to the readers as an excise.
\end{proof}

\begin{theorem}[Conservitivity of $qAPPTC_{G_{\tau}}$ with guarded linear recursion and shadow constant]
$qAPPTC_{G_{\tau}}$ with guarded linear recursion and shadow constant is a conservative extension of $qAPPTC_{G_{\tau}}$ with guarded linear recursion.
\end{theorem}

\begin{proof}
We leave the proof to the readers as an excise.
\end{proof}

\begin{theorem}[Congruence theorem of $qAPPTC_{G_{\tau}}$ with guarded linear recursion and shadow constant]
Probabilistic rooted branching truly concurrent bisimulation equivalences $\approx_{prbp}$, $\approx_{prbs}$, $\approx_{prbhp}$ and $\approx_{prbhhp}$ are all congruences with respect
to $qAPPTC_{G_{\tau}}$ with guarded linear recursion and shadow constant.
\end{theorem}

\begin{proof}
We leave the proof to the readers as an excise.
\end{proof}

\begin{theorem}[Soundness of $qAPPTC_{G_{\tau}}$ with guarded linear recursion and shadow constant]
Let $p$ and $q$ be closed $qAPPTC_{G_{\tau}}$ with guarded linear recursion and shadow constant terms. If $qAPPTC_{G_{\tau}}$ with guarded linear recursion and shadow constant$\vdash x=y$, then

\begin{enumerate}
  \item $x\approx_{rbs} y$;
  \item $x\approx_{rbp} y$;
  \item $x\approx_{rbhp} y$;
  \item $x\approx_{rbhhp} y$.
\end{enumerate}
\end{theorem}

\begin{proof}
We leave the proof to the readers as an excise.
\end{proof}

\begin{theorem}[Completeness of $qAPPTC_{G_{\tau}}$ with guarded linear recursion and shadow constant]
Let $p$ and $q$ are closed $qAPPTC_{G_{\tau}}$ with guarded linear recursion and shadow constant terms, then,

\begin{enumerate}
  \item if $p\approx_{rbs} q$ then $p=q$;
  \item if $p\approx_{rbp} q$ then $p=q$;
  \item if $p\approx_{rbhp} q$ then $p=q$;
  \item if $p\approx_{rbhhp} q$ then $p=q$.
\end{enumerate}
\end{theorem}

\begin{proof}
We leave the proof to the readers as an excise.
\end{proof}

\subsection{Unification of Quantum and Classical Computing for Closed Quantum Systems}\label{uni2}

We give the transition rules under quantum configuration for traditional atomic actions (events) $e'\in\mathbb{E}$ as Table \ref{TRForBPA5} shows.

\begin{center}
    \begin{table}
        $$\frac{}{\langle\epsilon,s,\varrho\rangle\rightsquigarrow\langle\breve{\epsilon},s,\varrho\rangle}$$
        $$\frac{}{\langle e',s,\varrho\rangle\rightsquigarrow\langle\breve{e},s,\varrho\rangle}$$
        $$\frac{}{\langle\phi,s,\varrho\rangle\rightsquigarrow\langle\breve{\phi},s,\varrho\rangle}$$
        $$\frac{\langle x,s,\varrho\rangle\rightsquigarrow \langle x',s,\varrho\rangle}{\langle x\cdot y,s,\varrho\rangle\rightsquigarrow \langle x'\cdot y,s,\varrho\rangle}$$
        $$\frac{\langle x,s,\varrho\rangle\rightsquigarrow \langle x',s,\varrho\rangle\quad \langle y,s,\varrho\rangle\rightsquigarrow \langle y',s,\varrho\rangle}{\langle x+y,s,\varrho\rangle\rightsquigarrow \langle x'+y',s,\varrho\rangle}$$
        $$\frac{\langle x,s,\varrho\rangle\rightsquigarrow \langle x',s,\varrho\rangle}{\langle x\boxplus_{\pi}y,s,\varrho\rangle\rightsquigarrow \langle x',s,\varrho\rangle}\quad \frac{\langle y,s,\varrho\rangle\rightsquigarrow \langle y',s,\varrho\rangle}{\langle x\boxplus_{\pi}y,s,\varrho\rangle\rightsquigarrow \langle y',s,\varrho\rangle}$$

        $$\frac{}{\langle\breve{\epsilon},s,\varrho\rangle\rightarrow\langle\surd,s,\varrho\rangle}$$
        $$\frac{}{\langle \breve{e'},s,\varrho\rangle\xrightarrow{e'}\langle\surd,s',\varrho\rangle}\textrm{ if }s'\in effect(e,s)$$
        $$\frac{}{\langle\breve{\phi},s,\varrho\rangle\rightarrow\langle\surd,s,\varrho\rangle}\textrm{ if }test(\phi,s)$$
        $$\frac{\langle x,s,\varrho\rangle\xrightarrow{e'}\langle\surd,s',\varrho\rangle}{\langle x+ y,s,\varrho\rangle\xrightarrow{e'}\langle\surd,s',\varrho\rangle} \quad\frac{\langle x,s,\varrho\rangle\xrightarrow{e'}\langle x',s',\varrho\rangle}{\langle x+ y,s,\varrho\rangle\xrightarrow{e'}\langle x',s',\varrho\rangle}$$
        $$\frac{\langle y,s,\varrho\rangle\xrightarrow{e'}\langle\surd,s',\varrho\rangle}{\langle x+ y,s,\varrho\rangle\xrightarrow{e'}\langle\surd,s',\varrho\rangle} \quad\frac{\langle y,s,\varrho\rangle\xrightarrow{e'}\langle y',s',\varrho\rangle}{\langle x+ y,s,\varrho\rangle\xrightarrow{e'}\langle y',s',\varrho\rangle}$$
        $$\frac{\langle x,s,\varrho\rangle\xrightarrow{e'}\langle\surd,s',\varrho\rangle}{\langle x\cdot y,s,\varrho\rangle\xrightarrow{e'} \langle y,s',\varrho\rangle} \quad\frac{\langle x,s,\varrho\rangle\xrightarrow{e'}\langle x',s',\varrho\rangle}{\langle x\cdot y,s,\varrho\rangle\xrightarrow{e'}\langle x'\cdot y,s',\varrho\rangle}$$
        \caption{Transition rules of BAPTC under quantum configuration}
        \label{TRForBPA5}
    \end{table}
\end{center}

And the axioms for traditional actions are the same as those of $qBAPTC_G$. And it is natural can be extended to $qAPPTC_G$, recursion and abstraction. So, quantum and classical computing
are unified under the framework of $qAPPTC_G$ for closed quantum systems.

\newpage\section{Applications of $qAPPTC_G$}\label{aqapptcg}

Quantum and classical computing in closed systems are unified with $qAPPTC_G$, which have the same equational logic and the same quantum configuration based operational semantics.
The unification can be used widely in verification for the behaviors of quantum and classical computing mixed systems. In this chapter, we show its usage in verification of the
quantum communication protocols.

\subsection{Verification of Quantum Teleportation Protocol}\label{VQT6}

\begin{figure}
  \centering
  \includegraphics{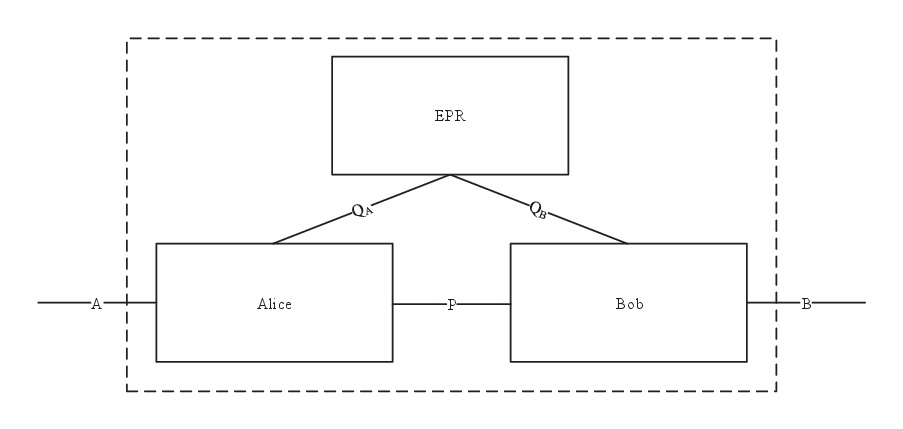}
  \caption{Quantum teleportation protocol.}
  \label{QT}
\end{figure}

Quantum teleportation \cite{QT} is a famous quantum protocol in quantum information theory to teleport an unknown quantum state by sending only classical information, provided that the sender and the receiver, Alice and Bob, shared an entangled state in advance. Firstly, we introduce the basic quantum teleportation protocol briefly, which is illustrated in Figure \ref{QT}. In this section, we show how to process quantum entanglement in an implicit way.

\begin{enumerate}
  \item EPR generates 2-qubits entangled EPR pair $q=q_1\otimes q_2$, and he sends $q_1$ to Alice through quantum channel $Q_A$ and $q_2$ to Bob through quantum channel $Q_B$;
  \item Alice receives $q_1$, after some preparations, she measures on $q_1$, and sends the measurement results $x$ to Bob through classical channel $P$;
  \item Bob receives $q_2$ from EPR, and also the classical information $x$ from Alice. According to $x$, he chooses specific Pauli transformation on $q_2$.
\end{enumerate}

We re-introduce the basic quantum teleportation protocol in an abstract way with more technical details as Figure \ref{QT} illustrates.

Now, we assume the generation of 2-qubits $q$ through two unitary operators $Set[q]$ and $H[q]$. EPR sends $q_1$ to Alice through the quantum channel $Q_A$ by quantum communicating action $send_{Q_A}(q_1)$ and Alice receives $q_1$ through $Q_A$ by quantum communicating action $receive_{Q_A}(q_1)$. Similarly, for Bob, those are $send_{Q_B}(q_2)$ and $receive_{Q_B}(q_2)$. After Alice receives $q_1$, she does some preparations, including a unitary transformation $CNOT$ and a Hadamard transformation $H$, then Alice do measurement $M=\sum^3_{i=0}M_i$, and sends measurement results $x$ to Bob through the public classical channel $P$ by classical communicating action $send_{P}(x)$, and Bob receives $x$ through channel $P$ by classical communicating action $receive_{P}(x)$. According to $x$, Bob performs specific Pauli transformations $\sigma_x$ on $q_2$. Let Alice, Bob and EPR be a system $ABE$ and let interactions between Alice, Bob and EPR be internal actions. $ABE$ receives external input $D_i$ through channel $A$ by communicating action $receive_A(D_i)$ and sends results $D_o$ through channel $B$ by communicating action $send_B(D_o)$. Note that the entangled EPR pair $q=q_1\otimes q_2$ is within $ABE$, so quantum entanglement can be processed implicitly.

Then the state transitions of EPR can be described by PQRA as follows.

\begin{eqnarray}
&&E=Set[q]\cdot E_1\nonumber\\
&&E_1=H[q]\cdot E_2\nonumber\\
&&E_2=send_{Q_A}(q_1)\cdot E_3\nonumber\\
&&E_3=send_{Q_B}(q_2)\cdot E\nonumber
\end{eqnarray}

And the state transitions of Alice can be described by PQRA as follows.

\begin{eqnarray}
&&A=\sum_{D_i\in \Delta_i}receive_A(D_i)\cdot A_1\nonumber\\
&&A_1=receive_{Q_A}(q_1)\cdot A_2\nonumber\\
&&A_2=CNOT\cdot A_3\nonumber\\
&&A_3=H\cdot A_4\nonumber\\
&&A_4=(M_0\cdot send_P(0)\boxplus_{\frac{1}{4}}M_1\cdot send_P(1)\boxplus_{\frac{1}{4}}M_2\cdot send_P(2)\boxplus_{\frac{1}{4}}M_3\cdot send_P(3))\cdot A\nonumber
\end{eqnarray}

where $\Delta_i$ is the collection of the input data.

And the state transitions of Bob can be described by PQRA as follows.

\begin{eqnarray}
&&B=receive_{Q_B}(q_2)\cdot B_1\nonumber\\
&&B_1=(receive_P(0)\cdot\sigma_0\boxplus_{\frac{1}{4}}receive_P(1)\cdot\sigma_1\boxplus_{\frac{1}{4}}receive_P(2) \cdot\sigma_2\boxplus_{\frac{1}{4}}receive_P(3)\cdot\sigma_3)\cdot B_2\nonumber\\
&&B_2=\sum_{D_o\in\Delta_o}send_B(D_o)\cdot B\nonumber
\end{eqnarray}

where $\Delta_o$ is the collection of the output data.

The send action and receive action of the same data through the same channel can communicate each other, otherwise, a deadlock $\delta$ will be caused. We define the following communication functions.

\begin{eqnarray}
&&\gamma(send_{Q_A}(q_1),receive_{Q_A}(q_1))\triangleq c_{Q_A}(q_1)\nonumber\\
&&\gamma(send_{Q_B}(q_2),receive_{Q_B}(q_2))\triangleq c_{Q_B}(q_2)\nonumber\\
&&\gamma(send_P(0),receive_P(0))\triangleq c_P(0)\nonumber\\
&&\gamma(send_P(1),receive_P(1))\triangleq c_P(1)\nonumber\\
&&\gamma(send_P(2),receive_P(2))\triangleq c_P(2)\nonumber\\
&&\gamma(send_P(3),receive_P(3))\triangleq c_P(3)\nonumber
\end{eqnarray}

Let $A$, $B$ and $E$ in parallel, then the system $ABE$ can be represented by the following process term.

$$\tau_I(\partial_H(\Theta(A\between B\between E)))$$

where $H=\{send_{Q_A}(q_1), receive_{Q_A}(q_1), send_{Q_B}(q_2), receive_{Q_B}(q_2),\\
send_P(0), receive_P(0), send_P(1), receive_P(1),\\
send_P(2), receive_P(2), send_P(3), receive_P(3)\}$ and $I=\{Set[q], H[q], CNOT, H, M_0, M_1,\\ M_2, M_3, \sigma_0, \sigma_1, \sigma_2, \sigma_3, \\ c_{Q_A}(q_1), c_{Q_B}(q_2), c_P(0), c_P(1), c_P(2), c_P(3)\}$.

Then we get the following conclusion.

\begin{theorem}
The basic quantum teleportation protocol $\tau_I(\partial_H(\Theta(A\between B\between E)))$ exhibits desired external behaviors.
\end{theorem}

\begin{proof}
We can get $\tau_I(\partial_H(\Theta(A\between B\between E)))=\sum_{D_i\in \Delta_i}\sum_{D_o\in\Delta_o}receive_A(D_i)\leftmerge send_B(D_o)\leftmerge 
\tau_I(\partial_H(\Theta(A\between B\between E)))$. So, the basic quantum teleportation protocol $\tau_I(\partial_H(\Theta(A\between B\between E)))$ exhibits desired external behaviors.
\end{proof}

\subsection{Verification of BB84 Protocol}\label{VBB86}

\begin{figure}
  \centering
  \includegraphics{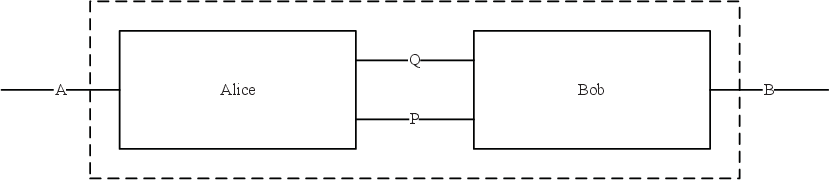}
  \caption{BB84 protocol.}
  \label{BB84}
\end{figure}

The BB84 protocol \cite{BB84} is used to create a private key between two parities, Alice and Bob. Firstly, we introduce the basic BB84 protocol briefly, which is illustrated in Figure \ref{BB84}.

\begin{enumerate}
  \item Alice create two string of bits with size $n$ randomly, denoted as $B_a$ and $K_a$;
  \item Alice generates a string of qubits $q$ with size $n$, and the $i$th qubit in $q$ is $|x_y\rangle$, where $x$ is the $i$th bit of $B_a$ and $y$ is the $i$th bit of $K_a$;
  \item Alice sends $q$ to Bob through a quantum channel $Q$ between Alice and Bob;
  \item Bob receives $q$ and randomly generates a string of bits $B_b$ with size $n$;
  \item Bob measures each qubit of $q$ according to a basis by bits of $B_b$. And the measurement results would be $K_b$, which is also with size $n$;
  \item Bob sends his measurement bases $B_b$ to Alice through a public channel $P$;
  \item Once receiving $B_b$, Alice sends her bases $B_a$ to Bob through channel $P$, and Bob receives $B_a$;
  \item Alice and Bob determine that at which position the bit strings $B_a$ and $B_b$ are equal, and they discard the mismatched bits of $B_a$ and $B_b$. Then the remaining bits of $K_a$ and $K_b$, denoted as $K_a'$ and $K_b'$ with $K_{a,b}=K_a'=K_b'$.
\end{enumerate}

We re-introduce the basic BB84 protocol in an abstract way with more technical details as Figure \ref{BB84} illustrates.

Now, we assume a special measurement operation $Rand[q;B_a]=\sum^{2n-1}_{i=0}Rand[q;B_a]_i$ which create a string of $n$ random bits $B_a$ from the $q$ quantum system, and the same as $Rand[q;K_a]=\sum^{2n-1}_{i=0}Rand[q;K_a]_i$, $Rand[q';B_b]=\sum^{2n-1}_{i=0}Rand[q';B_b]_i$. $M[q;K_b]=\sum^{2n-1}_{i=0}M[q;K_b]_i$ denotes the Bob's measurement on $q$. The generation of $n$ qubits $q$ through two unitary operators $Set_{K_a}[q]$ and $H_{B_a}[q]$. Alice sends $q$ to Bob through the quantum channel $Q$ by quantum communicating action $send_{Q}(q)$ and Bob receives $q$ through $Q$ by quantum communicating action $receive_{Q}(q)$. Bob sends $B_b$ to Alice through the public classical channel $P$ by classical communicating action $send_{P}(B_b)$ and Alice receives $B_b$ through channel $P$ by classical communicating action $receive_{P}(B_b)$, and the same as $send_{P}(B_a)$ and $receive_{P}(B_a)$. Alice and Bob generate the private key $K_{a,b}$ by a classical comparison action $cmp(K_{a,b},K_a,K_b,B_a,B_b)$. Let Alice and Bob be a system $AB$ and let interactions between Alice and Bob be internal actions. $AB$ receives external input $D_i$ through channel $A$ by communicating action $receive_A(D_i)$ and sends results $D_o$ through channel $B$ by communicating action $send_B(D_o)$.

Then the state transitions of Alice can be described by PQRA as follows.

\begin{eqnarray}
&&A=\sum_{D_i\in \Delta_i}receive_A(D_i)\cdot A_1\nonumber\\
&&A_1=\boxplus_{\frac{1}{2n},i=0}^{2n-1}Rand[q;B_a]_i\cdot A_2\nonumber\\
&&A_2=\boxplus_{\frac{1}{2n},i=0}^{2n-1}Rand[q;K_a]_i\cdot A_3\nonumber\\
&&A_3=Set_{K_a}[q]\cdot A_4\nonumber\\
&&A_4=H_{B_a}[q]\cdot A_5\nonumber\\
&&A_5=send_Q(q)\cdot A_6\nonumber\\
&&A_6=receive_P(B_b)\cdot A_7\nonumber\\
&&A_7=send_P(B_a)\cdot A_8\nonumber\\
&&A_8=cmp(K_{a,b},K_a,K_b,B_a,B_b)\cdot A_9\nonumber\\
&&A_9=\{B_{a_i}=B_{b_i}\}\cdot generate(K_a)\cdot A+\{B_{a_i}\neq B_{b_i}\}\cdot discard\cdot A\nonumber
\end{eqnarray}

where $\Delta_i$ is the collection of the input data.

And the state transitions of Bob can be described by PQRA as follows.

\begin{eqnarray}
&&B=receive_Q(q)\cdot B_1\nonumber\\
&&B_1=\boxplus_{\frac{1}{2n},i=0}^{2n-1}Rand[q';B_b]_i\cdot B_2\nonumber\\
&&B_2=\boxplus_{\frac{1}{2n},i=0}^{2n-1}M[q;K_b]_i\cdot B_3\nonumber\\
&&B_3=send_P(B_b)\cdot B_4\nonumber\\
&&B_4=receive_P(B_a)\cdot B_5\nonumber\\
&&B_5=cmp(K_{a,b},K_a,K_b,B_a,B_b)\cdot B_6\nonumber\\
&&B_6=\{B_{a_i}=B_{b_i}\}\cdot generate(K_b)\cdot B_7+\{B_{a_i}\neq B_{b_i}\}\cdot discard\cdot B_7\nonumber\\
&&B_7=\sum_{D_o\in\Delta_o}send_B(D_o)\cdot B\nonumber
\end{eqnarray}

where $\Delta_o$ is the collection of the output data.

The send action and receive action of the same data through the same channel can communicate each other, otherwise, a deadlock $\delta$ will be caused. We define the following communication functions.

\begin{eqnarray}
&&\gamma(send_Q(q),receive_Q(q))\triangleq c_Q(q)\nonumber\\
&&\gamma(send_P(B_b),receive_P(B_b))\triangleq c_P(B_b)\nonumber\\
&&\gamma(send_P(B_a),receive_P(B_a))\triangleq c_P(B_a)\nonumber
\end{eqnarray}

Let $A$ and $B$ in parallel, then the system $AB$ can be represented by the following process term.

$$\tau_I(\partial_H(\Theta(A\between B)))$$

where $H=\{send_Q(q),receive_Q(q),send_P(B_b),receive_P(B_b),send_P(B_a),receive_P(B_a)\}$ and 
$I=\{Rand[q;B_a]_i, Rand[q;K_a]_i, Set_{K_a}[q], H_{B_a}[q], Rand[q';B_b]_i, M[q;K_b]_i, \\c_Q(q), c_P(B_b), c_P(B_a), cmp(K_{a,b},K_a,K_b,B_a,B_b),\{B_{a_i}=B_{b_i}\},\{B_{a_i}\neq B_{b_i}\},\\
generate(K_a),generate(K_b),discard\}$.

Then we get the following conclusion.

\begin{theorem}
The basic BB84 protocol $\tau_I(\partial_H(\Theta(A\between B)))$ exhibits desired external behaviors.
\end{theorem}

\begin{proof}
We can get $\tau_I(\partial_H(\Theta(A\between B)))=\sum_{D_i\in \Delta_i}\sum_{D_o\in\Delta_o}receive_A(D_i)\leftmerge send_B(D_o)\leftmerge \tau_I(\partial_H(\Theta(A\between B)))$.
So, the basic BB84 protocol $\tau_I(\partial_H(\Theta(A\between B)))$ exhibits desired external behaviors.
\end{proof}

\subsection{Verification of E91 Protocol}\label{VE916}

\begin{figure}
  \centering
  \includegraphics{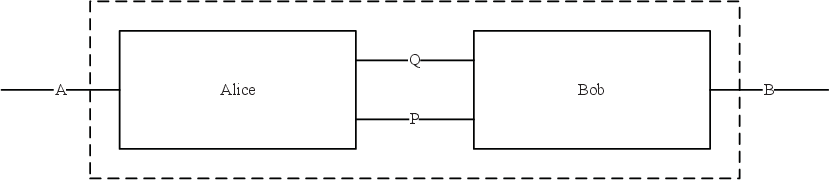}
  \caption{E91 protocol.}
  \label{E91}
\end{figure}

With support of Entanglement merge $\between$, PQRA can be used to verify quantum protocols utilizing entanglement explicitly. E91 protocol\cite{E91} is the first quantum protocol which utilizes entanglement. E91 protocol is used to create a private key between two parities, Alice and Bob. Firstly, we introduce the basic E91 protocol briefly, which is illustrated in Figure \ref{E91}.

\begin{enumerate}
  \item Alice generates a string of EPR pairs $q$ with size $n$, i.e., $2n$ particles, and sends a string of qubits $q_b$ from each EPR pair with $n$ to Bob through a quantum channel $Q$, remains the other string of qubits $q_a$ from each pair with size $n$;
  \item Alice create two string of bits with size $n$ randomly, denoted as $B_a$ and $K_a$;
  \item Bob receives $q_b$ and randomly generates a string of bits $B_b$ with size $n$;
  \item Alice measures each qubit of $q_a$ according to a basis by bits of $B_a$. And the measurement results would be $K_a$, which is also with size $n$;
  \item Bob measures each qubit of $q_b$ according to a basis by bits of $B_b$. And the measurement results would be $K_b$, which is also with size $n$;
  \item Bob sends his measurement bases $B_b$ to Alice through a public channel $P$;
  \item Once receiving $B_b$, Alice sends her bases $B_a$ to Bob through channel $P$, and Bob receives $B_a$;
  \item Alice and Bob determine that at which position the bit strings $B_a$ and $B_b$ are equal, and they discard the mismatched bits of $B_a$ and $B_b$. Then the remaining bits of $K_a$ and $K_b$, denoted as $K_a'$ and $K_b'$ with $K_{a,b}=K_a'=K_b'$.
\end{enumerate}

We re-introduce the basic E91 protocol in an abstract way with more technical details as Figure \ref{E91} illustrates.

Now, $M[q_a;K_a]=\sum_{i=0}^{2n-1}M[q_a;K_a]_i$ denotes the Alice's measurement operation of $q_a$, and $\circledS_{M[q_a;K_a]}=\sum_{i=0}^{2n-1}\circledS_{M[q_a;K_a]_i}$ denotes the responding shadow constant; $M[q_b;K_b]=\sum_{i=0}^{2n-1}M[q_b;K_b]_i$ denotes the Bob's measurement operation of $q_b$, and $\circledS_{M[q_b;K_b]}=\sum_{i=0}^{2n-1}\circledS_{M[q_b;K_b]_i}$ denotes the responding shadow constant. Alice sends $q_b$ to Bob through the quantum channel $Q$ by quantum communicating action $send_{Q}(q_b)$ and Bob receives $q_b$ through $Q$ by quantum communicating action $receive_{Q}(q_b)$. Bob sends $B_b$ to Alice through the public channel $P$ by classical communicating action $send_{P}(B_b)$ and Alice receives $B_b$ through channel $P$ by classical communicating action $receive_{P}(B_b)$, and the same as $send_{P}(B_a)$ and $receive_{P}(B_a)$. Alice and Bob generate the private key $K_{a,b}$ by a classical comparison action $cmp(K_{a,b},K_a,K_b,B_a,B_b)$. Let Alice and Bob be a system $AB$ and let interactions between Alice and Bob be internal actions. $AB$ receives external input $D_i$ through channel $A$ by communicating action $receive_A(D_i)$ and sends results $D_o$ through channel $B$ by communicating action $send_B(D_o)$.

Then the state transitions of Alice can be described by PQRA as follows.

\begin{eqnarray}
&&A=\sum_{D_i\in \Delta_i}receive_A(D_i)\cdot A_1\nonumber\\
&&A_1=send_Q(q_b)\cdot A_2\nonumber\\
&&A_2=\boxplus_{\frac{1}{2n},i=0}^{2n-1}M[q_a;K_a]_i\cdot A_3\nonumber\\
&&A_3=\boxplus_{\frac{1}{2n},i=0}^{2n-1}\circledS_{M[q_b;K_b]_i}\cdot A_4\nonumber\\
&&A_4=receive_P(B_b)\cdot A_5\nonumber\\
&&A_5=send_P(B_a)\cdot A_6\nonumber\\
&&A_6=cmp(K_{a,b},K_a,K_b,B_a,B_b)\cdot A_7\nonumber\\
&&A_7=\{B_{a_i}=B_{b_i}\}\cdot generate(K_a)\cdot A+\{B_{a_i}\neq B_{b_i}\}\cdot discard\cdot A\nonumber
\end{eqnarray}

where $\Delta_i$ is the collection of the input data.

And the state transitions of Bob can be described by PQRA as follows.

\begin{eqnarray}
&&B=receive_Q(q_b)\cdot B_1\nonumber\\
&&B_1=\boxplus_{\frac{1}{2n},i=0}^{2n-1}\circledS_{M[q_a;K_a]_i}\cdot B_2\nonumber\\
&&B_2=\boxplus_{\frac{1}{2n},i=0}^{2n-1}M[q_b;K_b]_i\cdot B_3\nonumber\\
&&B_3=send_P(B_b)\cdot B_4\nonumber\\
&&B_4=receive_P(B_a)\cdot B_5\nonumber\\
&&B_5=cmp(K_{a,b},K_a,K_b,B_a,B_b)\cdot B_6\nonumber\\
&&B_6=\{B_{a_i}=B_{b_i}\}\cdot generate(K_b)\cdot B_7+\{B_{a_i}\neq B_{b_i}\}\cdot discard\cdot B_7\nonumber\\
&&B_7=\sum_{D_o\in\Delta_o}send_B(D_o)\cdot B\nonumber
\end{eqnarray}

where $\Delta_o$ is the collection of the output data.

The send action and receive action of the same data through the same channel can communicate each other, otherwise, a deadlock $\delta$ will be caused. The quantum operation and its shadow constant pair will lead entanglement occur, otherwise, a deadlock $\delta$ will occur. We define the following communication functions.

\begin{eqnarray}
&&\gamma(send_Q(q_b),receive_Q(q_b))\triangleq c_Q(q_b)\nonumber\\
&&\gamma(send_P(B_b),receive_P(B_b))\triangleq c_P(B_b)\nonumber\\
&&\gamma(send_P(B_a),receive_P(B_a))\triangleq c_P(B_a)\nonumber
\end{eqnarray}

Let $A$ and $B$ in parallel, then the system $AB$ can be represented by the following process term.

$$\tau_I(\partial_H(\Theta(A\between B)))$$

where $H=\{send_Q(q_b),receive_Q(q_b),send_P(B_b),receive_P(B_b),send_P(B_a),receive_P(B_a),\\ M[q_a;K_a]_i, \circledS_{M[q_a;K_a]_i}, M[q_b;K_b]_i, \circledS_{M[q_b;K_b]_i}\}$ and 
$I=\{c_Q(q_b), c_P(B_b), c_P(B_a), M[q_a;K_a], M[q_b;K_b],\\ cmp(K_{a,b},K_a,K_b,B_a,B_b),\{B_{a_i}=B_{b_i}\},\{B_{a_i}\neq B_{b_i}\},\\
generate(K_a),generate(K_b),discard\}$.

Then we get the following conclusion.

\begin{theorem}
The basic E91 protocol $\tau_I(\partial_H(\Theta(A\between B)))$ exhibits desired external behaviors.
\end{theorem}

\begin{proof}
We can get $\tau_I(\partial_H(\Theta(A\between B)))=\sum_{D_i\in \Delta_i}\sum_{D_o\in\Delta_o}receive_A(D_i)\leftmerge send_B(D_o)\leftmerge \tau_I(\partial_H(\Theta(A\between B)))$.
So, the basic E91 protocol $\tau_I(\partial_H(\Theta(A\between B)))$ exhibits desired external behaviors.
\end{proof}

\subsection{Verification of B92 Protocol}\label{VB926}

The famous B92 protocol\cite{B92} is a quantum key distribution protocol, in which quantum information and classical information are mixed. We take an example of the B92 protocol to illustrate the usage of probabilistic quantum process algebra in verification of quantum protocols.

\begin{figure}
  \centering
  \includegraphics{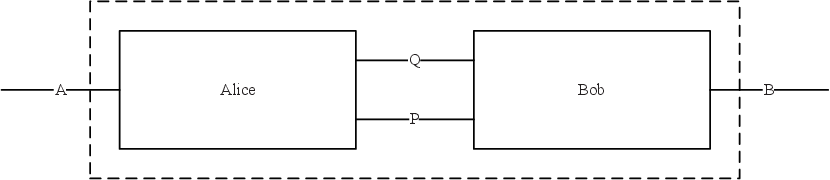}
  \caption{The B92 protocol.}
  \label{B92}
\end{figure}

The B92 protocol is used to create a private key between two parities, Alice and Bob. B92 is a protocol of quantum key distribution (QKD) which uses polarized photons as information carriers. Firstly, we introduce the basic B92 protocol briefly, which is illustrated in Figure \ref{B92}.

\begin{enumerate}
  \item Alice create a string of bits with size $n$ randomly, denoted as $A$.
  \item Alice generates a string of qubits $q$ with size $n$, carried by polarized photons. If $A_i=0$, the ith qubit is $|0\rangle$; else if $A_i=1$, the ith qubit is $|+\rangle$.
  \item Alice sends $q$ to Bob through a quantum channel $Q$ between Alice and Bob.
  \item Bob receives $q$ and randomly generates a string of bits $B$ with size $n$.
  \item If $B_i=0$, Bob chooses the basis $\oplus$; else if $B_i=1$, Bob chooses the basis $\otimes$. Bob measures each qubit of $q$ according to the above basses. And Bob builds a String of bits $T$, if the measurement produces $|0\rangle$ or $|+\rangle$, then $T_i=0$; else if the measurement produces $|1\rangle$ or $|-\rangle$, then $T_i=1$, which is also with size $n$.
  \item Bob sends $T$ to Alice through a public channel $P$.
  \item Alice and Bob determine that at which position the bit strings $A$ and $B$ are remained for which $T_i=1$. In absence of Eve, $A_i=1-B_i$, a shared raw key $K_{a,b}$ is formed by $A_i$.
\end{enumerate}

We re-introduce the basic B92 protocol in an abstract way with more technical details as Figure \ref{B92} illustrates.

Now, we assume a special measurement operation $Rand[q;A]=\sum^{2n-1}_{i=0}Rand[q;A]_i$ which create a string of $n$ random bits $A$ from the $q$ quantum system, and the same as $Rand[q';B]=\sum^{2n-1}_{i=0}Rand[q';B]_i$. $M[q;T]=\sum^{2n-1}_{i=0}M[q;T]_i$ denotes the Bob's measurement operation of $q$. The generation of $n$ qubits $q$ through a unitary operator $Set_{A}[q]$. Alice sends $q$ to Bob through the quantum channel $Q$ by quantum communicating action $send_{Q}(q)$ and Bob receives $q$ through $Q$ by quantum communicating action $receive_{Q}(q)$. Bob sends $T$ to Alice through the public channel $P$ by classical communicating action $send_{P}(T)$ and Alice receives $T$ through channel $P$ by classical communicating action $receive_{P}(T)$. Alice and Bob generate the private key $K_{a,b}$ by a classical comparison action $cmp(K_{a,b},T,A,B)$. Let Alice and Bob be a system $AB$ and let interactions between Alice and Bob be internal actions. $AB$ receives external input $D_i$ through channel $A$ by communicating action $receive_A(D_i)$ and sends results $D_o$ through channel $B$ by communicating action $send_B(D_o)$.

Then the state transition of Alice can be described by probabilistic quantum process algebra as follows.

\begin{eqnarray}
&&A=\sum_{D_i\in \Delta_i}receive_A(D_i)\cdot A_1\nonumber\\
&&A_1=\boxplus_{\frac{1}{2n},i=0}^{2n-1}Rand[q;A]_i\cdot A_2\nonumber\\
&&A_2=Set_{A}[q]\cdot A_3\nonumber\\
&&A_3=send_Q(q)\cdot A_4\nonumber\\
&&A_4=receive_P(T)\cdot A_5\nonumber\\
&&A_5=cmp(K_{a,b},T,A,B)\cdot A_6\nonumber\\
&&A_6=\{A_{i}=B_{i}\}\cdot generate(K_a)\cdot A+\{A_{i}\neq B_{i}\}\cdot discard\cdot A\nonumber
\end{eqnarray}

where $\Delta_i$ is the collection of the input data.

And the state transition of Bob can be described by probabilistic quantum process algebra as follows.

\begin{eqnarray}
&&B=receive_Q(q)\cdot B_1\nonumber\\
&&B_1=\boxplus_{\frac{1}{2n},i=0}^{2n-1}Rand[q';B]_i\cdot B_2\nonumber\\
&&B_2=\boxplus_{\frac{1}{2n},i=0}^{2n-1}M[q;T]_i\cdot B_3\nonumber\\
&&B_3=send_P(T)\cdot B_4\nonumber\\
&&B_4=cmp(K_{a,b},T,A,B)\cdot B_5\nonumber\\
&&B_5=\{A_{i}=B_{b_i}\}\cdot generate(K_b)\cdot B_6+\{A_{i}\neq B_{i}\}\cdot discard\cdot B_6\nonumber\\
&&B_6=\sum_{D_o\in\Delta_o}send_B(D_o)\cdot B\nonumber
\end{eqnarray}

where $\Delta_o$ is the collection of the output data.

The send action and receive action of the same data through the same channel can communicate each other, otherwise, a deadlock $\delta$ will be caused. We define the following communication functions.

\begin{eqnarray}
&&\gamma(send_Q(q),receive_Q(q))\triangleq c_Q(q)\nonumber\\
&&\gamma(send_P(T),receive_P(T))\triangleq c_P(T)\nonumber\\
\end{eqnarray}

Let $A$ and $B$ in parallel, then the system $AB$ can be represented by the following process term.

$$\tau_I(\partial_H(\Theta(A\between B)))$$

where $H=\{send_Q(q),receive_Q(q),send_P(T),receive_P(T)\}$ and $I=\{\boxplus_{\frac{1}{2n},i=0}^{2n-1}Rand[q;A]_i, \\Set_{A}[q], \boxplus_{\frac{1}{2n},i=0}^{2n-1}Rand[q';B]_i, \boxplus_{\frac{1}{2n},i=0}^{2n-1}M[q;T]_i, c_Q(q), c_P(T), cmp(K_{a,b},T,A,B)\},\{A_{i}=B_{i}\},\{A_{i}\neq B_{i}\},\\
generate(K_a),generate(K_b),discard\}$.

Then we get the following conclusion.

\begin{theorem}
The basic B92 protocol $\tau_I(\partial_H(\Theta(A\between B)))$ exhibits desired external behaviors.
\end{theorem}

\begin{proof}
We can get $\tau_I(\partial_H(\Theta(A\between B)))=\sum_{D_i\in \Delta_i}\sum_{D_o\in\Delta_o}receive_A(D_i)\leftmerge send_B(D_o)\leftmerge \tau_I(\partial_H(\Theta(A\between B)))$.
So, the basic B92 protocol $\tau_I(\partial_H(\Theta(A\between B)))$ exhibits desired external behaviors.
\end{proof}

\subsection{Verification of DPS Protocol}\label{VDPS6}

The famous DPS protocol\cite{DPS} is a quantum key distribution protocol, in which quantum information and classical information are mixed. We take an example of the DPS protocol to illustrate the usage of probabilistic quantum process algebra in verification of quantum protocols.

\begin{figure}
  \centering
  \includegraphics{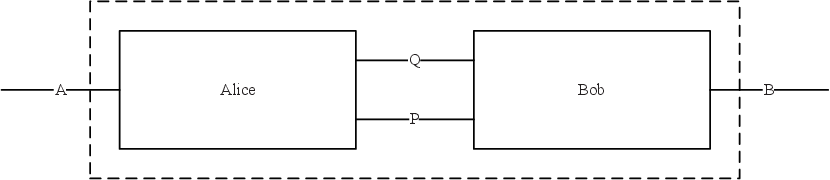}
  \caption{The DPS protocol.}
  \label{DPS}
\end{figure}

The DPS protocol is used to create a private key between two parities, Alice and Bob. DPS is a protocol of quantum key distribution (QKD) which uses pulses of a photon which has nonorthogonal four states. Firstly, we introduce the basic DPS protocol briefly, which is illustrated in Figure \ref{DPS}.

\begin{enumerate}
  \item Alice generates a string of qubits $q$ with size $n$, carried by a series of single photons possily at four time instances.
  \item Alice sends $q$ to Bob through a quantum channel $Q$ between Alice and Bob.
  \item Bob receives $q$ by detectors clicking at the second or third time instance, and records the time into $T$ with size $n$ and which detector clicks into $D$ with size $n$.
  \item Bob sends $T$ to Alice through a public channel $P$.
  \item Alice receives $T$. From $T$ and her modulation data, Alice knows which detector clicked in Bob's site, i.e. $D$.
  \item Alice and Bob have an identical bit string, provided that the first detector click represents "0" and the other detector represents "1", then a shared raw key $K_{a,b}$ is formed.
\end{enumerate}

We re-introduce the basic DPS protocol in an abstract way with more technical details as Figure \ref{DPS} illustrates.

Now, we assume $M[q;T]=\sum^{2n-1}_{i=0}M[q;T]_i$ denotes the Bob's measurement operation of $q$. The generation of $n$ qubits $q$ through a unitary operator $Set_{A}[q]$. Alice sends $q$ to Bob through the quantum channel $Q$ by quantum communicating action $send_{Q}(q)$ and Bob receives $q$ through $Q$ by quantum communicating action $receive_{Q}(q)$. Bob sends $T$ to Alice through the public channel $P$ by classical communicating action $send_{P}(T)$ and Alice receives $T$ through channel $P$ by classical communicating action $receive_{P}(T)$. Alice and Bob generate the private key $K_{a,b}$ by a classical comparison action $cmp(K_{a,b},D)$. Let Alice and Bob be a system $AB$ and let interactions between Alice and Bob be internal actions. $AB$ receives external input $D_i$ through channel $A$ by communicating action $receive_A(D_i)$ and sends results $D_o$ through channel $B$ by communicating action $send_B(D_o)$.

Then the state transition of Alice can be described by probabilistic quantum process algebra as follows.

\begin{eqnarray}
&&A=\sum_{D_i\in \Delta_i}receive_A(D_i)\cdot A_1\nonumber\\
&&A_1=Set_{A}[q]\cdot A_2\nonumber\\
&&A_2=send_Q(q)\cdot A_3\nonumber\\
&&A_3=receive_P(T)\cdot A_4\nonumber\\
&&A_4=cmp(K_{a,b},D)\cdot A\nonumber
\end{eqnarray}

where $\Delta_i$ is the collection of the input data.

And the state transition of Bob can be described by probabilistic quantum process algebra as follows.

\begin{eqnarray}
&&B=receive_Q(q)\cdot B_1\nonumber\\
&&B_1=\boxplus_{\frac{1}{2n},i=0}^{2n-1}M[q;T]_i\cdot B_2\nonumber\\
&&B_2=send_P(T)\cdot B_3\nonumber\\
&&B_3=cmp(K_{a,b},D)\cdot B_4\nonumber\\
&&B_4=\sum_{D_o\in\Delta_o}send_B(D_o)\cdot B\nonumber
\end{eqnarray}

where $\Delta_o$ is the collection of the output data.

The send action and receive action of the same data through the same channel can communicate each other, otherwise, a deadlock $\delta$ will be caused. We define the following communication functions.

\begin{eqnarray}
&&\gamma(send_Q(q),receive_Q(q))\triangleq c_Q(q)\nonumber\\
&&\gamma(send_P(T),receive_P(T))\triangleq c_P(T)\nonumber\\
\end{eqnarray}

Let $A$ and $B$ in parallel, then the system $AB$ can be represented by the following process term.

$$\tau_I(\partial_H(\Theta(A\between B)))$$

where $H=\{send_Q(q),receive_Q(q),send_P(T),receive_P(T)\}$ and $I=\{Set_{A}[q], \\ \boxplus_{\frac{1}{2n},i=0}^{2n-1}M[q;T]_i, c_Q(q), c_P(T), cmp(K_{a,b},D)\}$.

Then we get the following conclusion.

\begin{theorem}
The basic DPS protocol $\tau_I(\partial_H(\Theta(A\between B)))$ exhibits desired external behaviors.
\end{theorem}

\begin{proof}
We can get $\tau_I(\partial_H(\Theta(A\between B)))=\sum_{D_i\in \Delta_i}\sum_{D_o\in\Delta_o}receive_A(D_i)\leftmerge send_B(D_o)\leftmerge \tau_I(\partial_H(\Theta(A\between B)))$. 
So, the basic DPS protocol $\tau_I(\partial_H(\Theta(A\between B)))$ exhibits desired external behaviors.
\end{proof}

\subsection{Verification of BBM92 Protocol}\label{VBBM926}

The famous BBM92 protocol\cite{BBM92} is a quantum key distribution protocol, in which quantum information and classical information are mixed. We take an example of the BBM92 protocol to illustrate the usage of probabilistic quantum process algebra in verification of quantum protocols.

\begin{figure}
  \centering
  \includegraphics{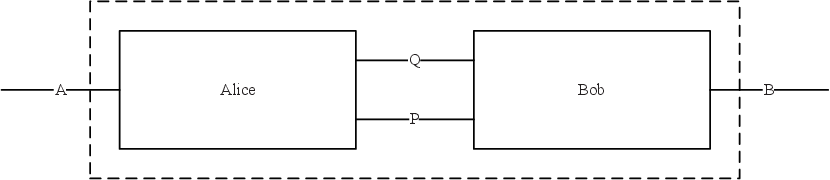}
  \caption{The BBM92 protocol.}
  \label{BBM92}
\end{figure}

The BBM92 protocol is used to create a private key between two parities, Alice and Bob. BBM92 is a protocol of quantum key distribution (QKD) which uses EPR pairs as information carriers. Firstly, we introduce the basic BBM92 protocol briefly, which is illustrated in Figure \ref{BBM92}.

\begin{enumerate}
  \item Alice generates a string of EPR pairs $q$ with size $n$, i.e., $2n$ particles, and sends a string of qubits $q_b$ from each EPR pair with $n$ to Bob through a quantum channel $Q$, remains the other string of qubits $q_a$ from each pair with size $n$.
  \item Alice create a string of bits with size $n$ randomly, denoted as $B_a$.
  \item Bob receives $q_b$ and randomly generates a string of bits $B_b$ with size $n$.
  \item Alice measures each qubit of $q_a$ according to bits of $B_a$, if $B_{a_i}=0$, then uses $x$ axis ($\rightarrow$); else if $B_{a_i}=1$, then uses $z$ axis ($\uparrow$).
  \item Bob measures each qubit of $q_b$ according to bits of $B_b$, if $B_{b_i}=0$, then uses $x$ axis ($\rightarrow$); else if $B_{b_i}=1$, then uses $z$ axis ($\uparrow$).
  \item Bob sends his measurement axis choices $B_b$ to Alice through a public channel $P$.
  \item Once receiving $B_b$, Alice sends her axis choices $B_a$ to Bob through channel $P$, and Bob receives $B_a$.
  \item Alice and Bob agree to discard all instances in which they happened to measure along different axes, as well as instances in which measurements fails because of imperfect quantum efficiency of the detectors. Then the remaining instances can be used to generate a private key $K_{a,b}$.
\end{enumerate}

We re-introduce the basic BBM92 protocol in an abstract way with more technical details as Figure \ref{BBM92} illustrates.

Now, $M[q_a;B_a]=\sum_{i=0}^{2n-1}M[q_a;K_a]_i$ denotes the Alice's measurement operation of $q_a$, and $\circledS_{M[q_a;B_a]}=\sum_{i=0}^{2n-1}\circledS_{M[q_a;B_a]_i}$ denotes the responding shadow constant; $M[q_b;B_b]=\sum_{i=0}^{2n-1}M[q_b;B_b]_i$ denotes the Bob's measurement operation of $q_b$, and $\circledS_{M[q_b;B_b]}=\sum_{i=0}^{2n-1}\circledS_{M[q_b;B_n]_i}$ denotes the responding shadow constant. Alice sends $q_b$ to Bob through the quantum channel $Q$ by quantum communicating action $send_{Q}(q_b)$ and Bob receives $q_b$ through $Q$ by quantum communicating action $receive_{Q}(q_b)$. Bob sends $B_b$ to Alice through the public channel $P$ by classical communicating action $send_{P}(B_b)$ and Alice receives $B_b$ through channel $P$ by classical communicating action $receive_{P}(B_b)$, and the same as $send_{P}(B_a)$ and $receive_{P}(B_a)$. Alice and Bob generate the private key $K_{a,b}$ by a classical comparison action $cmp(K_{a,b},B_a,B_b)$. Let Alice and Bob be a system $AB$ and let interactions between Alice and Bob be internal actions. $AB$ receives external input $D_i$ through channel $A$ by communicating action $receive_A(D_i)$ and sends results $D_o$ through channel $B$ by communicating action $send_B(D_o)$.

Then the state transition of Alice can be described by probabilistic quantum process algebra as follows.

\begin{eqnarray}
&&A=\sum_{D_i\in \Delta_i}receive_A(D_i)\cdot A_1\nonumber\\
&&A_1=send_Q(q_b)\cdot A_2\nonumber\\
&&A_2=\boxplus_{\frac{1}{2n},i=0}^{2n-1}M[q_a;B_a]_i\cdot A_3\nonumber\\
&&A_3=\boxplus_{\frac{1}{2n},i=0}^{2n-1}\circledS_{M[q_b;B_b]_i}\cdot A_4\nonumber\\
&&A_4=receive_P(B_b)\cdot A_5\nonumber\\
&&A_5=send_P(B_a)\cdot A_6\nonumber\\
&&A_6=cmp(K_{a,b},B_a,B_b)\cdot A_7\nonumber\\
&&A_7=\{B_{a_i}=B_{b_i}\}\cdot generate(K_a)\cdot A+\{B_{a_i}\neq B_{b_i}\}\cdot discard\cdot A\nonumber
\end{eqnarray}

where $\Delta_i$ is the collection of the input data.

And the state transition of Bob can be described by probabilistic quantum process algebra as follows.

\begin{eqnarray}
&&B=receive_Q(q_b)\cdot B_1\nonumber\\
&&B_1=\boxplus_{\frac{1}{2n},i=0}^{2n-1}\circledS_{M[q_a;B_a]_i}\cdot B_2\nonumber\\
&&B_2=\boxplus_{\frac{1}{2n},i=0}^{2n-1}M[q_b;B_b]_i\cdot B_3\nonumber\\
&&B_3=send_P(B_b)\cdot B_4\nonumber\\
&&B_4=receive_P(B_a)\cdot B_5\nonumber\\
&&B_5=cmp(K_{a,b},B_a,B_b)\cdot B_6\nonumber\\
&&B_6=\{B_{a_i}=B_{b_i}\}\cdot generate(K_b)\cdot B_7+\{B_{a_i}\neq B_{b_i}\}\cdot discard\cdot B_7\nonumber\\
&&B_7=\sum_{D_o\in\Delta_o}send_B(D_o)\cdot B\nonumber
\end{eqnarray}

where $\Delta_o$ is the collection of the output data.

The send action and receive action of the same data through the same channel can communicate each other, otherwise, a deadlock $\delta$ will be caused. The quantum measurement and its shadow constant pair will lead entanglement occur, otherwise, a deadlock $\delta$ will occur. We define the following communication functions.

\begin{eqnarray}
&&\gamma(send_Q(q_b),receive_Q(q_b))\triangleq c_Q(q_b)\nonumber\\
&&\gamma(send_P(B_b),receive_P(B_b))\triangleq c_P(B_b)\nonumber\\
&&\gamma(send_P(B_a),receive_P(B_a))\triangleq c_P(B_a)\nonumber
\end{eqnarray}

Let $A$ and $B$ in parallel, then the system $AB$ can be represented by the following process term.

$$\tau_I(\partial_H(\Theta(A\between B)))$$

where $H=\{send_Q(q_b),receive_Q(q_b),send_P(B_b),receive_P(B_b),send_P(B_a),receive_P(B_a),\\ \boxplus_{\frac{1}{2n},i=0}^{2n-1}M[q_a;B_a]_i, \boxplus_{\frac{1}{2n},i=0}^{2n-1}\circledS_{M[q_a;B_a]_i}, \boxplus_{\frac{1}{2n},i=0}^{2n-1}M[q_b;B_b]_i, \boxplus_{\frac{1}{2n},i=0}^{2n-1}\circledS_{M[q_b;B_b]_i}\}$

and $I=\{c_Q(q_b), c_P(B_b), c_P(B_a), M[q_a;B_a], M[q_b;B_b],\\ cmp(K_{a,b},B_a,B_b),\{B_{a_i}=B_{b_i}\},\{B_{a_i}\neq B_{b_i}\},\\
generate(K_a),generate(K_b),discard\}$.

Then we get the following conclusion.

\begin{theorem}
The basic BBM92 protocol $\tau_I(\partial_H(\Theta(A\between B)))$ exhibits desired external behaviors.
\end{theorem}

\begin{proof}
We can get $\tau_I(\partial_H(\Theta(A\between B)))=\sum_{D_i\in \Delta_i}\sum_{D_o\in\Delta_o}receive_A(D_i)\leftmerge send_B(D_o)\leftmerge \tau_I(\partial_H(\Theta(A\between B)))$.
So, the basic BBM92 protocol $\tau_I(\partial_H(\Theta(A\between B)))$ exhibits desired external behaviors.
\end{proof}

\subsection{Verification of SARG04 Protocol}\label{VSARG046}

The famous SARG04 protocol\cite{SARG04} is a quantum key distribution protocol, in which quantum information and classical information are mixed. We take an example of the SARG04 protocol to illustrate the usage of probabilistic quantum process algebra in verification of quantum protocols.

\begin{figure}
  \centering
  \includegraphics{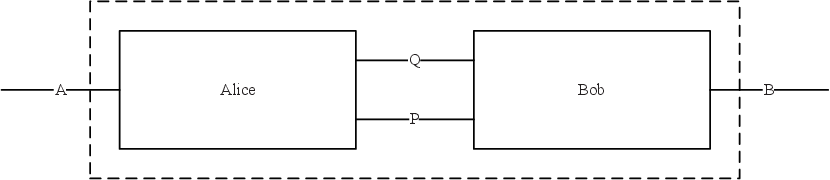}
  \caption{The SARG04 protocol.}
  \label{SARG04}
\end{figure}

The SARG04 protocol is used to create a private key between two parities, Alice and Bob. SARG04 is a protocol of quantum key distribution (QKD) which refines the BB84 protocol against PNS (Photon Number Splitting) attacks. The main innovations are encoding bits in nonorthogonal states and the classical sifting procedure. Firstly, we introduce the basic SARG04 protocol briefly, which is illustrated in Figure \ref{SARG04}.

\begin{enumerate}
  \item Alice create a string of bits with size $n$ randomly, denoted as $K_a$.
  \item Alice generates a string of qubits $q$ with size $n$, and the $i$th qubit of $q$ has four nonorthogonal states, it is $|\pm x\rangle$ if $K_a=0$; it is $|\pm z\rangle$ if $K_a=1$. And she records the corresponding one of the four pairs of nonorthogonal states into $B_a$ with size $2n$.
  \item Alice sends $q$ to Bob through a quantum channel $Q$ between Alice and Bob.
  \item Alice sends $B_a$ through a public channel $P$.
  \item Bob measures each qubit of $q$ $\sigma_x$ or $\sigma_z$. And he records the unambiguous discriminations into $K_b$ with a raw size $n/4$, and the unambiguous discrimination information into $B_b$ with size $n$.
  \item Bob sends $B_b$ to Alice through the public channel $P$.
  \item Alice and Bob determine that at which position the bit should be remained. Then the remaining bits of $K_a$ and $K_b$ is the private key $K_{a,b}$.
\end{enumerate}

We re-introduce the basic SARG04 protocol in an abstract way with more technical details as Figure \ref{SARG04} illustrates.

Now, we assume a special measurement operation $Rand[q;K_a]=\sum^{2n-1}_{i=0}Rand[q;K_a]_i$ which create a string of $n$ random bits $K_a$ from the $q$ quantum system. $M[q;K_b]=\sum^{2n-1}_{i=0}M[q;K_b]_i$ denotes the Bob's measurement operation of $q$. The generation of $n$ qubits $q$ through a unitary operator $Set_{K_a}[q]$. Alice sends $q$ to Bob through the quantum channel $Q$ by quantum communicating action $send_{Q}(q)$ and Bob receives $q$ through $Q$ by quantum communicating action $receive_{Q}(q)$. Bob sends $B_b$ to Alice through the public channel $P$ by classical communicating action $send_{P}(B_b)$ and Alice receives $B_b$ through channel $P$ by classical communicating action $receive_{P}(B_b)$, and the same as $send_{P}(B_a)$ and $receive_{P}(B_a)$. Alice and Bob generate the private key $K_{a,b}$ by a classical comparison action $cmp(K_{a,b},K_a,K_b,B_a,B_b)$. Let Alice and Bob be a system $AB$ and let interactions between Alice and Bob be internal actions. $AB$ receives external input $D_i$ through channel $A$ by communicating action $receive_A(D_i)$ and sends results $D_o$ through channel $B$ by communicating action $send_B(D_o)$.

Then the state transition of Alice can be described by probabilistic quantum process algebra as follows.

\begin{eqnarray}
&&A=\sum_{D_i\in \Delta_i}receive_A(D_i)\cdot A_1\nonumber\\
&&A_1=\boxplus_{\frac{1}{2n},i=0}^{2n-1}Rand[q;K_a]_i\cdot A_2\nonumber\\
&&A_2=Set_{K_a}[q]\cdot A_3\nonumber\\
&&A_3=send_Q(q)\cdot A_4\nonumber\\
&&A_4=send_P(B_a)\cdot A_5\nonumber\\
&&A_5=receive_P(B_b)\cdot A_6\nonumber\\
&&A_6=cmp(K_{a,b},K_a,K_b,B_a,B_b)\cdot A_7\nonumber\\
&&A_7=\{B_{a_i}=B_{b_i}\}\cdot generate(K_a)\cdot A+\{B_{a_i}\neq B_{b_i}\}\cdot discard\cdot A\nonumber
\end{eqnarray}

where $\Delta_i$ is the collection of the input data.

And the state transition of Bob can be described by probabilistic quantum process algebra as follows.

\begin{eqnarray}
&&B=receive_Q(q)\cdot B_1\nonumber\\
&&B_1=receive_P(B_a)\cdot B_2\nonumber\\
&&B_2=\boxplus_{\frac{1}{2n},i=0}^{2n-1}M[q;K_b]_i\cdot B_3\nonumber\\
&&B_3=send_P(B_b)\cdot B_4\nonumber\\
&&B_4=cmp(K_{a,b},K_a,K_b,B_a,B_b)\cdot B_5\nonumber\\
&&B_5=\{B_{a_i}=B_{b_i}\}\cdot generate(K_b)\cdot B_6+\{B_{a_i}\neq B_{b_i}\}\cdot discard\cdot B_6\nonumber\\
&&B_6=\sum_{D_o\in\Delta_o}send_B(D_o)\cdot B\nonumber
\end{eqnarray}

where $\Delta_o$ is the collection of the output data.

The send action and receive action of the same data through the same channel can communicate each other, otherwise, a deadlock $\delta$ will be caused. We define the following communication functions.

\begin{eqnarray}
&&\gamma(send_Q(q),receive_Q(q))\triangleq c_Q(q)\nonumber\\
&&\gamma(send_P(B_b),receive_P(B_b))\triangleq c_P(B_b)\nonumber\\
&&\gamma(send_P(B_a),receive_P(B_a))\triangleq c_P(B_a)\nonumber
\end{eqnarray}

Let $A$ and $B$ in parallel, then the system $AB$ can be represented by the following process term.

$$\tau_I(\partial_H(\Theta(A\between B)))$$

where $H=\{send_Q(q),receive_Q(q),send_P(B_b),receive_P(B_b),send_P(B_a),receive_P(B_a)\}$ and 
$I=\{\boxplus_{\frac{1}{2n},i=0}^{2n-1}Rand[q;K_a]_i, Set_{K_a}[q], \boxplus_{\frac{1}{2n},i=0}^{2n-1}M[q;K_b]_i, c_Q(q), c_P(B_b),\\ c_P(B_a), cmp(K_{a,b},K_a,K_b,B_a,B_b),\{B_{a_i}=B_{b_i}\},\{B_{a_i}\neq B_{b_i}\},\\
generate(K_a),generate(K_b),discard\}$.
Then we get the following conclusion.

\begin{theorem}
The basic SARG04 protocol $\tau_I(\partial_H(\Theta(A\between B)))$ exhibits desired external behaviors.
\end{theorem}

\begin{proof}
We can get $\tau_I(\partial_H(\Theta(A\between B)))=\sum_{D_i\in \Delta_i}\sum_{D_o\in\Delta_o}receive_A(D_i)\leftmerge send_B(D_o)\leftmerge \tau_I(\partial_H(\Theta(A\between B)))$.
So, the basic SARG04 protocol $\tau_I(\partial_H(\Theta(A\between B)))$ exhibits desired external behaviors.
\end{proof}

\subsection{Verification of COW Protocol}\label{VCOW6}

The famous COW protocol\cite{COW} is a quantum key distribution protocol, in which quantum information and classical information are mixed. We take an example of the COW protocol to illustrate the usage of probabilistic quantum process algebra in verification of quantum protocols.

\begin{figure}
  \centering
  \includegraphics{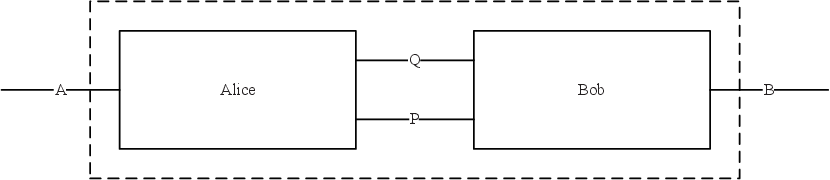}
  \caption{The COW protocol.}
  \label{COW}
\end{figure}

The COW protocol is used to create a private key between two parities, Alice and Bob. COW is a protocol of quantum key distribution (QKD) which is practical. Firstly, we introduce the basic COW protocol briefly, which is illustrated in Figure \ref{COW}.

\begin{enumerate}
  \item Alice generates a string of qubits $q$ with size $n$, and the $i$th qubit of $q$ is "0" with probability $\frac{1-f}{2}$, "1" with probability $\frac{1-f}{2}$ and the decoy sequence with probability $f$.
  \item Alice sends $q$ to Bob through a quantum channel $Q$ between Alice and Bob.
  \item Alice sends $A$ of the items corresponding to a decoy sequence through a public channel $P$.
  \item Bob removes all the detections at times $2A-1$ and $2A$ from his raw key and looks whether detector $D_{2M}$ has ever fired at time $2A$.
  \item Bob sends $B$ of the times $2A+1$ in which he had a detector in $D_{2M}$ to Alice through the public channel $P$.
  \item Alice receives $B$ and verifies if some of these items corresponding to a bit sequence "1,0".
  \item Bob sends $C$ of the items that he has detected through the public channel $P$.
  \item Alice and Bob run error correction and privacy amplification on these bits, and the private key $K_{a,b}$ is established.
\end{enumerate}

We re-introduce the basic COW protocol in an abstract way with more technical details as Figure \ref{COW} illustrates.

Now, we assume The generation of $n$ qubits $q$ through a unitary operator $Set[q]$. $M[q]=\sum^{2n-1}_{i=0}M[q]_i$ denotes the Bob's measurement operation of $q$.  Alice sends $q$ to Bob through the quantum channel $Q$ by quantum communicating action $send_{Q}(q)$ and Bob receives $q$ through $Q$ by quantum communicating action $receive_{Q}(q)$. Alice sends $A$ to Alice through the public channel $P$ by classical communicating action $send_{P}(A)$ and Alice receives $A$ through channel $P$ by classical communicating action $receive_{P}(A)$, and the same as $send_{P}(B)$ and $receive_{P}(B)$, and $send_{P}(C)$ and $receive_{P}(C)$. Alice and Bob generate the private key $K_{a,b}$ by a classical comparison action $cmp(K_{a,b})$. Let Alice and Bob be a system $AB$ and let interactions between Alice and Bob be internal actions. $AB$ receives external input $D_i$ through channel $A$ by communicating action $receive_A(D_i)$ and sends results $D_o$ through channel $B$ by communicating action $send_B(D_o)$.

Then the state transition of Alice can be described by probabilistic quantum process algebra as follows.

\begin{eqnarray}
&&A=\sum_{D_i\in \Delta_i}receive_A(D_i)\cdot A_1\nonumber\\
&&A_1=Set[q]\cdot A_2\nonumber\\
&&A_2=send_Q(q)\cdot A_3\nonumber\\
&&A_3=send_P(A)\cdot A_4\nonumber\\
&&A_4=receive_P(B)\cdot A_5\nonumber\\
&&A_5=receive_P(C)\cdot A_6\nonumber\\
&&A_6=cmp(K_{a,b})\cdot A\nonumber
\end{eqnarray}

where $\Delta_i$ is the collection of the input data.

And the state transition of Bob can be described by probabilistic quantum process algebra as follows.

\begin{eqnarray}
&&B=receive_Q(q)\cdot B_1\nonumber\\
&&B_1=receive_P(A)\cdot B_2\nonumber\\
&&B_2=\boxplus_{\frac{1}{2n},i=0}^{2n-1}M[q]_i\cdot B_3\nonumber\\
&&B_3=send_P(B)\cdot B_4\nonumber\\
&&B_4=send_P(C)\cdot B_5\nonumber\\
&&B_5=cmp(K_{a,b})\cdot B_6\nonumber\\
&&B_6=\sum_{D_o\in\Delta_o}send_B(D_o)\cdot B\nonumber
\end{eqnarray}

where $\Delta_o$ is the collection of the output data.

The send action and receive action of the same data through the same channel can communicate each other, otherwise, a deadlock $\delta$ will be caused. We define the following communication functions.

\begin{eqnarray}
&&\gamma(send_Q(q),receive_Q(q))\triangleq c_Q(q)\nonumber\\
&&\gamma(send_P(A),receive_P(A))\triangleq c_P(A)\nonumber\\
&&\gamma(send_P(B),receive_P(B))\triangleq c_P(B)\nonumber\\
&&\gamma(send_P(C),receive_P(C))\triangleq c_P(C)\nonumber
\end{eqnarray}

Let $A$ and $B$ in parallel, then the system $AB$ can be represented by the following process term.

$$\tau_I(\partial_H(\Theta(A\between B)))$$

where $H=\{send_Q(q),receive_Q(q),send_P(A),receive_P(A),send_P(B),receive_P(B),send_P(C),\\receive_P(C)\}$ and $I=\{Set[q], \boxplus_{\frac{1}{2n},i=0}^{2n-1}M[q]_i, c_Q(q), c_P(A),\\ c_P(B),c_P(C), cmp(K_{a,b})\}$.

Then we get the following conclusion.

\begin{theorem}
The basic COW protocol $\tau_I(\partial_H(\Theta(A\between B)))$ exhibits desired external behaviors.
\end{theorem}

\begin{proof}
We can get $\tau_I(\partial_H(\Theta(A\between B)))=\sum_{D_i\in \Delta_i}\sum_{D_o\in\Delta_o}receive_A(D_i)\leftmerge send_B(D_o)\leftmerge \tau_I(\partial_H(\Theta(A\between B)))$. 
So, the basic COW protocol $\tau_I(\partial_H(\Theta(A\between B)))$ exhibits desired external behaviors.
\end{proof}

\subsection{Verification of SSP Protocol}\label{VSSP6}

The famous SSP protocol\cite{SSP} is a quantum key distribution protocol, in which quantum information and classical information are mixed. We take an example of the SSP protocol to illustrate the usage of probabilistic quantum process algebra in verification of quantum protocols.

\begin{figure}
  \centering
  \includegraphics{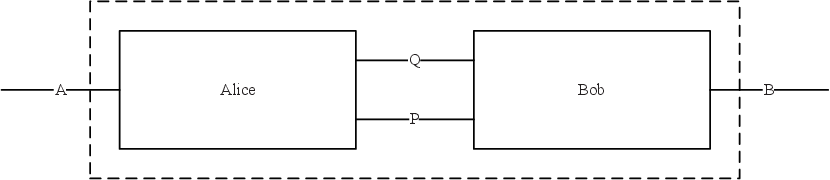}
  \caption{The SSP protocol.}
  \label{SSP}
\end{figure}

The SSP protocol is used to create a private key between two parities, Alice and Bob. SSP is a protocol of quantum key distribution (QKD) which uses six states. Firstly, we introduce the basic SSP protocol briefly, which is illustrated in Figure \ref{SSP}.

\begin{enumerate}
  \item Alice create two string of bits with size $n$ randomly, denoted as $B_a$ and $K_a$.
  \item Alice generates a string of qubits $q$ with size $n$, and the $i$th qubit in $q$ is one of the six states $\pm x$, $\pm y$ and $\pm z$.
  \item Alice sends $q$ to Bob through a quantum channel $Q$ between Alice and Bob.
  \item Bob receives $q$ and randomly generates a string of bits $B_b$ with size $n$.
  \item Bob measures each qubit of $q$ according to a basis by bits of $B_b$, i.e., $x$, $y$ or $z$ basis. And the measurement results would be $K_b$, which is also with size $n$.
  \item Bob sends his measurement bases $B_b$ to Alice through a public channel $P$.
  \item Once receiving $B_b$, Alice sends her bases $B_a$ to Bob through channel $P$, and Bob receives $B_a$.
  \item Alice and Bob determine that at which position the bit strings $B_a$ and $B_b$ are equal, and they discard the mismatched bits of $B_a$ and $B_b$. Then the remaining bits of $K_a$ and $K_b$, denoted as $K_a'$ and $K_b'$ with $K_{a,b}=K_a'=K_b'$.
\end{enumerate}

We re-introduce the basic SSP protocol in an abstract way with more technical details as Figure \ref{SSP} illustrates.

Now, we assume a special measurement operation $Rand[q;B_a]=\sum^{2n-1}_{i=0}Rand[q;B_a]_i$ which create a string of $n$ random bits $B_a$ from the $q$ quantum system, and the same as $Rand[q;K_a]=\sum^{2n-1}_{i=0}Rand[q;K_a]_i$, $Rand[q';B_b]=\sum^{2n-1}_{i=0}Rand[q';B_b]_i$. $M[q;K_b]=\sum^{2n-1}_{i=0}M[q;K_b]_i$ denotes the Bob's measurement operation of $q$. The generation of $n$ qubits $q$ through two unitary operators $Set_{K_a}[q]$ and $H_{B_a}[q]$. Alice sends $q$ to Bob through the quantum channel $Q$ by quantum communicating action $send_{Q}(q)$ and Bob receives $q$ through $Q$ by quantum communicating action $receive_{Q}(q)$. Bob sends $B_b$ to Alice through the public channel $P$ by classical communicating action $send_{P}(B_b)$ and Alice receives $B_b$ through channel $P$ by classical communicating action $receive_{P}(B_b)$, and the same as $send_{P}(B_a)$ and $receive_{P}(B_a)$. Alice and Bob generate the private key $K_{a,b}$ by a classical comparison action $cmp(K_{a,b},K_a,K_b,B_a,B_b)$. Let Alice and Bob be a system $AB$ and let interactions between Alice and Bob be internal actions. $AB$ receives external input $D_i$ through channel $A$ by communicating action $receive_A(D_i)$ and sends results $D_o$ through channel $B$ by communicating action $send_B(D_o)$.

Then the state transition of Alice can be described by probabilistic quantum process algebra as follows.

\begin{eqnarray}
&&A=\sum_{D_i\in \Delta_i}receive_A(D_i)\cdot A_1\nonumber\\
&&A_1=\boxplus_{\frac{1}{2n},i=0}^{2n-1}Rand[q;B_a]_i\cdot A_2\nonumber\\
&&A_2=\boxplus_{\frac{1}{2n},i=0}^{2n-1}Rand[q;K_a]_i\cdot A_3\nonumber\\
&&A_3=Set_{K_a}[q]\cdot A_4\nonumber\\
&&A_4=H_{B_a}[q]\cdot A_5\nonumber\\
&&A_5=send_Q(q)\cdot A_6\nonumber\\
&&A_6=receive_P(B_b)\cdot A_7\nonumber\\
&&A_7=send_P(B_a)\cdot A_8\nonumber\\
&&A_8=cmp(K_{a,b},K_a,K_b,B_a,B_b)\cdot A_9\nonumber\\
&&A_9=\{B_{a_i}=B_{b_i}\}\cdot generate(K_a)\cdot A+\{B_{a_i}\neq B_{b_i}\}\cdot discard\cdot A\nonumber
\end{eqnarray}

where $\Delta_i$ is the collection of the input data.

And the state transition of Bob can be described by probabilistic quantum process algebra as follows.

\begin{eqnarray}
&&B=receive_Q(q)\cdot B_1\nonumber\\
&&B_1=\boxplus_{\frac{1}{2n},i=0}^{2n-1}Rand[q';B_b]_i\cdot B_2\nonumber\\
&&B_2=\boxplus_{\frac{1}{2n},i=0}^{2n-1}M[q;K_b]_i\cdot B_3\nonumber\\
&&B_3=send_P(B_b)\cdot B_4\nonumber\\
&&B_4=receive_P(B_a)\cdot B_5\nonumber\\
&&B_5=cmp(K_{a,b},K_a,K_b,B_a,B_b)\cdot B_6\nonumber\\
&&B_6=\{B_{a_i}=B_{b_i}\}\cdot generate(K_b)\cdot B_7+\{B_{a_i}\neq B_{b_i}\}\cdot discard\cdot B_7\nonumber\\
&&B_7=\sum_{D_o\in\Delta_o}send_B(D_o)\cdot B\nonumber
\end{eqnarray}

where $\Delta_o$ is the collection of the output data.

The send action and receive action of the same data through the same channel can communicate each other, otherwise, a deadlock $\delta$ will be caused. We define the following communication functions.

\begin{eqnarray}
&&\gamma(send_Q(q),receive_Q(q))\triangleq c_Q(q)\nonumber\\
&&\gamma(send_P(B_b),receive_P(B_b))\triangleq c_P(B_b)\nonumber\\
&&\gamma(send_P(B_a),receive_P(B_a))\triangleq c_P(B_a)\nonumber
\end{eqnarray}

Let $A$ and $B$ in parallel, then the system $AB$ can be represented by the following process term.

$$\tau_I(\partial_H(\Theta(A\between B)))$$

where $H=\{send_Q(q),receive_Q(q),send_P(B_b),receive_P(B_b),send_P(B_a),receive_P(B_a)\}$ and
$I=\{\boxplus_{\frac{1}{2n},i=0}^{2n-1}Rand[q;B_a]_i, \boxplus_{\frac{1}{2n},i=0}^{2n-1}Rand[q;K_a]_i, Set_{K_a}[q],
\\ H_{B_a}[q], \boxplus_{\frac{1}{2n},i=0}^{2n-1}Rand[q';B_b]_i, \boxplus_{\frac{1}{2n},i=0}^{2n-1}M[q;K_b]_i, c_Q(q), c_P(B_b),\\ c_P(B_a), cmp(K_{a,b},K_a,K_b,B_a,B_b),\{B_{a_i}=B_{b_i}\},\{B_{a_i}\neq B_{b_i}\},\\
generate(K_a),generate(K_b),discard\}$.

Then we get the following conclusion.

\begin{theorem}
The basic SSP protocol $\tau_I(\partial_H(\Theta(A\between B)))$ exhibits desired external behaviors.
\end{theorem}

\begin{proof}
We can get $\tau_I(\partial_H(\Theta(A\between B)))=\sum_{D_i\in \Delta_i}\sum_{D_o\in\Delta_o}receive_A(D_i)\leftmerge send_B(D_o)\leftmerge \tau_I(\partial_H(\Theta(A\between B)))$.
So, the basic SSP protocol $\tau_I(\partial_H(\Theta(A\between B)))$ exhibits desired external behaviors.
\end{proof}

\subsection{Verification of S09 Protocol}\label{VS096}

The famous S09 protocol\cite{S09} is a quantum key distribution protocol, in which quantum information and classical information are mixed. We take an example of the S09 protocol to illustrate the usage of probabilistic quantum process algebra in verification of quantum protocols.

\begin{figure}
  \centering
  \includegraphics{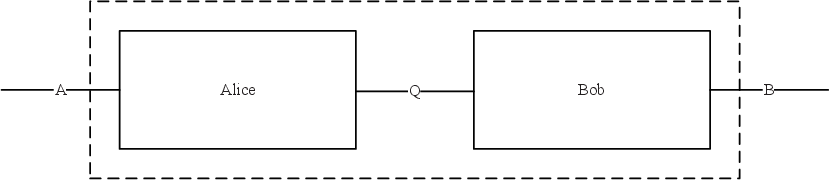}
  \caption{The S09 protocol.}
  \label{S09}
\end{figure}

The S09 protocol is used to create a private key between two parities, Alice and Bob, by use of pure quantum information. Firstly, we introduce the basic S09 protocol briefly, which is illustrated in Figure \ref{S09}.

\begin{enumerate}
  \item Alice create two string of bits with size $n$ randomly, denoted as $B_a$ and $K_a$.
  \item Alice generates a string of qubits $q$ with size $n$, and the $i$th qubit in $q$ is $|x_y\rangle$, where $x$ is the $i$th bit of $B_a$ and $y$ is the $i$th bit of $K_a$.
  \item Alice sends $q$ to Bob through a quantum channel $Q$ between Alice and Bob.
  \item Bob receives $q$ and randomly generates a string of bits $B_b$ with size $n$.
  \item Bob measures each qubit of $q$ according to a basis by bits of $B_b$. After the measurement, the state of $q$ evolves into $q'$.
  \item Bob sends $q'$ to Alice through the quantum channel $Q$.
  \item Alice measures each qubit of $q'$ to generate a string $C$.
  \item Alice sums $C_i\oplus B_{a_i}$ to get the private key $K_{a,b}=B_b$.
\end{enumerate}

We re-introduce the basic S09 protocol in an abstract way with more technical details as Figure \ref{S09} illustrates.

Now, we assume a special measurement operation $Rand[q;B_a]=\sum^{2n-1}_{i=0}Rand[q;B_a]_i$ which create a string of $n$ random bits $B_a$ from the $q$ quantum system, and the same as $Rand[q;K_a]=\sum^{2n-1}_{i=0}Rand[q;K_a]_i$, $Rand[q';B_b]=\sum^{2n-1}_{i=0}Rand[q';B_b]_i$. $M[q;B_b]=\sum^{2n-1}_{i=0}M[q;B_b]_i$ denotes the Bob's measurement operation of $q$, and the same as $M[q';C]=\sum^{2n-1}_{i=0}Rand[q';C]_i$. The generation of $n$ qubits $q$ through two unitary operators $Set_{K_a}[q]$ and $H_{B_a}[q]$. Alice sends $q$ to Bob through the quantum channel $Q$ by quantum communicating action $send_{Q}(q)$ and Bob receives $q$ through $Q$ by quantum communicating action $receive_{Q}(q)$, and the same as $send_{Q}(q')$ and $receive_{Q}(q')$. Alice and Bob generate the private key $K_{a,b}$ by a classical comparison action $cmp(K_{a,b},B_b)$. We omit the sum classical $\oplus$ actions without of loss of generality. Let Alice and Bob be a system $AB$ and let interactions between Alice and Bob be internal actions. $AB$ receives external input $D_i$ through channel $A$ by communicating action $receive_A(D_i)$ and sends results $D_o$ through channel $B$ by communicating action $send_B(D_o)$.

Then the state transition of Alice can be described by probabilistic quantum process algebra as follows.

\begin{eqnarray}
&&A=\sum_{D_i\in \Delta_i}receive_A(D_i)\cdot A_1\nonumber\\
&&A_1=\boxplus_{\frac{1}{2n},i=0}^{2n-1}Rand[q;B_a]_i\cdot A_2\nonumber\\
&&A_2=\boxplus_{\frac{1}{2n},i=0}^{2n-1}Rand[q;K_a]_i\cdot A_3\nonumber\\
&&A_3=Set_{K_a}[q]\cdot A_4\nonumber\\
&&A_4=H_{B_a}[q]\cdot A_5\nonumber\\
&&A_5=send_Q(q)\cdot A_6\nonumber\\
&&A_6=receive_Q(q')\cdot A_{7}\nonumber\\
&&A_7=\boxplus_{\frac{1}{2n},i=0}^{2n-1}M[q';C]_i\cdot A_8\nonumber\\
&&A_{8}=cmp(K_{a,b},B_b)\cdot A\nonumber
\end{eqnarray}

where $\Delta_i$ is the collection of the input data.

And the state transition of Bob can be described by probabilistic quantum process algebra as follows.

\begin{eqnarray}
&&B=receive_Q(q)\cdot B_1\nonumber\\
&&B_1=\boxplus_{\frac{1}{2n},i=0}^{2n-1}Rand[q';B_b]_i\cdot B_2\nonumber\\
&&B_2=\boxplus_{\frac{1}{2n},i=0}^{2n-1}M[q;B_b]_i\cdot B_3\nonumber\\
&&B_3=send_Q(q')\cdot B_4\nonumber\\
&&B_4=cmp(K_{a,b},B_b)\cdot B_{5}\nonumber\\
&&B_{5}=\sum_{D_o\in\Delta_o}send_B(D_o)\cdot B\nonumber
\end{eqnarray}

where $\Delta_o$ is the collection of the output data.

The send action and receive action of the same data through the same channel can communicate each other, otherwise, a deadlock $\delta$ will be caused. We define the following communication functions.

\begin{eqnarray}
&&\gamma(send_Q(q),receive_Q(q))\triangleq c_Q(q)\nonumber\\
&&\gamma(send_Q(q'),receive_Q(q'))\triangleq c_Q(q')\nonumber
\end{eqnarray}

Let $A$ and $B$ in parallel, then the system $AB$ can be represented by the following process term.

$$\tau_I(\partial_H(\Theta(A\between B)))$$

where $H=\{send_Q(q),receive_Q(q),send_Q(q'),receive_Q(q')\}$ and $I=\{\boxplus_{\frac{1}{2n},i=0}^{2n-1}Rand[q;B_a]_i, \\ \boxplus_{\frac{1}{2n},i=0}^{2n-1}Rand[q;K_a]_i, Set_{K_a}[q], H_{B_a}[q], \boxplus_{\frac{1}{2n},i=0}^{2n-1}Rand[q';B_b]_i, \boxplus_{\frac{1}{2n},i=0}^{2n-1}M[q;B_b]_i,  \\ \boxplus_{\frac{1}{2n},i=0}^{2n-1}Rand[q';C]_i, c_Q(q), c_Q(q'), cmp(K_{a,b},B_b)\}$.

Then we get the following conclusion.

\begin{theorem}
The basic S09 protocol $\tau_I(\partial_H(\Theta(A\between B)))$ exhibits desired external behaviors.
\end{theorem}

\begin{proof}
We can get $\tau_I(\partial_H(\Theta(A\between B)))=\sum_{D_i\in \Delta_i}\sum_{D_o\in\Delta_o}receive_A(D_i)\leftmerge send_B(D_o)\leftmerge \tau_I(\partial_H(\Theta(A\between B)))$. 
So, the basic S09 protocol $\tau_I(\partial_H(\Theta(A\between B)))$ exhibits desired external behaviors.
\end{proof}

\subsection{Verification of KMB09 Protocol}\label{VKMB096}

The famous KMB09 protocol\cite{KMB09} is a quantum key distribution protocol, in which quantum information and classical information are mixed. We take an example of the KMB09 protocol to illustrate the usage of probabilistic quantum process algebra in verification of quantum protocols.

\begin{figure}
  \centering
  \includegraphics{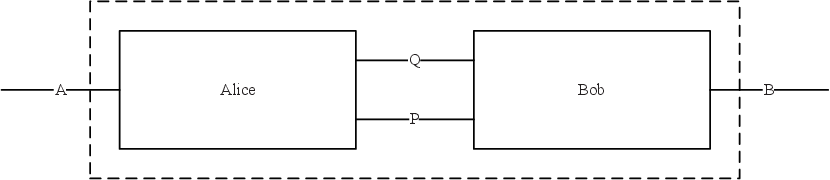}
  \caption{The KMB09 protocol.}
  \label{KMB09}
\end{figure}

The KMB09 protocol is used to create a private key between two parities, Alice and Bob. KMB09 is a protocol of quantum key distribution (QKD) which refines the BB84 protocol against PNS (Photon Number Splitting) attacks. The main innovations are encoding bits in nonorthogonal states and the classical sifting procedure. Firstly, we introduce the basic KMB09 protocol briefly, which is illustrated in Figure \ref{KMB09}.

\begin{enumerate}
  \item Alice create a string of bits with size $n$ randomly, denoted as $K_a$, and randomly assigns each bit value a random index $i=1,2,...,N$ into $B_a$.
  \item Alice generates a string of qubits $q$ with size $n$, accordingly either in $|e_i\rangle$ or $|f_i\rangle$.
  \item Alice sends $q$ to Bob through a quantum channel $Q$ between Alice and Bob.
  \item Alice sends $B_a$ through a public channel $P$.
  \item Bob measures each qubit of $q$ by randomly switching the measurement basis between $e$ and $f$. And he records the unambiguous discriminations into $K_b$, and the unambiguous discrimination information into $B_b$.
  \item Bob sends $B_b$ to Alice through the public channel $P$.
  \item Alice and Bob determine that at which position the bit should be remained. Then the remaining bits of $K_a$ and $K_b$ is the private key $K_{a,b}$.
\end{enumerate}

We re-introduce the basic KMB09 protocol in an abstract way with more technical details as Figure \ref{KMB09} illustrates.

Now, we assume a special measurement operation $Rand[q;K_a]=\sum^{2n-1}_{i=0}Rand[q;K_a]_i$ which create a string of $n$ random bits $K_a$ from the $q$ quantum system. $M[q;K_b]=\sum^{2n-1}_{i=0}M[q;K_b]_i$ denotes the Bob's measurement operation of $q$. The generation of $n$ qubits $q$ through a unitary operator $Set_{K_a}[q]$. Alice sends $q$ to Bob through the quantum channel $Q$ by quantum communicating action $send_{Q}(q)$ and Bob receives $q$ through $Q$ by quantum communicating action $receive_{Q}(q)$. Bob sends $B_b$ to Alice through the public channel $P$ by classical communicating action $send_{P}(B_b)$ and Alice receives $B_b$ through channel $P$ by classical communicating action $receive_{P}(B_b)$, and the same as $send_{P}(B_a)$ and $receive_{P}(B_a)$. Alice and Bob generate the private key $K_{a,b}$ by a classical comparison action $cmp(K_{a,b},K_a,K_b,B_a,B_b)$. Let Alice and Bob be a system $AB$ and let interactions between Alice and Bob be internal actions. $AB$ receives external input $D_i$ through channel $A$ by communicating action $receive_A(D_i)$ and sends results $D_o$ through channel $B$ by communicating action $send_B(D_o)$.

Then the state transition of Alice can be described by probabilistic quantum process algebra as follows.

\begin{eqnarray}
&&A=\sum_{D_i\in \Delta_i}receive_A(D_i)\cdot A_1\nonumber\\
&&A_1=\boxplus_{\frac{1}{2n},i=0}^{2n-1}Rand[q;K_a]_i\cdot A_2\nonumber\\
&&A_2=Set_{K_a}[q]\cdot A_3\nonumber\\
&&A_3=send_Q(q)\cdot A_4\nonumber\\
&&A_4=send_P(B_a)\cdot A_5\nonumber\\
&&A_5=receive_P(B_b)\cdot A_6\nonumber\\
&&A_6=cmp(K_{a,b},K_a,K_b,B_a,B_b)\cdot A_7\nonumber\\
&&A_7=\{B_{a_i}=B_{b_i}\}\cdot generate(K_a)\cdot A+\{B_{a_i}\neq B_{b_i}\}\cdot discard\cdot A\nonumber
\end{eqnarray}

where $\Delta_i$ is the collection of the input data.

And the state transition of Bob can be described by probabilistic quantum process algebra as follows.

\begin{eqnarray}
&&B=receive_Q(q)\cdot B_1\nonumber\\
&&B_1=receive_P(B_a)\cdot B_2\nonumber\\
&&B_2=\boxplus_{\frac{1}{2n},i=0}^{2n-1}M[q;K_b]_i\cdot B_3\nonumber\\
&&B_3=send_P(B_b)\cdot B_4\nonumber\\
&&B_4=cmp(K_{a,b},K_a,K_b,B_a,B_b)\cdot B_5\nonumber\\
&&B_5=\{B_{a_i}=B_{b_i}\}\cdot generate(K_b)\cdot B_6+\{B_{a_i}\neq B_{b_i}\}\cdot discard\cdot B_6\nonumber\\
&&B_6=\sum_{D_o\in\Delta_o}send_B(D_o)\cdot B\nonumber
\end{eqnarray}

where $\Delta_o$ is the collection of the output data.

The send action and receive action of the same data through the same channel can communicate each other, otherwise, a deadlock $\delta$ will be caused. We define the following communication functions.

\begin{eqnarray}
&&\gamma(send_Q(q),receive_Q(q))\triangleq c_Q(q)\nonumber\\
&&\gamma(send_P(B_b),receive_P(B_b))\triangleq c_P(B_b)\nonumber\\
&&\gamma(send_P(B_a),receive_P(B_a))\triangleq c_P(B_a)\nonumber
\end{eqnarray}

Let $A$ and $B$ in parallel, then the system $AB$ can be represented by the following process term.

$$\tau_I(\partial_H(\Theta(A\between B)))$$

where $H=\{send_Q(q),receive_Q(q),send_P(B_b),receive_P(B_b),send_P(B_a),receive_P(B_a)\}$ and
$I=\{\boxplus_{\frac{1}{2n},i=0}^{2n-1}Rand[q;K_a]_i, Set_{K_a}[q], \boxplus_{\frac{1}{2n},i=0}^{2n-1}M[q;K_b]_i, c_Q(q), c_P(B_b),\\ c_P(B_a), cmp(K_{a,b},K_a,K_b,B_a,B_b),\{B_{a_i}=B_{b_i}\},\{B_{a_i}\neq B_{b_i}\},\\
generate(K_a),generate(K_b),discard\}$.
Then we get the following conclusion.

\begin{theorem}
The basic KMB09 protocol $\tau_I(\partial_H(\Theta(A\between B)))$ exhibits desired external behaviors.
\end{theorem}

\begin{proof}
We can get $\tau_I(\partial_H(\Theta(A\between B)))=\sum_{D_i\in \Delta_i}\sum_{D_o\in\Delta_o}receive_A(D_i)\leftmerge send_B(D_o)\leftmerge \tau_I(\partial_H(\Theta(A\between B)))$.
So, the basic KMB09 protocol $\tau_I(\partial_H(\Theta(A\between B)))$ exhibits desired external behaviors.
\end{proof}

\subsection{Verification of S13 Protocol}\label{VS136}

The famous S13 protocol\cite{S13} is a quantum key distribution protocol, in which quantum information and classical information are mixed. We take an example of the S13 protocol to illustrate the usage of probabilistic quantum process algebra in verification of quantum protocols.

\begin{figure}
  \centering
  \includegraphics{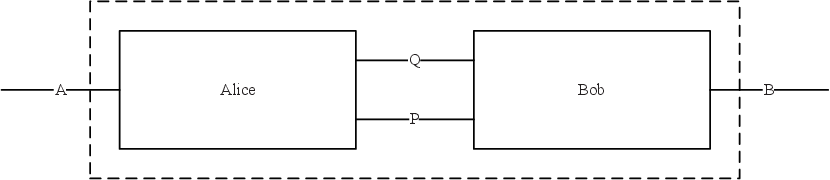}
  \caption{The S13 protocol.}
  \label{S13}
\end{figure}

The S13 protocol is used to create a private key between two parities, Alice and Bob. Firstly, we introduce the basic S13 protocol briefly, which is illustrated in Figure \ref{S13}.

\begin{enumerate}
  \item Alice create two string of bits with size $n$ randomly, denoted as $B_a$ and $K_a$.
  \item Alice generates a string of qubits $q$ with size $n$, and the $i$th qubit in $q$ is $|x_y\rangle$, where $x$ is the $i$th bit of $B_a$ and $y$ is the $i$th bit of $K_a$.
  \item Alice sends $q$ to Bob through a quantum channel $Q$ between Alice and Bob.
  \item Bob receives $q$ and randomly generates a string of bits $B_b$ with size $n$.
  \item Bob measures each qubit of $q$ according to a basis by bits of $B_b$. And the measurement results would be $K_b$, which is also with size $n$.
  \item Alice sends a random binary string $C$ to Bob through the public channel $P$.
  \item Alice sums $B_{a_i}\oplus C_i$ to obtain $T$ and generates other random string of binary values $J$. From the elements occupying a concrete position, $i$, of the preceding strings, Alice get the new states of $q'$, and sends it to Bob through the quantum channel $Q$.
  \item Bob sums $1\oplus B_{b_i}$ to obtain the string of binary basis $N$ and measures $q'$ according to these bases, and generating $D$.
  \item Alice sums $K_{a_i}\oplus J_i$ to obtain the binary string $Y$ and sends it to Bob through the public channel $P$.
  \item Bob encrypts $B_b$ to obtain $U$ and sends to Alice through the public channel $P$.
  \item Alice decrypts $U$ to obtain $B_b$. She sums $B_{a_i}\oplus B_{b_i}$ to obtain $L$ and sends $L$ to Bob through the public channel $P$.
  \item Bob sums $B_{b_i}\oplus L_i$ to get the private key $K_{a,b}$.
\end{enumerate}

We re-introduce the basic S13 protocol in an abstract way with more technical details as Figure \ref{S13} illustrates.

Now, we assume a special measurement operation $Rand[q;B_a]=\sum^{2n-1}_{i=0}Rand[q;B_a]_i$ which create a string of $n$ random bits $B_a$ from the $q$ quantum system, and the same as $Rand[q;K_a]=\sum^{2n-1}_{i=0}Rand[q;K_a]_i$, $Rand[q';B_b]=\sum^{2n-1}_{i=0}Rand[q';B_b]_i$. $M[q;K_b]=\sum^{2n-1}_{i=0}M[q;K_b]_i$ denotes the Bob's measurement operation of $q$, and the same as $M[q';D]=\sum^{2n-1}_{i=0}M[q';D]_i$. The generation of $n$ qubits $q$ through two unitary operators $Set_{K_a}[q]$ and $H_{B_a}[q]$, and the same as $Set_{T}[q']$. Alice sends $q$ to Bob through the quantum channel $Q$ by quantum communicating action $send_{Q}(q)$ and Bob receives $q$ through $Q$ by quantum communicating action $receive_{Q}(q)$, and the same as $send_{Q}(q')$ and $receive_{Q}(q')$. Bob sends $B_b$ to Alice through the public channel $P$ by classical communicating action $send_{P}(B_b)$ and Alice receives $B_b$ through channel $P$ by classical communicating action $receive_{P}(B_b)$, and the same as $send_{P}(B_a)$ and $receive_{P}(B_a)$, $send_{P}(C)$ and $receive_{P}(C)$, $send_{P}(Y)$ and $receive_{P}(Y)$, $send_{P}(U)$ and $receive_{P}(U)$, $send_{P}(L)$ and $receive_{P}(L)$. Alice and Bob generate the private key $K_{a,b}$ by a classical comparison action $cmp(K_{a,b},K_a,K_b,B_a,B_b)$. We omit the sum classical $\oplus$ actions without of loss of generality. Let Alice and Bob be a system $AB$ and let interactions between Alice and Bob be internal actions. $AB$ receives external input $D_i$ through channel $A$ by communicating action $receive_A(D_i)$ and sends results $D_o$ through channel $B$ by communicating action $send_B(D_o)$.

Then the state transition of Alice can be described by probabilistic quantum process algebra as follows.

\begin{eqnarray}
&&A=\sum_{D_i\in \Delta_i}receive_A(D_i)\cdot A_1\nonumber\\
&&A_1=\boxplus_{\frac{1}{2n},i=0}^{2n-1}Rand[q;B_a]_i\cdot A_2\nonumber\\
&&A_2=\boxplus_{\frac{1}{2n},i=0}^{2n-1}Rand[q;K_a]_i\cdot A_3\nonumber\\
&&A_3=Set_{K_a}[q]\cdot A_4\nonumber\\
&&A_4=H_{B_a}[q]\cdot A_5\nonumber\\
&&A_5=send_Q(q)\cdot A_6\nonumber\\
&&A_6=send_P(C)\cdot A_7\nonumber\\
&&A_7=send_Q(q')\cdot A_8\nonumber\\
&&A_8=send_P(Y)\cdot A_9\nonumber\\
&&A_9=receive_P(U)\cdot A_{10}\nonumber\\
&&A_{10}=send_P(L)\cdot A_{11}\nonumber\\
&&A_{11}=cmp(K_{a,b},K_a,K_b,B_a,B_b)\cdot A_{12}\nonumber\\
&&A_{12}=\{B_{a_i}=B_{b_i}\}\cdot generate(K_a)\cdot A+\{B_{a_i}\neq B_{b_i}\}\cdot discard\cdot A\nonumber
\end{eqnarray}

where $\Delta_i$ is the collection of the input data.

And the state transition of Bob can be described by probabilistic quantum process algebra as follows.

\begin{eqnarray}
&&B=receive_Q(q)\cdot B_1\nonumber\\
&&B_1=\boxplus_{\frac{1}{2n},i=0}^{2n-1}Rand[q';B_b]_i\cdot B_2\nonumber\\
&&B_2=\boxplus_{\frac{1}{2n},i=0}^{2n-1}M[q;K_b]_i\cdot B_3\nonumber\\
&&B_3=receive_P(C)\cdot B_4\nonumber\\
&&B_4=receive_Q(q')\cdot B_5\nonumber\\
&&B_5=\boxplus_{\frac{1}{2n},i=0}^{2n-1}M[q';D]_i\cdot B_6\nonumber\\
&&B_6=receive_P(Y)\cdot B_7\nonumber\\
&&B_7=send_P(U)\cdot B_8\nonumber\\
&&B_8=receive_P(L)\cdot B_9\nonumber\\
&&B_9=cmp(K_{a,b},K_a,K_b,B_a,B_b)\cdot B_{10}\nonumber\\
&&B_{10}=\{B_{a_i}=B_{b_i}\}\cdot generate(K_b)\cdot B_{11}+\{B_{a_i}\neq B_{b_i}\}\cdot discard\cdot B_{11}\nonumber\\
&&B_{11}=\sum_{D_o\in\Delta_o}send_B(D_o)\cdot B\nonumber
\end{eqnarray}

where $\Delta_o$ is the collection of the output data.

The send action and receive action of the same data through the same channel can communicate each other, otherwise, a deadlock $\delta$ will be caused. We define the following communication functions.

\begin{eqnarray}
&&\gamma(send_Q(q),receive_Q(q))\triangleq c_Q(q)\nonumber\\
&&\gamma(send_Q(q'),receive_Q(q'))\triangleq c_Q(q')\nonumber\\
&&\gamma(send_P(C),receive_P(C))\triangleq c_P(C)\nonumber\\
&&\gamma(send_P(Y),receive_P(Y))\triangleq c_P(Y)\nonumber\\
&&\gamma(send_P(U),receive_P(U))\triangleq c_P(U)\nonumber\\
&&\gamma(send_P(L),receive_P(L))\triangleq c_P(L)\nonumber
\end{eqnarray}

Let $A$ and $B$ in parallel, then the system $AB$ can be represented by the following process term.

$$\tau_I(\partial_H(\Theta(A\between B)))$$

where $H=\{send_Q(q),receive_Q(q),send_Q(q'),receive_Q(q'),send_P(C),receive_P(C),send_P(Y),\\receive_P(Y),send_P(U),receive_P(U),send_P(L),receive_P(L)\}$

 and $I=\{\boxplus_{\frac{1}{2n},i=0}^{2n-1}Rand[q;B_a]_i, \boxplus_{\frac{1}{2n},i=0}^{2n-1}Rand[q;K_a]_i, Set_{K_a}[q], \\H_{B_a}[q], \boxplus_{\frac{1}{2n},i=0}^{2n-1}Rand[q';B_b]_i, \boxplus_{\frac{1}{2n},i=0}^{2n-1}M[q;K_b]_i, \boxplus_{\frac{1}{2n},i=0}^{2n-1}M[q';D]_i, c_Q(q), c_P(C),\\c_Q(q'), c_P(Y), c_P(U), c_P(L), cmp(K_{a,b},K_a,K_b,B_a,B_b),\{B_{a_i}=B_{b_i}\},\{B_{a_i}\neq B_{b_i}\},\\
generate(K_a),generate(K_b),discard\}$.

Then we get the following conclusion.

\begin{theorem}
The basic S13 protocol $\tau_I(\partial_H(\Theta(A\between B)))$ exhibits desired external behaviors.
\end{theorem}

\begin{proof}
We can get $\tau_I(\partial_H(\Theta(A\between B)))=\sum_{D_i\in \Delta_i}\sum_{D_o\in\Delta_o}receive_A(D_i)\leftmerge send_B(D_o)\leftmerge \tau_I(\partial_H(\Theta(A\between B)))$.
So, the basic S13 protocol $\tau_I(\partial_H(\Theta(A\between B)))$ exhibits desired external behaviors.
\end{proof}


\end{document}